%% file: Thesis_Sharmila_Final.tex
\documentclass[PhD]{iitmdiss}
\usepackage{times}
 \usepackage{t1enc}
 
\usepackage{graphicx}
\usepackage{epstopdf}
\usepackage{hyperref} 
\usepackage{amsmath,amssymb} 
\usepackage{tikz}
\usetikzlibrary{arrows}
\usepackage{xcolor}
\usepackage{braket}
\graphicspath{{./figs_p1/},{./figs_p2/},{./figs_p3/},{./figs_p4/},{./figs_misc/},{./figs_pdf/}}

\newcommand{\delim}{,}
\newcommand{\vn}{\mathbf{n}}
\newcommand{\aver}[1]{\langle #1 \rangle}
\newcommand{\thetaa}{\theta_{\textsc{a}}}
\newcommand{\thetaat}{\theta_{\textsc{a}}}
\newcommand{\thetab}{\theta_{\textsc{b}}}
\newcommand{\thetabt}{\theta_{\textsc{b}}}
\newcommand{\eref}[1]{(\ref{#1})}
\newcommand{\rmd}{\text{d}}
\newcommand{\af}{\textsc{af}}

\newcommand{\neweps}{\varepsilon}
\newcommand{\epsarg}[1]{\varepsilon_{\small\textsc{#1}}}
\newcommand{\Tr}{\text{Tr}}

\newcommand{\vardel}{\Delta}
\usepackage{enumitem}
\usepackage{mathtools}
\makeatletter
\providecommand*{\input@path}{}
\g@addto@macro\input@path{{preamble/}{chapters/}}
\makeatother

\begin{document}

\input{title.tex}
\maketitle

\input{cert.tex}

\input{acknowledge.tex}

\input{abstract.tex}

\input{contents.tex}

\input{abbr.tex}

\input{notation.tex}


\pagenumbering{arabic}

\input{chapter1.tex} 
\input{chapter2.tex} 
\input{chapter3.tex} 
\input{chapter4.tex} 
\input{chapter5.tex} 
\input{chapter6.tex} 



\appendix

\input{appendix_TomoSinModExpr.tex} 
\input{appendix_NormalOrderedMom.tex} 
\input{appendix_DensMat.tex} 
\input{appendix_TimeSeries.tex} 
\input{appendix_2states.tex} 
\input{appendix_ChronoTomo.tex} 


\begin{singlespace}
\bibliography{references}
\end{singlespace}

\listofpapers
~~~~{\Large {\bf Papers in refereed journals}}
\begin{enumerate}	

\item Signatures of nonclassical effects in optical tomograms,\\
{\bf B. Sharmila}, K. Saumitran, S. Lakshmibala, and V. Balakrishnan \\
{\it J. Phys. B.: At. Mol. Opt. Phys.}  {\bf 50}, 045501 (2017). 

\item Estimation of entanglement in bipartite systems directly from tomograms,\\
{\bf B. Sharmila}, S. Lakshmibala, and V. Balakrishnan \\
{\it Quantum Inf. Process.} {\bf 18}, 236 (2019). 

\item Tomographic entanglement indicators in multipartite systems,\\
{\bf B. Sharmila}, S. Lakshmibala, and V. Balakrishnan \\
{\it Quantum Inf. Process.} {\bf 19}, 127 (2020).

\end{enumerate}
~~~~{\Large {\bf Manuscript submitted for publication}}
\begin{enumerate}	

\item Signatures of avoided energy-level crossings in entanglement indicators obtained from quantum tomograms,\\
{\bf B. Sharmila}, S. Lakshmibala, and V. Balakrishnan \\
{\it ArXiv} {\bf arXiv:2006.13536} [quant-ph], (2020).
	

\end{enumerate}
%
\newpage
\begin{center}
	{\Large {\bf CURRICULUM VITAE}}
\end{center}
\begin{tabular}{l l l l} \\
	Name           &: B. Sharmila & &\\
	Date of Birth &: October 7, 1991 & &\\
	Nationality         &: Indian & &\\
	Permanent Address&: Nachiar Kudil, 32, Sundaresan Nagar & &\\
	                     &~~Chennai, 600095.& &\\
	                     &~~Tamil Nadu, India&&\\
\end{tabular}
~\\
{\bf Education: }\\
\\
\begin{tabular}{ l  l }
    2014- & Ph.D. Physics, Department of Physics\\
	      &Indian Institute of Technology Madras, Chennai.  \\
	      &Date of registration:  July 2014.\\
	      & \\
	      2012-2014    & M.Sc. Physics,\\
	      & Indian Institute of Technology Madras, Chennai.\\
	     & \\
	     2009-2012   & B.Sc. Physics, Women's Christian College(Autonomous)\\
	     &Tamil Nadu, India.  \\
\end{tabular}
%

\newpage
\begin{center}
{\Large {\bf DOCTORAL COMMITTEE}}
\end{center}
\begin{tabular}{l l}
Supervisor:     & Dr. S.~Lakshmi Bala \vspace{-1ex}\\
	                 &Professor\vspace{-1ex} \\
	                 &Dept. of Physics, IIT Madras \vspace{-1ex}\\
\\
Members: & Dr. Suresh Govindarajan \vspace{-1ex}\\
                & Professor \vspace{-1ex}\\
                & Dept. of Physics,  IIT Madras \vspace{-1ex}\\
\\
                & Dr. Sunethra Ramanan \vspace{-1ex}\\
 	       &Assistant Professor \vspace{-1ex}\\
	       & Dept. of Physics,  IIT Madras \vspace{-1ex}\\
\\
               & Dr. R.~Radha \vspace{-1ex}\\
               &Professor \vspace{-1ex} \\
	      & Dept. of Mathematics,  IIT Madras
\end{tabular}


\end{document}

%% file: title.tex
\title{SIGNATURES OF NONCLASSICAL EFFECTS IN TOMOGRAMS}

\author{B. SHARMILA}

\date{SEPTEMBER 2020}
\department{PHYSICS}

%% file: cert.tex
\certificate

\vspace*{0.5in}

\noindent This is to certify that the thesis entitled {\bf Signatures of nonclassical effects in tomograms}, submitted by {\bf B. Sharmila} 
  to the Indian Institute of Technology Madras for
the award of the degree of {\bf Doctor of Philosophy}, is a bona fide
record of the research work done by her under my supervision.  The
contents of this thesis, in full or in part, have not been submitted
to any other Institute or University for the award of any degree or
diploma.

\vspace*{1.5in}

\begin{minipage}{\textwidth}
\begin{minipage}{0.64\textwidth}
\end{minipage}%
\hfill
\begin{minipage}{0.35\textwidth}
\noindent {\bf Prof. S. Lakshmi Bala} \\
\noindent Thesis Guide and Professor \\
\noindent Dept. of Physics\\
\noindent IIT Madras \\
\end{minipage}%
\end{minipage}%

\vspace*{0.25in}
\noindent Place: Chennai, India.\\
Date: \today

%% file: acknowledge.tex
\acknowledgements
I shall remain eternally grateful to my guide Dr. S. Lakshmi Bala, for teaching me to think independently and analyse critically anything presented to me, both academically and otherwise. She has been a pillar of support through my tough times. I aspire to achieve her level of dedication, expertise in a variety of topics, and attention to detail. She has been a source of inspiration, encouragement and advice. Her constructive criticisms have always helped me understand my limitations and build upon my strengths. There will never be enough words or actions to express my gratitude. 

I am blessed to have had the opportunity to work with Dr. V. Balakrishnan. Every interaction with him has been  an enriching experience. I have benefitted immensely from his teaching and meticulous attention to detail. I shall cherish every one of his witty remarks and his kind words of advice. I sincerely hope that some day I will be able to inspire at least one student as much as he has inspired me.  It has been  my great privilege to have been taught and mentored by him.

I am extremely grateful to my doctoral committee members - Dr. Suresh Govindarajan, Dr. Sunethra Ramanan and Dr. R. Radha, for their discussions, feedback and constant support. I especially gained a lot from regular interactions with Dr. Suresh Govindarajan and Dr. Sunethra Ramanan. I am grateful for all the encouragement and advice from my doctoral committee chairman, Dr. M. V. Satyanarayana. Since the time I had first met him as an undergraduate student, he had always encouraged me to work harder. My learning as a research scholar
has been enhanced by  all the inputs that I have  received from Dr. Arul Lakshminarayan, Dr. Prabha Mandayam and Dr. Rajesh Narayanan. It is my pleasure to have known Dr. M. S. Sriram and Dr. Radha Balakrishnan, and I thank them for their support and encouragement.

I am grateful to Dr. K. Sethupathi, Head of the Department of Physics, and Dr. M. S. R. Rao and Dr. P. B. Sunil Kumar, former Heads of the Department of Physics, for their kind help and advice during my doctoral programme. I am thankful to IIT Madras for providing generous financial support and excellent departmental and central computing facilities. I also thank Dr. Sunethra Ramanan for allowing me access to the computing facility of her research group, whenever it was necessary. I thank the IITM Alumni Association for providing partial financial support to attend the conference `Quantum Information and Measurement (QIM V) 2019' in Rome, Italy.

I take this opportunity to thank Prof. Prasanta Panigrahi, IISER Kolkata, for useful discussions and suggestions regarding my work. In Chapter 5 of this thesis, I have used experimental data supplied by the NMR-QIP group, IISER Pune, India. I acknowledge with thanks that these  raw experimental data helped me in my study of spin systems. I thank Prof. T. S. Mahesh and the NMR-QIP group for their help. I also personally thank Mr. Soham Pal for discussions and clarifications regarding the experiment. 

I have gained a fantastic set of friends in my office mates. I am especially thankful to Pradip Laha for all the help and discussions, academic and otherwise. As a senior, Pradip is the best one could ask for, and as a friend, he and his wife were most supportive and encouraging. I had the best fun and learnt a lot at the same time with Sashi, Sarath, Swaminath and Malayaja. Be it coding, networking in a conference, or binge-watching a sitcom series, I will always remember their tips and encouragement. I also thank Athreya Shankar for discussions and feedback on my work. I take this opportunity to thank Sutapa, Tanya, Akshaya, Vasumathi, Sai Smruti and Madhuparna for all our academic discussions and chit-chat that kept me going through thick and thin. I also thank my juniors, Soumyabrata Paul and  Vishnu, for discussions. 
 
I also take this opportunity  to express my gratitude to all my teachers. Specifically, I shall always be grateful to my science teacher Mrs. Hemalatha Murthy, (Late) Dr. Nirupama Raghavan, and Dr. N. Lakshminarayanan, for inspiring me to choose physics and work in it. 

I thank my best friend Shakthi, for her unconditional support and presence in the darkest hours of need. 
A `Thank you' is insufficient  and, according to my parents, unnecessary, for all their support, encouragement, love and frank criticisms. It is not in my power to ever finish thanking my parents for all that they have done for me. I  thank God for having blessed me with all these people who have made this journey not only possible but incredibly enjoyable. 


%% file: abstract.tex

\abstract

\noindent KEYWORDS: \hspace*{0.5em} \parbox[t]{4.4in}{Tomograms; full and fractional revivals; super-revivals; quadrature, higher-order, entropic and spin squeezing; entanglement; tomographic entanglement indicators; mutual information; participation ratio; Bhattacharyya distance; Pearson correlation coefficient; avoided energy-level crossings; Bose-Einstein condensate; atom-field interaction models; IBM quantum computer; NMR; chronocyclic tomograms; local Lyapunov exponents.}

\vspace*{24pt}

\noindent  Investigations on nonclassical effects such as revivals, squeezing and entanglement in quantum systems require, in general, knowledge of the state of the system as it evolves in time. The state (equivalently, the density matrix) is reconstructed from a tomogram obtained from experiments. A tomogram is a set of histograms of appropriately chosen observables. Reconstruction of the quantum state from a tomogram typically involves statistical procedures that could be cumbersome and inherently error-prone. It is therefore desirable to extract as much information as possible about the properties of the state {\em  directly} from the tomogram. The theme of this thesis is the identification and quantification of nonclassical effects from appropriate tomograms. We have examined continuous-variable (CV) systems, hybrid quantum (HQ) systems and spin systems. The program is two-fold: (a) To compute tomograms of known states numerically at various instants during temporal evolution under specific Hamiltonians, and to examine their revival, squeezing and entanglement properties at these instants; (b) to compute tomograms from available experimental data, and to investigate their nonclassical features. The CV systems considered are primarily (i) a Bose-Einstein condensate (BEC) trapped in a double-well potential, and (ii) a radiation field interacting with a nonlinear multi-level atomic medium. The HQ systems considered are two-level atoms interacting with radiation fields. This allows for the possibility of examining both optical tomograms and qubit tomograms. Several  initial states have  been considered,  including the standard coherent state (CS), the two-mode squeezed state, the binomial state, and states which display quantifiable departure from coherence, such as the photon-added coherent state and the boson-added coherent state.

Wave packet revival phenomena including full, fractional and super-revivals have been examined tomographically in the context of a single-mode field system and the bipartite BEC system. Squeezing properties such as quadrature, higher-order Hong-Mandel and Hillery-type squeezing, and entropic squeezing have been investigated in detail in CV systems by evaluating appropriate moments of observables, from tomograms. A major part of the thesis is devoted to a comparison between the performance of different entanglement indicators (computed from tomograms) that we have proposed. 

We have also obtained and examined tomograms directly related to experiments reported in the literature. (i) We have analysed appropriate tomograms to distinguish between two $2$-photon states produced in an experiment on a CV bipartite system. While photon coincidence count measurements were  used for this purpose in the experiment, our investigation demonstrates that tomograms provide another powerful tool for examining differences between various quantum states. (ii) We have computed spin tomograms from data, obtained using liquid-state NMR techniques, from an experimental group. (iii) Equivalent circuits of multipartite HQ systems were provided by us to the IBM quantum computing platform. Based on these circuits, tomograms were generated by the platform by experiment as well as  simulation. Corresponding tomograms were computed by us using the HQ model. The entanglement indicators calculated from these three sources of tomograms have been compared and contrasted.

The thesis highlights and elaborates upon the very useful and significant role played by tomograms in assessing nonclassical effects displayed by quantum systems, without resorting to detailed state reconstruction procedures.

\pagebreak


%% file: contents.tex

\begin{singlespace}
\tableofcontents
\thispagestyle{empty}

\listoffigures
\addcontentsline{toc}{chapter}{LIST OF FIGURES}
\end{singlespace}


%% file: abbr.tex
\abbreviations

\noindent 
\begin{tabbing}
xxxxxxxxxxx \= xxxxxxxxxxxxxxxxxxxxxxxxxxxxxxxxxxxxxxxxxxxxxxxx \kill
\textbf{CV}   \> continuous-variable \\
\textbf{HQ} \> hybrid quantum \\
\textbf{BEC} \> Bose-Einstein condensate\\
\textbf{CS} \> coherent state \\
\textbf{PACS} \> photon-added coherent state \\
\textbf{QED} \> quantum electrodynamics \\
\textbf{IPR} \> inverse participation ratio \\
\textbf{PCC} \> Pearson correlation coefficient \\
\textbf{GKP} \> Gottesman-Kitaev-Preskill \\
\textbf{DJC} \> double Jaynes-Cummings \\
\textbf{DTC} \> double Tavis-Cummings \\
\textbf{QASM} \> quantum assembly language\\
\textbf{NMR} \> nuclear magnetic resonance \\

\end{tabbing}

\pagebreak


%% file: notation.tex

\chapter*{\centerline{NOTATION}}
\addcontentsline{toc}{chapter}{GLOSSARY OF SYMBOLS}

\begin{singlespace}
\begin{tabbing}
xxxxxxxxxxxxxxxx \= xxxxxxxxxxxxxxxxxxxxxxxxxxxxxxxxxxxxxxxxxxxxxxxx \kill
$\theta$ \> angle (similarly, $\thetaa$, $\thetab$, $\theta_{\textsc{a}p}$, $\theta_{\textsc{b}q}$, $\vartheta$, $\varphi$, $\theta^{\prime}$, $\varphi^{\prime}$, $\upsilon$)\\
$\mathbb{X}_{\theta}$ \> rotated quadrature operator (similarly, $\mathbb{X}_{\thetaa}$, $\mathbb{X}_{\thetab}$) \\
$(a, \, a^{\dagger})$ \> ladder operators (similarly, $(b,\,b^{\dagger})$, $(a_{\textsc{a}},\,a_{\textsc{a}}^{\dagger})$, $(a_{\textsc{b}},\,a_{\textsc{b}}^{\dagger})$) \\
$(\sigma_{+}, \sigma_{-})$ \> spin ladder operators (additional subscripts denote subsystems labels) \\
$\ket{X_{\theta},\theta}$ \> eigenstate of rotated quadrature operator (similarly, $\ket{X_{\thetaa},\thetaa}$, $\ket{X_{\thetab},\thetab}$, $\ket{X}$)\\
$\ket{X_{\thetaa},\thetaa;X_{\thetab},\thetab}$ \> factored product of $\ket{X_{\thetaa},\thetaa}$ and $\ket{X_{\thetab},\thetab}$ \\
$X_{\theta}$ \> eigenvalue of rotated quadrature operator (similarly, $X_{\thetaa}$, $X_{\thetab}$, $X_{\textsc{a}0}$, $X_{\textsc{b}0}$,\\
\> $X_{\theta_{\textsc{a}p}}$, $X_{\theta_{\textsc{b}q}}$, $X$)\\
$\rho$ \> density matrix (similarly, $\varrho$, $\overline{\varrho}$, $\overline{\rho}$, $\rho_{\textsc{ab}}$, $\rho_{\textsc{s}}$, $\rho_{\textsc{a}}$, $\rho_{\textsc{b}}$, $\rho_{m_{1},m_{2}}$, $\rho_{\textsc{mab}}$, $\rho_{+}$, $\rho_{-}$) \\
$w(X_{\theta},\theta)$ \> single-mode tomogram \\
$w(X_{\thetaa},\thetaa;X_{\thetab},\thetab)$ \> bipartite tomogram (similarly,
 $w^{\alpha}(t_{\textsc{s}};t_{\textsc{i}})$, $w^{\beta}(t_{\textsc{s}};t_{\textsc{i}})$) \\
$w_{\textsc{a}}(X_{\thetaat},\thetaa)$ \> reduced tomogram (similarly, $w_{\textsc{b}}(X_{\thetabt},\thetab)$) \\
$\alpha$ \> a complex number (similarly, $\zeta$, $\alpha_{a}$, $\alpha_{b}$, $\alpha(t)$, $\beta(t)$) \\
$\nu$ \> $|\alpha|^{2}$\\
$\ket{\alpha}$ \> coherent state of a single-mode continuous-variable system \\ 
$\ket{\alpha}_{tcs}$ \> truncated coherent state\\
$\ket{n}$ \> a Fock state/ an $n$-photon state/ an $n$-boson state \\
$\ket{\alpha,m}$ \> $m$-photon-added coherent state/ $m$-boson-added coherent state\\
$\ket{\psi_{\text{bin}}}$ \> binomial state \\
$\ket{\zeta}$ \> two-mode squeezed state \\
$\ket{\ell \delim m}$ \> factored product of number states $\ket{\ell}$ and $\ket{m}$ \\
$L_{m}$ \> Laguerre polynomial of order $m$\\
$H_{n}$ \> Hermite polynomial of order $n$\\
$T_{rev}$ \> revival time\\
$H$ \> Hamiltonian of a single-mode radiation field (similarly, $H'$) \\
$\chi_{1}$ \> strength of nonlinearity (similarly, $\chi_{2}$, $U$, $\gamma$, $\chi$)\\
$\mathcal{N}$ \> photon number operator $a^{\dagger}a$ \\
$F_{q}(\mathcal{N})$ \> $[Z_{1},Z_{2}]$, a polynomial of order $q-1$ in $\mathcal{N}$ \\
$\ket{\psi(t)}$ \> state of the radiation field at instant $t$\\
$\text{LCM}\left(\frac{1}{\chi_{1}},\frac{1}{\chi_{2}}\right)$ \> least common multiple of $(1/\chi_{1})$ and $(1/\chi_{2})$\\
$\Gamma$ \> rate of decoherence (similarly, $\Gamma_{p}$)\\
$x$  \> position quadrature (similarly, two-mode position quadrature $\eta$) \\
$p$ \> momentum quadrature \\
$\aver{(\Delta x)^{2 q}}$ \> $2 q$-order central moment (similarly, $\aver{(\Delta \eta)^{2 q}}$, $\aver{(\Delta Z_{1})^{2 q}}$)\\
$(Z_{1}$, $Z_{2})$ \> conjugate pair in Hillery-type squeezing \\
$D_{q}$ \> squeezing parameter \\
$H_{\textsc{bec}}$ \> Hamiltonian of the double-well BEC system (on transformation, $\widetilde{H}_{\textsc{bec}}$) \\
$N_{\text{tot}}$ \> total number operator that commutes with the Hamiltonian (similarly, $\mathcal{N}_{\text{tot}}$) \\
$\omega_{0}$ \> frequency (similarly, $\omega_{1}$, $\lambda_{1}$, $\omega_{\textsc{a}}$, $\omega_{\textsc{f}}$, $\Omega_{\textsc{f}}$, $\epsilon$, $\omega_{p}$, $\overline{\omega}$, $\omega_{\textsc{s}}$, $\omega_{\textsc{i}}$, $\Omega$, $\Omega_{0}$, $\chi_{\textsc{f}}$, $\chi_{0}$, $\chi_{\textsc{s}}$,\\
\> $\Omega_{i}$, $\Delta_{i}$ where $i=1,2,\dots,M$)\\
$\lambda$ \> hopping frequency (similarly, $g$, $\Lambda$, $\Lambda_{s}$, $g_{0}$) \\
$\ket{\Psi_{m_{1}m_{2}}(t)}$ \> state of BEC system for initial $\ket{\alpha_{a},m_{1}}\otimes \ket{\alpha_{b},m_{2}}$ at instant $t$ \\
$M_{m_{1},m_{2}}$ \> unitary operator (similarly, $\mathcal{U}$, $V$, $U(\mathbf{n})$, $U_{3}$, $S^{\dagger}$, $C^{\prime}$, $Z_{t_{\textsc{s}}}$); identity operator $\mathbb{I}$.\\
$\xi_{\textsc{svne}}$ \> subsystem von Neumann entropy (similarly, $\xi^{(\textsc{a})}_{\textsc{svne}}$, $\xi^{(\textsc{b})}_{\textsc{svne}}$, $\xi^{(\textsc{ab})}_{\textsc{svne}}$) \\
$\xi_{\textsc{sle}}$ \> subsystem linear entropy \\
$\xi_{\textsc{qmi}}$ \> quantum mutual information \\
$\epsarg{tei}$ \> single-slice entanglement indicator (similarly, $\epsarg{ipr}$, $\epsarg{bd}$, $\epsarg{pcc}$)\\
$\xi_{\textsc{tei}}$ \> averaged entanglement indicator (similarly, $\xi^{\prime}_{\textsc{tei}}$, $\xi_{\textsc{ipr}}$, $\xi_{\textsc{bd}}$, $\xi_{\textsc{pcc}}$)\\
$S(\thetaa)$ \> subsystem tomographic entropy (similarly, $S(\thetab)$, $S_{0}$)\\
$S(\thetaa,\thetab)$ \>  two-mode tomographic entropy\\
$\eta_{\textsc{a}}(\thetaa)$ \> inverse participation ratio (similarly, $\eta_{\textsc{b}}(\thetab)$, $\eta_{\textsc{ab}}(\thetaa,\thetab)$)\\
$D_{\textsc{kl}}$ \> Kullback-Leibler divergence \\
$D_{\textsc{b}}$ \> Bhattacharyya distance \\
$PCC(X,Y)$ \> Pearson correlation coefficient between $X$ and $Y$ \\
$\text{Cov}(X,Y)$ \> covariance of $X$ and $Y$ \\
$(\Delta X)^{2}$ \> variance (similarly, $(\Delta Y)^{2}$, $(\Delta \mathbf{J \cdot v}_{\perp})^{2}$, $(\Delta J_{\text{min}})^{2}$, $(\Delta \mathcal{J})^{2}$, $(\Delta \mathcal{J}_{min})^{2}$,\\
\> $(\Delta \omega)^{2}$, $(\Delta \Omega)^{2}$) \\
$(\sigma_{x}$, $\sigma_{y}$, $\sigma_{z})$ \> spin operators\\
$\ket{\mathbf{n},m}$ \> eigenstate of spin operators for $\mathbf{n}$ chosen appropriately\\
\> (for instance, $\mathbf{n}=\mathbf{e}_{z}$ corresponds to eigenstate of $\sigma_{z}$)\\
$w(\mathbf{n},m)$ \> spin tomogram \\
$\ket{\psi_{N,k}}$ \> eigenstate of $H_{\textsc{bec}}$ ($N=0,1,2\dots$ and $k=0,1,\dots,N$) \\
$E(N,k)$ \> eigenvalue of $H_{\textsc{bec}}$ \\
$H_{\textsc{af}}$ \> Hamiltonian of the atom-field interaction system\\
$\ket{\phi_{N,k}}$ \> eigenstate of $H_{\textsc{af}}$ \\
$E_{\textsc{af}}(N,k)$ \> eigenvalue of $H_{\textsc{af}}$ \\
$H_{\textsc{tc}}$ \> Tavis-Cummings Hamiltonian for $M$ two-level systems interacting with a \\
\> radiation field \\
$\ket{\psi_{M,N,k}}$ \> eigenstate of $H_{\textsc{tc}}$\\
$d_{1}(t)$ \> difference between entanglement indicators at instant $t$ \\
\>(similarly, $d_{2}(t)$, $d_{3}(t)$, $\Delta(t)$)\\
$\tau_{d}$ \> time delay\\
$d_{\text{emb}}$ \> embedding dimension \\
$I(T)$ \> mutual information \\
$C(r)$ \> correlation integral \\
$\mathbf{DF}$ \> Jacobian obtained from time-series (similarly, product of Jacobians $\mathbf{DF}^{L}$)\\
$\mathbf{M}_{\text{os}}$ \> Oseledec matrix\\
$\Lambda_{L}$ \> maximum local Lyapunov exponent over $L$ time steps\\
$\Lambda_{\infty}$ \> maximum Lyapunov exponent \\
$PS(f)$ \> power spectrum as a function of frequency $f$\\
$\ket{\Psi_{\alpha}}$ \> frequency combs (similarly, $\ket{\Psi_{\beta}}$, $\ket{\psi^{\prime}}$, $\ket{\widetilde{+}}_{\omega_{\textsc{s}}}$, $\ket{\widetilde{-}}_{\omega_{\textsc{s}}}$, $\ket{\widetilde{+}}_{\omega_{\textsc{i}}}$, $\ket{\widetilde{-}}_{\omega_{\textsc{i}}}$)\\
$f_{+}$ \> Gaussian function (similarly, $f_{-}$)\\
$f_{\text{cav}}$ \>  superposition of Gaussian functions (similarly, $g_{\text{cav}}$, $\mathcal{F}$, $\mathcal{G}$)\\
$t$ \> time (simialrly, $\tau$, $\delta t$, $\tau_{\textsc{p}}$, $T$, $t_{\textsc{s}}$, $t_{\textsc{i}}$)\\
$\ket{\psi_{+}}$ \> maximally entangled state (similarly, $\ket{\phi_{+}}$)\\
$H_{\textsc{djc}}$ \> double Jaynes-Cummings Hamiltonian \\
$H_{\textsc{dtc}}$ \> double Tavis-Cummings Hamiltonian \\
$H_{\textsc{s}}$ \> effective Hamiltonian of $3$-qubit system \\
$(\ket{\uparrow}$, $\ket{\downarrow})$ \> eigenstates of $\sigma_{z}$ (similarly, in two-level atoms $(\ket{e}$, $\ket{g})$)\\
$(\ket{+}$, $\ket{-})$ \> eigenstates of $\sigma_{x}$ \\
$\mathbf{J}$ \> total spin operator with dyad $\mathbf{JJ}$\\
$\mathbf{v}_{\textsc{s}}$ \> vector (similarly, $\mathbf{v}_{\perp}$, $\mathbf{v}_{1}$, $\mathbf{v}_{2}$) \\
$\mathcal{J}$ \> spin operator defined in terms of $\mathbf{JJ}$, $\mathbf{v}_{1}$ and $\mathbf{v}_{2}$\\
$N(\rho_{\textsc{ab}})$ \> negativity\\
$\rho_{\textsc{ab}}^{T_{\textsc{a}}}$ \> partial transpose with respect to subsystem A (similarly, $\rho_{\textsc{ab}}^{T_{\textsc{b}}}$)\\
$\mu$ \> normalisation constants (similarly, $d_{10}$, $d_{11}$, $\mathcal{N}_{\alpha}$, $\mathcal{N}_{\beta}$, $\mathcal{M}_{\alpha}$, $\mathcal{M}_{\beta}$) \\

\end{tabbing}
\end{singlespace}

\pagebreak
\clearpage

%% file: chapter1.tex

\chapter{Introduction}
\label{ch:intro}


Measurement of any  observable in  a quantum mechanical system yields a histogram of the state of the system  in the basis of that observable. Measurements of a judiciously chosen \textit{quorum} of appropriate observables of a system that are informationally complete, yield a set of histograms called a tomogram. In the context of atoms interacting with radiation fields, both the optical tomogram and the tomogram pertaining to atomic observables would 
yield,  in principle, information about the full system and its subsystems.
 Quantum state reconstruction seeks to obtain the density matrix from the tomogram. 
 However, even in the simple case of a bipartite system comprising two $2$-level atoms (two qubits), state reconstruction from relevant tomograms typically employs statistical tools that are inherently error-prone~\cite{photonnumbertomo2}. The reconstruction procedure is significantly more difficult in the case of entangled multipartite qubit states~\cite{QubitRecon2016}. Attempts at scalable reconstruction programs for systems with a large number of qubits, and the challenges faced in this context, have been reported in the literature (see, for instance,~\cite{ReconTech2013,ReconTech2018}). 
It is therefore desirable to extract information about the state {\em  directly from the tomogram}, avoiding  the reconstruction procedure. This has been demonstrated in bipartite qubit systems by estimating state fidelity with respect to a specific target state directly from the tomogram, and comparing the errors that arise with the corresponding errors in procedures involving detailed state reconstruction~\cite{HuiKhoon}. Further, efficient methods have been proposed to estimate entanglement entropies directly from experimental data in the context of qubit systems
~\cite{DSPE1,DSPE2,DSPE3,DSPE4}.

Reconstruction of the state of a radiation field from optical tomograms is more challenging.
Even in the case of a single-mode radiation field, the Hilbert space is infinite-dimensional.  Thus, for optical tomography, the infinite set of rotated quadrature operators~\cite{VogelRisken, ibort} given by
 \begin{equation}
\label{eqn:quadop}
\mathbb{X}_{\theta} = \frac{1}{\sqrt{2}} (a^\dagger e^{i \theta} + a e^{-i \theta}), 
\;\;\theta \in [0,\pi)
\end{equation}
constitutes the quorum of observables that carries complete information  about the state.
 Here $(a,a^{\dagger})$ are the photon annihilation and creation operators  satisfying the commutation relation $[a, a^{\dagger}] = 1$. We note that $\theta=0$ corresponds to the $x$-quadrature and $\theta=\tfrac{1}{2}\pi$ corresponds to the conjugate 
 $p$-quadrature. The eigenvalue equation 
 for $\mathbb{X}_{\theta}$ 
 is given by $\mathbb{X}_{\theta} 
 \ket{X_\theta, \theta} = X_{\theta} \ket{X_\theta, \theta}$,  and the optical tomogram~\cite{VogelRisken, LvovskyRaymer} is
\begin{equation}
\label{eqn:tomogdef}
w(X_\theta, \theta) = \bra{X_\theta, \theta} \rho \ket{X_\theta, \theta}
\end{equation}
where $\rho$ is the field density matrix.
It is worth noting that even in the simple case of a two-level atom interacting with a radiation field, the state of the field subsystem was experimentally reconstructed from the corresponding tomogram at various instants of temporal evolution only as recently as 2017~\cite{ReconJCM2017}. With an increase in the number of field modes interacting with an atomic system, the inevitable entanglement that arises during dynamical evolution makes state reconstruction a more formidable task. The problem posed by the size  of the Hilbert space holds for other continuous-variable (CV) systems,  such as a Bose-Einstein condensate (BEC) in a double-well potential and hybrid quantum (HQ) systems such as an atomic array interacting with many radiation fields. It would 
 therefore  be efficient to \textit{read off} information about a state, wherever possible, directly from the tomogram. In particular, identifying signatures of nonclassical effects through simple manipulations of the relevant tomograms  {\it alone} becomes an interesting and important exercise. Investigations  carried out in this regard  employ the `inverse procedure' of starting with a known state, and obtaining the corresponding tomogram with the purpose of understanding how tomographic patterns carry signatures of nonclassical effects such as revivals, squeezing and entanglement. Such as  exercise also helps to build a dictionary of signatures that can be captured in tomograms corresponding to known states. This would be  the first step in applying the tomographic approach in experiments where the state is not known \textit{a priori}.  The programme just described is essentially the central theme of this thesis. We have identified and quantified nonclassical effects in both CV and HQ systems for known states, and applied the lessons learnt to extract information from experimental data when the state is not initially known. We now  give a brief outline of the nonclassical effects examined in this thesis.
  
We first consider the revival phenomena. An initial wave packet  $\ket{\psi(0)}$  governed by a nonlinear Hamiltonian  is said to revive fully  at an instant $T_{rev}$ during its dynamical evolution if the wave packet $\ket{\psi(T_{rev})}$  differs from $\ket{\psi(0)}$ only by an overall phase. The state revives due to very specific quantum interference between the basis states that comprise the wave packet.  In certain systems, revivals occur {\it periodically} at integer multiples of $T_{rev}$.  Under certain circumstances, $\ell$-subpacket fractional revivals of the wave packet (where $\ell$ is  a positive integer) can occur at specific instants between two successive revivals~\cite{aver,robi}. At these instants the initial wave packet becomes $\ell$ superposed copies of itself, each with 
an amplitude less than that of the initial state.
For instance, a radiation field governed by the Kerr Hamiltonian  $ \hbar \chi_{1} a^{\dagger 2}a^{2}$ can exhibit periodic revivals and fractional revivals, with $T_{rev}=\pi/\chi_{1}$. (Here $\chi_{1}$ is the third-order nonlinear susceptibility of the medium). For an initial coherent state (CS) $\ket{\alpha}$ ($\alpha \in \mathbb{C}$), expressed in  the photon number basis $\lbrace \ket{p} \rbrace$  as
\begin{equation}
\ket{\alpha}=e^{-|\alpha|^{2}/2} \sum_{p=0}^{\infty} \frac{\alpha^{p}}{\sqrt{p!}} \ket{p},
\label{eqn:CSdefn}
\end{equation}
the time evolved state is a superposition of $\ell$ coherent states~\cite{TaraAgarwal} at instants $m T_{rev}/\ell$ (where $m= 1,2,\dotsc, \ell -1$).
The optical tomogram corresponding to an $\ell$-subpacket fractional revival of the CS is an $\ell$-strand pattern~\cite{sudhrohithrev}. 
With the addition of higher-order nonlinearities (and hence more than one time scale) to the Hamiltonian, this 
simple picture gets modified. 
Super-revivals then occur in the system, when a delicate balance is struck between two or more time scales~\cite{agarwalbanerji,bluhm}. Straightforward correlations between tomograms and fractional revivals are therefore not to be expected. Investigations on these lines gain impetus because super-revivals have been experimentally detected in systems of alkali atoms subject to an external field (see, for instance, \cite{suprev_expt}). In this thesis, we have examined tomograms of CV systems subject to Hamiltonians with more than one time scale both in the case of a single-mode field and the double-well BEC system.

We now turn to the squeezing properties of the states considered in this thesis. We have used the tomographic approach to examine quadrature, higher-order and entropic squeezing in CV systems, and spin squeezing from experimental data obtained by employing NMR techniques on a $3$-qubit system. Quadrature squeezing is quantified by the numerical value of the variance of the corresponding observable. For instance, the state of the field is squeezed in $x$ if the corresponding variance 
$\aver{(\Delta x)^{2}}$ is less than the variance of $x$ 
in a CS $\ket{\alpha}$. Generalisation of this definition to include higher-order squeezing allows for two possibilities, namely, Hong-Mandel~\cite{hong} and Hillery-type~\cite{Hillery} higher-order squeezing.  Hong-Mandel squeezing of order $q$ in $x$ requires that the $(2q)^{\rm th}$ central moment of $x$ in the given state be  less than the corresponding moment for the CS. Hillery-type squeezing of order $q$ refers to squeezing in either $Z_{1} =( a^{q} + {a^{\dagger}}^{q})/ \sqrt{2}$ or $Z_{2} =( a^{q} - {a^{\dagger}}^{q})/ (\sqrt{2} \, i)$ ($q = 2,3,\dotsc$). We note that the tomogram slice corresponding to a specific value of $\theta$ gives the probability distribution in that quadrature basis. Thus the central moments calculated from appropriate slices of the tomogram can be used to estimate quadrature and Hong-Mandel higher-order squeezing of the state. An elegant method has been proposed~\cite{wunsche} for estimating Hillery-type higher-order squeezing in single-mode systems. In this thesis, we have extended this procedure to two-mode squeezing. 

Further, the information entropy $S(\theta)=-\int_{-\infty}^{\infty} \text{d}X_{\theta} \, w(X_{\theta},\theta) \, \ln\, w(X_{\theta},\theta)$ corresponding to a given value of 
$\theta$ can be computed readily from the tomogram. The information entropies in conjugate quadrature bases, such as $x$ and $p$  ($\theta=0$ and 
$\tfrac{1}{2}\pi$, respectively), satisfy an entropic uncertainty relation~\cite{bialynicki} $S(0)+
S(\tfrac{1}{2}\pi) \geqslant (1+\ln\, \pi)$. If the entropy in either quadrature is less than 
$\tfrac{1}{2}(1+\ln\,\pi)$, entropic squeezing occurs in that quadrature. We have examined entropic squeezing in the double-well BEC system. 

A major part of this thesis is devoted to finding an efficient entanglement indicator \textit{directly} from relevant tomograms in CV, HQ and spin systems.
Entanglement is an essential resource in quantum information processing. Interesting phenomena such as sudden death and birth of entanglement~\cite{eberly}, and its collapse to a constant non-zero value over a significant interval of time~\cite{pradip1} have been found in model systems. A standard measure of entanglement between the two subsystems A and B of a bipartite system is the subsystem von Neumann entropy $\xi_{\textsc{svne}}=-\mathrm{Tr}\,(\rho_{i} \,\log_{2} \,\rho_{i})$ where $\rho_{i}$ ($i=\text{A,\,B}$) is the reduced density matrix of the subsystem concerned. Computation of this measure, however, requires a knowledge of the density matrix. Qualitative signatures of entanglement in the output state of a  quantum beamsplitter have been identified solely from tomograms, and reported in the literature~\cite{sudhrohithbs}. We have carried out detailed \textit{quantitative} analysis of entanglement, using indicators obtained directly from tomograms in a variety of multipartite systems of experimental interest. We have assessed their efficacy by comparing them with 
$\xi_{\textsc{svne}}$ and other standard measures of entanglement. In particular, we have undertaken this study on two experimentally viable bipartite CV systems, namely, the double-well BEC and a multi-level atom modelled as an oscillator interacting with the radiation field. We have examined the performance of these entanglement indicators in multipartite HQ systems of two-level atoms interacting with radiation fields. 

We have also examined how the entanglement indicator can be used to distinguish between two $2$-photon states produced in a recent experiment~\cite{perola} using ultrashort light pulses. In this experiment, photon coincidence count measurements were used to distinguish between the two states, and further, a quantum logic operation was implemented. The relevant tomogram in this experiment is a chronocyclic tomogram~\cite{paye} (i.e., where time and frequency are the relevant observables analogous to the $x$ and 
$p$ quadratures in an optical tomogram). We have demonstrated that entanglement indicators computed from the chronocyclic tomograms corresponding to these two states distinguish unambiguously  between them. The purpose of this investigation was to provide an alternative method for distinguishing between two optical states.

Of particular interest and relevance is the performance of the tomographic entanglement indicators computed directly from experimental data. In this context, we have examined HQ systems using the IBM quantum computer and also the spin system mentioned earlier. In the former case, equivalent circuits that mimic the atomic subsystem of the multipartite HQ system considered, were provided to the IBM quantum computer for generation of the tomogram. The purpose of this investigation was to assess the extent to which experimental losses affected entanglement indicators. In the latter case, the NMR-QIP group in IISER Pune, India, provided us with experimentally reconstructed density matrices from an NMR spectroscopy experiment~\cite{nmrExpt}. We have computed the corresponding tomograms  and from these,  the  
entanglement indicators. The purpose of this investigation was to assess,  using a simple experimentally viable entangled system, the limitations that could possibly arise by neglecting the off-diagonal elements of the density matrix. A significant outcome of this thesis is the identification of \textit{useful and reliable entanglement indicators directly from tomograms} in generic quantum systems. The thesis showcases the advantages of extracting information from the tomogram. These outweigh the limitations that this approach entails  in certain contexts.

A variety of initial states of the radiation field have been considered in our investigation. This facilitates understanding of the sensitivity of the nonclassical effects to the specific initial state considered. Apart from the standard CS (Eq. \eref{eqn:CSdefn}), states such as the photon-added coherent states, which exhibit quantifiable departure from coherence, have been used. The $m$-photon-added coherent state $\ket{\alpha,m}$ ($m$-PACS) is given by~\cite{zavatta}
\begin{equation}
\ket{\alpha,m}=\frac{a^{\dagger m}}{\sqrt{L_{m}(-|\alpha|^{2}) m!}} \ket{\alpha},
\label{eqn:mPACSDefn}
\end{equation}
where $L_{m}$ is the Laguerre polynomial of order $m$.  In bipartite systems comprising two field modes, we have considered both unentangled and entangled initial states. The former are factored products of combinations of coherent and photon-added coherent states.  The latter are the binomial state 
$\ket{\psi_{\rm{bin}}}$ and the two-mode squeezed state $\ket{\zeta}$. For a non-negative integer $N$, the binomial state $\ket{\psi_{\rm{bin}}}$ is given by~\cite{stolerBS} 
\begin{equation}
\ket{\psi_{\rm{bin}}} = 2^{-N/2} \sum_{n=0}^{N} 
{\tbinom{N}{n}}^{1/2}
 \ket{N-n \delim n},
\label{eqn:BS_defn}
\end{equation}
where $\ket{N-n \delim n} \equiv  \ket{N-n}\otimes\ket{n}$.
 The two-mode squeezed state is given by~\cite{cavesTMS}
\begin{equation}
\ket{\zeta} = e^{\zeta^{*} a b - \zeta a^{\dagger} b^{\dagger}} \ket{0 \delim 0},
\label{eqn:sq_state}
\end{equation} 
where $\zeta \in \mathbb{C}$,  
$\zeta^{*}$ is its complex conjugate, 
and  $\ket{0 \delim 0}$ is the 
 product state corresponding to 
 $N = 0, n = 0$. 

\noindent The contents of the rest of this thesis are as follows:

In {\bf Chapter 2}, we summarise the salient features of the tomograms corresponding to single-mode radiation fields and bipartite CV systems that are relevant for our purpose. This is followed by an illustration of how full, fractional and super-revivals are identified from  tomograms corresponding to the state of a single-mode radiation field governed by a Hamiltonian with more than one time scale, at appropriate instants of temporal evolution. We comment on the decoherence properties of states at the instant $\tfrac{1}{2} T_{rev}$. We have investigated the (quadrature and higher-order) squeezing properties of states of the radiation field directly from their tomograms. We have extended our investigations on the revival and squeezing phenomena to the double-well BEC system. The results have been published in Ref.~\cite{sharmila}.

In {\bf Chapter 3},  we propose several entanglement indicators that can be obtained directly from tomograms. We compare the performance of these indicators and assess them quantitatively in bipartite CV systems (the double-well BEC,  and the radiation field interacting with a multi-level atom modelled as an oscillator) and in a multipartite HQ system (two-level atoms interacting with a radiation field). The contents of this chapter are based on Ref.~\cite{arxiv4}.

In {\bf Chapter 4}, we assess the performance of the tomographic entanglement indicators in comparison with standard measures of entanglement during temporal evolution in CV systems, using tools of nonlinear time-series analysis. These results  have been  published in  Ref.~\cite{sharmila2}. We also give details pertaining to the manner in which we have used tomograms to distinguish between the two optical states reported in the experiment~\cite{perola} mentioned earlier.

In {\bf Chapter 5},  we have examined entanglement indicators in the context of the double Jaynes-Cummings and the double Tavis-Cummings models. The IBM quantum computing platform has been used to simulate and execute equivalent circuits of these HQ models. Quantitative comparison of the extent of entanglement obtained from experimental runs, simulation, and direct calculation from the HQ models, has been carried out. These results  have been  published in  Ref.~\cite{sharmila3}. We have also assessed the performance of the entanglement indicator in the spin system studied in~\cite{nmrExpt}. 

Finally in {\bf Chapter 6}, the results of the thesis are summarised, emphasising the novel aspects, and indicating avenues for future research. Appendices \ref{appen:TomoSinModExpr}  to  \ref{appen:ChronoTomo} augment the material presented in the main text.


%% file: chapter2.tex

\chapter{Signatures of wave packet revivals and squeezing in optical tomograms}
\label{ch:RevSqueezeOptTomo}

\section{Introduction}
\label{sec:Ch2Intro}

In this chapter, we examine the revival phenomena and the squeezing properties of quantum wave packets corresponding to both single-mode and bipartite systems. Our investigations are carried out by examining tomograms obtained at specific instants during temporal evolution of the system of interest. The effect of decoherence on the system state is also examined. Full, fractional and super-revivals are investigated in the case of a single-mode radiation field propagating through a nonlinear medium. Quantitative estimates of quadrature and higher-order squeezing of the field state have been obtained using procedures to compute the central moments and expectation values of appropriate observables \textit{directly} from tomograms. The bipartite continuous-variable system that we have examined is a Bose-Einstein condensate (BEC) trapped in a double-well with nonlinear interactions. We identify and assess the manner in which the revival phenomena and two-mode amplitude squeezing properties manifest in this case, choosing initial states which are factored products of the subsystem states. We have determined the extent of entropic squeezing of the condensate in one of the wells from the corresponding tomogram. 

To facilitate the discussion, we first review relevant properties of tomograms of single-mode and bipartite systems in Section \ref{sec:Ch2Sec2}. In Section \ref{sec:Ch2SM}, we consider single-mode systems governed by Hamiltonians with one or more time scales, and we identify full, fractional and super-revivals from tomograms. The manner in which a specific state of the system decoheres under amplitude and phase damping is studied. The quadrature and higher-order squeezing properties of the single-mode system are assessed tomographically. In Section \ref{sec:Ch2Sec3}, we carry out our investigations on the revival, quadrature and entropic squeezing phenomena in the double-well BEC system for initial states that are unentangled coherent states or boson-added coherent states. We conclude with brief remarks. 

\section{Tomograms: A brief review\label{sec:Ch2Sec2}}
We recall from Chapter \ref{ch:intro} that a tomogram is obtained by measuring a quorum of observables which, in principle, gives complete information about the state of the system. In the case of a single-mode radiation field the tomogram (Eq. \eref{eqn:tomogdef}) is given by 
\begin{equation}
\nonumber w(X_\theta, \theta) = \bra{X_\theta, \theta} \rho \ket{X_\theta, \theta}.
\end{equation}
Here $\rho$ is the density matrix, and $\mathbb{X}_{\theta} \ket{X_\theta, \theta} = X_{\theta} \ket{X_\theta, \theta} $. It is evident that for a pure state $\ket{\psi}$, $w(X_{\theta},\theta)=|\aver{X_{\theta},\theta|\psi}|^{2}$.
We see that the tomogram is a collection of probability distributions corresponding to the rotated quadrature operators $\mathbb{X}_{\theta}=(a e^{-i \theta} +a^{\dagger} e^{i \theta})/\sqrt{2}$. For each $\theta$,
\begin{equation}
\int_{-\infty}^{\infty}  \rmd X_{\theta} \,  w(X_\theta, \theta) =  1.
\end{equation}
The tomogram $w(X_{\theta}, \theta)$ is plotted with $X_{\theta}$ on the $x$-axis and $\theta$ on the $y$-axis.

We can show that 
\begin{equation}
\mathbb{X}_{\theta} = e^{i \theta a^{\dagger}a} \frac{(a+a^{\dagger})}{\sqrt{2}} e^{-i \theta a^{\dagger}a},
\label{eqn:rotatedQuadOps_BCH}
\end{equation}
 using the Baker-Campbell-Hausdorff identity for any operator $\mathcal{O}$, given by 
\begin{equation}
e^{i \mathcal{Q}}\mathcal{O} e^{- i \mathcal{Q}}= \mathcal{O} + i [\mathcal{Q},\mathcal{O}] + \frac{i^{2}}{2!} [\mathcal{Q},[\mathcal{Q},\mathcal{O}]]+ \frac{i^{3}}{3!} [\mathcal{Q},[\mathcal{Q},[\mathcal{Q},\mathcal{O}]]+\cdots,
\label{eqn:BCHidentity}
\end{equation}
where $\mathcal{Q}$ is a Hermitian operator. Substituting $\mathcal{O}=(a+a^{\dagger})/\sqrt{2}$ and $\mathcal{Q}=\theta a^{\dagger} a$ in Eq. \eref{eqn:BCHidentity}, we get Eq. \eref{eqn:rotatedQuadOps_BCH}.

From Eq. \eref{eqn:rotatedQuadOps_BCH} and $\mathbb{X}_{\theta} \ket{X_\theta, \theta} = X_{\theta} \ket{X_\theta, \theta} $, we get
\begin{equation} 
\ket{X_\theta, \theta} = e^{i \theta a^{\dagger} a} \ket{X},
\label{eqn:phaseshift}
\end{equation}
 where $\ket{X}$ is the eigenstate of the operator $\left((a+a^{\dagger})/\sqrt{2}\right)$. 
We note that $e^{i \pi a^{\dagger} a}$ is the parity operator since $\mathbb{X}_{\theta+\pi}=e^{i \pi a^{\dagger} a} \mathbb{X}_{\theta} e^{-i \pi a^{\dagger} a} = -\mathbb{X}_{\theta}$, which in turn implies that $e^{i \pi a^{\dagger} a} \ket{X_{\theta},\theta}=\ket{-X_\theta, \theta}$.
Using this and Eq. \eref{eqn:phaseshift}, we get
\begin{equation}
w(X_{\theta+\pi}, \theta + \pi ) = w(-X_{\theta}, \theta).
\end{equation}
For state reconstruction, although it is sufficient to work with the range $0 \leq \theta < \pi$, the tomogram plotted for $0 \leq \theta < 2\pi$ helps visualise various features better.
 
The tomogram of a normalised  pure state $\ket{\psi}$, which can be expanded in the photon number basis $\lbrace \ket{p} \rbrace$ as $\sum\limits_{p=0}^{\infty} c_p \ket{p}$, is given by~\cite{MankoFockStates} 
\begin{equation}
\label{eqn:tomogfock}
w(X_{\theta}, \theta) = \frac{e^{-X_{\theta}^{2}}}{\sqrt{\pi}} \left| \sum_{n=0}^{\infty} \frac{c_{n} e^{-i n \theta} }{\sqrt{n!} 2^{\frac{n}{2}} } H_{n}(X_\theta)  \right|^{2},
\end{equation}
where $H_{n}(X_{\theta})$ is the Hermite polynomial.
In this expression (derived in Appendix \ref{appen:TomoSinModExpr}), $c_{n}$ alone is a function of time. Hence Eq. \eref{eqn:tomogfock} is useful for numerical computation of tomograms at various instants of time. 
 
These ideas can be extended in a straightforward manner to multimode systems. In particular, for a bipartite system we define rotated quadrature operators,
\begin{align*}
\mathbb{X}_{\thetaat} = (a^\dagger e^{i \thetaat} + a e^{-i \thetaat})/\sqrt{2}, \qquad \text{and} \qquad \mathbb{X}_{\thetabt} = (b^\dagger e^{i \thetabt} + b e^{-i \thetabt})/\sqrt{2}.
\end{align*}
Here $(a, a^{\dagger})$ and $(b, b^{\dagger})$ are the particle annihilation and creation operators corresponding to subsystems A and B respectively, of the bipartite system. The bipartite tomogram
\begin{equation}
w(X_{\thetaat}, \thetaa; X_{\thetabt}, \thetab)= \aver{X_{\thetaat}, \thetaa; X_{\thetabt}, \thetab|\rho_{\textsc{ab}}|X_{\thetaat}, \thetaa; X_{\thetabt}, \thetab},
\label{eqn:2modeTomoDefn}
\end{equation}
where $\rho_{\textsc{ab}}$ is the bipartite density matrix and $\mathbb{X}_{\theta_{i}} \ket{X_{\theta_{i}},\theta_{i}} = X_{\theta_{i}} \ket{X_{\theta_{i}},\theta_{i}}$ ($i=\text{A,B}$). Here, and in the rest of the thesis, $\ket{X_{\thetaat},\thetaat} \otimes \ket{X_{\thetabt},\thetabt}$ is denoted by $\ket{X_{\thetaat},\thetaat;X_{\thetabt},\thetabt}$. The normalisation condition is given by 
\begin{equation}
\int_{-\infty}^{\infty}\! \rmd X_{\thetaat} \int_{-\infty}^{\infty} \!\rmd X_{\thetabt} w(X_{\thetaat},\thetaa;X_{\thetabt},\thetab) = 1
\label{eqn:tomoNorm}
\end{equation}
for each $\thetaa$ and $\thetab$. 

Analogous to the single-mode example, we now consider a pure state $\ket{\psi}$  expanded in the Fock bases $\{\ket{m}\}$, $\{\ket{n}\}$  corresponding to subsystems A and B respectively as  
$\ket{\psi}  = \sum\limits_{m,n=0}^{\infty} c_{mn} \ket{m \delim n}$ . Here, $\ket{m \delim n}$ is a short-hand notation 
for $\ket{m} \otimes \ket{n}$ and $c_{mn}$ are the expansion coefficients.

It is straightforward to extend the procedure (Appendix \ref{appen:TomoSinModExpr}) used in deriving Eq. \eref{eqn:tomogfock} to generic bipartite systems whose subsystems are infinite-dimensional.  We then get
\begin{align}
\nonumber w(X_{\thetaat} , \thetaa;& X_{\thetabt}, \thetab)
= \frac{\exp(-X_{\thetaat}^{2}-X_{\thetabt} ^{2})}{\pi} \\
&\times \left| \sum_{m,n=0}^{\infty} \frac{c_{mn} e^{- i (m \thetaat + n \thetabt)}}{(m! n! 2^{m+n})^{1/2}} H_{m} (X_{\thetaat}) H_{n}(X_{\thetabt}) \right|^{2}.
\label{eqn:2_mode_opt_tomo}
\end{align}
The tomograms corresponding to the subsystems A and B (reduced tomograms) are respectively given  by 
\begin{align}
w_{\textsc{a}}(X_{\thetaat}, \thetaa)= \aver{X_{\thetaat}, \thetaa|\rho_{\textsc{a}}|X_{\thetaat}, \thetaa} = \int_{-\infty}^{\infty} w(X_{\thetaat}, \thetaa; X_{\thetabt}, \thetab) \rmd X_{\thetabt},
\label{eqn:tomo_2_mode_sub_A}
\end{align}
for any fixed value of $\thetabt$ and
\begin{align}
w_{\textsc{b}}(X_{\thetabt}, \thetab)= \aver{X_{\thetabt}, \thetab|\rho_{\textsc{b}}|X_{\thetabt}, \thetab} = \int_{-\infty}^{\infty} w(X_{\thetaat}, \thetaa; X_{\thetabt}, \thetab) \rmd X_{\thetaat}
\label{eqn:tomo_2_mode_sub_B}
\end{align}
for any fixed value of $\thetaa$. The reduced density matrix $\rho_{\textsc{a}}$ (respectively, $\rho_{\textsc{b}}$) is given by $\text{Tr}_{\textsc{b}}(\rho_{\textsc{ab}})$ (resp., $\text{Tr}_{\textsc{a}}(\rho_{\textsc{ab}})$). 

\section{\label{sec:Ch2SM}Single-mode system: A tomographic approach}
\subsection{Revivals and fractional revivals}

As stated in Chapter \ref{ch:intro}, when an initial quantum wave packet $\ket{\psi(0)}$ evolves under a nonlinear Hamiltonian, it could revive at a later instant of time $T_{rev}$ under certain circumstances. The fidelity $|\aver{\psi(0)|\psi(T_{rev})}|^{2}=1$. When an initial coherent state of light $\ket{\alpha}$ ($\alpha \in \mathbb{C}$) propagates through a nonlinear optical medium, a simple choice for the effective Hamiltonian is $ \hbar \chi_{1} {a^{\dagger}}^{2}a^{2}$ (Kerr Hamiltonian) where $(a,a^{\dagger})$ are photon destruction and creation operators respectively. In this example, the state of the radiation field exhibits both periodic revivals and fractional revivals. Here $T_{rev}=\pi/\chi_{1}$ and an $\ell$-subpacket fractional revival occurs at $T_{rev}/\ell$ ($l=2,3,\dots$). 

   In this system, it has been shown earlier~\cite{sudhrohithrev}, that 
the optical tomogram at instants of revivals and fractional revivals is composed of distinct strands in contrast to blurred patterns at other generic instants during temporal evolution of the field. By using the method of `strand-counting', the authors infer that corresponding to an $\ell$-subpacket fractional revival the field tomogram has $\ell$ strands. Thus for instance, at $T_{rev}/3$, the tomogram is comprised of only 3 distinct strands. A limitation in this method is that  individual  strands in the tomogram corresponding to an $\ell$-subpacket fractional revival will not be distinct for $\ell$  sufficiently large (5 or more for $|\alpha|\sim 3$) due to quantum interference effects. However, to understand the broad features of the revival phenomena it suffices to employ this procedure  without resorting to detailed quantitative methods. 

 We now proceed to apply this tomographic approach to a system whose dynamics is governed by the Hamiltonian $H =( \hbar \chi_{1} {a^{\dagger}}^{2} a^{2} + \hbar \chi_{2} {a^{\dagger}}^{3} a^{3})$. In what follows we set $\hbar = 1$ for convenience.
To facilitate the discussion, we first consider an initial CS evolving in an effective cubic Hamiltonian $H^{'} =  \chi_{2} {a^{\dagger}}^{3} a^{3} = \chi_{2} \mathcal{N} (\mathcal{N}-1) (\mathcal{N} - 2)$, where $\mathcal{N} = a^{\dagger}a$ and $\chi_{2}$ is a constant with appropriate dimensions.  We will then move on to consider the effects of the full Hamiltonian $H$ using the tomographic approach. 
 We corroborate our numerical findings with  analytical explanations wherever possible for the revival patterns that we observe. 

\begin{figure*}
	\centering    
	\includegraphics[width=0.3\textwidth]{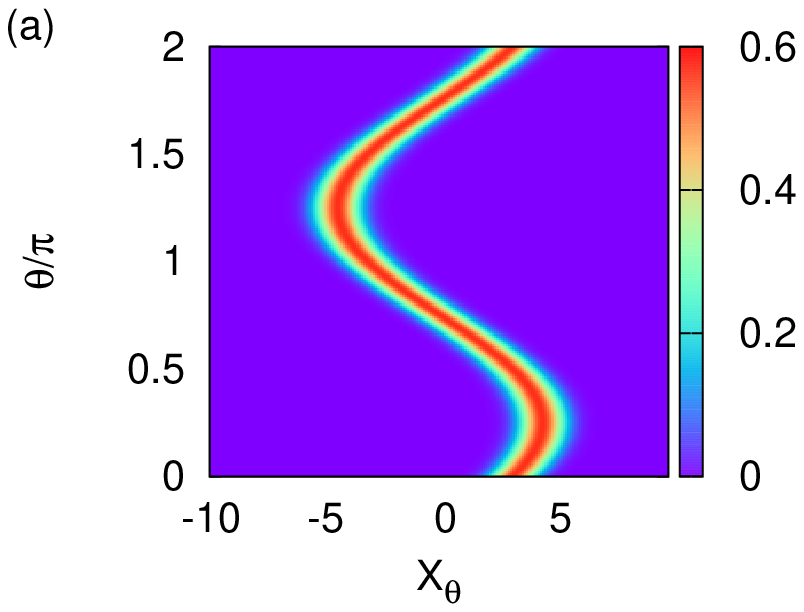}
	\includegraphics[width=0.3\textwidth]{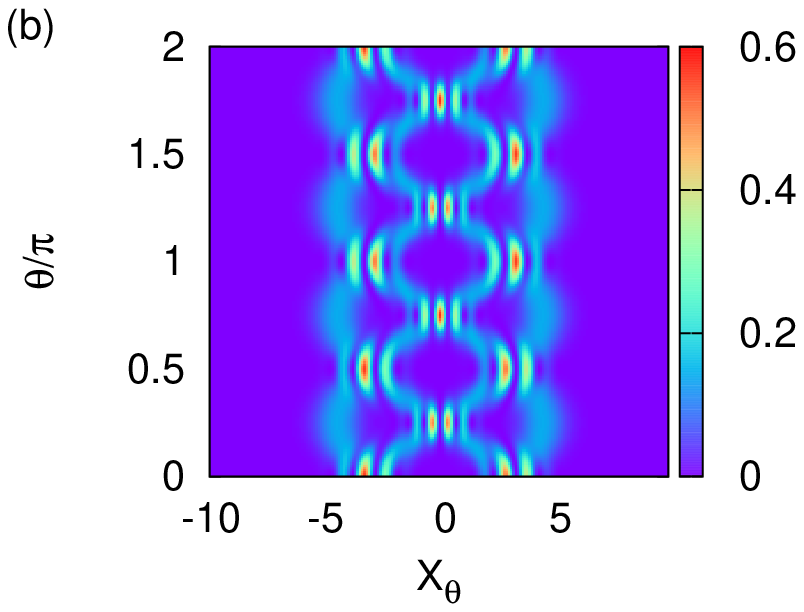}
	\includegraphics[width=0.3\textwidth]{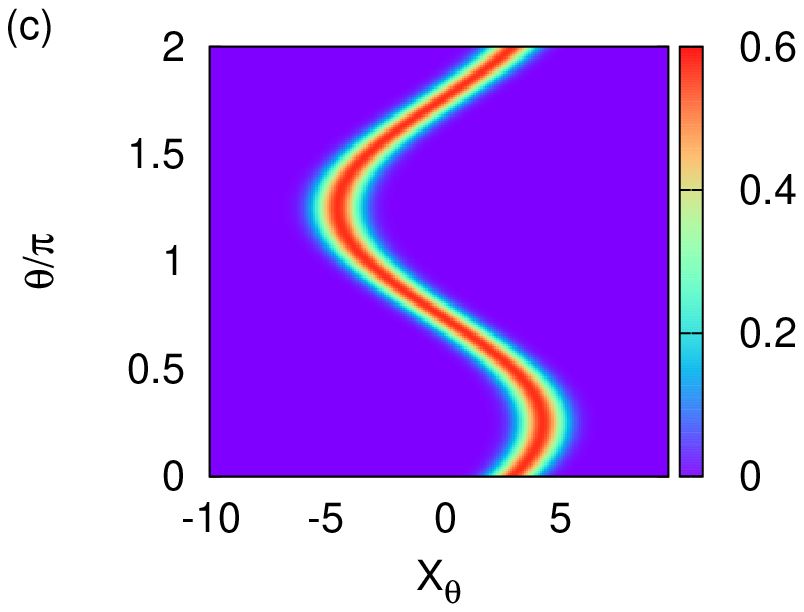}
	\includegraphics[width=0.3\textwidth]{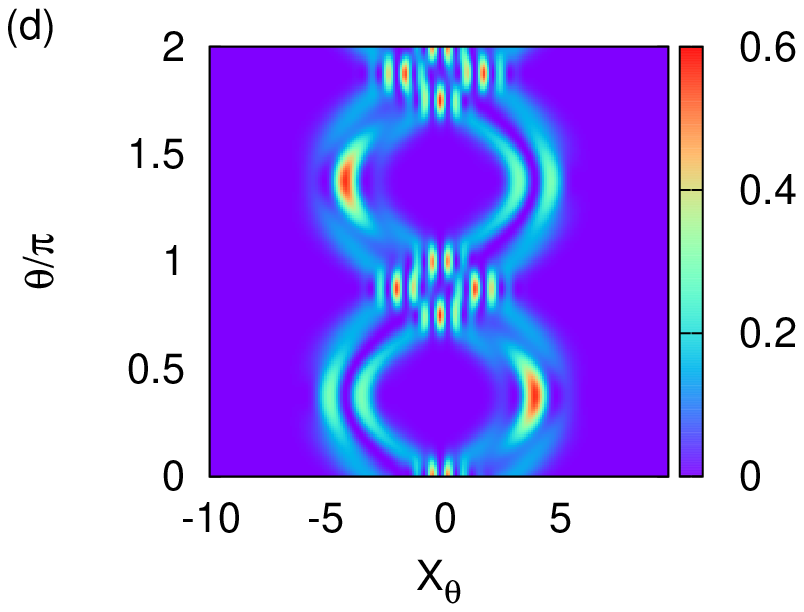}
	\includegraphics[width=0.3\textwidth]{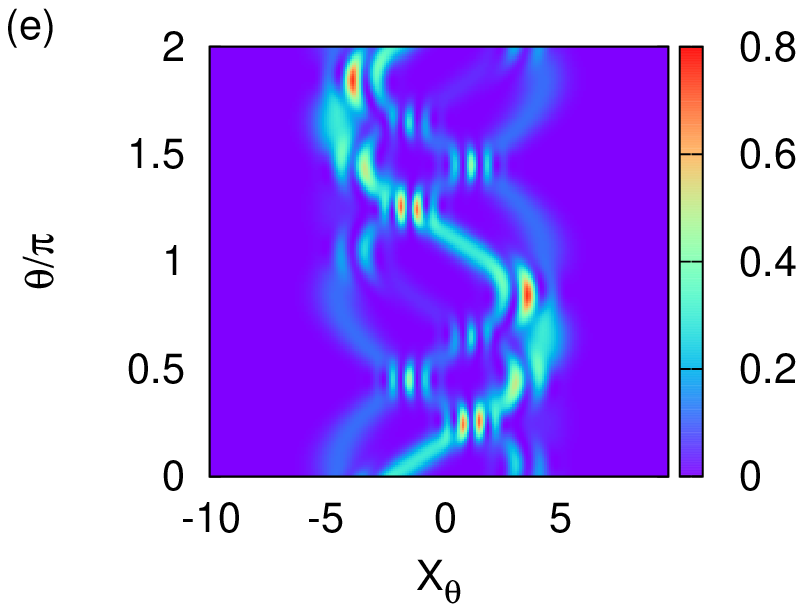}
	\includegraphics[width=0.3\textwidth]{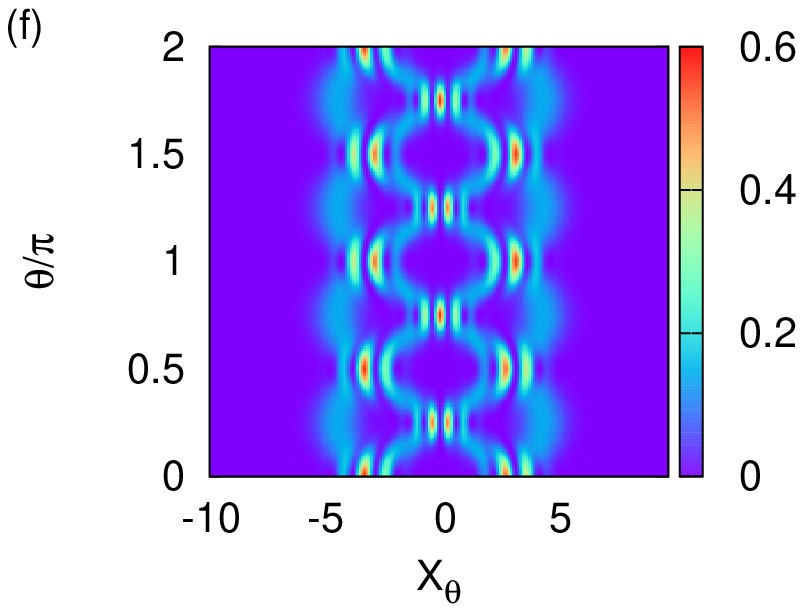}
	\includegraphics[width=0.3\textwidth]{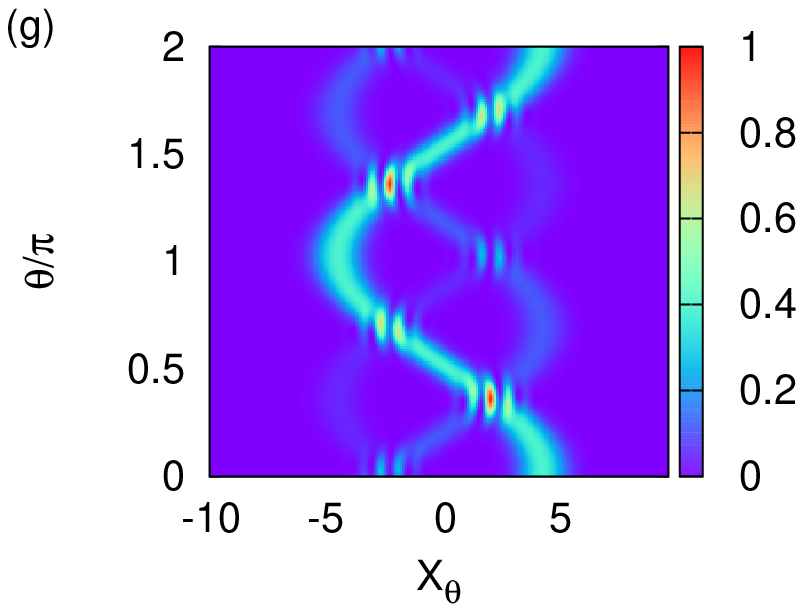}
	\includegraphics[width=0.3\textwidth]{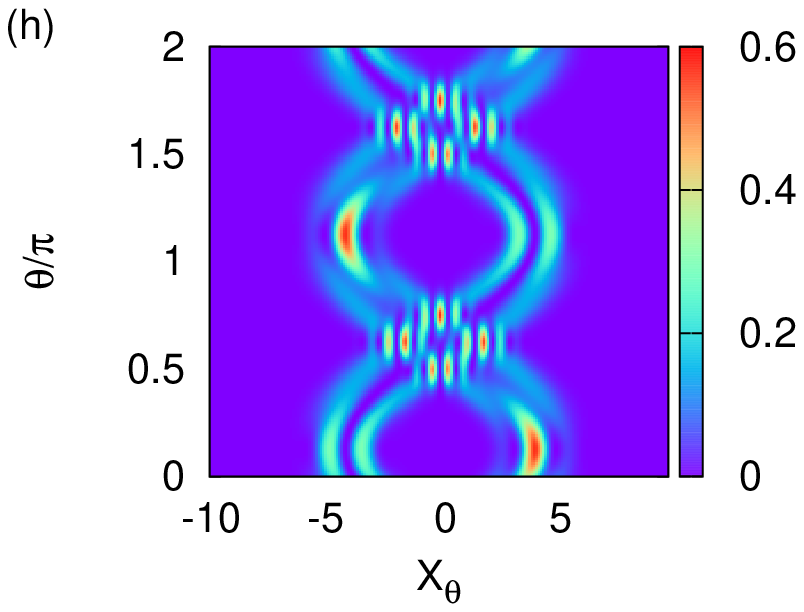}
	\includegraphics[width=0.3\textwidth]{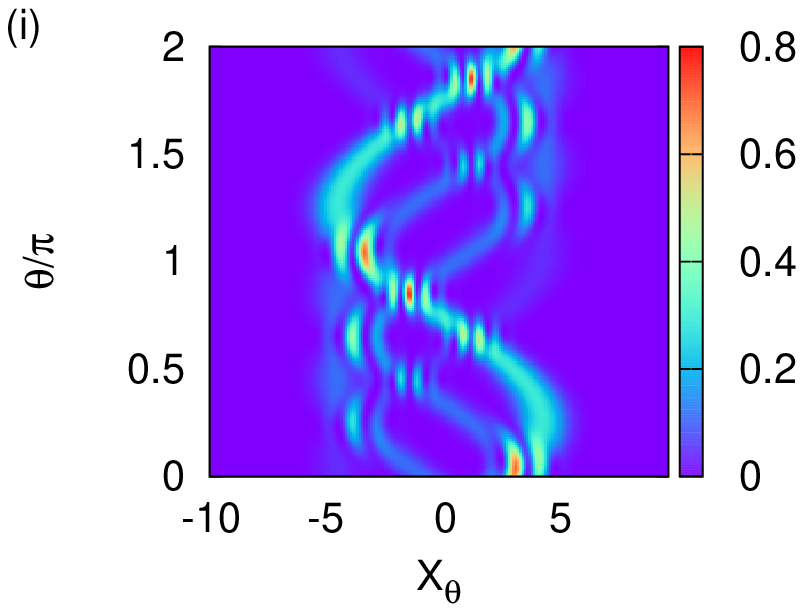}
	\caption{ Tomograms of an initial CS for a cubic Hamiltonian and $\alpha=\sqrt{10}e^{i\pi/4}$ at  instants (a) $0$ and $T_{rev}$, (b) $T_{rev}/2$, (c) $T_{rev}/3$, (d) $T_{rev}/4$, (e) $T_{rev}/5$, (f) $T_{rev}/6$, (g) $T_{rev}/9$, (h) $T_{rev}/12$, and (i) $T_{rev}/15$.}
	\label{fig:tomogrevivalcubic}
\end{figure*}

It can be easily seen that an initial CS or $1$-PACS governed by $H^{'}$ revives fully at instants $ T_{rev} = \pi/\chi_{2}$. Of immediate interest to us is the nature of tomograms at instants $T_{rev}/\ell$
where $\ell$ is a positive  integer. The optical tomograms for different values of $\ell$ are shown in Figs.~\ref{fig:tomogrevivalcubic} (a)-(i). The following observations are in order. (i) At initial time (equivalently at instant $T_{rev}$) and at $T_{rev}/3$, the tomograms look similar. This is in sharp contrast to the case of the system governed by the Kerr Hamiltonian mentioned earlier. While the system at hand is more complicated, the full revival at $T_{rev}/3$  is  a simple consequence of the fact that $n(n-1)(n-2)/3$ is even $\forall\, n\epsilon\mathbb{N}$. Here, $\mathcal{N} \ket{n}  = n \ket{n}$, with $ \{\ket{n} \}$ denoting the photon number basis. Hence corresponding to  an initial state $\ket{\psi(0)} = \sum_{m=0}^{\infty} c_{n} \ket{n}$, the state at instant $T_{rev}/3$ is
\begin{equation}
\nonumber\ket{\psi(T_{rev}/3)} = \mathcal{U}(T_{rev}/3) \ket{\psi(0)} = \sum_{n=0}^{\infty} e^{-i\pi n(n-1)(n-2)/3} c_n \ket{n} = \ket{\psi(0)}.
\end{equation}
(ii) At  instants $T_{rev}/2$ and $T_{rev}/6$,  the tomograms are similar  and have four strands each, in contrast to what is reported in \cite{sudhrohithrev}, where at $T_{rev}/2$ the tomogram has two strands and at $T_{rev}/6$ it has six strands.

The new  features in our system obviously follow from the properties of the unitary time evolution operator corresponding to $H^{'}$. 
An analysis of the properties of the time evolution operator would, in principle,  explain the appearance of a specific number of strands in the tomogram at different instants $T_{rev}/\ell$. 
 
\begin{figure}
	\centering
	\includegraphics[width=0.4\textwidth]{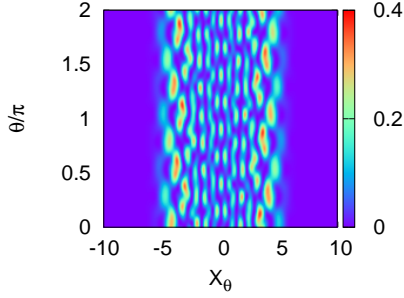}
	\caption{ Tomogram of an initial CS at $t=\pi / \chi_{2}$ for $\alpha=\sqrt{10}e^{i\pi/4}$, $\chi _{1}= 1$ and $\chi_{2}= 10^{-7}/\sqrt{2}$.}
	\label{fig:tomogcsquadcubirrat}
\end{figure}

 We are now in a position to investigate tomogram patterns for the full Hamiltonian 
\begin{equation}
H =  (\chi_{1} a^{\dagger 2} a^2 + \chi_{2} a^{\dagger 3} a^3).
\label{eqn:full_single_mode_H}
\end{equation}
 In this case, 
\begin{equation}
 T_{rev} = \pi \,  \mathrm {LCM}\left(\frac{1}{\chi_{1}}, \frac{1}{\chi_{2}}\right).
\end{equation}
 It is easy to see that  
 if the ratio $\chi_{1}/\chi_{2}$ is irrational,  revivals are absent and the generic tomogram at any instant is blurred.  This is illustrated in Fig. \ref{fig:tomogcsquadcubirrat}
 for an initial CS with $\alpha = \sqrt{10} \exp(i \pi /4)$  with $\chi_{1} = 1$, and $\chi_{2} = 10^{-7}/\sqrt{2}$ at $t = \pi/\chi_{2}$.

\begin{figure}
	\centering    
	\includegraphics[width=0.4\textwidth]{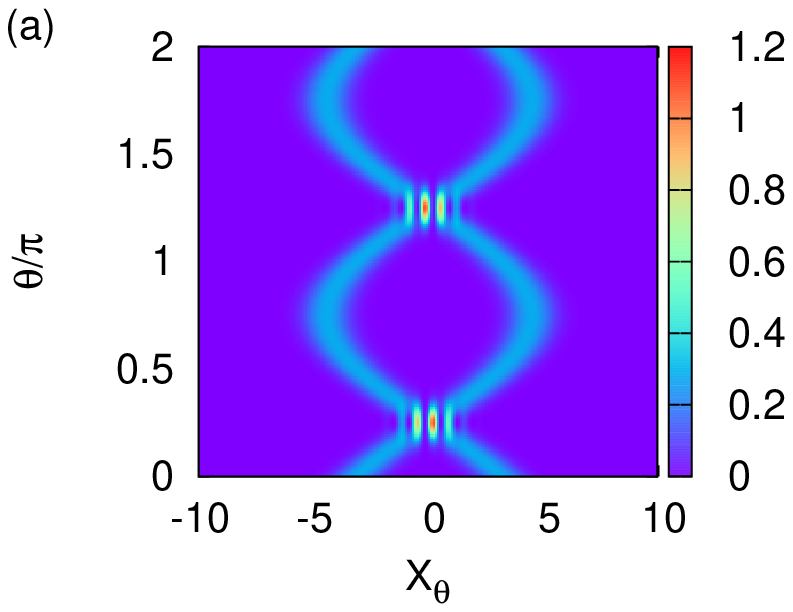}
	\includegraphics[width=0.4\textwidth]{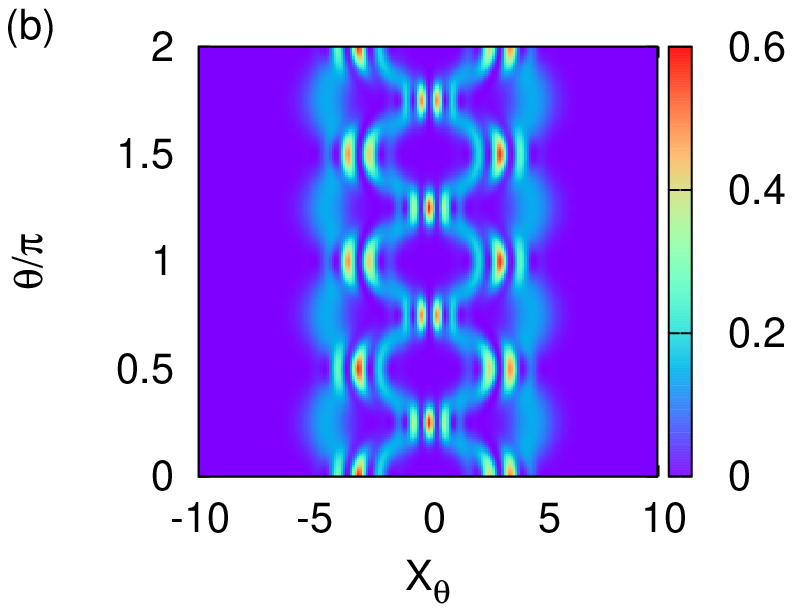}
	\caption{ Tomograms of an initial CS at $ t = T_{rev}/2$ for $\alpha=\sqrt{10}e^{i\pi/4}$, $\chi_{1}=1$ and (a) $\chi_{2}=2.048 \times 10^{-7}$, (b)  $\chi_{2}=1.024 \times 10^{-7}$.}
	\label{fig:tomogrevivalquadcubtrev2}
\end{figure}

\begin{figure}
	\centering    
	\includegraphics[width=0.4\textwidth]{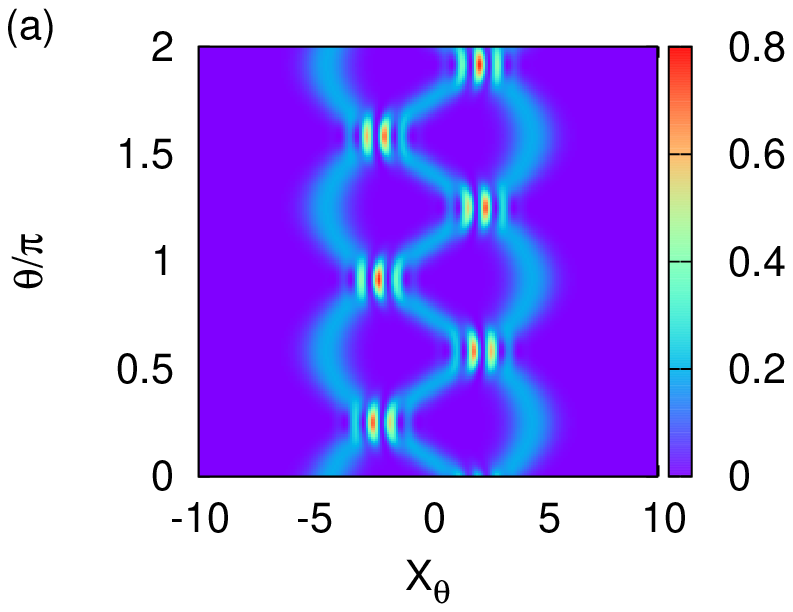}
	\includegraphics[width=0.4\textwidth]{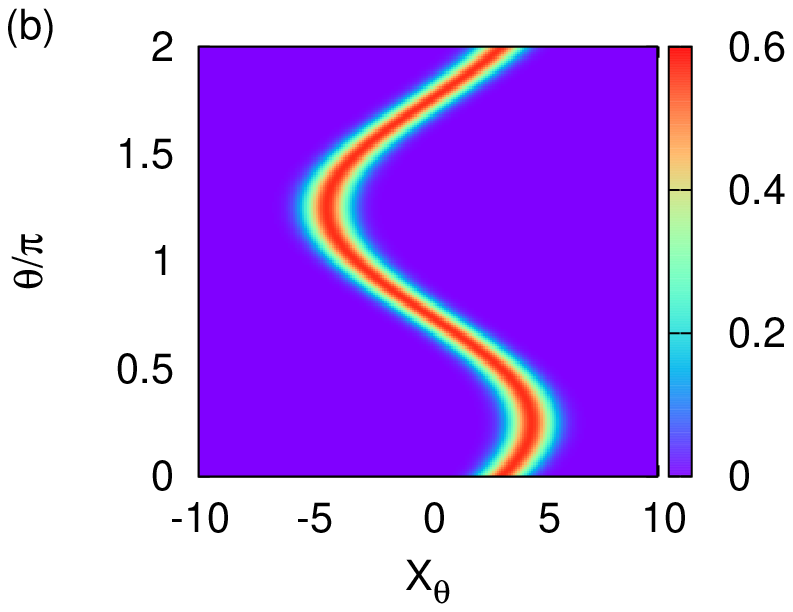}
	\caption{ Tomograms of an initial CS  at $t = T_{rev}/3$ for $\alpha=\sqrt{10}e^{i\pi/4}$, $\chi_{1}=1$ and (a) $\chi_{2}=2.048 \times 10^{-7}$, (b) $\chi_{2}=(4/3) \times 10^{-7}$.}
	\label{fig:tomogrevivalquadcubtrev3}
\end{figure}

\begin{figure*}
	\centering    
	\includegraphics[width=0.32\textwidth]{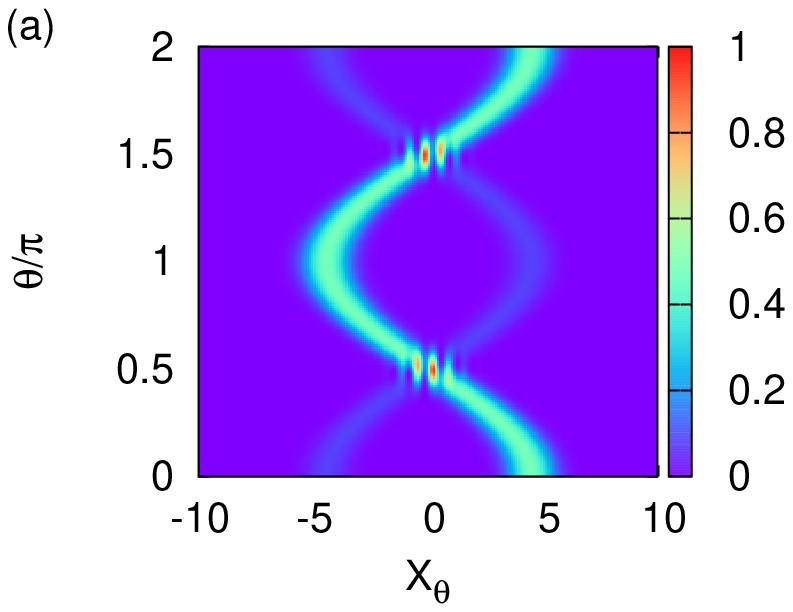}
	\includegraphics[width=0.32\textwidth]{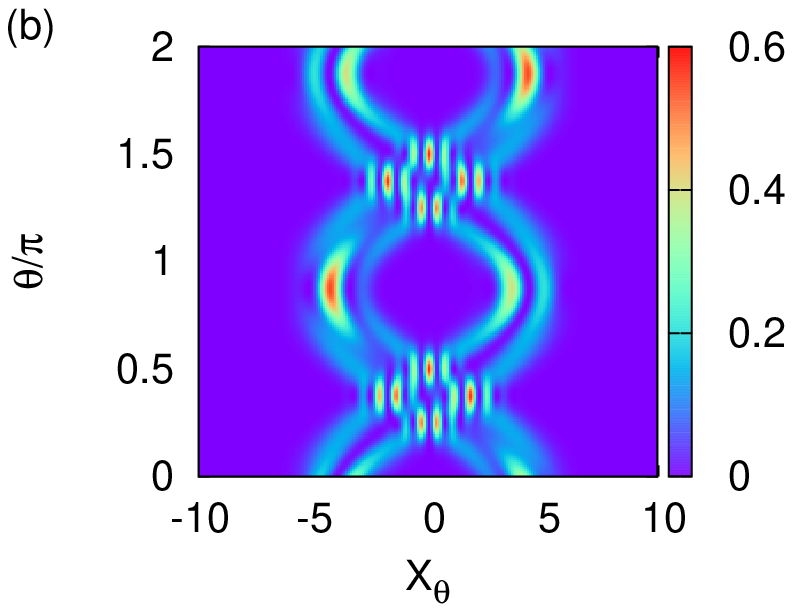}
	\includegraphics[width=0.32\textwidth]{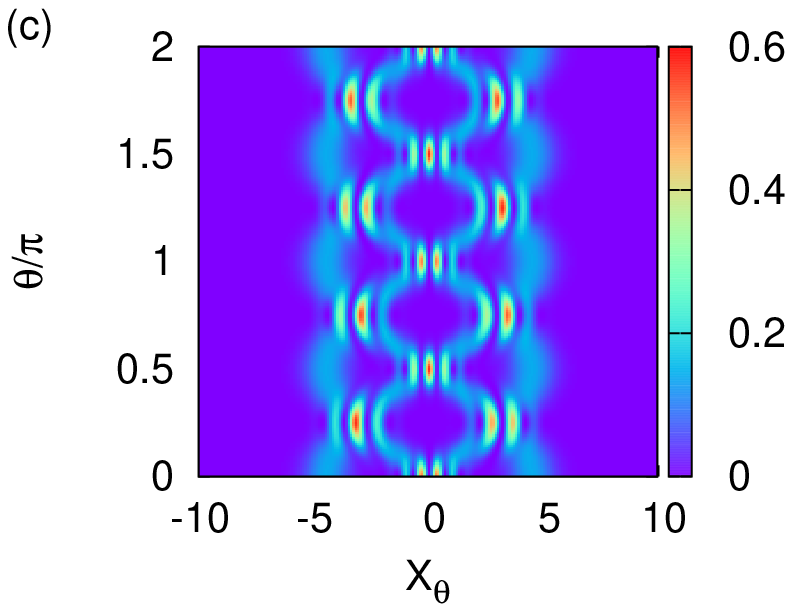}
	\caption{ Tomograms of an initial CS  at $t=T_{rev}/4$ for $\alpha=\sqrt{10}e^{i\pi/4}$, $\chi_{1}=1$ and (a) $\chi_{2}=2.048 \times 10^{-7}$, (b) $\chi_{2}=1.024 \times 10^{-7}$, and (c) $\chi_{2}=4.096 \times 10^{-7}$.}
	\label{fig:tomogrevivalquadcubtrev4}
\end{figure*}

For rational  $\chi_{1}/\chi_{2}$,  revivals and fractional revivals are seen.  Fractional revivals  occur at instants $T_{rev}/\ell$ as before, but the corresponding tomogram patterns  are sensitive to the ratio $\chi_{1}/\chi_{2}$.  
  We expect that  for a given $\ell$, the tomogram will have $\ell$ strands as a consequence of the Kerr term in $H$.  The cubic term in $H$ however allows for other possibilities. We illustrate this for an initial CS with Hamiltonian $H$  and 
$\alpha = \sqrt{10} \exp(i \pi /4)$  in  Figs. \ref{fig:tomogrevivalquadcubtrev2} -- \ref{fig:tomogrevivalquadcubtrev4}.  
 At the instant $T_{rev}/2$,  apart from the two-strand tomogram for $\chi_{1} = 1$ and $\chi_{2} = 2.048 \times 10^{-7}$, 
 one of the other possibilities is a four-strand tomogram for $\chi_{1} = 1$ and $\chi_{2} = 1.024 \times 10^{-7}$ (Figs. \ref{fig:tomogrevivalquadcubtrev2} (a),(b)). Similarly at $t = T_{rev}/3$, the tomogram has three strands for $\chi_{1} = 1$ and $\chi_{2} = 2.048 \times 10^{-7}$  and a single strand  similar to the tomogram of a CS for 
$\chi_{1} = 1$ and $\chi_{2} = (4/3) \times 10^{-7}$ (Figs. \ref{fig:tomogrevivalquadcubtrev3} (a),(b)). At $t = T_{rev}/4$ three specimen tomograms which are distinctly different from each other are shown in Figs.~\ref{fig:tomogrevivalquadcubtrev4} (a)-(c) with $\chi_{1} = 1$ and $\chi_{2} = 2.048 \times 10^{-7}$,  $1.024 \times 10^{-7}$ and $4.096 \times 10^{-7}$ respectively.

These features can be explained on a case by case basis as before, by examining the periodicity properties of the unitary time evolution operator at appropriate instants. It is however evident that the simple inference that an $\ell$-subpacket fractional revival is associated with an $\ell$-strand tomogram alone, does not hold when more than one time scale is involved in the Hamiltonian, and there can be several tomograms possible at a given instant depending on the interplay between the different time scales in the system.

Thus distinct signatures of higher-order nonlinearities are captured in tomograms corresponding to specific instants of fractional revivals.  
This observation could  be of considerable use in understanding the role of higher-order susceptibilities through experiments.  Practical  difficulties arise because the  susceptibility parameter is numerically small for the Kerr medium. (In our discussions we have scaled $\chi_{1}$ to unity). Typically, the loss rate of the field is such that damping occurs much faster than $T_{rev}$. It is therefore considerably difficult to  sustain field collapses and revivals till $T_{rev}$ in experiments.  However a  forerunner to such studies  \cite{kirchmair} has been successfully carried out by engineering an artificial Kerr medium with sufficiently  high susceptibility in which collapses and revivals of a coherent state have been observed. 
Since then, several experiments (see, for instance, \cite{circQEDcatSt,circQEDcrossKerr}) have been performed in circuit QED to implement Kerr-type nonlinearities in single-mode and bipartite systems. The coherence time of the microwave cavity in circuit QED has also significantly increased to orders of milliseconds~\cite{circQEDcohtime}.
While effects of higher-order nonlinearities are more subtle as we have shown, it should still be possible in future to capture the tomographic signatures reported above, particularly because they can be seen even at the instant $T_{rev}/4$. 

We now examine the role played by interaction of the  single-mode system with an external environment, and the manner in which the purity of the system gets affected. For illustrative purposes, we  consider the state at the instant $T_{rev}/2$ corresponding to an initial CS governed by the Hamiltonian \eref{eqn:full_single_mode_H}.  This is allowed to interact with the environment. The procedure used is  similar to that employed for the Kerr Hamiltonian~\cite{sudhrohithrev} so as to facilitate comparison of the two cases.  

We consider two models of decoherence, namely, amplitude decay and phase damping of the state. First, dissipation is modelled using the master equation for amplitude decay
\begin{equation}
\frac{\rmd \varrho}{\rmd \tau} = - \Gamma (a^{\dagger} a \varrho - 2 a \varrho a^{\dagger} + \varrho a^{\dagger} a).
\label{eqn:master_eqn}
\end{equation}
Here $\varrho$ denotes the density matrix, $\Gamma$ is the rate of loss of photons and $\tau$ the time parameter is reckoned from the instant $T_{rev}/2$. 

The solution to this master equation is ~\cite{agarwalmaster}
\begin{equation}
\varrho(\tau) = \sum_{n,n'=0}^{\infty}\varrho_{n,n'}(\tau) \ket{n}\bra{n'},
\label{eqn:soln_master}
\end{equation}
with matrix elements
\begin{align}
\nonumber \varrho_{n,n'}(\tau)=e^{-\Gamma\tau(n+n')}&\sum_{r=0}^{\infty} \sqrt{{n+r \choose r}{n'+r \choose r}} \\
&(1-e^{-2\Gamma\tau})^{r} \varrho_{n+r,n'+r}(\tau=0).
\label{eqn:rho_soln_master}
\end{align}

\begin{figure*}
 \includegraphics[width=0.32\textwidth]{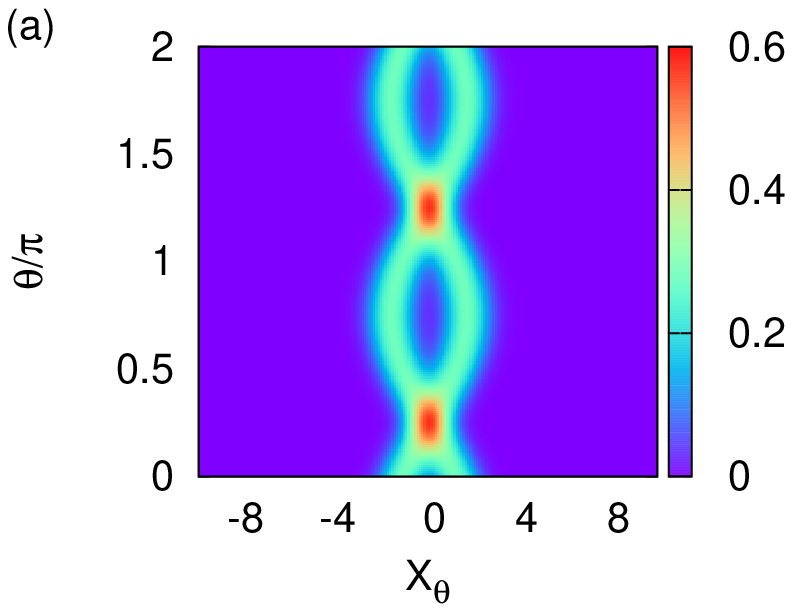}
 \includegraphics[width=0.32\textwidth]{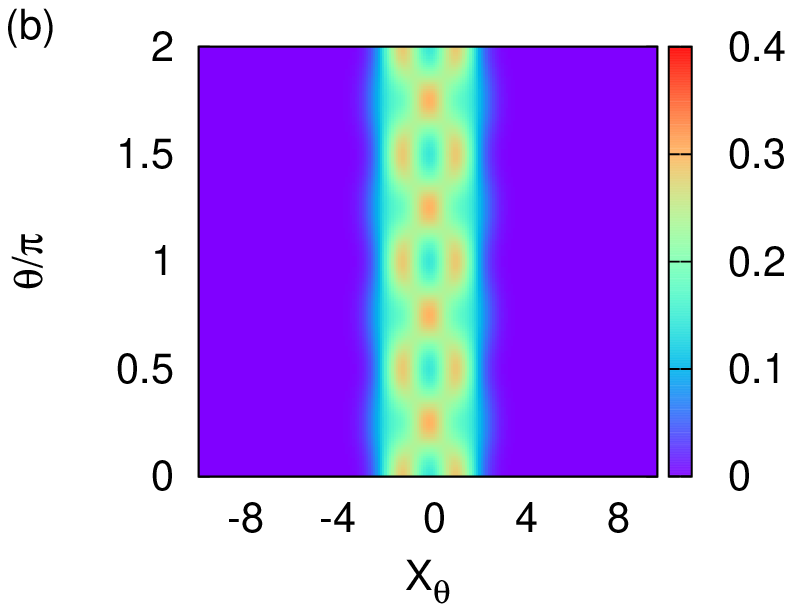}
 \includegraphics[width=0.32\textwidth]{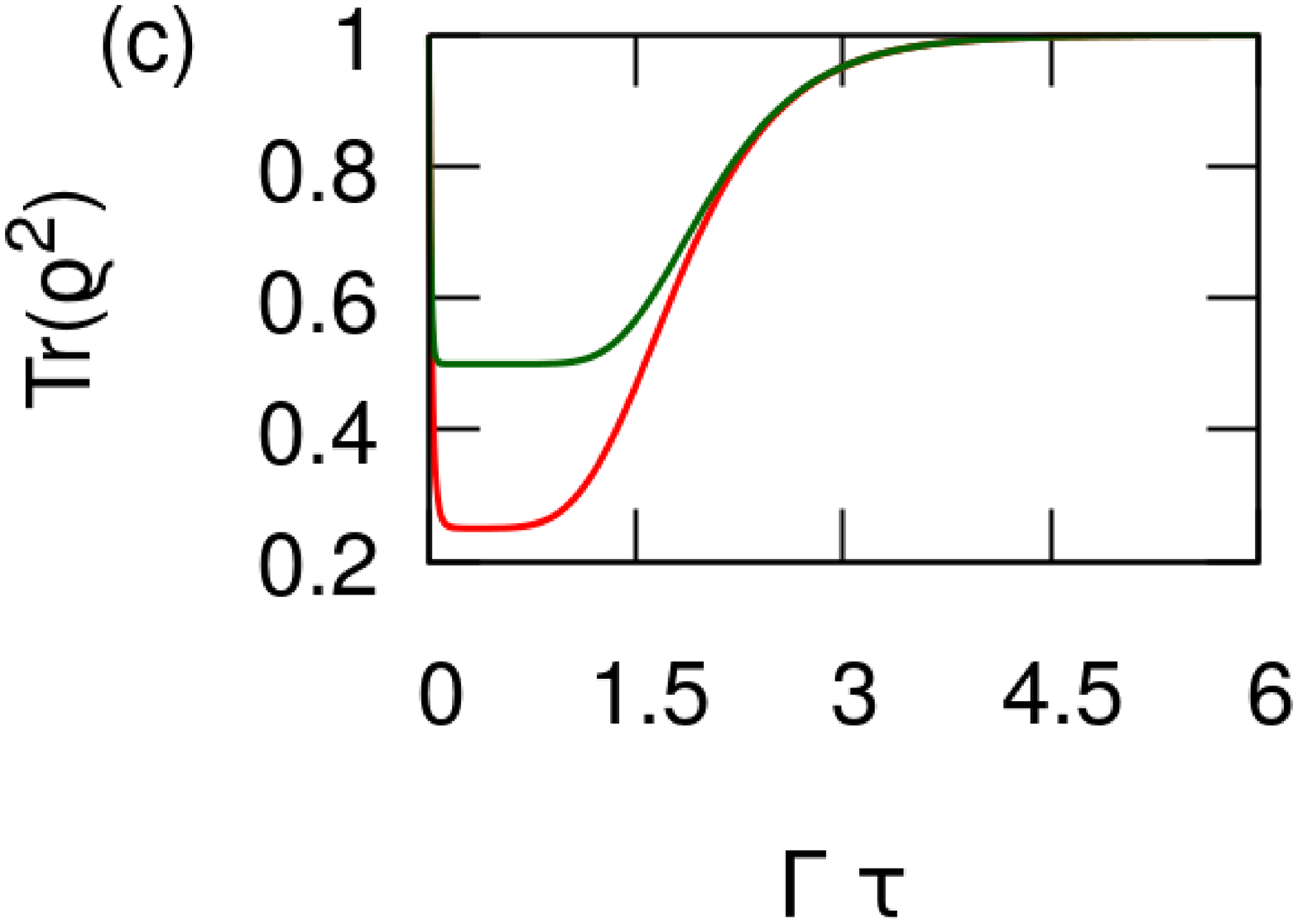}
 \caption{ Tomograms of an initial CS at $T_{rev}/2$ in the amplitude decay model for $\alpha=\sqrt{10}e^{i\pi/4}$, $\Gamma\tau=1$, $\chi_{1}=1$ and (a) $\chi_{2}=2.048 \times 10^{-7}$, (b) $\chi_{2}=1.024 \times 10^{-7}$. (c) $\mathrm{Tr}(\varrho^{2})$ vs. $\Gamma\tau$ for $\chi_{1}=1$ and $\chi_{2}=2.048 \times 10^{-7}$ (green), $\chi_{2}=1.024 \times 10^{-7}$ (red).}
 \label{fig:decoherence}
 \end{figure*}

 As is to be expected from Eqs. \eref{eqn:soln_master} and \eref{eqn:rho_soln_master} as $\Gamma\tau\rightarrow\infty$  the state evolves to $\ket{0}\bra{0}$. The purity of the state $\left(\mathrm{Tr}(\varrho^{2})\right)$ initially decreases from $1$ corresponding to the initial pure state, and subsequently increases back to $1$  when $\Gamma\tau\approx4.5$  (Fig. \ref{fig:decoherence} (c)). Depending on the ratio $\chi_{1}/\chi_{2}$ we see  different extents of loss of purity of the state. The red (green) curve corresponds to $\chi_{1}=1$ and $\chi_{2}=1.024 \times 10^{-7}$ ($\chi_{2}=2.048 \times 10^{-7}$). Further, new aspects of decoherence in the system considered by us arise as a consequence of the possibility of more than one distinctly different tomograms at $T_{rev}/2$,  depending on the ratio $\chi_{1}/\chi_{2}$ (Figs. \ref{fig:decoherence} (a),(b)).

\begin{figure}
 \includegraphics[width=0.4\textwidth]{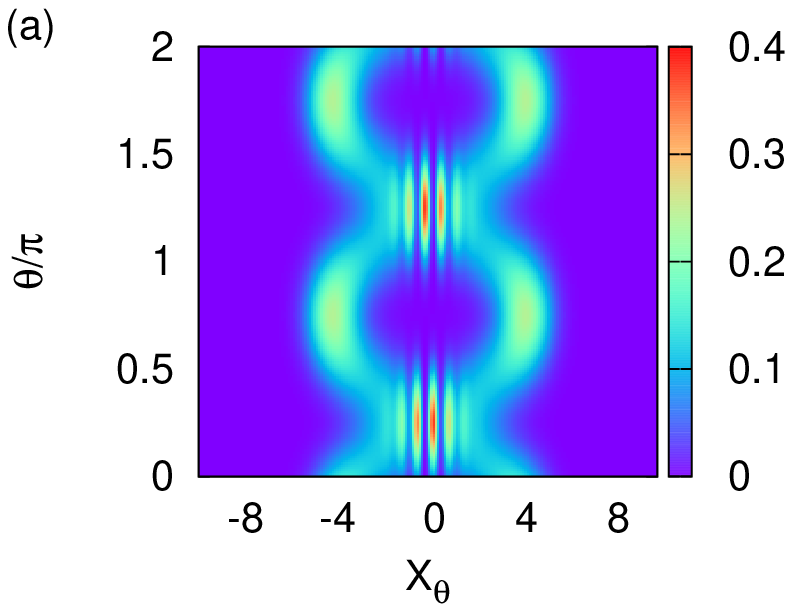}
 \includegraphics[width=0.4\textwidth]{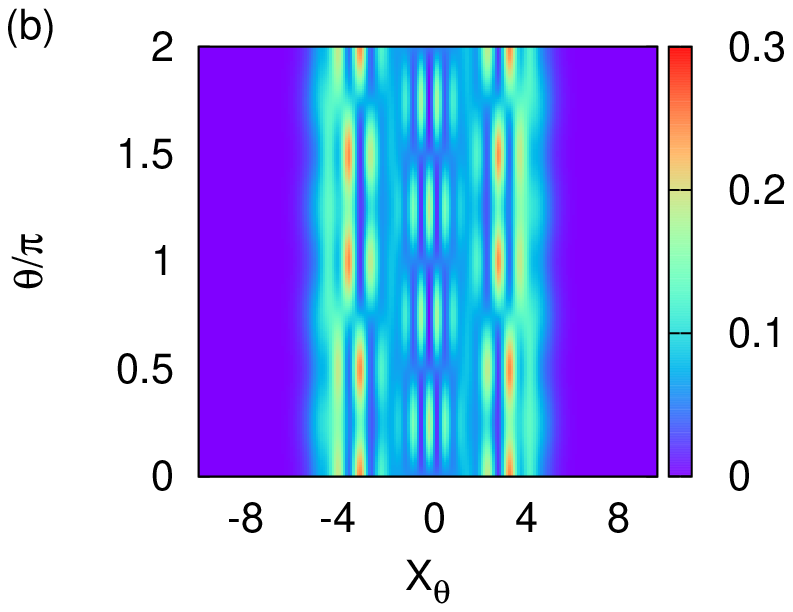}
 \caption{ Tomograms of an initial CS at $T_{rev}/2$ in the phase damping model for $\alpha=\sqrt{10}e^{i\pi/4}$, $\Gamma_{p}\tau=0.1$, $\chi_{1}=1$ and (a) $\chi_{2}=2.048 \times 10^{-7}$, 
 (b)  $\chi_{2}=1.024 \times 10^{-7}$.}
 \label{fig:decoherence_phase}
 \end{figure}

Dissipation through phase damping is modelled using the master equation~\cite{agarwalmaster},
\begin{equation}
\frac{\rmd \overline{\varrho}}{\rmd \tau} = - \Gamma_{p} (\mathcal{N}^{2} \overline{\varrho} - 2 \mathcal{N} \overline{\varrho} \mathcal{N} + \overline{\varrho} \mathcal{N}^{2}),
\label{eqn:master_eqn_phase}
\end{equation}
where $\overline{\varrho}$ is the density matrix in the phase damping model and $\Gamma_{p}$ is the rate of decoherence. The solution to this master equation also can be  expressed in the form given in Eq. \eref{eqn:soln_master}, with matrix elements~\cite{agarwalmaster}
\begin{align}
\overline{\varrho}_{n,n'}(\tau)=e^{-\Gamma_{p}\tau(n-n')^{2}} \overline{\varrho}_{n,n'}(\tau=0).
\label{eqn:rho_soln_master_phase}
\end{align}
As $\Gamma_{p}\tau\rightarrow\infty$, it is evident that the off-diagonal terms of the density matrix vanish while the diagonal terms of $\overline{\varrho}_{n,n'}(\tau=0)$ remain unchanged due to  phase damping. In contrast to amplitude damping, the state does not go back to a pure state as $\Gamma_{p}\tau\rightarrow\infty$ but remains a mixed state. Further, the differences between tomograms with different strand structures (consequent  to different ratios of $\chi_{1}/\chi_{2}$) disappear faster than in the case of amplitude damping. Figures \ref{fig:decoherence_phase} (a) and (b) show the effect of phase damping at $\Gamma_{p}\tau=0.1$ for two such ratios. However, for $\Gamma_{p}\tau =1$  the differences are barely visible.

\subsection{Squeezing and higher-order squeezing properties}
We now proceed to examine the squeezing and higher-order squeezing properties of the state of the system with Hamiltonian $H$. Once again, our aim is to identify and assess  these nonclassical effects directly from the tomogram without attempting to reconstruct the state at any instant of time. As mentioned in Chapter \ref{ch:intro}, we will quantify the extent of quadrature, Hong-Mandel and Hillery-type higher-order squeezing.
The extent of Hong-Mandel squeezing is obtained by calculating the central moments of the probability distribution corresponding to the appropriate  quadrature. Thus for instance, to quantify the extent of second-order Hong-Mandel squeezing in the $x$-quadrature, we calculate the fourth central moment of a horizontal cut of the tomogram
 at $\theta=0$. (For $q=1$, Hong-Mandel squeezing is identical to quadrature squeezing). 

  We recall that Hillery-type squeezing of order $q$ corresponds  to squeezing in either $Z_{1} =( a^{q} + {a^{\dagger}}^{q})/ \sqrt{2}$ or $Z_{2} =( a^{q} - {a^{\dagger}}^{q})/ (\sqrt{2} \, i)$ ($q = 2,3,\dots$).  A useful quantifier $D_{q}$ of $q$th-power squeezing in $Z_{1}$ for instance, is defined  \cite{Hillery} in terms of the commutator 
 $[Z_{1}, Z_{2}] = F_{q}(\mathcal{N})$  as 
\begin{equation}
D_{q}= \frac{2 \aver{(\Delta Z_{1})^{2}} - |\aver{F_{q}(\mathcal{N})}|}{|\aver{F_{q}(\mathcal{N})}|}.
\label{eqn:squeeze_param}
\end{equation}
where $\aver{(\Delta Z_{1})^{2}}$ is the variance in $Z_{1}$. A similar definition holds for $q$th-power squeezing in $Z_{2}$.  
We note that $F_{q}(\mathcal{N})$ is a polynomial function of  order $(q - 1)$ in $\mathcal{N}$. A state is $q$th-power squeezed if $-1 \leq D_{q} < 0$.
 It is clear that  $Z_1$ and $Z_2$ cannot be obtained in a straightforward manner from the tomogram as they involve terms with products of powers of different rotated  quadrature operators and hence cannot be assigned probability distributions directly from a set of tomograms.

 However an illustrative treatment~\cite{wunsche} of the problem of expressing the expectation value of a product of moments of creation and destruction operators in terms of the tomogram $w$ and Hermite polynomials leads to the result (Appendix \ref{appen:NormOrdMom})
\begin{align}
	\label{eqn:wunsche}
\nonumber \braket{a^{\dagger k} a^l} &= C_{k l} \sum_{m=0}^{k+l} \exp \left(-i(k-l)\left(\frac{m\pi}{k+l+1}\right)\right) \\
& \int_{-\infty}^{\infty} \rmd X_\theta  \ w\left(X_\theta, \ \frac{m\pi}{k+l+1}\right) H_{k+l}\left( X_{\theta}  \right) ,
\end{align}
where
\begin{equation}
\nonumber C_{k l}=\frac{k! l!}{(k+l+1)! \sqrt{2^{k+l}}}.
\end{equation}

This form is useful for computing $D_{q}$ numerically.
We therefore need to consider $(k+l+1)$ values of $\theta$  in order to calculate a moment of order $(k+l)$.  In a single tomogram, this amounts to using  $(k+l+1)$  probability distributions 
$w(X_{\theta}$) corresponding to  these chosen values of $\theta$,  in order to compute $D_{q}$.  As the system evolves in time, the extent of squeezing at various instants is determined from the instantaneous tomograms in this manner.

\begin{figure}
	\centering    
	\includegraphics[width=0.48\textwidth]{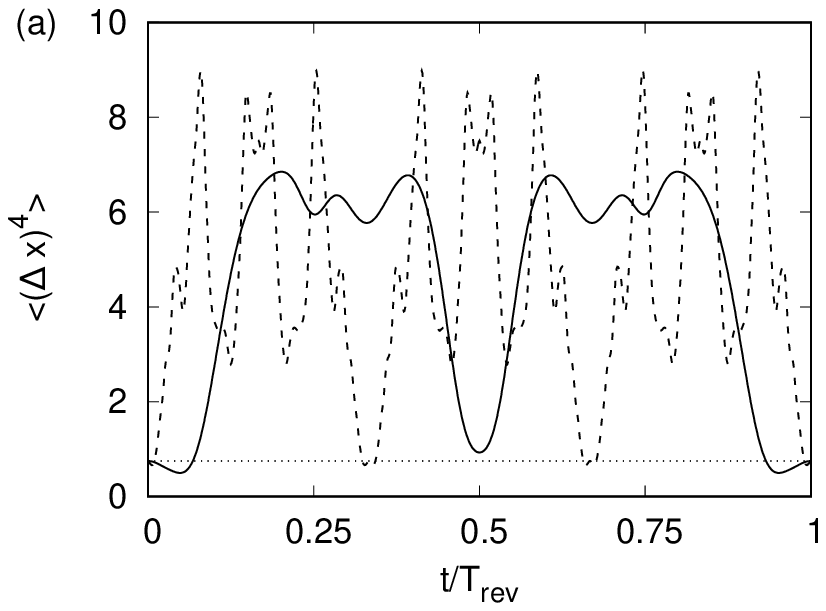}
	\includegraphics[width=0.48\textwidth]{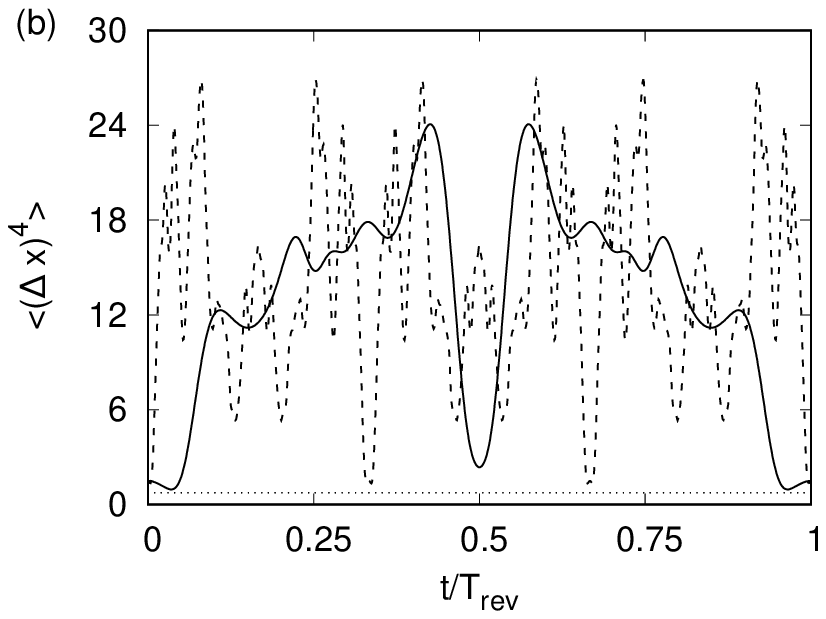}
	\caption{Hong-Mandel squeezing  as a function of scaled time $t/T_{rev}$ for initial  states (a)$\ket{\alpha}$ and (b)$\ket{\alpha,1}$ for $\alpha = 1$, corresponding to the Kerr (solid) and cubic (dashed) Hamiltonians. The horizontal line at $0.75$ denotes the value below which the state is squeezed.}
	\label{fig:squeezingcspacshongmandel}
\end{figure}

For the squeezed vacuum, $\ket{\alpha}$ and $\ket{\alpha,1}$  we have verified that the variance and hence the Hong-Mandel (equivalently Hillery-type squeezing) properties inferred directly from tomograms are in excellent agreement with  corresponding results obtained analytically from the state. 
We have also computed $\aver{(\Delta x)^{4}}$ (equivalently the  higher-order Hong-Mandel squeezing parameter)  directly from the tomogram for initial states $\ket{\alpha}$ and $\ket{\alpha,1}$ evolving under the Kerr Hamiltonian and the  cubic Hamiltonian (Figs.~\ref{fig:squeezingcspacshongmandel} (a), (b)). Without loss of generality, we have set $\alpha = 1$. 
From these figures it is evident that independent of the precise nature of the initial field state, $\aver{(\Delta x)^{4}}$ oscillates more rapidly in the case of the cubic Hamiltonian compared to the Kerr Hamiltonian. Earlier these squeezing properties were investigated  for an initial CS and $1$-PACS evolving under a Kerr Hamiltonian alone~\cite{sudhsqueezing} by calculating $\aver{(\Delta x)^{4}}$ explicitly for the state at different instants of time. Our results from the corresponding tomograms are in excellent agreement with these.
 
\begin{figure*}
	\centering
	\includegraphics[width=0.3\textwidth]{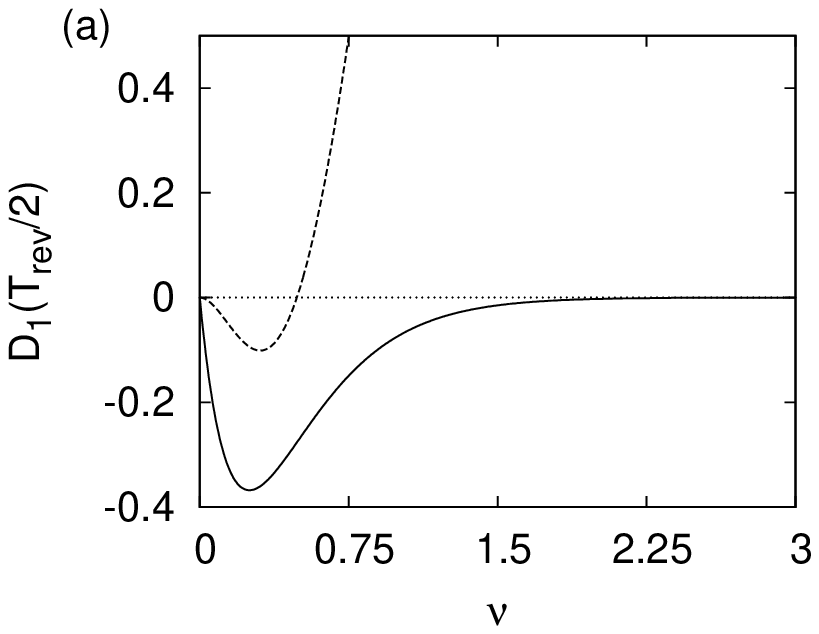}
	\includegraphics[width=0.3\textwidth]{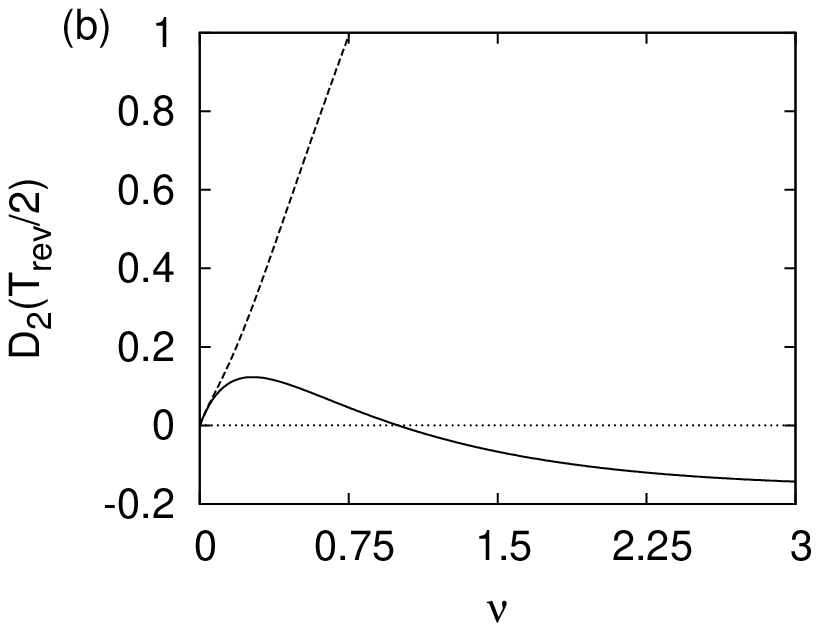}
	\includegraphics[width=0.3\textwidth]{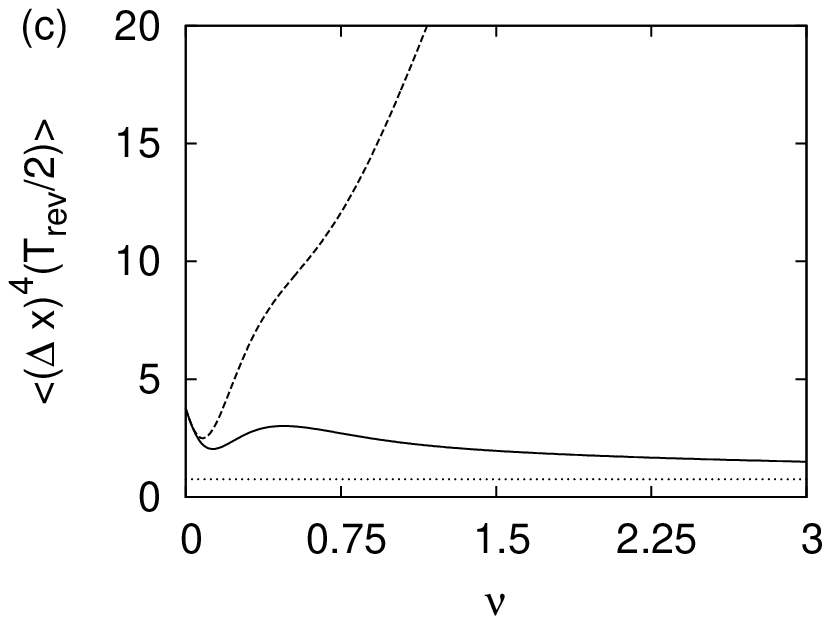}
	\caption{ (a) $D_{1}$ for initial $\ket{\alpha}$, (b) $D_{2}$ for initial $\ket{\alpha,1}$, and (c) $\langle (\Delta x)^{4} \rangle$ for initial $\ket{\alpha,1}$ vs. $\nu$ at instant $T_{rev}/2$, corresponding to the Kerr (solid) and cubic (dashed) Hamiltonians. The horizontal line at $0$ in (a) and (b) ($3/4$ in (c)) denotes the value below which the state is squeezed.}
	\label{fig:squeezingcspacshillery}
\end{figure*}

Figure \ref{fig:squeezingcspacshillery} (a) is a plot of $D_{1}$ versus $\nu = \alpha^{2}$ ($\alpha$ real) at $T_{rev}/2$ for an initial CS evolving under both the Kerr and cubic Hamiltonians for $0\leq \alpha < \sqrt{3}$. It is evident that the state is squeezed for a larger range of  values of $\nu$, and that the numerical value of $D_{1}$ is larger for a given $\nu$ in the case of the Kerr Hamiltonian as compared with the cubic Hamiltonian.

For the same parameter values, we have also considered an initial state $\ket{\alpha,1}$ and obtained $D_{2}$ from the tomogram at $T_{rev}/2$. While $D_{2}$ is not negative for any $\nu$ in this range for the cubic Hamiltonian, it becomes negative for $\nu \geq 0.8$ approximately for the Kerr Hamiltonian (Fig. \ref{fig:squeezingcspacshillery} (b)). In contrast $\aver{(\Delta x)^{4}}$  is never less than $3/4$  in both cases over this range of values of $\nu$ (Fig. \ref{fig:squeezingcspacshillery} (c)).

\section{The double-well BEC: A tomographic approach \label{sec:Ch2Sec3}}
 \subsection{The model}
We now proceed to examine nonclassical effects in a BEC condensed in a double-well potential. The effective Hamiltonian of this bipartite system, setting $\hbar=1$, is ~\cite{sanz}
\begin{equation}
H_{\textsc{bec}} =\omega_{0} N_{\text{tot}} + \omega_{1} (a^{\dagger}a - b^{\dagger}b)
+ U N_{\text{tot}}^{2} - \lambda (a^{\dagger} b + a b^{\dagger}),
\label{eqn:HBEC}
\end{equation}
where $\omega_{0}$, $\omega_{1}$ and $\lambda$ are constants with appropriate dimensions.
In this case, $(a, a^{\dagger})$ and $(b, b^{\dagger})$ are the boson annihilation and creation operators corresponding to the two subsystems A and B, which comprise the atomic species condensed in the two wells.
Here, $[a,a^{\dagger}] = 1$, $[b,b^{\dagger}] = 1$ and all other commutators are equal to zero. $N_{\text{tot}}=(a^{\dagger}a+b^{\dagger}b)$ and $U$ is the strength of the nonlinearity. 
It is easy to see that  $[H_{\textsc{bec}},N_{\text{tot}}]=0$. 

We denote by $\ket{\alpha_{a}}$ (respectively $\ket{\alpha_{b}}$) the CS  formed from the condensate corresponding to subsystem A [resp. B] 
and by $\ket{\alpha_{a},1}$ [$\ket{\alpha_{b},1}$] a 1-boson-added CS  corresponding to subsystem A [B]. 
The initial states considered  by us are  factored product states  of the form 
  $\ket{\alpha_{a}} \otimes \ket{\alpha_{b}}$ (denoted by $\ket{\Psi_{00}}$), 
  $\ket{\alpha_{a},1} \otimes \ket{\alpha_{b},1}$( denoted by $\ket{\Psi_{11}}$) and
 $ \ket{\alpha_{a},1} \otimes \ket{\alpha_{b}}$ (denoted by $\ket{\Psi_{10}}$).

In general, the state at a subsequent time $t$ is  entangled. It has been shown~\cite{sanz} that corresponding to the initial state $\ket{\Psi_{00}}$ we have
\begin{align}
\nonumber \ket{\Psi_{00} (t)} &= \exp (-({|\alpha_{a}|}^{2} + {|\alpha_{b}|}^{2})/2) \sum_{p,q=0}^{\infty} \frac{(\alpha(t))^{p} (\beta(t))^{q}}{\sqrt{p! q!}}\\
& \exp (- i t (\omega_{0} (p+q)+ U (p + q)^{2})) \ket{p \delim q}.
\label{eqn:Psi_t}
\end{align}
We recall that $\ket{p \delim q}$ denotes $\ket{p}\otimes \ket{q}$ where $\lbrace \ket{p} \rbrace$, $\lbrace \ket{q}\rbrace$ are the boson number basis states of subsystems A and B respectively, and, 
\begin{equation}
\alpha (t) = \alpha_{a} \cos (\lambda_{1} t) + i \frac{\sin (\lambda_{1} t)}{\lambda_{1}} \left( \lambda \alpha_{b} - \omega_{1} \alpha_{a} \right),\\
\label{eqn:alpha_t}
\end{equation}
\begin{equation}
\beta (t) = \alpha_{b} \cos (\lambda_{1} t) + i \frac{\sin (\lambda_{1} t)}{\lambda_{1}} \left( \lambda \alpha_{a} + \omega_{1} \alpha_{b} \right),
\label{eqn:beta_t}
\end{equation}
and $\lambda_{1}=\sqrt{\omega_{1}^{2}+\lambda^{2}}$. 

We have used a similar procedure (outlined in Appendix \ref{appen:DensMat}) to obtain the states $\ket{\Psi_{10}(t)}$, $\ket{\Psi_{11}(t)}$ and the expression for the density matrix corresponding to the factored product of generic boson-added coherent states $\ket{\alpha_{a}, m_{1}} \otimes \ket{\alpha_{b}, m_{2}}$ ($m_{1}$,$m_{2}$: positive integers). We denote this density matrix by $\rho_{m_{1},m_{2}}(t)$.
We give the final expressions below.
 We have
\begin{align}
\nonumber \ket{\Psi_{10} (t)} = \frac{1}{d_{10}}  &\left(a^{\dagger} \lambda_{1} \cos (\lambda_{1} t) +  i (\lambda b^{\dagger} - \omega_{1} a^{\dagger}) \sin (\lambda_{1} t) \right)\\
& \times \exp(- i U (2 N_{\text{tot}} + 1) t) \ket{\Psi_{00} (t)},
\label{eqn:Psi_10_t}
\end{align}
and
\begin{align}
\nonumber \ket{\Psi_{11} (t)} = &\frac{1}{d_{11}} \biggl( 2 \omega_{1}^{2} \, a^{\dagger} b^{\dagger} +\omega_{1} \lambda \left( a^{\dagger 2}- b^{\dagger 2} \right) \\
\nonumber &+ \cos (2 \lambda_{1} t) \left(2 \lambda^{2} \, a^{\dagger} b^{\dagger} - \omega_{1} \lambda \left( a^{\dagger 2}-b^{\dagger 2} \right) \right)\\
\nonumber &+ i \sin (2 \lambda_{1} t) \lambda \lambda_{1} (a^{\dagger 2} +  b^{\dagger 2}) \biggr)\\
& \times \exp (- 4 i U ( N_{\text{tot}} + 1) t) \ket{\Psi_{00} (t)}.
\label{eqn:Psi_11_t}
\end{align}
Here $d_{10}=  \lambda_{1} \exp (i \omega_{0} t) \sqrt{1+{|\alpha_{a}|}^{2}}$ and $d_{11} = 2 \lambda_{1}^{2} \exp (2 i \omega_{0} t) \sqrt{1+ {|\alpha_{a}|}^{2}}\sqrt{1+{|\alpha_{b}|}^{2}}$. 

To obtain an expression for $\rho_{m_{1},m_{2}}(t)$, we first define $p_{max}=(k+m_{2}-l)$ and $q_{max}=(l+m_{1}-k)$. Then,
\begin{align}
\nonumber &M_{m_{1},m_{2}}(t)= \frac{1}{\mu} \biggl[\sum_{k=0}^{m_{1}}\sum_{l=0}^{m_{2}}\sum_{p=0}^{p_{max}}\sum_{q=0}^{q_{max}}(-1)^{k-p} {m_{1} \choose k} {m_{2} \choose l} \\
\nonumber &\hspace{2 em} {p_{max} \choose p} {q_{max} \choose q} \exp(-i\lambda_{1} t (2(k-l)+m_{2}-m_{1}))\\ 
\nonumber &\hspace{2 em} (\cos(\kappa/2))^{(k+l+p+q)} (\sin(\kappa/2))^{(2(m_{1}+m_{2})-(k+l+p+q))}\\
\nonumber &\hspace{2 em} a^{\dagger (p+q_{max}-q)} b^{\dagger (q+p_{max}-p)} \biggr] \exp(-i\omega_{0} t(m_{1}+m_{2}))\\
& \times \exp(-i U t (m_{1}+m_{2}) (2 N_{\text{tot}} + m_{1}+m_{2})).
\label{eqn:intermed_rho_numerics}
\end{align}
Here $\mu=\sqrt{m_{1}! L_{m_{1}}(-{|\alpha_{a}|}^{2}) m_{2}! L_{m_{2}}(-{|\alpha_{b}|}^{2})}$ and  $\kappa=\tan^{-1}(\lambda/\omega_{1})$. $L_{m}(-|\alpha|^{2})$ are the Laguerre polynomials which appear in the normalisation of an $m$-boson-added CS $\ket{\alpha,m}$.
 Then,
\begin{equation}
\rho_{m_{1},m_{2}} (t) = M_{m_{1},m_{2}}(t) \ket{\Psi_{00}(t)}\bra{\Psi_{00}(t)} M^{\dagger}_{m_{1},m_{2}}(t).
\label{rho_numerics}
\end{equation}
This expression for the general density matrix can be easily seen to reduce to the forms needed in our case where the states are $\ket{\Psi_{00}(t)}$, $\ket{\Psi_{11}(t)}$ and  $\ket{\Psi_{10}(t)}$.

 \subsection{The Revival Phenomena}
A straightforward calculation reveals that full and fractional revivals occur provided $\omega_{0}=m U$, $\lambda_{1}=m' U$, $m,m' \in \mathbb{Z}$, and $(m+m')$ is odd. Here the revival time $T_{rev}=\pi /U$.  This follows from the periodicity property of $\exp(-i U N_{\text{tot}}^{2})$. For instance, we can easily show that for an initial state $\ket{\Psi_{00}}$, the state at the instant $\pi/(s U)$  ($s$: even integer) is
\begin{align}
\nonumber \ket{\Psi_{00} (\pi / s U)} = &\sum_{j=0}^{s-1}  f_{j} \ket{\alpha(\pi/s U) e^{- i \pi(m + 2 j)/s}} \\
& \otimes \ket{\beta(\pi/s U) e^{- i \pi (m + 2 j)/s}}.
\label{eqn:frac_rev_even}
\end{align}
If $s$ is an odd integer,
\begin{align}
\nonumber \ket{\Psi_{00} (\pi/s U)} = &\sum_{j=0}^{s-1} g_{j} \ket{\alpha(\pi/s U) e^{- i \pi(m + 2 j + 1)/s}} \\
& \otimes \ket{\beta(\pi/s U) e^{- i \pi(m + 2 j + 1) / s}}.
\label{eqn:frac_rev_odd}
\end{align}
We note that $\alpha\left( \pi/(s U) \right)$ and $\beta\left( \pi/(s U) \right)$ are obtained from Eqs. \eref{eqn:alpha_t} and  \eref{eqn:beta_t}.

 We recall that in the single-mode system the occurrence of full and fractional revivals are related to the presence of distinct strands in the tomogram. In this bipartite system however, the tomogram is a 4-dimensional hypersurface. Hence we need to consider appropriate sections to identify and examine nonclassical effects. The 2-dimensional section ($X_{\thetabt}$-$X_{\thetaat}$) obtained by choosing a specific value for $\thetaa$ and $\thetab$, is a natural choice for investigating  not only the revival phenomena but also the state's squeezing properties. In contrast to the single-mode case where strands appear in the tomograms, these sections are characterised by  `blobs'  at instants of fractional revivals. The number of blobs gives the number of subpackets in the wave packet. 
 
\begin{figure}
\centering
\includegraphics[scale=0.7]{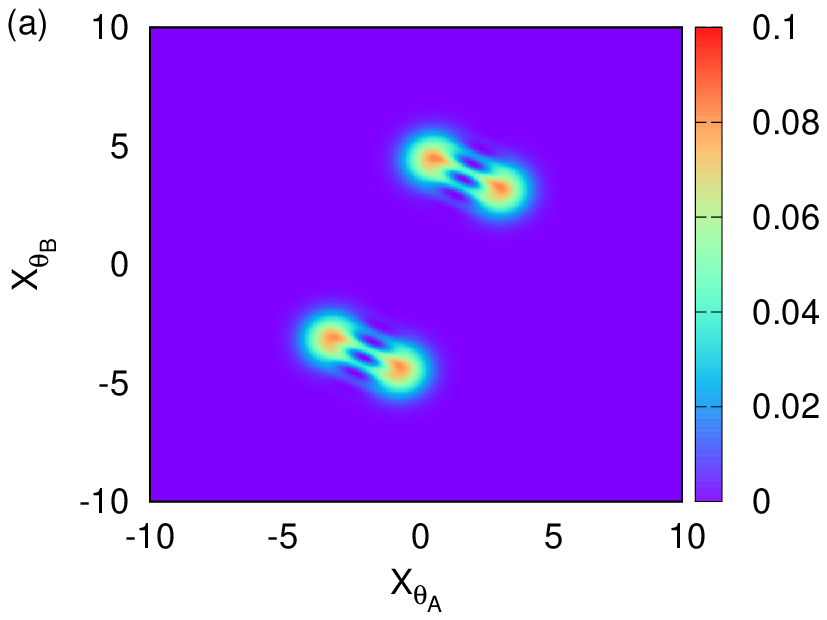}
\includegraphics[scale=0.7]{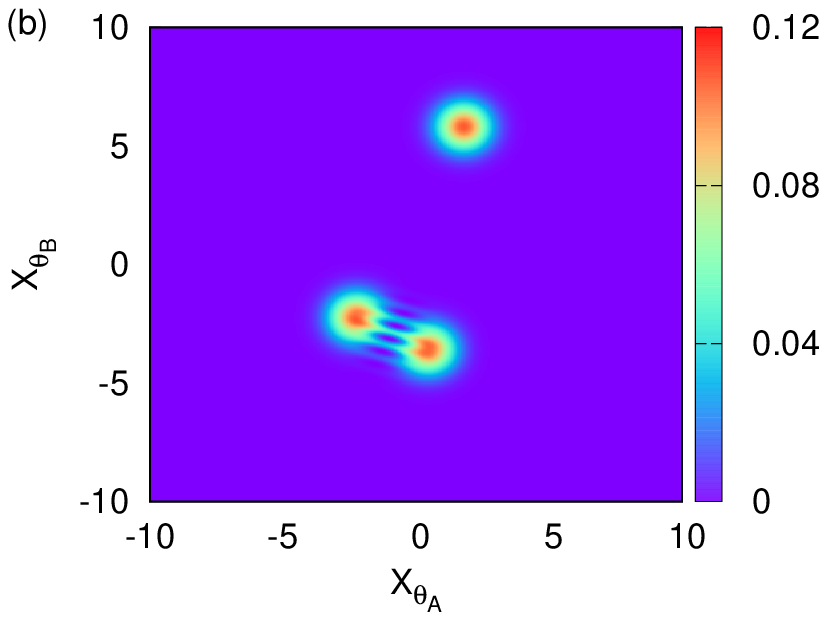}
\includegraphics[scale=0.7]{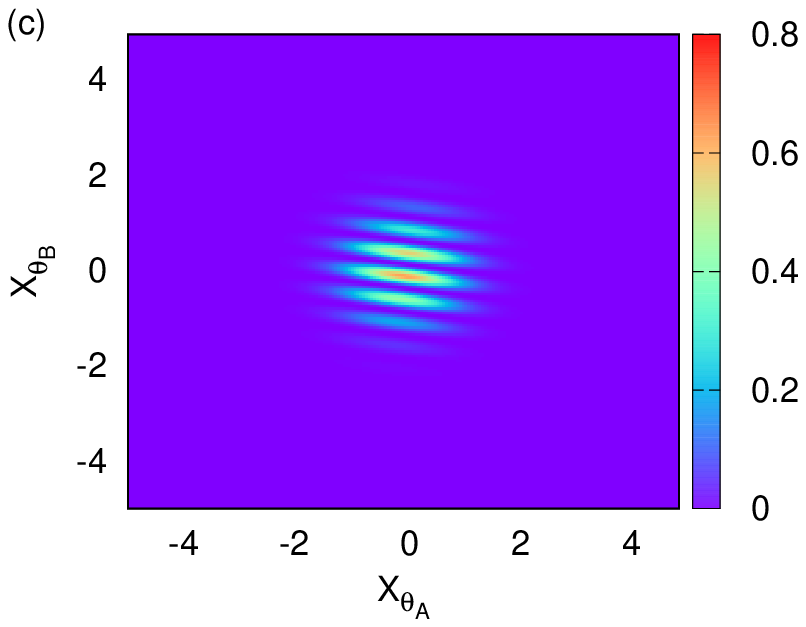}
\includegraphics[scale=0.7]{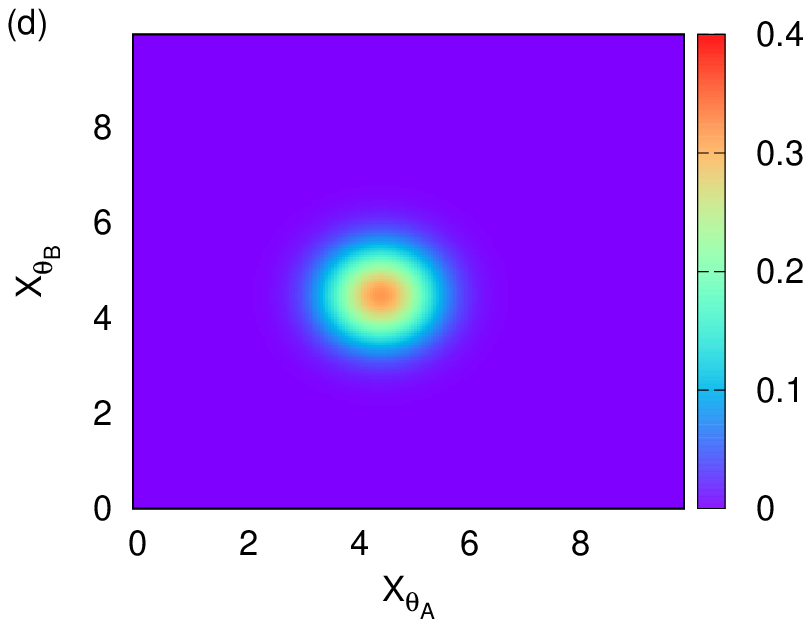}
\caption{ Sections of the optical tomogram for $\thetaa=\thetab=0$ at instants (a) $T_{rev}/4$, (b)$T_{rev}/3$, (c) $T_{rev}/2$, and (d) $T_{rev}$. $\alpha_{a}=\alpha_{b}= \sqrt{10}$, and initial state $\ket{\Psi_{00}}$.}
\label{fig:2_mode_frac_rev}
\end{figure}

 Although $\ket{\alpha}$ is expanded as an infinite superposition of boson number states, in practice numerical computations can  be carried  out using only a large but finite sum of these basis states. While this is the procedure that we have used, an alternative is to use truncated coherent state(TCS)  \cite{tcs} instead of the standard CS. The latter are  defined as 
\begin{equation}
{\ket{\alpha}}_{tcs} = \frac{1}{\left(\sum_{n=0}^{N_{max}}( |\alpha|^{2 n}/n!)\right)^{1/2}} \sum_{p=0}^{N_{max}} \frac{\alpha^{p}}{\sqrt{p!}} \ket{p}.
\label{eqn:TCS}
\end{equation}
 where $N_{max}$ is a sufficiently large but finite integer. We have verified that 
the revival phenomena and the squeezing properties as observed from appropriate tomograms in these two approaches agree remarkably well. We have set $\thetaa = \thetab = 0$, $\omega_{0}=10$, $\omega_{1}=3$, $\lambda=4$, $U=1$, and $\alpha_{a} = \alpha_{b} = \sqrt{10}$ in our numerical computation pertaining to revivals and fractional revivals. We note that with this choice of values, the necessary conditions for the revival phenomena to occur are satisfied, namely, $\omega_{0}=m U$, $\lambda_{1}=m' U$, $m,m' \in \mathbb{Z}$, and $(m+m')$ is odd. Figures \ref{fig:2_mode_frac_rev}  (a)-(d) are tomograms corresponding to different fractional revivals for the initial state $\ket{\Psi_{00}}$.  At instants $T_{rev}/4$,  $T_{rev}/3$ and $T_{rev}$  (Figs. \ref{fig:2_mode_frac_rev} (a), (b) and (d) respectively), there are 4, 3 and a single blob respectively in the corresponding tomograms along with interference patterns. 
However in Fig. \ref{fig:2_mode_frac_rev} (c)  corresponding to the instant $T_{rev}/2$  blobs are absent  and we  merely see interference patterns.
 This is primarily due to the specific choice of  {\it  real} values of  $\alpha_{a}$ and $\alpha_{b}$  as explained below. At the instant $T_{rev}/2$ it follows from Eq. \eref{eqn:Psi_t} that the state of the system can be expanded in terms of superpositions of factored products of CS  corresponding to A and B as 
\begin{align}
\nonumber \ket{\Psi_{00}(T_{rev}/2)}= &\frac{(1-i)}{2} \ket{{- 2 i}/{\sqrt{10}}} \otimes \ket{{- 14 i}/{\sqrt{10}}} \\
&+ \frac{(1+i)}{2} \ket{{2 i}/{\sqrt{10}}} \otimes \ket{{14 i}/{\sqrt{10}}}.
\label{eqn:Psi_half_rev_sp_param}
\end{align}
It is  now straightforward to see why the interference patterns alone appear in the tomogram (Fig. \ref{fig:2_mode_frac_rev} (c)), as a simple calculation gives 
\begin{align}
\nonumber |\Psi_{00}(X_{\textsc{a}0},X_{\textsc{b}0})|^{2} = |\langle X_{\textsc{a}0},0;X_{\textsc{b}0},0 | \Psi_{00}(T_{rev}/2)\rangle|^{2}\\
=\frac{1}{\pi} e^{-(X_{\textsc{a}0}^{2} + X_{\textsc{b}0}^{2})} \left(1-\sin(4 (X_{\textsc{a}0} + 7 X_{\textsc{b}0})/ \sqrt{5} )\right). 
\label{eqn:2_mode_Trevby2_interference}
\end{align}
Here  $X_{\textsc{a}0}$ denotes $X_{\thetaa}$ for $\thetaa=0$ and $X_{\textsc{b}0}$  denotes $X_{\thetab}$ for $\thetab=0$.
In contrast, it can be seen that if $\alpha_{a}$ and $\alpha_{b}$  are generic complex numbers, two blobs together with the interference pattern would appear at $T_{rev}/2$ also.   
Similar results hold for initial states $\ket{\Psi_{11}}$ and $\ket{\Psi_{10}}$.
  
\subsection{Squeezing and higher-order squeezing properties of the condensate}

\begin{figure}
\centering
\includegraphics[width=0.49\textwidth]{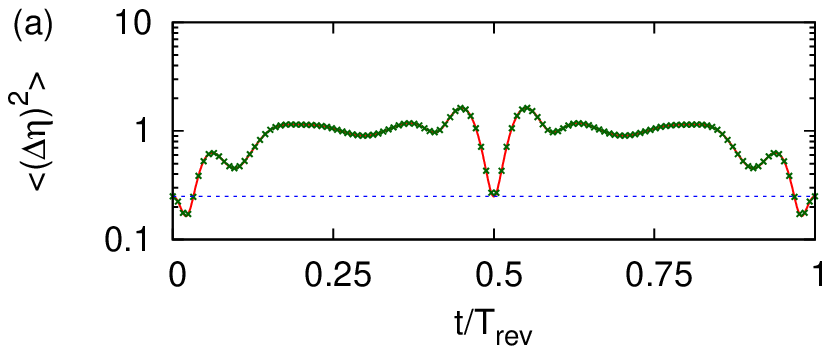}
\includegraphics[width=0.49\textwidth]{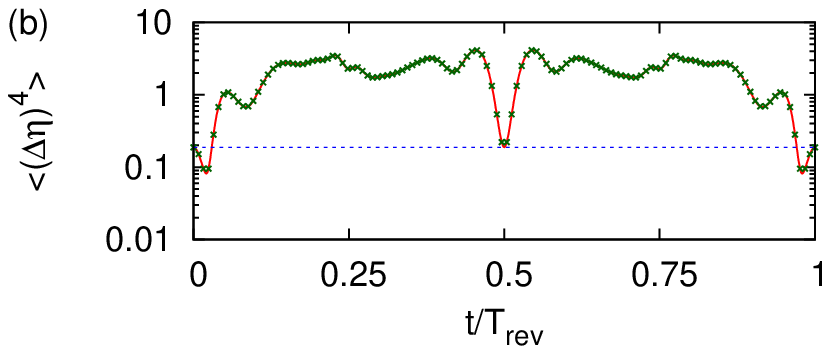}
\includegraphics[width=0.49\textwidth]{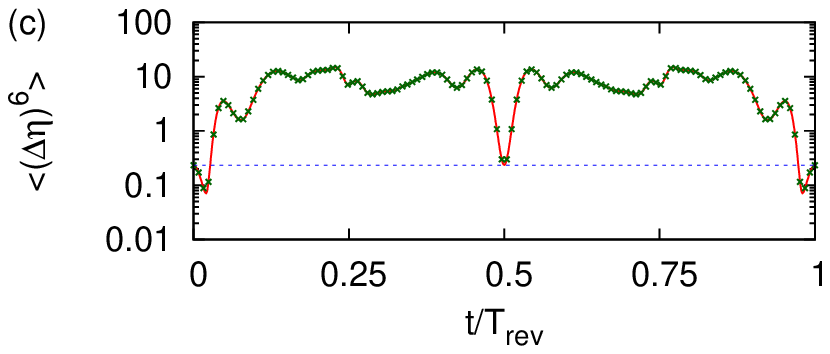}
\includegraphics[width=0.49\textwidth]{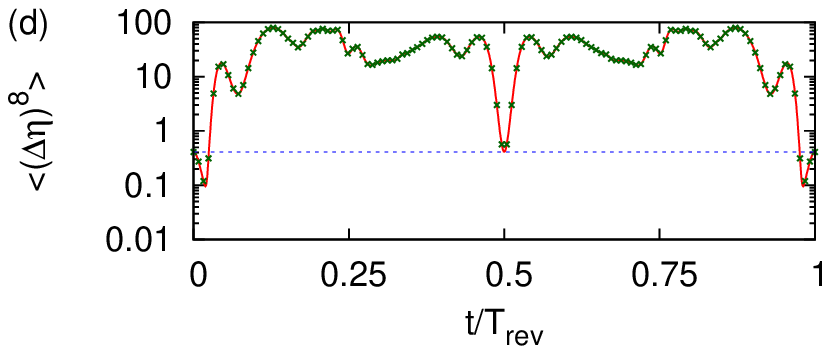}
\caption{ $2 q$-order moment of $\eta=(a+a^{\dagger}+b+b^{\dagger})/2\sqrt{2}$ 
vs. $t/T_{rev}$ for initial state $\ket{\Psi_{00}}$, $|\alpha_{a}|= |\alpha_{b}|=1$, and $q=$ (a) $1$, (b) $2$, (c) $3$, and (d) $4$. Moments calculated from the tomogram (green) and moments from unitarily evolved state (red). The horizontal line denotes the value below which $q$-order squeezing occurs ($1/4$, $3/16$, $15/64$, and $105/256$ for $q=1,2,3,4$ respectively).}\label{fig:2_mode_hong_time_dep}
\end{figure}

The extent of Hong-Mandel squeezing is obtained  as in the single-mode example,  by calculating the central moments of the probability distribution corresponding to a given quadrature. We examine two-mode squeezing  by evaluating appropriate moments of the quadrature variable $\eta=(a+a^{\dagger}+b+b^{\dagger})/(2\sqrt{2})$.  These are obtained from the $\thetaa=\thetab=0$ section of the tomogram as the system evolves in time. We have also obtained these moments by explicit calculation of the relevant expectation values of $\eta$ in the state of the system at different instants of time. In both cases the initial state is $\ket{\Psi_{00}}$.
In Figs.~\ref{fig:2_mode_hong_time_dep} (a)-(d), $2 q$-order moments for $q = 1,2,3$ and $4$ obtained  both from the states and directly from the tomogram are plotted as functions of time. It is evident that they are in excellent agreement with each other at all instants. The horizontal line in each figure denotes the value below which the state is squeezed ($1/4$, $3/16$, $15/64$, and $105/256$ for $q=1,2,3,4$ respectively). For all values of $q$ considered,  the state is squeezed (higher-order squeezed)  in the neighborhood of revivals, and the extent of squeezing at various instants is considerably sensitive to the values of $q$, $\alpha_{a}$ and $\alpha_{b}$ as expected. 

Turning to Hillery-type higher-order squeezing, we recall that in the single-mode case we used Eq. \eref{eqn:wunsche} for this calculation.
A straightforward extension to the two-mode system gives us the expression
\begin{align}
\nonumber \langle a^{\dagger k} a^{l} &b^{\dagger m} b^{n} \rangle =  c_{k l m n} \sum_{p=0}^{k+l} \sum_{q=0}^{m+n} \exp \left(-i (k-l)\theta_{\textsc{a} p} \right)\\
\nonumber & \exp\left( - i (m-n)\theta_{\textsc{b} q}\right) \int_{-\infty}^{+\infty} \rmd X_{\theta_{\textsc{a} p}} \int_{-\infty}^{+\infty} \rmd X_{\theta_{\textsc{b} q}} \\
&\hspace*{-1.5 em}w\left(X_{\theta_{\textsc{a} p}}, \theta_{\textsc{a} p}; X_{\theta_{\textsc{b} q}}, \theta_{\textsc{b} q}\right) H_{k+l}(X_{\theta_{\textsc{a} p}}) H_{m+n}(X_{\theta_{\textsc{b} q}}).
\label{eqn:2_mode_Wunsche}
\end{align}
Here
$$c_{k l m n}=\frac{k! l! m! n!}{(k+l+1)! (m+n+1)! \sqrt{2^{k+l+m+n}}},$$ $\theta_{\textsc{a} p}= \frac{p \pi}{k+l+1}$, 
and $\theta_{\textsc{b} q} = \frac{q \pi}{m+n+1}$. We note that $(k+l+1)(m+n+1)$ gives the number of 2- dimensional slices of the tomogram that are required to calculate $\langle a^{\dagger k} a^{l} b^{\dagger m} b^{n} \rangle$.

\begin{figure}
\centering
\includegraphics[width=0.49\textwidth]{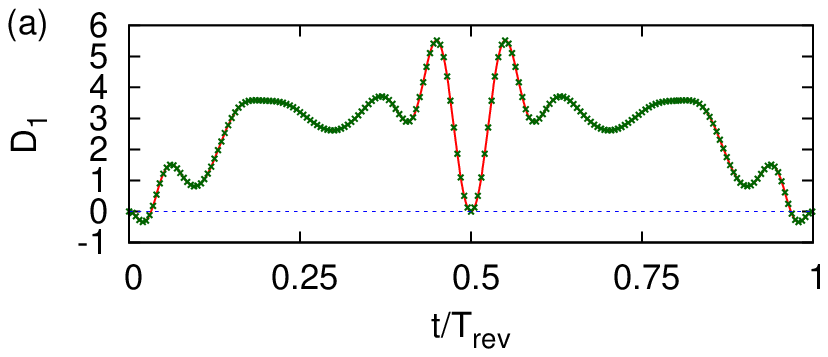}
\includegraphics[width=0.49\textwidth]{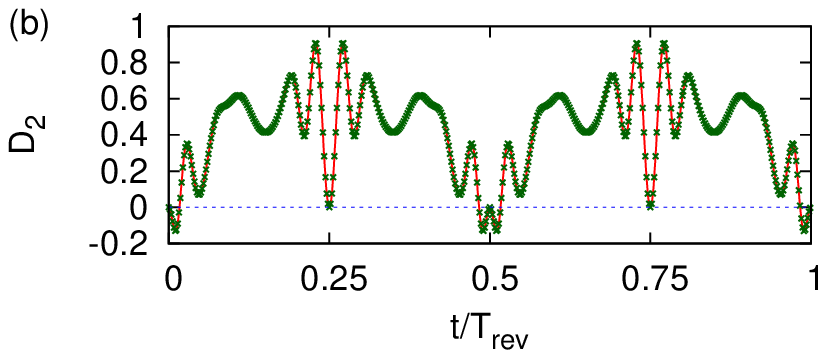}
\includegraphics[width=0.49\textwidth]{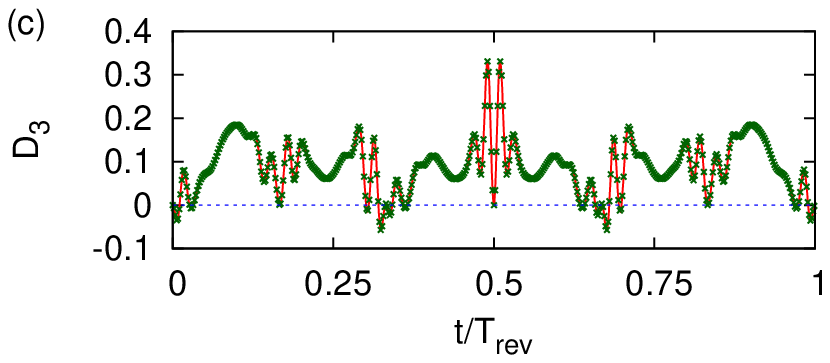}
\includegraphics[width=0.49\textwidth]{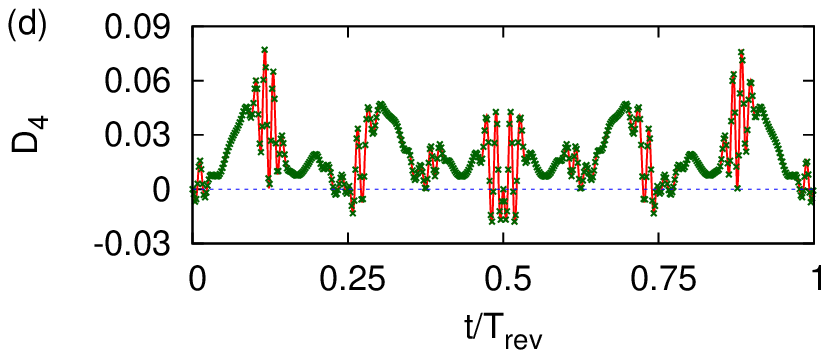}
\caption{ $D_{q}(t)$ vs. $t/T_{rev}$ for initial state $\ket{\Psi_{00}}$, $|\alpha_{a}|=|\alpha_{b}|=1$, and $q=$ (a) $1$, (b) $2$, 
(c) $3$, and (d) $4$. Squeezing parameter from the tomogram (green) and from unitarily evolved state (red). The horizontal line corresponds to $D_{q}(t) = 0$.}
\label{fig:2_mode_hillery_time_dep}
\end{figure}

From Figs.~\ref{fig:2_mode_hillery_time_dep} (a)-(d), we see that $D_{q}(t)$ obtained both from the states  and directly from the tomogram  for an initial state $\ket{\Psi_{00}}$ are in excellent agreement. (The expression in Eq. \eref{eqn:squeeze_param} holds in this case also, with $Z_{1} =( a^{q} + {a^{\dagger}}^{q} + b^{q} + {b^{\dagger}}^{q})/ (2 \sqrt{2})$ and $Z_{2} =( a^{q} - {a^{\dagger}}^{q} + b^{q} - {b^{\dagger}}^{q})/ (2 \sqrt{2} \, i)$). It is clear that for higher values of $q$ there are more instants of time when higher-order squeezing occurs as expected from the fact that more cross-terms involving the creation and destruction operators arise with increase in $q$. 
We have also verified that  both the Hong-Mandel and Hillery-type squeezing and higher-order squeezing parameters obtained from tomograms and from expectation values of appropriate operators for the initial states $\ket{\Psi_{10}}$ and $\ket{\Psi_{11}}$ are equal to each other at all instants between $t =0$ and $T_{rev}$.  

\subsection{Subsystem entropies from tomograms}

We now proceed to examine the subsystem's entropic squeezing properties by computing the subsystem information entropy $S(\thetaa)$ at various instants of time.  (Here, without loss of generality we have considered subsystem A by setting $\thetab = 0$, and integrating over the full range of $X_{\thetabt}$ to obtain $w_{\textsc{a}}(X_{\thetaat}, \thetaa)$). Expressed in terms of the subsystem tomogram, 
\begin{equation}
S(\thetaa)=-\int_{-\infty}^{\infty} \text{d}X_{\thetaat} \; w_{\textsc{a}}(X_{\thetaat}, \thetaa) \; \ln \; w_{\textsc{a}}(X_{\thetaat}, \thetaa).
\label{eqn:subsysEntropy_A}
\end{equation}

\begin{figure}
\centering
\includegraphics[width=0.49\textwidth]{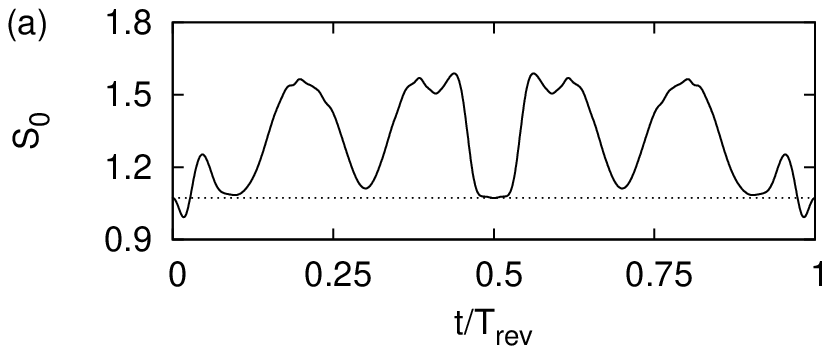}
\includegraphics[width=0.49\textwidth]{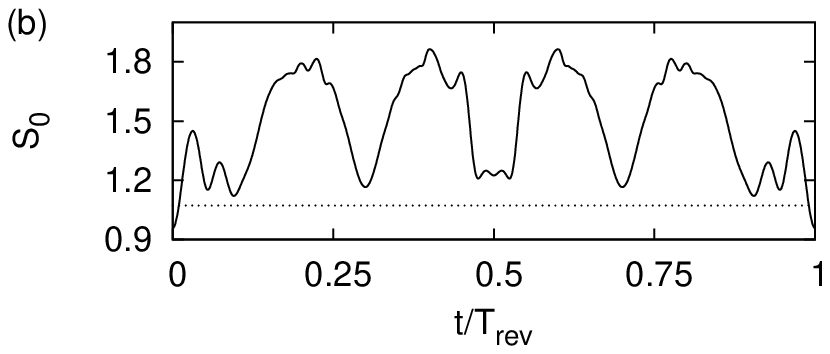}
\includegraphics[width=0.49\textwidth]{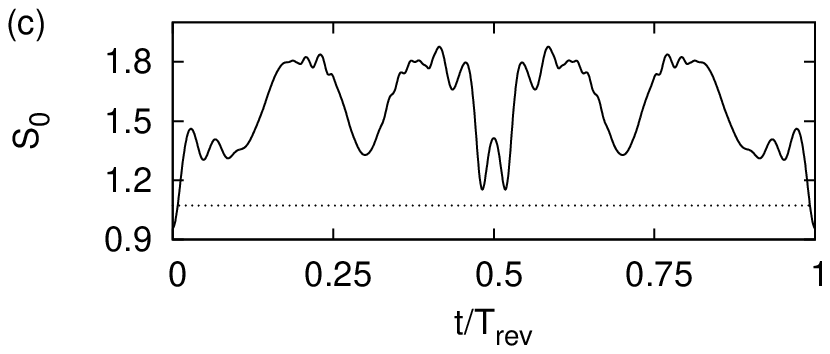}
\caption{Subsystem entropy $S_{0}$ vs. $t/T_{rev}$ for initial states (a) $\ket{\Psi_{00}}$, (b) $\ket{\Psi_{10}}$, and (c) $\ket{\Psi_{11}}$ with $\alpha_{a} = \alpha_{b} = 1$.}
\label{fig:2_mode_entropy_squeezing_time_dep}
\end{figure}

For our purpose, we set $\thetaa = 0$ (i.e., the $x$-quadrature). Denoting the corresponding tomogram by $w_{\textsc{a}}(x)$, the subsystem entropy $S_{0}$ is now equal to $- \int_{-\infty}^{\infty}  \rmd x \,w_{\textsc{a}}(x) \, \ln w_{\textsc{a}}(x)$. It can be shown that for a given state, $S_{0}  \leq (1/2)[ 1 + \ln {\pi} + \ln(2 \aver{(\Delta x)^{2}}) ]$, where $ \aver{(\Delta x)^{2}}$ is the variance  in $x$~\cite{orlowski}. The state exhibits entropic squeezing if $S_{0}$ is less than $(1/2) (1 + \ln {\pi})$. 
Plots of $S_{0}$ versus scaled time $t/T_{rev}$ are shown in Figs. \ref{fig:2_mode_entropy_squeezing_time_dep} (a)-(c)  for  initial states $\ket{\Psi_{00}}$,  $\ket{\Psi_{10}}$ and $\ket{\Psi_{11}}$ respectively.  The horizontal line in these figures denotes the numerical value below which entropic squeezing occurs.  We see that entropic squeezing occurs close to $t = 0$ and $T_{rev}$.   At other instants the subsystem entropy is significantly higher for initial states $\ket{\Psi_{10}}$ and $\ket{\Psi_{11}}$  compared to initial ideal coherence.  This feature is very prominent close to $T_{rev}/2$. Further, comparing Figs. \ref{fig:2_mode_entropy_squeezing_time_dep} (b) and (c) it is clear that $S_{0}$ is larger at all instants if both subsystems  depart from coherence initially,  compared to the case where one of them displays initial coherence.

\begin{figure}
\centering
\includegraphics[scale=0.7]{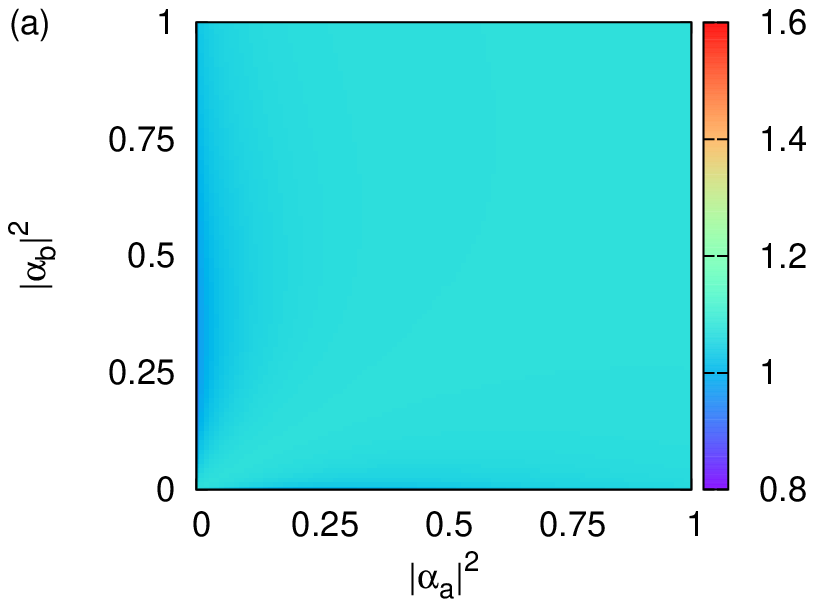}
\includegraphics[scale=0.7]{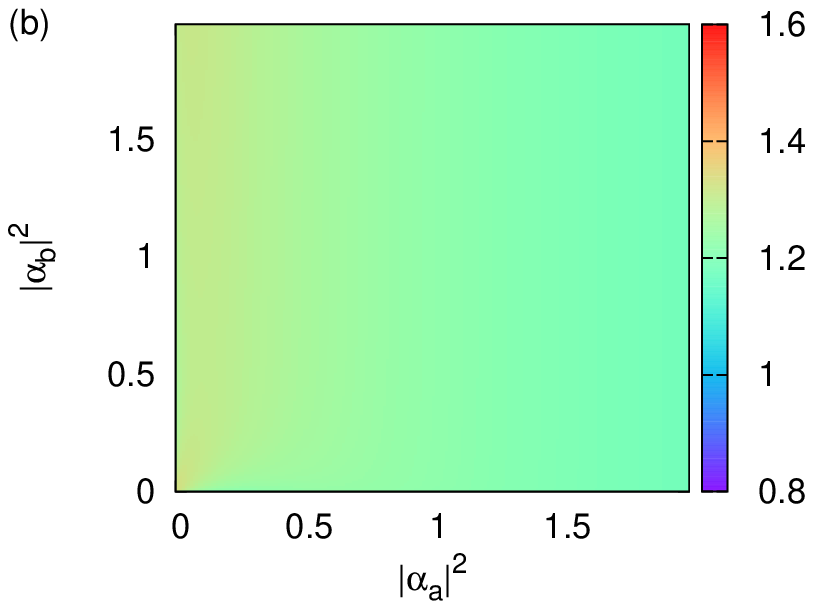}
\includegraphics[scale=0.7]{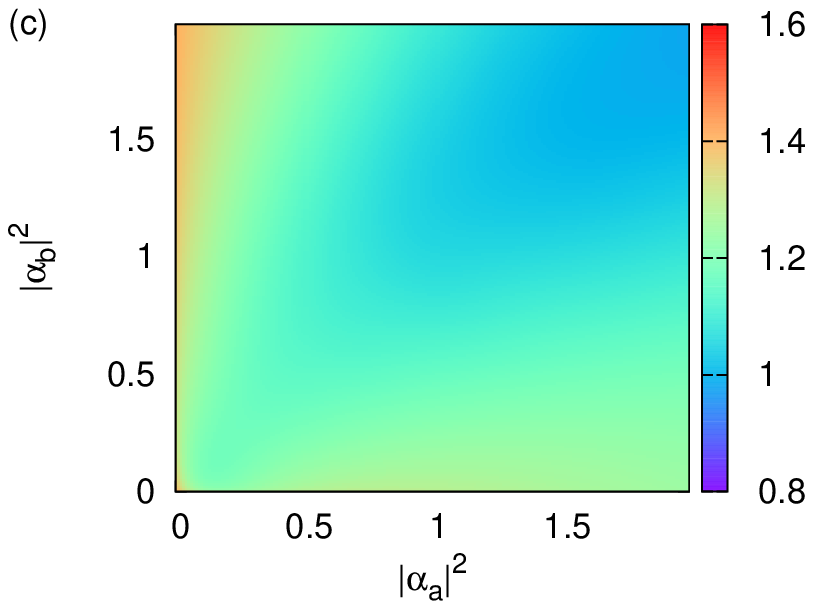}
\includegraphics[scale=0.7]{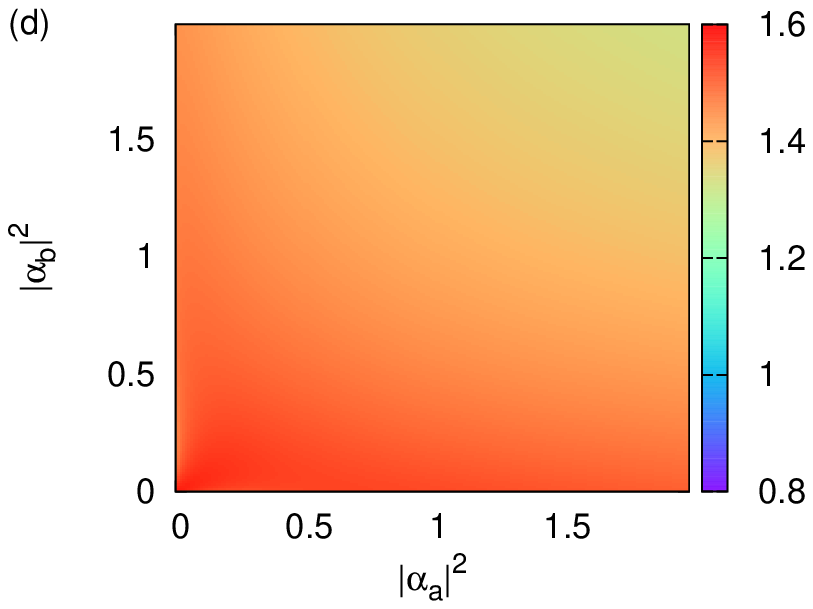}
\caption{ Variation of $S_{0}$ with $|\alpha_{a}|^{2}$ and $|\alpha_{b} |^{2}$  at $T_{rev}/2$ for initial states (a) $\ket{\Psi_{00}}$, (b) and (c) $\ket{\Psi_{10}}$, and (d) $\ket{\Psi_{11}}$. Figures (a), (b) and (d) correspond to subsystem A and (c) to B.}
\label{fig:2_mode_entropy_squeezing_alpha_dep}
\end{figure}

The variation of $S_{0}$ with $|\alpha_{a}|^{2}$ and $|\alpha_{b}|^{2}$, at $T_{rev}/2$,  for initial states $\ket{\Psi_{00}}$, $\ket{\Psi_{10}}$ and $\ket{\Psi_{11}}$ are shown in Figs. \ref{fig:2_mode_entropy_squeezing_alpha_dep} (a), (b) and (d) respectively. Figure \ref{fig:2_mode_entropy_squeezing_alpha_dep} (c) corresponds to the entropy of subsystem B for an initial state $\ket{\Psi_{10}}$. This facilitates comparison of the features in Figs.~\ref{fig:2_mode_entropy_squeezing_alpha_dep} (b) and (c)  where for the same asymmetric initial state the two subsystems examined are different. It is evident that $S_{0}$ corresponding to A is not squeezed while that corresponding to B exhibits squeezing for certain values of $\alpha_{a}$ and $\alpha_{b}$. The role played by the asymmetry in the initial states of the two subsystems is thus clearly brought out in these figures. It is also clear that entropic squeezing is more if the initial states of the subsystems are coherent.

\section{Concluding remarks}
 We have established how tomograms can be exploited to identify and characterise a variety of nonclassical effects such as the wave packet revival phenomena and squeezing and higher-order squeezing in both single-mode and bipartite systems. While a simple relation has been shown to exist between the number of strands in tomogram patterns  and the nature of fractional revivals when a single-mode radiation field propagates in a Kerr medium \cite{sudhrohithrev}, our investigations reveal that this no longer holds even in single-mode systems which display super-revivals during temporal evolution. The role played by decoherence due to amplitude decay and phase damping of the state has also been discussed. We have also analysed the revival phenomena in bipartite systems such as the double-well BEC evolving in time, \textit{solely} from tomograms. 
We have obtained the extent of squeezing, higher-order Hong-Mandel and Hillery-type squeezing from tomograms. 
  We have also investigated entropic squeezing from the tomograms in the case of the double-well BEC. 
  
  In the next chapter, we discuss procedures for extracting the extent of entanglement directly from quantum state tomograms in the context of both continuous-variable and hybrid quantum systems. We propose several entanglement indicators for this purpose. We compare the extent of entanglement as captured by these indicators with $\xi_{\textsc{svne}}$ which is a standard measure of entanglement. 
  

%% file: chapter3.tex

\chapter{Tomographic entanglement indicators and avoided energy-level crossings}
\label{ch:EIatAC}

\section{\label{sec:Ch3intro}Introduction}

In this chapter, we introduce various entanglement indicators that can be obtained directly from appropriate tomograms. We assess their performance as quantifiers of the extent of entanglement in both bipartite and multipartite systems. We have considered both CV and HQ systems. Our reference for comparison is set by $\xi_{\textsc{svne}}$. We recall that $\xi_{\textsc{svne}}=-\Tr\,(\rho \,\log_{2} \,\rho)$, where $\rho$ is the subsystem density matrix~\cite{niel}.

Some of the entanglement indicators we examine are inspired by classical tomography. Indicators of correlations  between different parts of a {\em classical} system  have been used extensively in various applications such  as automated image processing. While these correlators are obtained from classical tomograms, their definitions however,  are not intrinsically classical in nature. It is therefore worth examining their applicability in quantum contexts. Since correlations are inherently present in entangled states of quantum systems, a natural question that arises is whether the  performance of entanglement quantifiers obtained from classical correlators is comparable to that of standard indicators such as  $\xi_{\textsc{svne}}$. 

We focus on quantum systems whose eigenspectrums exhibit avoided energy-level crossings. In these systems, the spacings between energy levels change significantly with changes in the parameters, with two or more levels moving close to each other for specific values of the parameters, and then moving away as these values change. Our interest stems from the fact that extensive studies~\cite{AClinkPT,ent_change,ent_max} have established that entanglement as measured by standard indicators such as $\xi_{\textsc{svne}}$ is generically at an extremum  at an avoided crossing. Typically, the energy spectrum and the spacing between the energy levels depend on the strengths of the nonlinearity and the coupling between subsystems. With changes in the values of  these  parameters, the spacing between adjacent  levels can decrease, and even tend to zero, resulting in an energy-level crossing. According to the von Neumann-Wigner no-crossing theorem, energy levels within a multiplet generically avoid crossing, provided only one of the parameters is varied in the Hamiltonian governing the system. In what follows, we  investigate how effectively some of these entanglement indicators mimic the behaviour of $\xi_{\textsc{svne}}$ close to avoided crossings.

Energy-level  crossings display other interesting features. Since they affect the level spacings and their probability distribution ~\cite{BGSconj},  they  are
also important from the point of view of non-integrability and quantum chaos (see, for instance,~\cite{haake}). In addition, avoided crossings point to phase transitions which trigger a change in the quantum correlations in the system ~\cite{cejnar,heiss,AClinkPT,ent_change}. This aspect has been investigated extensively both  theoretically and in experiments~\cite{santhAC1,santhAC2,ACtheor1,ACexpt1,ACexpt2,ACexpt3}.

The bipartite CV systems that we examine here are a BEC in a double-well trap~\cite{sanz} introduced in Chapter \ref{ch:RevSqueezeOptTomo}, and a multi-level atom interacting with a radiation field~\cite{agarwalpuri}. We also investigate a multipartite HQ system~\cite{QuantMeta2,QuantMeta} comprising qubits interacting with a microwave field that is effectively described by the Tavis-Cummings model~\cite{tavis}. 

In the next section we introduce the entanglement indicators to be employed.  In Section \ref{sec:Ch3work}, we investigate how these indicators behave close to avoided crossings in the two  bipartite CV models mentioned above.  In Section \ref{sec:Ch3work2}, we extend our analysis to the multipartite HQ model. We conclude with brief remarks.

\section{\label{sec:Ch3review}Entanglement indicators from tomograms}

A typical example of a bipartite CV system  is two coupled oscillators (equivalently, a 
single-mode  radiation field interacting with a multi-level atom modelled as an oscillator) with $(a, a^{\dagger})$ (respectively, $(b, b^{\dagger})$) being the oscillator annihilation and creation operators corresponding to the two subsystems  A and B. We recall that Eqs. \eref{eqn:2modeTomoDefn}, \eref{eqn:tomo_2_mode_sub_A} and \eref{eqn:tomo_2_mode_sub_B} define the tomograms corresponding to the full system and the two subsystems respectively. 
In order to estimate the degree of  correlation between the subsystems, we use the following tomographic entropies.
The bipartite tomographic entropy is given by 
\begin{align}
\nonumber S(\thetaa,\thetab) = - & \int_{-\infty}^{\infty} \, \rmd X_{\thetaat} \int_{-\infty}^{\infty} \, \rmd X_{\thetabt} w(X_{\thetaat},\thetaa;X_{\thetabt},\thetab)\,\times \\[4pt] 
 & \log_{2} \, w(X_{\thetaat},\thetaa;X_{\thetabt},\thetab).
\label{eqn:2modeEntropy}
\end{align}
The subsystem tomographic entropy (generalizing Eq. \eref{eqn:subsysEntropy_A}) is
\begin{align}
S(\theta_{i}) = - &\int_{-\infty}^{\infty} \!\rmd X_{\theta_{i}} w_{i}(X_{\theta_{i}},\theta_{i}) \log_{2} \,[w_{i}(X_{\theta_{i}},\theta_{i})] \;\; (i= \textsc{A,B}).
\label{eqn:1modeEntropy}
\end{align}
Some of the correlators that we examine are obtained from a section of the tomogram
 corresponding to specific values of $\thetaat$ and $\thetabt$. The efficacy  of such a correlator  
 as a measure  of entanglement is therefore sensitive to the choice of  the tomographic section. We now define these  correlators, and the corresponding entanglement indicators.  

The mutual information 
$\epsarg{tei}(\thetaa,\thetab)$  which we get from the tomogram of a quantum system 
can carry  signatures of entanglement. This  
quantity is expressed in terms of  the  tomographic entropies defined above  as
\begin{equation}
\epsarg{tei}(\thetaa,\thetab)=S(\thetaa) + S(\thetab) - S(\thetaa,\thetab).
\label{eqn:epsTEI}
\end{equation}
Indicators based on the inverse participation ratio (IPR) are also found to be good candidates for estimating  the extent of entanglement ~\cite{ViolaBrown,sharmila2}. The participation ratio is a measure of delocalisation in a given basis. The IPR corresponding to a bipartite system in the basis of the rotated quadrature operators  is defined as
\begin{equation}
\eta_{\textsc{ab}}(\thetaa,\thetab) = \int_{-\infty}^{\infty} \!\rmd X_{\thetaat} \int_{-\infty}^{\infty} \!\rmd X_{\thetabt} [w(X_{\thetaat},\thetaa;X_{\thetabt},\thetab)]^{2}.
\label{eqn:IPRab}
\end{equation}
The IPR for each subsystem is given by 
\begin{equation}
\eta_{i}(\theta_{i}) = \int_{-\infty}^{\infty} \!\rmd X_{\theta_{i}} [w_{i}(X_{\theta_{i}},\theta_{i})]^{2} \;\; 
(i= \textsc{A,B}).
\label{eqn:IPRi}
\end{equation}
The entanglement indicator  in this case is given by 
\begin{equation}
\epsarg{ipr}(\thetaa,\thetab)=1+\eta_{\textsc{ab}}(\thetaa,\thetab)-\eta_{\textsc{a}}(\thetaa)-\eta_{\textsc{b}}(\thetab). 
\label{eqn:epsIPR}
\end{equation}  

Apart from these, we have examined two  
other correlators which are familiar in the context of classical tomograms. The first of these is the Pearson correlation coefficient~\cite{SmithMIcorr}   between two random variables $X$ and $Y$, given by 
\begin{equation}
\textsc{PCC}(X,Y)
 = \frac{\mathrm{Cov}(X,Y)}{(\Delta X) (\Delta Y)}.
\label{eqn:PCC}
\end{equation}
Here $\Delta X, \, \Delta Y$ are the standard deviations of $X$ and $Y$ respectively, and  
$\mathrm{Cov}(X,Y)$ is their covariance.
Of direct relevance to us is 
$\textsc{PCC}(X_{\thetaat},X_{\thetabt})$ 
calculated  for  fixed values of $\thetaa$ and $\thetab$.
Since the quantifier  of entanglement between two subsystems must be non-negative, a simple definition of the entanglement indicator in this case would be 
\begin{equation}
\epsarg{pcc}(\thetaa,\thetab)=
\vert \textsc{PCC}(X_{\thetaat},X_{\thetabt})\vert.
\label{eqn:epsPCC}
\end{equation}
This indicator captures  the effect of linear correlations.  Our motivation for assessing this indicator arises from the fact that,   in recent experiments on generating and testing the extent of entanglement in CV systems, the variances of suitably chosen conjugate observables and the corresponding standard quantum limit alone are used ~\cite{ExptCVentang}. 
We reiterate that these merely capture the extent of linear correlations between two states. 

The second indicator (to be denoted by 
$\epsarg{bd}$) that we introduce and use is arrived at as follows. 
In probability theory, 
the mutual information~\cite{ITCoverThomas}  
 between two continuous random variables 
$X$ and $Y$ 
can be expressed in terms of the Kullback-Leibler divergence
$D_{\textsc{kl}}$~\cite{kullback} 
between their joint probability density  
$p_{XY}(x,y)$ 
and the product of the corresponding  marginal 
densities  $p_{X}(x)= \int  p_{XY}(x,y) dy$ 
and $p_{Y}(y) = \int p_{XY}(x,y) dx$,    
as~\cite{MI_KL_link} 
\begin{equation}
D_{\textsc{kl}}[p_{XY}\!:\!p_{X}p_{Y}] 
= \int \!dx \int\!dy \,p_{XY}(x,y)\,\log_{2}
\frac{p_{XY}(x,y)}{p_{X}(x) p_{Y}(y)},
\label{eqn:KLmutualinfo}
\end{equation}
The quantity  $\epsarg{tei}(\thetaa,\thetab)$ 
defined in  Eq. \eref{eqn:epsTEI} 
is precisely the mutual information in the case of 
optical tomograms (which are continuous probability distributions): 
\begin{equation}
\epsarg{tei}(\thetaa,\thetab)= 
D_{\textsc{kl}}
\big[w(X_{\thetaat},\thetaa;X_{\thetabt},\thetab)\!:\! w_{\textsc{a}}(X_{\thetaat},\theta) w_{\textsc{b}}(X_{\thetabt},\thetab)\big]. 
\label{eqn:KL_MI_link}
\end{equation}
 A simpler  alternative for our purposes is 
provided by the Bhattacharyya distance
$D_{\textsc{b}}$~\cite{KL_BD_link} between $p_{XY}$ and 
$p_{X}p_{Y}$, defined as  
\begin{equation}
D_{\textsc{b}}[p_{XY}\!:\!p_{X}p_{Y}] =  
 -\log_{2}\,\Big\{\int \!dx\int \!dy \,\big[p_{XY}(x,y) 
p_{X}(x) p_{Y}(y)\big]^{1/2}\Big\}. 
\label{eqn:Bhattmutualinfo}
\end{equation}
Using Jensen's inequality, it is easily shown that 
$D_{\textsc{b}} \leqslant \frac{1}{2} D_{\textsc{kl}}$. 
$D_{\textsc{b}}$ thus gives us an approximate estimate (that is an underestimate) 
of the mutual information. Based on this quantity,  
 we have 
 an entanglement indicator that is the 
 analogue of Eq. \eref{eqn:KL_MI_link}, namely, 
\begin{align}
\epsarg{bd}(\thetaa,\thetab)=& D_{\textsc{b}}[w(X_{\thetaat},\thetaa;X_{\thetabt},\thetab)\!:\! w_{\textsc{a}}(X_{\thetaat},\theta) w_{\textsc{b}}(X_{\thetabt},\thetab)].
\label{eqn:epsBD}
\end{align}

The dependence on 
$\thetaa$ and $\thetab$ of each of the foregoing entanglement 
 indicators $\varepsilon$ is removed by  
  averaging over a representative set of values of  
  those variables. We denote the 
corresponding averaged value by $\xi$. 
 In the context  of 
 bipartite CV models,     
 we find~\cite{sharmila,sharmila2} that  averaging 
 $\epsarg{tei}(\thetaa,\thetab)$ 
over  $25$ different values of 
 $(\thetaa,\thetab)$ selected at equal intervals in the range $[0,\pi)$ yields a reliable  
entanglement indicator  $\xi_{\textsc{tei}}$.
 A similar averaging of each of the quantities 
$\epsarg{ipr}, \epsarg{pcc}$ and 
$\epsarg{bd}$ yields 
  $\xi_{\textsc{ipr}}, \xi_{\textsc{pcc}}$ and   
 $\xi_{\textsc{bd}}$, respectively. 

Next, we turn  to  hybrid systems of field-atom interactions. For a two-level atom with ground state $\ket{g}$ and excited state $\ket{e}$, the quorum of observables  is ~\cite{thew} 
\begin{align}
\label{eqn:atomops}
\nonumber \sigma_{x}=
\textstyle{\frac{1}{2}} (\ket{e}\bra{g}&+\ket{g}\bra{e}), 
\;\sigma_{y}=\textstyle{\frac{1}{2}} i(\ket{g}\bra{e}-\ket{e}\bra{g}), \\
& \sigma_{z}=\textstyle{\frac{1}{2}} (\ket{e}\bra{e}-\ket{g}\bra{g}). 
\end{align}
Let $\sigma_{z}\ket{m}=m\ket{m}$. Then $U(\vartheta,\varphi) \ket{m} =  \ket{\vartheta,\varphi,m}$, where $U(\vartheta,\varphi)$ 
 is a general SU(2) transformation parametrised by $(\vartheta, \varphi)$. Denoting $(\vartheta, \varphi)$ by the unit vector $\vn$, the  qubit tomogram is  given by 
\begin{equation}
\label{eqn:spintomogram}
w(\vn,m)=\bra{\vn,m}\rho_{\textsc{s}}\ket{\vn,m}
\end{equation}
where $\rho_{\textsc{s}}$ is the qubit density matrix. Corresponding to each value of $\vn$ there exists a complete basis set and hence $\sum\limits_{m} w(\vn,m)=1, \; \forall \; \vn$.  The atomic tomograms 
are  obtained from these,  and the corresponding 
entanglement properties are quantified using  
appropriate adaptations of the indicators (suitably replacing the integrals by sums) described above.
Extension of the foregoing to the multipartite 
case is straightforward ~\cite{ibort}, and the tomograms 
obtained can be examined on similar lines.  

\section{\label{sec:Ch3work}Avoided energy-level crossings in bipartite CV models}

\subsection{\label{subsec:Ch3BEC}The double-well BEC model}

The effective Hamiltonian (Eq. \eref{eqn:HBEC}) for the system and its diagonalisation are as follows~\cite{sanz}. 
Setting $\hbar = 1$,
\begin{equation}
\nonumber H_{\textsc{bec}} =\omega_{0} N_{\mathrm{tot}} + \omega_{1} (a^{\dagger} a - b^{\dagger} b) + U  N_{\mathrm{tot}}^{2} - \lambda (a^{\dagger} b + a b^{\dagger}).
\end{equation}
We recall that $(a,a^{\dagger})$ and $(b,b^{\dagger})$ are the  respective boson annihilation and creation operators of the atoms in wells A and B (the two subsystems),  
and $N_{\mathrm{tot}} 
= (a^{\dagger} a + b^{\dagger} b)$.
 $U$ is the strength of  nonlinear interactions between atoms within each well, and also between  the two wells. $U>0$, ensuring that the energy spectrum is bounded from below. $\lambda$ is the linear interaction strength and $\omega_{1}$ is the strength of
 the population imbalance between the two wells.   
 The Hamiltonian is diagonalised by the unitary 
 transformation 
$V= e^{\kappa (a^{\dagger} b - b^{\dagger} a)/2}$ (Eq. \eref{eqn:VHVdag} of Appendix \ref{appen:DensMat}) where $\kappa
=\tan^{-1} (\lambda/\omega_{1})$, to yield 
\begin{equation} 
 V^{\dagger} H_{\textsc{bec}} V = 
 \widetilde{H}_{\textsc{bec}}
= \omega_{0} N_{\mathrm{tot}} + \lambda_{1} (a^{\dagger} a - b^{\dagger} b) + 
U  N_{\mathrm{tot}}^{2},
\label{eqn:VrotatedH}
\end{equation}
with $\lambda_{1}= (\lambda^{2} + 
\omega_{1}^{2})^{1/2}$.  
$\widetilde{H}_{\textsc{bec}}$ and $N_{\mathrm{tot}}$ 
commute with each other.
Their common eigenstates   are 
the  product states $\ket{k} \otimes \ket{N-k} \equiv 
\ket{k \delim N-k}$. Here  
$N = 0,1,2,\dots$ 
is the eigenvalue of 
$N_{\mathrm{tot}}$, 
and $\ket{k}$ is a 
 boson number state, with $k$ running from 
 $0$ to $N$ for a given $N$.
  The eigenstates and 
 eigenvalues of 
$H_{\textsc{bec}}$ are given by
\begin{equation}
\ket{\psi_{N,k}} = V \ket{k \delim N-k}
\label{eqn:EigvecBEC}
\end{equation}
and
\begin{equation}
E(N,k)= \omega_{0} N + \lambda_{1} (2 k - N) + U  N^{2}.
\label{eqn:EigvalBEC}
\end{equation}
For numerical analysis we set $\omega_{0} = 1, 
 U =1$.

\begin{figure}
\includegraphics[width=0.45\textwidth]{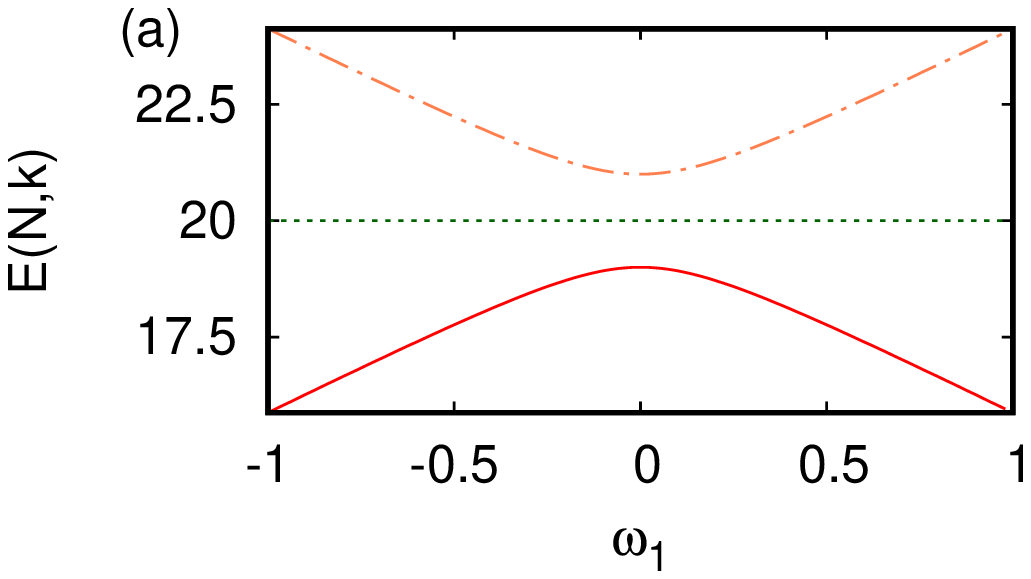}
\includegraphics[width=0.45\textwidth]{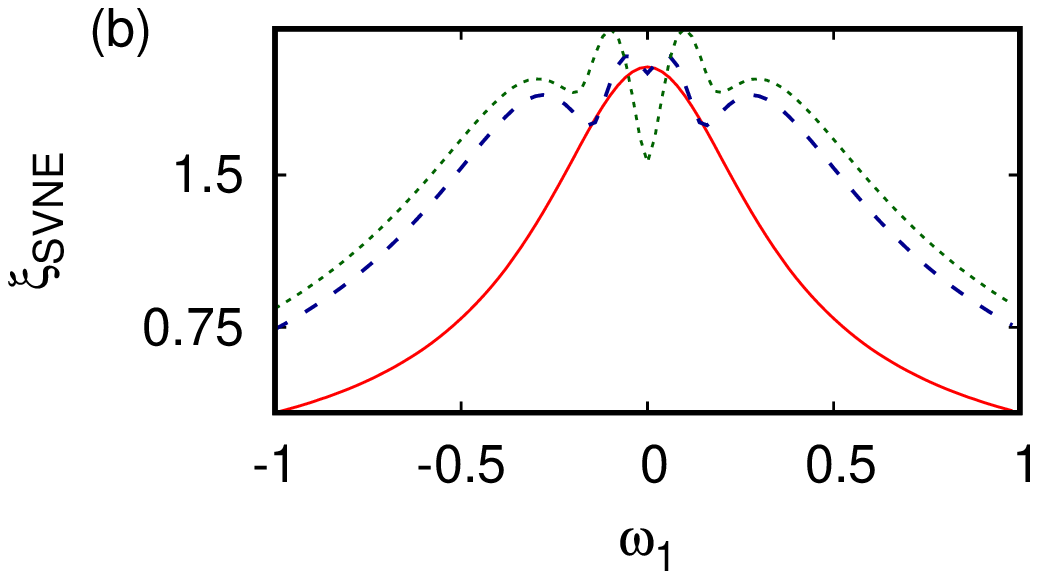}
\caption{(a) $E(N,k)$ vs.  $\omega_{1}$ for $N=4$ 
and $k=0,2,4$ in the BEC model. (b) $\xi_{\textsc{svne}}$ vs. $\omega_{1}$ for $N=4, \,k=0,1,2$.  The curves correspond to $k=0$ (red), $1$ (blue), $2$ (green) and $4$ (orange). $\lambda=0.25$.}
\label{fig:energySpectrumSVNE}
\end{figure}

In Fig. \ref{fig:energySpectrumSVNE}(a), 
 $E(N=4,k)$ is  plotted against  
$\omega_{1}$ for $k=0,2,4$, with 
$\lambda=0.25$. 
$E(N, N - k)$ is the reflection of  
$E(N, k)$  about the value 
$\omega_{0}N + UN^{2}$.  
Avoided  energy-level 
crossings are seen at $\omega_{1}=0$. In order to  
 set the reference level for the extent of entanglement between the two wells, we compute 
 $\xi_{\textsc{svne}} 
 = -\mathrm{Tr}\,
 (\rho_{\textsc{a}} \log_{2} \rho_{\textsc{a}})$, 
  where  $\rho_{\textsc{a}}$ 
is the reduced density matrix of the subsystem A. 
($\xi_{\textsc{svne}}$  is also equal to 
$-\mathrm{Tr}\,(\rho_{\textsc{b}} \log_{2} \rho_{\textsc{b}})$, since  
   $\ket{\psi_{N,k}}$ is a bipartite pure state.) 
Plots of $\xi_{\textsc{svne}}$ corresponding to the state $\ket{\psi_{4,k}}$ for $k = 0,1,2$ 
 are shown in 
 Fig. \ref{fig:energySpectrumSVNE}(b). 
 The states $\ket{\psi_{4,3}}$ and  
$\ket{\psi_{4,1}}$ have the same 
$\xi_{\textsc{svne}}$, (as do the states  
$\ket{\psi_{4,4}}$ and  $\ket{\psi_{4,0}}$), 
owing to the $k\leftrightarrow  N-k$ symmetry. 
It is evident that there is a significant extent of entanglement close to the avoided crossing,  and   $\omega_{1}=0$ is marked by a local maximum or minimum in $\xi_{\textsc{svne}}$. 

\begin{figure*}
\includegraphics[width=0.3\textwidth]{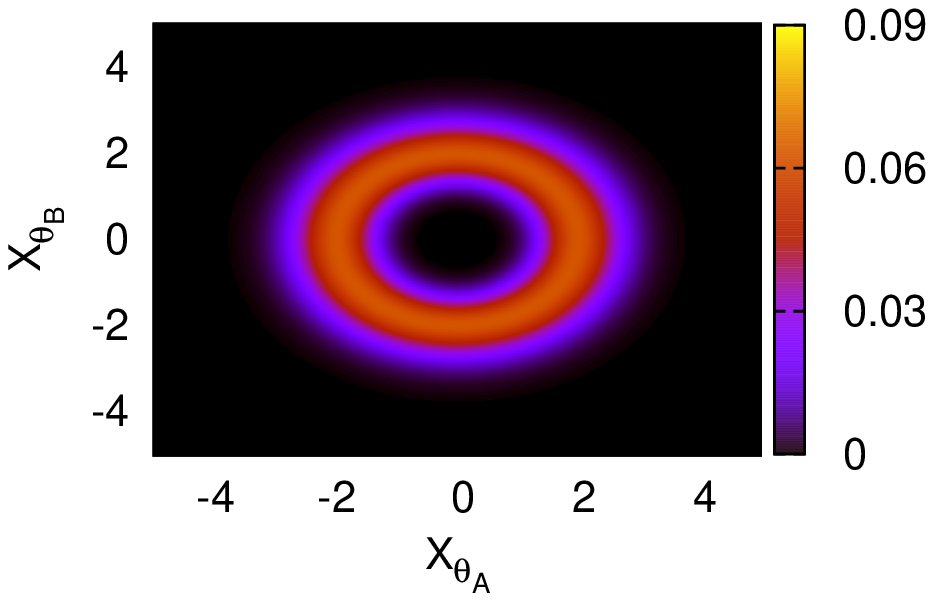}
\includegraphics[width=0.3\textwidth]{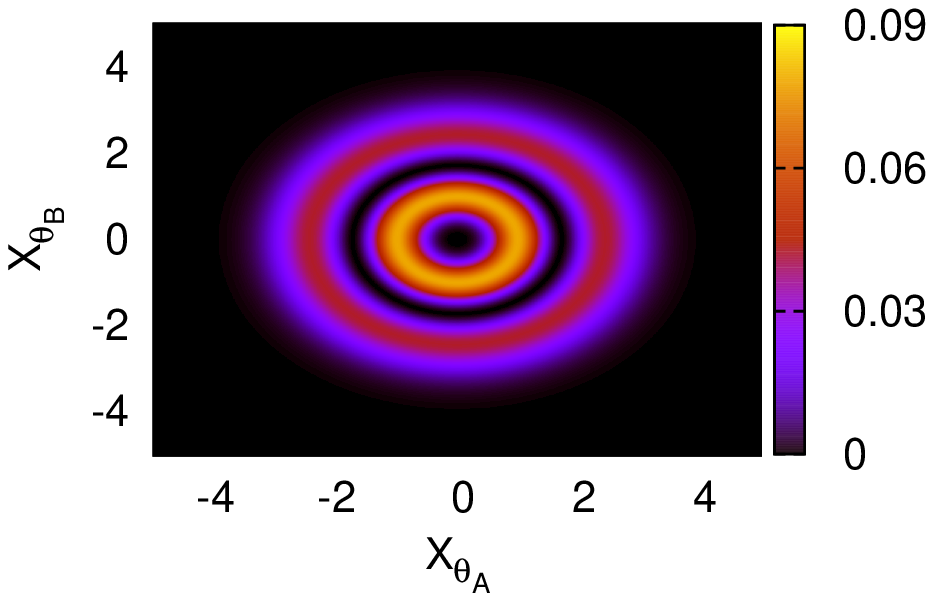}
\includegraphics[width=0.3\textwidth]{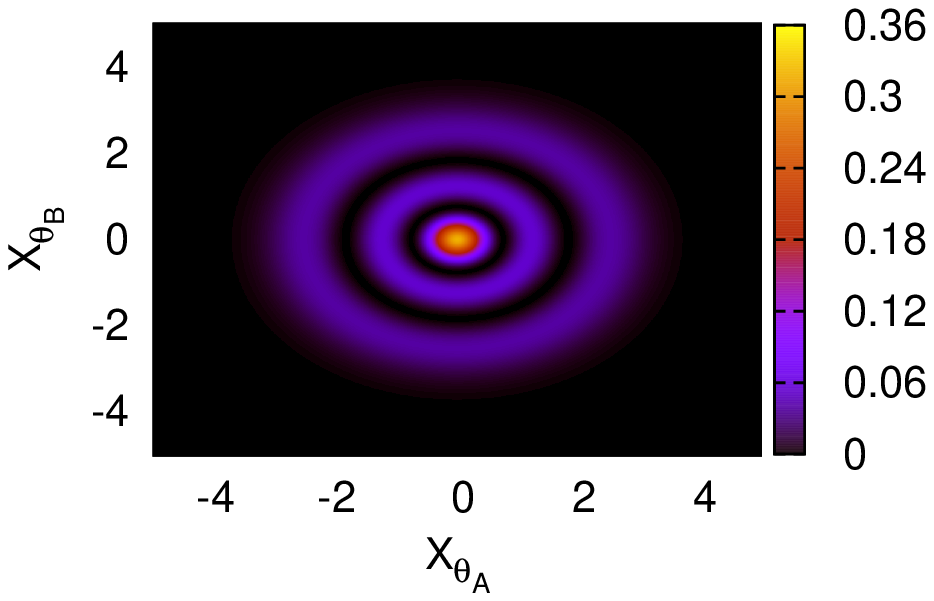}\\
\includegraphics[width=0.3\textwidth]{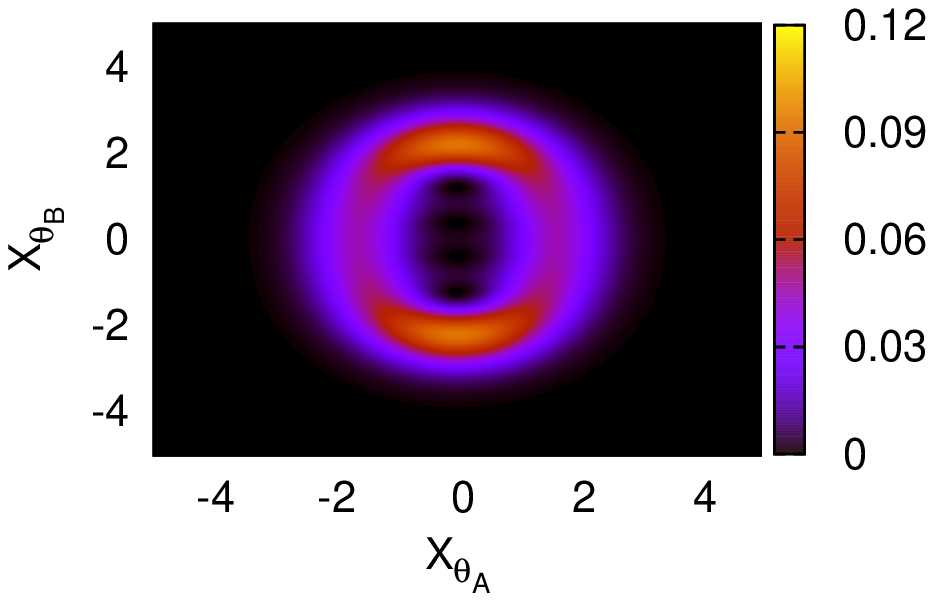}
\includegraphics[width=0.3\textwidth]{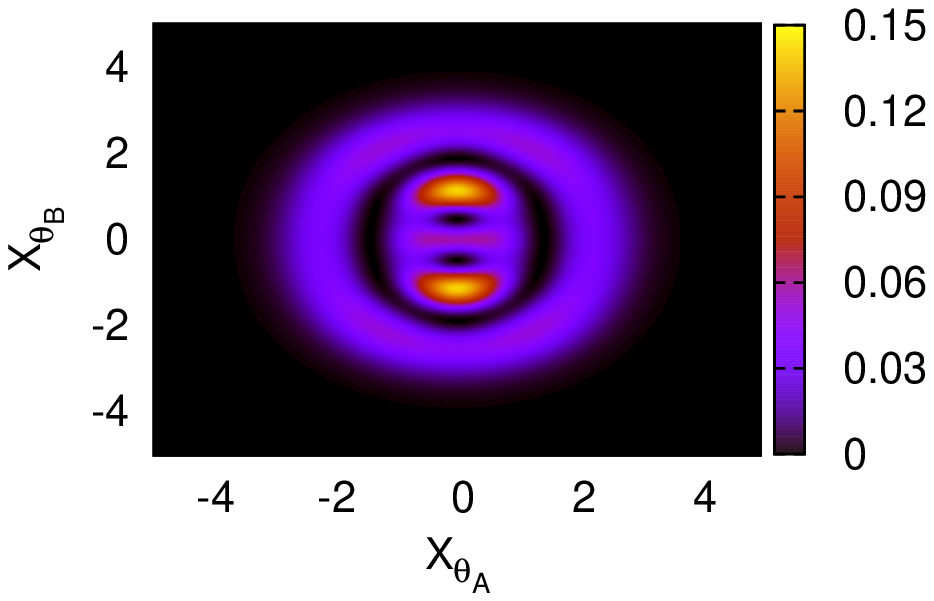}
\includegraphics[width=0.3\textwidth]{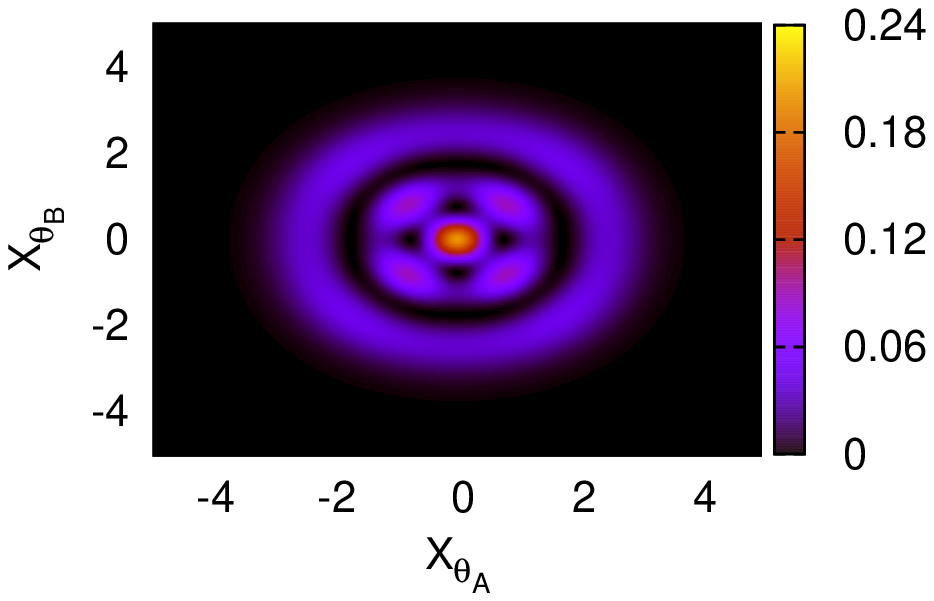}\\
\includegraphics[width=0.3\textwidth]{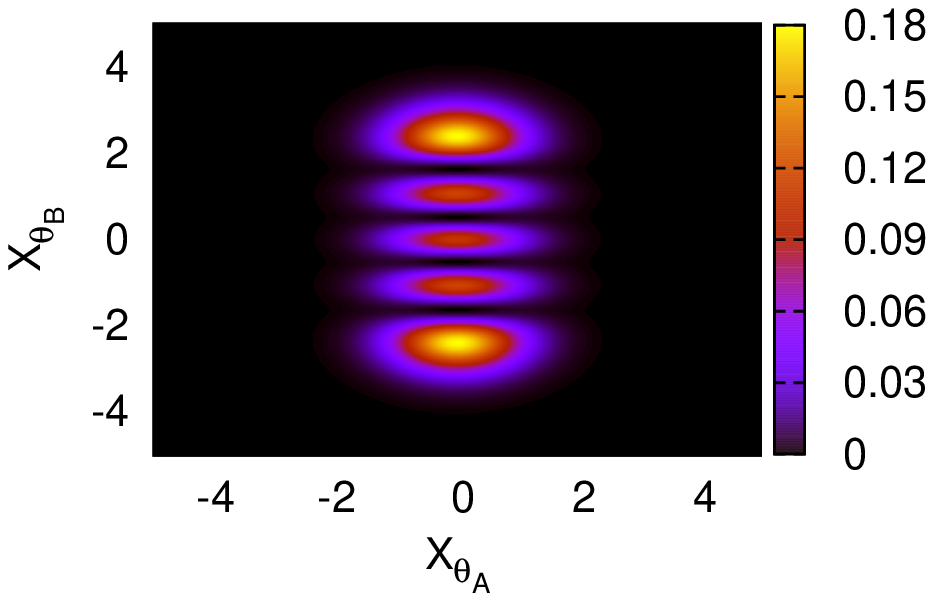}
\includegraphics[width=0.3\textwidth]{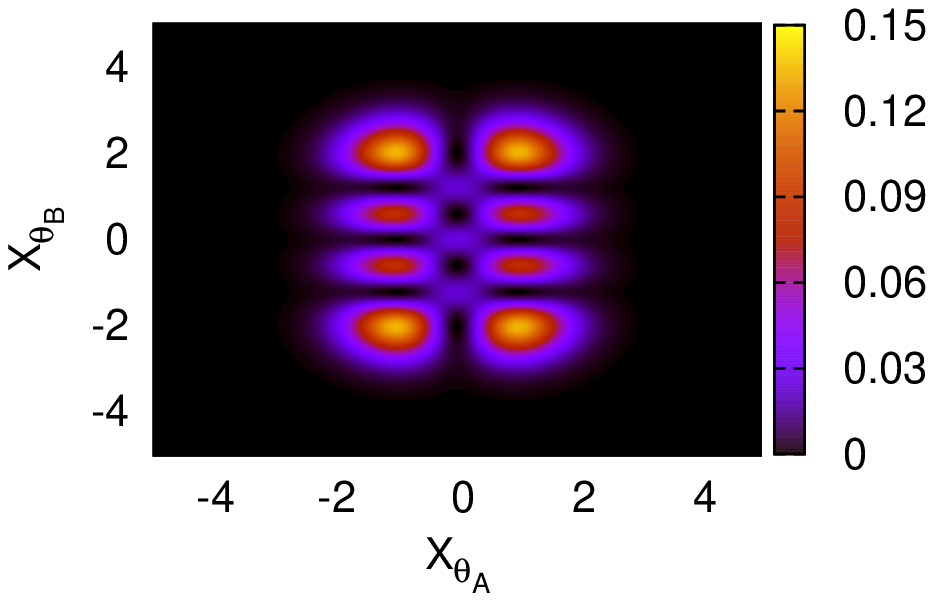}
\includegraphics[width=0.3\textwidth]{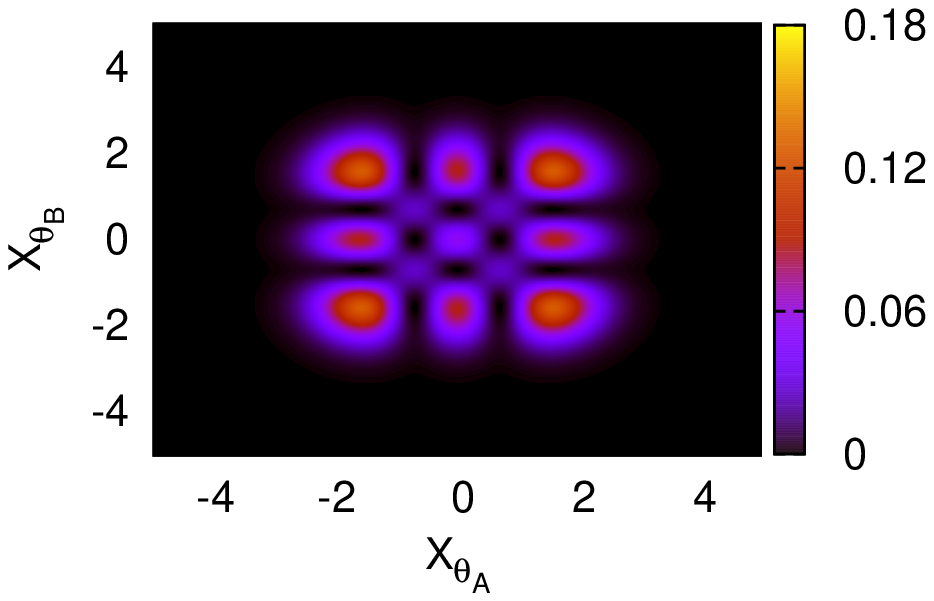}
\caption{$\thetaa=0,\thetab=\pi/2$ slice of 
the tomogram for $N=4$ in the BEC model. Left to right,  $k=0, 1$ and 
$2$. Top to bottom,  $\omega_{1}=0, 0.1$ and $1$.}
\label{fig:tomosAC}
\end{figure*}

Figure \ref{fig:tomosAC} depicts   $\thetaa = 0, \thetab = \frac{1}{2}\pi$ sections of the tomograms corresponding to the states 
$\ket{\psi_{4,k}}$ for  $k=0,1, 2$ and   
 $\omega_{1}=0, 0.1, 1$.
It is clear that, for a given value of $\omega_{1}$, the qualitative features of the tomograms are altered considerably as $k$ is varied. 
The patterns in the tomograms also reveal   nonlinear correlations between the quadrature variables $X_{\thetaat}$ and $X_{\thetabt}$ (top panel). For  instance, the tomogram slice on the top right 
 shows a probability distribution that is essentially  unimodal and symmetric about the origin with the annular structures diminished in magnitude. It is 
clear that this case is less correlated than the tomogram  in the top left corner. This conforms to the observed trend in the extent of entanglement (compare $\xi_{\textsc{svne}}$ corresponding to $k=0$ and $k=2$ at $\omega_{1}=0$ in Fig. \ref{fig:energySpectrumSVNE} (b)). Again,  in the bottom panel of the figure, the sub-structures in the patterns increase with increasing $k$, signifying 
a higher degree of nonlinear correlation. This is in 
consonance  with the trend in the  entanglement at 
$\omega_{1}=1$ (Fig. \ref{fig:energySpectrumSVNE} (b)). We therefore expect $\epsarg{tei}$ 
and its 
averaged version   
$\xi_{\textsc{tei}}$  to be 
 much better entanglement indicators  than 
  $\epsarg{pcc}$ and $\xi_{\textsc{pcc}}$. 
We also  mention here that the current experimental techniques of testing CV entanglement based on the variances and covariances of suitably chosen observables~\cite{ExptCVentang} are  
not as effective as calculating nonlinear 
correlators,  for the same reason.

\begin{figure}
\includegraphics[width=0.5\textwidth]{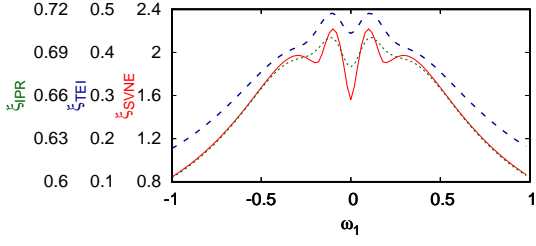}
\caption{$\xi_{\textsc{svne}}$ (red), $\xi_{\textsc{tei}}$ (blue) and $\xi_{\textsc{ipr}}$ (green) vs. 
$\omega_{1}$, for the state 
$\ket{\psi_{4,2}}$ in the BEC model.}
\label{fig:N4k2VaryOmega1}
\end{figure}

Our detailed investigations reveal that 
$\xi_{\textsc{tei}}$ and $\xi_{\textsc{ipr}}$ 
follow the trends in 
$\xi_{\textsc{svne}}$ 
reasonably well for generic eigenstates of $H_{\textsc{bec}}$.
This is illustrated in 
Fig. \ref{fig:N4k2VaryOmega1}, which shows plots of these indicators as functions of $\omega_{1}$. 
Apart from examining the suitability of  
$\epsarg{pcc}$ as an  entanglement indicator, 
 we have also assessed the extent of linear correlation between any two indicators based on 
 the corresponding PCC, as follows. 
  We have obtained  $100$ values each of $\xi_{\textsc{tei}}$ and $\xi_{\textsc{svne}}$ for different values of $\omega_{1}$ in the range 
  $(-1, 1)$ 
  in steps of $0.02$. Treating the two sets of values    
  as two sets of random numbers, we obtain the PCC between them, as defined 
   in Eq. \eref{eqn:PCC}. The  PCC between $\xi_{\textsc{tei}}$ and $\xi_{\textsc{svne}}$ (respectively,  $\xi_{\textsc{ipr}}$ and $\xi_{\textsc{svne}}$) estimates the extent of  linear correlation between the two indicators, and 
is found to be $0.97$ (resp., $0.99$) in the case shown in Fig. \ref{fig:N4k2VaryOmega1} corresponding to $\ket{\psi_{4,2}}$.  
 (In general, the  PCC ranges from $1$ for complete correlation,  to $-1$ for  maximal anti-correlation. 
 Its vanishing indicates the absence of  linear correlation).  

\begin{figure}
\includegraphics[width=0.3\textwidth]{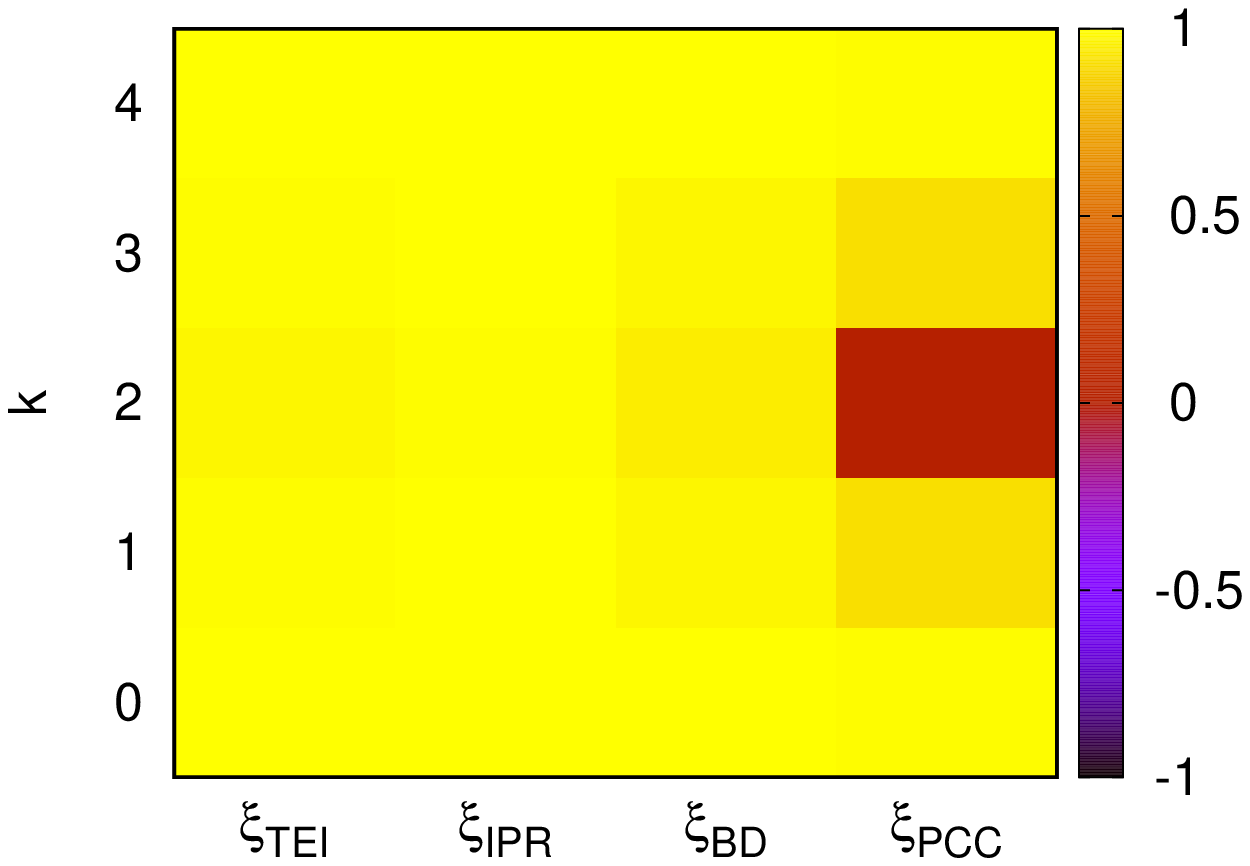}
\includegraphics[width=0.3\textwidth]{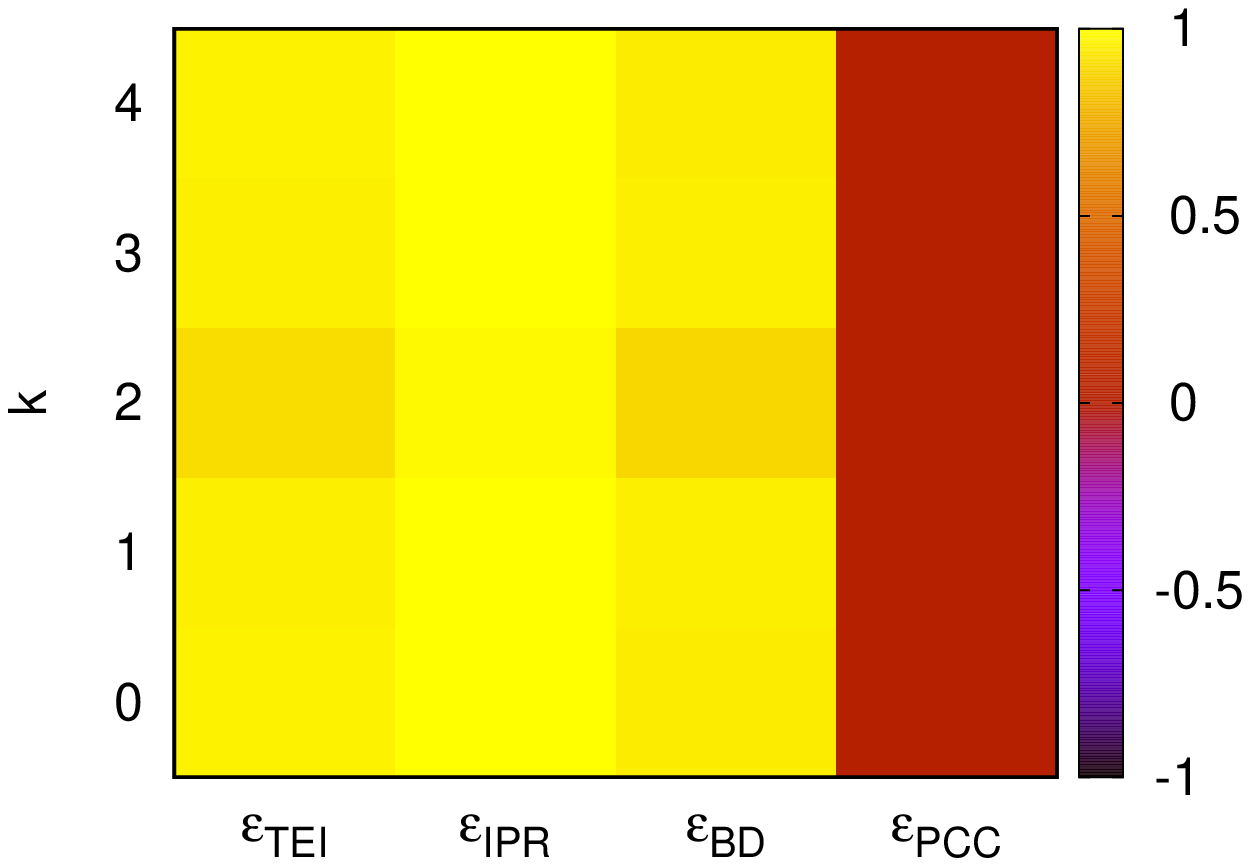}
\includegraphics[width=0.3\textwidth]{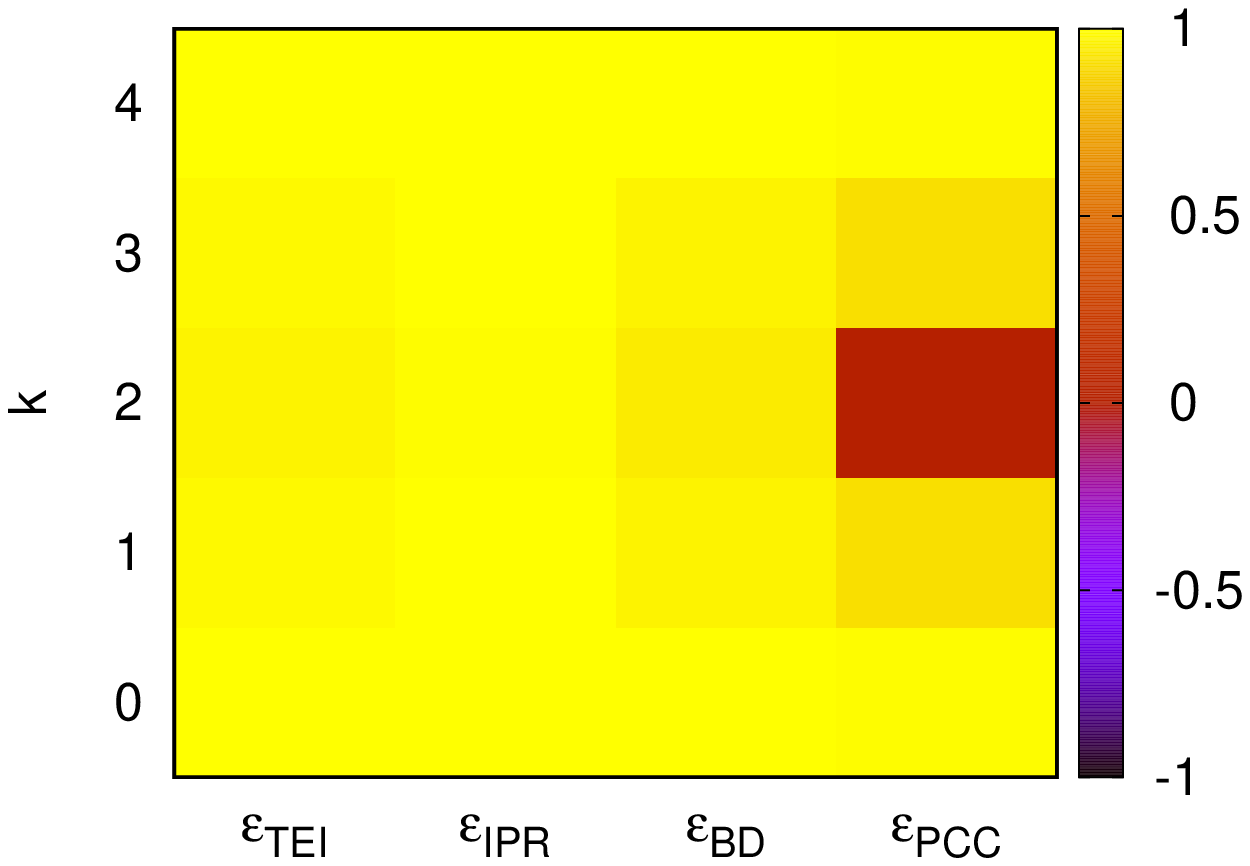}
\caption{Correlation of $\xi_{\textsc{svne}}$ 
with 
 $\xi$-indicators (left),   
 with $\neweps$-indicators for the slice 
 $\thetaa=0,\thetab=\pi/2$ (centre), 
 and with 
 $\neweps$-indicators  for the slice 
 $\thetaa=0,\thetab=0$ (right), for  the 
 eigenstates $\ket{\psi_{4, k}}, 
 \,0 \leqslant k \leqslant 4$ in the BEC model.}
\label{fig:CorrPlotN4VaryOmega1}
\end{figure}

Figure \ref{fig:CorrPlotN4VaryOmega1} 
shows 
the PCC  between 
$\xi_{\textsc{svne}}$ and various indicators, 
for the eigenstates 
$\ket{\psi_{4,k}}$ where $k=0,1,2,3,4$.
From Fig. \ref{fig:CorrPlotN4VaryOmega1}(a), we see that $\xi_{\textsc{ipr}}$,  $\xi_{\textsc{tei}}$ 
and $\xi_{\textsc{bd}}$ are very good  entanglement indicators. We have also found that 
all these indicators improve with increasing  $N$.
The performance of the $\neweps$-indicators depends, of course,  on the specific choice of the tomographic section. For instance, 
$\epsarg{tei}$ and $\epsarg{bd}$ perform marginally better for the slice  
$\thetaa = 0, \,\thetab = 0$ 
than for the slice
$\thetaa = 0, \,\thetab = \frac{1}{2}\pi$. 
It is also evident that 
$\xi_{\textsc{pcc}}$ does not fare as well as the other indicators. This is to be  expected, since 
 $\xi_{\textsc{pcc}}$  only captures  linear correlations, as already emphasised.

   We have verified that the sensitivity  of all the indicators decreases with an increase in 
   $\lambda$, the strength of  
 the coupling between the two  subsystems   
 (as in Eq. \eref{eqn:HBEC}).   
$\xi_{\textsc{ipr}}$, however, 
 remains closer to 
 $\xi_{\textsc{svne}}$ 
than the other indicators. 
This fact is consistent with inferences~\cite{sharmila2} drawn   
about  the relation between 
the Hamming distance~\cite{HammingInQM}  
and the efficacy of $\xi_{\textsc{ipr}}$. 
We recall that the 
Hamming distance between two bipartite qudits  
$\ket{u_{1}}\otimes \ket{u_{2}}$ and $\ket{v_{1}}\otimes\ket{v_{2}}$ attains its maximum value of 
$2$ when $\aver{u_{1}|v_{1}}=0$ and $\aver{u_{2}|v_{2}}=0$. 
A straightforward extension to CV systems implies  that the Hamming distance between $\ket{k_{1} \delim N-k_{1}}$ and $\ket{k_{2} \delim N-k_{2}}$ is 2 (so that these states are Hamming-uncorrelated), if $k_{1}\neq k_{2}$. Participation ratios are valid measures of entanglement for superpositions of Hamming-uncorrelated states in spin systems~\cite{ViolaBrown}. We will show in this and the next chapter that $\xi_{\textsc{ipr}}$ effectively mimics standard measures of entanglement in  CV systems as well. In the present instance, the eigenstates $\ket{\psi_{N,k}}$ are superpositions of the states $\lbrace \ket{j \delim N-j} \rbrace$ which are Hamming-uncorrelated for different values of 
$j$. This is the reason for the usefulness   
of $\xi_{\textsc{ipr}}$ as an entanglement indicator even for larger values of $\lambda$. 

\begin{figure}
\includegraphics[width=0.45\textwidth]{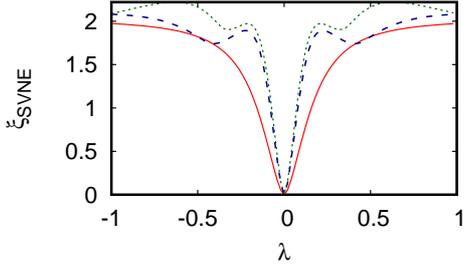}
\caption{$\xi_{\textsc{svne}}$ vs. $\lambda$ for $N=4, \,k=0,1,2$, in the BEC model.  The curves correspond to $k=0$ (red), $1$ (blue) and $2$ (green). $\omega_{1}=0.25$. }
\label{fig:energySpectrumSVNEvaryLambda}
\end{figure}

We now proceed to examine 
quantitatively the efficacy of the entanglement indicators  as functions of  $\lambda$. For numerical computation we have set $\omega_{1}=0.25$.  
Consider, as an illustration, plots of the eigenvalues 
$E(4,k)$ ($k=0,2,4$) as functions of  $\lambda$.  
These plots are  exactly the 
same as those in 
Fig. \ref{fig:energySpectrumSVNE}(a), with 
$\omega_{1}$ replaced by $\lambda$ on the 
horizontal axis, since 
$E(N, k)$ only depends on the parameters 
$\omega_{1}$ and $\lambda$ in the 
symmetric combination $\lambda_{1} = (\lambda^{2} 
+ \omega_{1}^{2})^{1/2}$. 
The avoided crossing of energy levels now occurs at 
$\lambda = 0$.   
But this symmetry between 
$\omega_{1}$ and 
$\lambda$ does not extend to 
the unitary transformation $V$, and hence 
to  the eigenstates 
of $H_{\textsc{bec}}$. 
(Recall that $V$ involves the parameter 
$\kappa = \tan^{-1}(\lambda/\omega_{1})$.) 
When  $\lambda=0$ 
there is  no linear interaction between the two modes.  $V$ then  reduces to the identity operator, and 
$H_{\textsc{bec}}$ is 
 diagonal in the basis 
$\lbrace \ket{k \delim N-k} \rbrace$. 
We therefore  expect the entanglement to vanish  at the avoided crossing. This is borne out in Fig. \ref{fig:energySpectrumSVNEvaryLambda} 
 in which  
$\xi_{\textsc{svne}}$   
for the state $\ket{\psi(4,k)}$ is plotted  
for different values of $k$. 
As before,  it suffices to depict the cases  
$k = 0,1$ and $2$ because of the $k \leftrightarrow N-k$ 
symmetry. We observe that, in the 
case $k = 0$, 
 while there is a  minimum 
in $\xi_{\textsc{svne}}$ at $\lambda = 0$, 
there is a maximum in this quantity at 
$\omega_{1} = 0$  
(Fig. \ref{fig:energySpectrumSVNE}(b)).  

We have also calculated the 
PCC between various indicators and $\xi_{\textsc{svne}}$ for the set of states $\ket{\psi_{4,k}}, \,0\leqslant k \leqslant 4$. For this purpose, we have used 
$100$ values of each of 
the $\xi$-indicators 
calculated for each 
$\lambda$ in the range $[-1, 1]$ with a step size of 
$0.02$. The results are   very similar to those 
already found 
(see Fig. \ref{fig:CorrPlotN4VaryOmega1})
using  $\omega_{1}$ as the variable 
parameter instead of $\lambda$.

\subsection{Atom-field interaction model}

We turn next to the case of a multi-level atom (modelled by an anharmonic oscillator) that is  linearly coupled with strength $g$ to a radiation field 
of  frequency $\omega_{\textsc{f}}$. The effective Hamiltonian (setting $\hbar = 1$ )  is given by~\cite{agarwalpuri}
\begin{equation}
H_{\textsc{af}}=\omega_{\textsc{f}} a^{\dagger} a +
 \omega_{\textsc{a}} b^{\dagger} b + \gamma b^{\dagger \,2} b^{2} + g ( a^{\dagger} b + a b^{\dagger}).
\label{eqn:HAF}
\end{equation}
 $\omega_{\textsc{a}}$ and 
 $\gamma$ ($> 0$ for stability) are 
 constants. 
  $(a,a^{\dagger})$ 
 and $(b,b^{\dagger})$ are the annihilation and creation operators for the field mode  and the oscillator mode, 
 respectively. As before, 
  $N_{\mathrm{tot}} = a^{\dagger}a+b^{\dagger}b$ and $[H_{\textsc{af}},N_{\mathrm{tot}}]=0$.  
  As in the BEC model 
  of the preceding section, the eigenvalues $E_{\af}(N,k)$ and the common eigenstates $\ket{\phi_{N,k}}$ of these two operators are labelled by  $N = 0,1,\ldots$ (the eigenvalue of 
$N_{\mathrm{tot}}$) and, within each $(N+1)$-dimensional 
subspace for a given $N$, by the index $k$ that runs from $0$ to $N$. 

\begin{figure}
\includegraphics[width=0.45\textwidth]{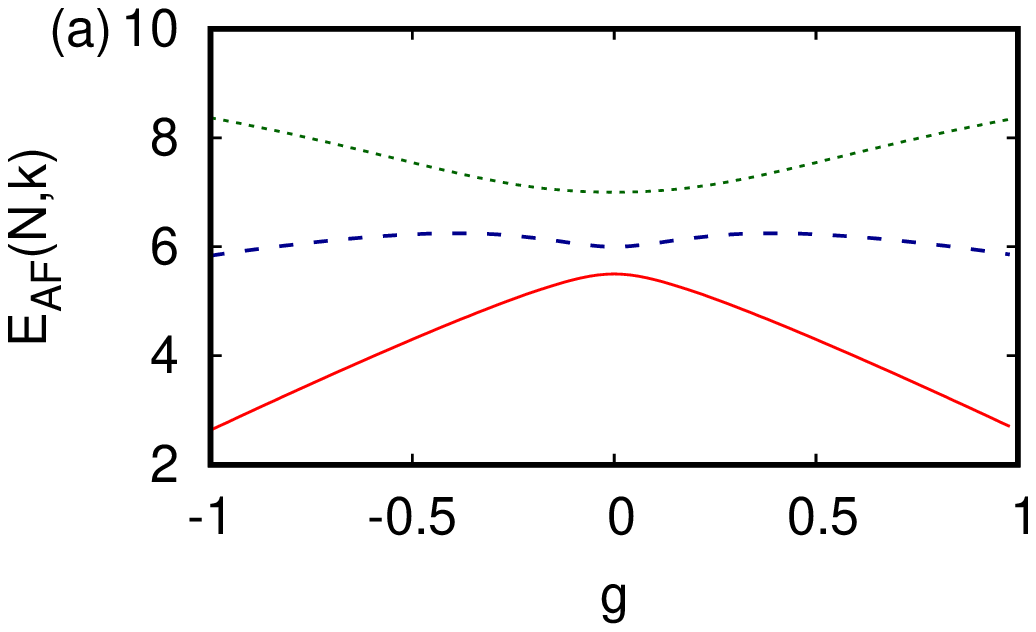}
\includegraphics[width=0.45\textwidth]{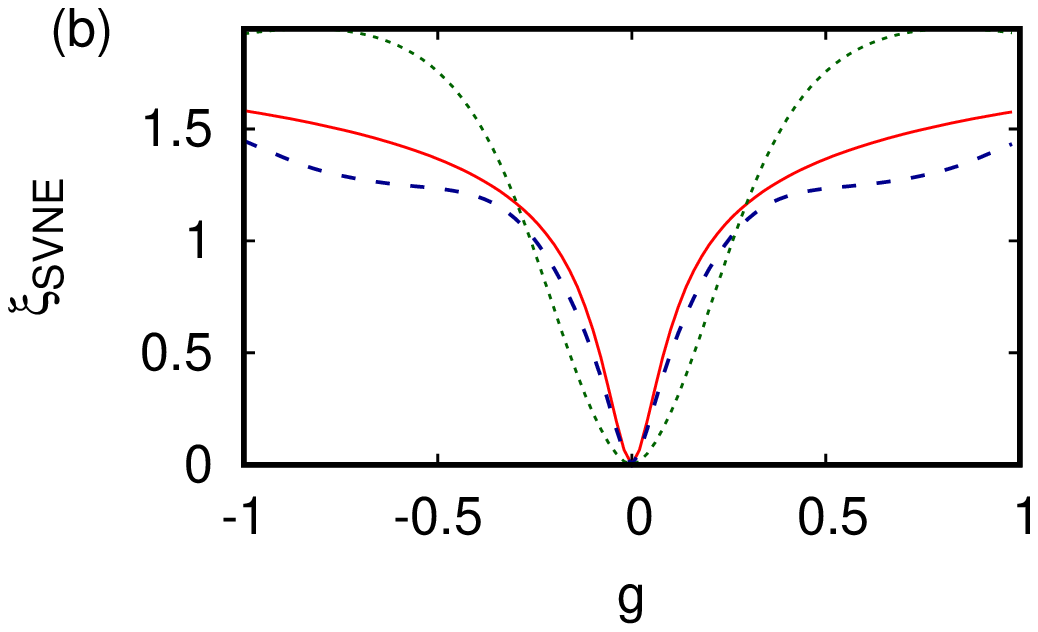}
\caption{(a) $E_{\af}(N,k)$ and (b) $\xi_{\textsc{svne}}$ vs. $g$ for $N=4$, $k=0,1,2$ in the atom-field interaction model. The curves correspond to $k=0$ (red), $1$ (blue) and $2$ (green). $\omega_{\textsc{f}}=1.5, \,\omega_{\textsc{a}}=1, \,
\gamma=1$.}
\label{fig:energySpectrumSVNEagarwalVaryLambdaDetuned}
\end{figure}
 We find 
 $\ket{\phi_{N,k}}$ and 
 $E_{\af}(N,k)$ numerically. 
Figures \ref{fig:energySpectrumSVNEagarwalVaryLambdaDetuned}(a) and (b) are plots of  
$E_{\af}(N,k)$ and $\xi_{\textsc{svne}}$ versus $g$  for $N=4$  and  $k=0,1,2$ in the case   
$\omega_{\textsc{f}}=1.5, \,\omega_{\textsc{a}}=1$. 
Avoided crossings occur at $g=0$, 
with a corresponding  minimum in  
$\xi_{\textsc{svne}}$ 
that drops down to zero 
for each of the three states 
$\ket{\phi_{4,0}},  \ket{\phi_{4,1}}$ and 
$\ket{\phi_{4, 2}}$. These states are 
therefore unentangled at $g=0$, i.e., 
in the absence of interaction between the 
two modes of the bipartite system, as one might expect.  

In order to examine what happens when there 
{\em  is} a 
crossing of energy levels, we introduce a degeneracy 
by setting  $\omega_{\textsc{f}}= \omega_{\textsc{a}}$.
\begin{figure}
\includegraphics[width=0.45\textwidth]{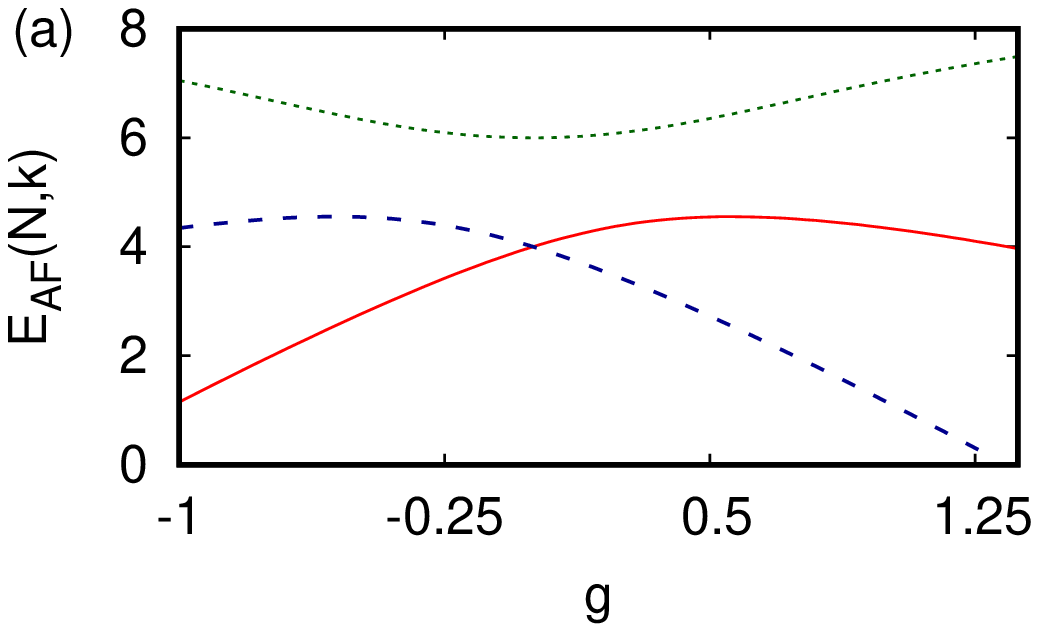}
\includegraphics[width=0.45\textwidth]{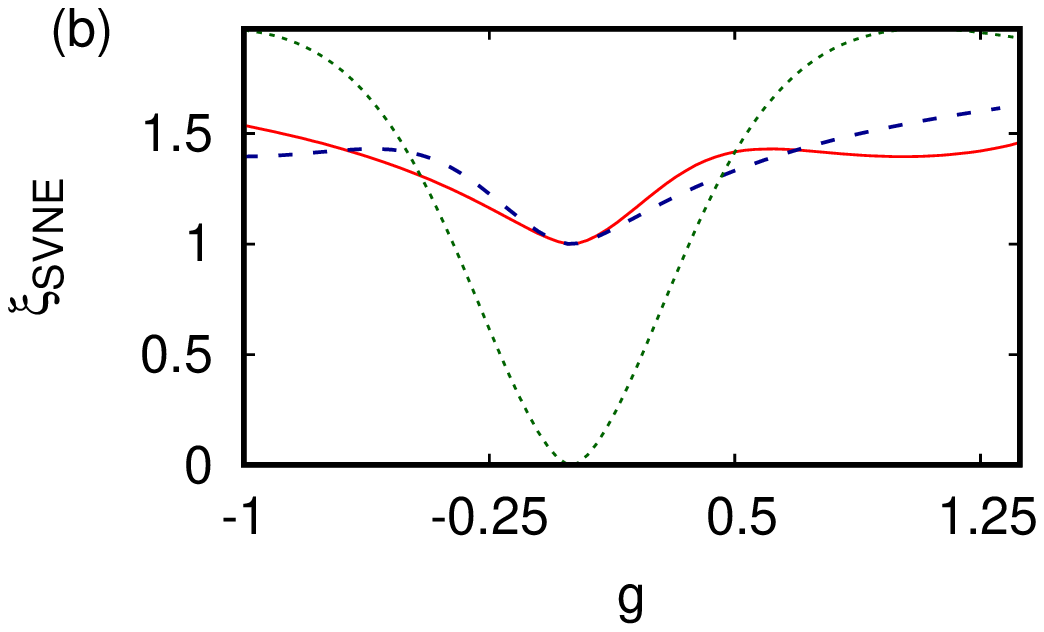}
\caption{(a) $E_{\af}(N,k)$ and (b) $\xi_{\textsc{svne}}$ vs. $g$ for $N=4$, $k=0,1,2$ in the atom-field interaction model, in the degenerate case  
$\omega_{\textsc{f}}=\omega_{\textsc{a}}
= 1$. The curves correspond to $k=0$ (red), $1$ (blue) and $2$ (green). 
$\gamma=1$.}
\label{fig:energySpectrumSVNEagarwalVaryLambda}
\end{figure}
Figures \ref{fig:energySpectrumSVNEagarwalVaryLambda} (a) and (b) are plots of  $E_{\af}(N,k)$ and  $\xi_{\textsc{svne}}$  versus $g$ for $N=4$ and 
$k = 0,1, 2$, with  
$\gamma$, $\omega_{\textsc{a}}$ and $\omega_{\textsc{f}}$  set equal to $1$.
Both a level crossing and an avoided crossing are seen to occur at $g=0$, signalled by a 
minimum in $\xi_{\textsc{svne}}$ for 
each of the three states concerned.  The crossing 
of $E_{\af}(4,0)$ and $E_{\af}(4,1)$ arises as follows. 
Let  $\ket{p \delim 4-p}$ denote the product state  $\ket{p}_{\textsc{f}}\otimes\ket{4-p}$, where  $\ket{p}_{\textsc{f}}$ is a photon number state of the field mode and $\ket{4-p}$ is an oscillator 
state of the atom mode. In the rest of this chapter, we drop the subscript F in the field state for ease of notation.
When $\gamma = \omega_{\textsc{a}} 
= \omega_{\textsc{f}} = 1$ and $g = 0$, the Hamiltonian 
reduces to $a^{\dagger} a + (b^{\dagger} b)^{2}$. 
The energy levels
$E_{\af}(4,0)$ and $E_{\af}(4,1)$ 
become degenerate 
at the value $4$. 
The degeneracy occurs because the 
operator $\ket{4 \delim 0}\bra{3 \delim 1} + \ket{3 \delim 1}\bra{4 \delim 0}$ 
commutes with 
$H_{\textsc{af}}$ when  $\omega_{\textsc{a}}=\omega_{\textsc{f}}$ and $g=0$.
Mixing of the states  
$\ket{4 \delim 0}$ and  $\ket{3 \delim 1}$
occurs, and the corresponding energy eigenstates are given by 
the symmetric linear combination 
$\ket{\phi_{4,0}} =
(\ket{4 \delim 0}+\ket{3 \delim 1})/\sqrt{2}$ 
and the antisymmetric linear combination 
$\ket{\phi_{4,1}} =  (\ket{4 \delim 0}-\ket{3 \delim 1})/\sqrt{2}$.  
As the symmetries of the two states are different, 
 the level crossing does not violate the 
von Neumann-Wigner no-crossing theorem. 
At the crossing,  each of the  states 
$\ket{\phi_{4,0}}$ and 
$\ket{\phi_{4,1}}$ 
remains a manifestly  
 entangled state that is,  in fact, a Bell state.  
This is why the corresponding 
$\xi_{\textsc{svne}}$ does not vanish at that point, but merely dips to a local minimum with value $1$, 
characteristic of a Bell state.
It is interesting to note that  the degeneracy  
that occurs when 
$\omega_{\textsc{f}}=\omega_{\textsc{a}}$ 
ensures entanglement even in the absence of 
any interaction between the two modes.  

The level
 $E_{\af}(4,2)$, on the other hand,  is repelled and has the value 
 $6$ at $g =  0$. The corresponding eigenstate 
 $\ket{\phi_{4,2}}$ becomes 
 the unentangled product state $\ket{2 \delim 2}$ at the 
 avoided crossing, and 
 $\xi_{\textsc{svne}}$ drops to zero in this case, as expected.

\begin{figure}
\includegraphics[width=0.3\textwidth]{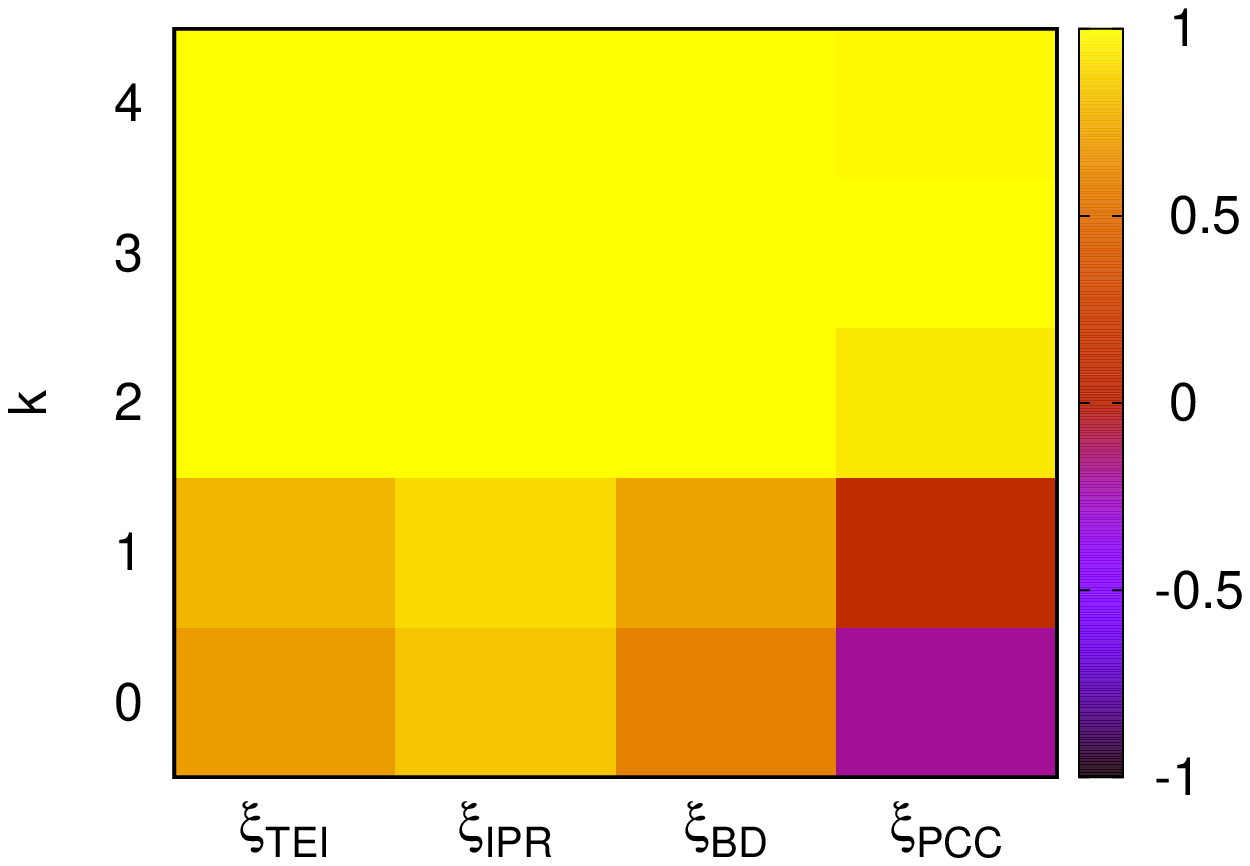}
\includegraphics[width=0.3\textwidth]{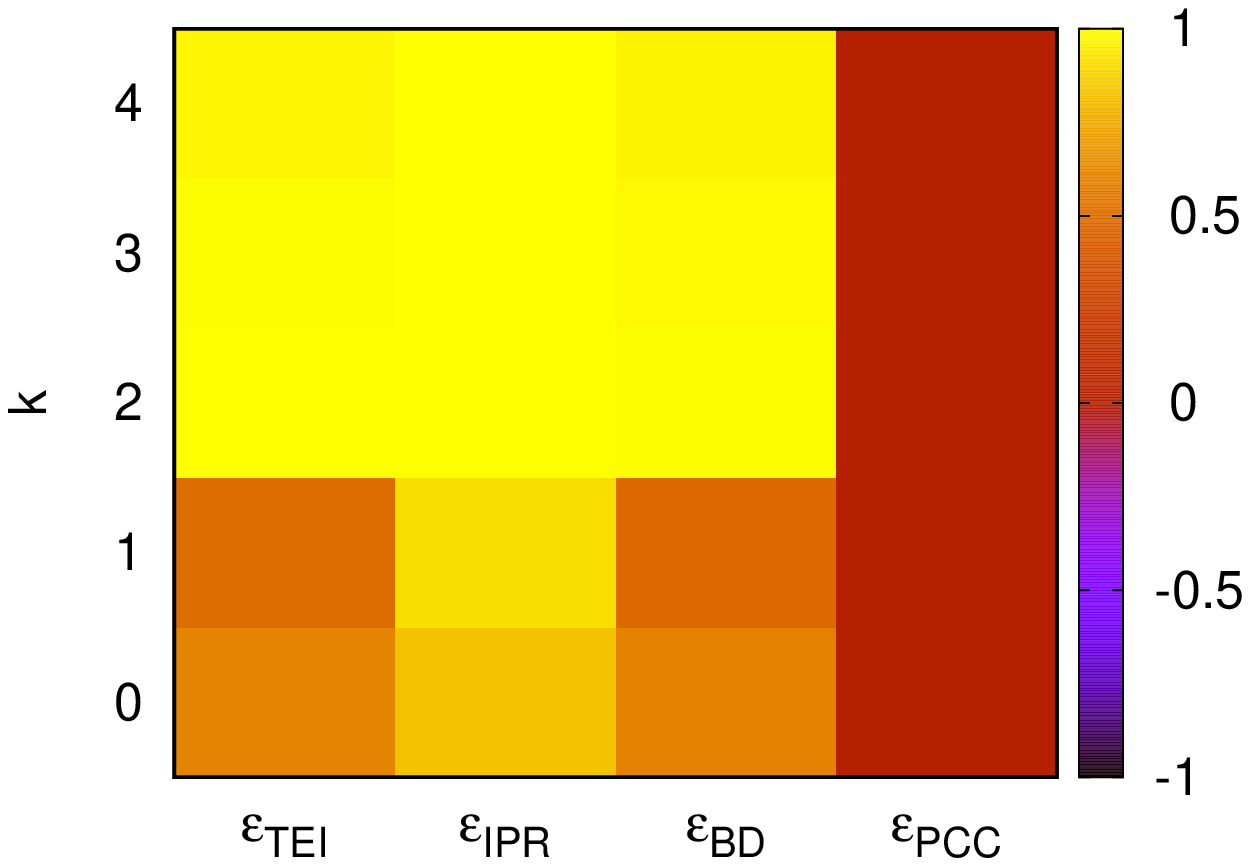}
\includegraphics[width=0.3\textwidth]{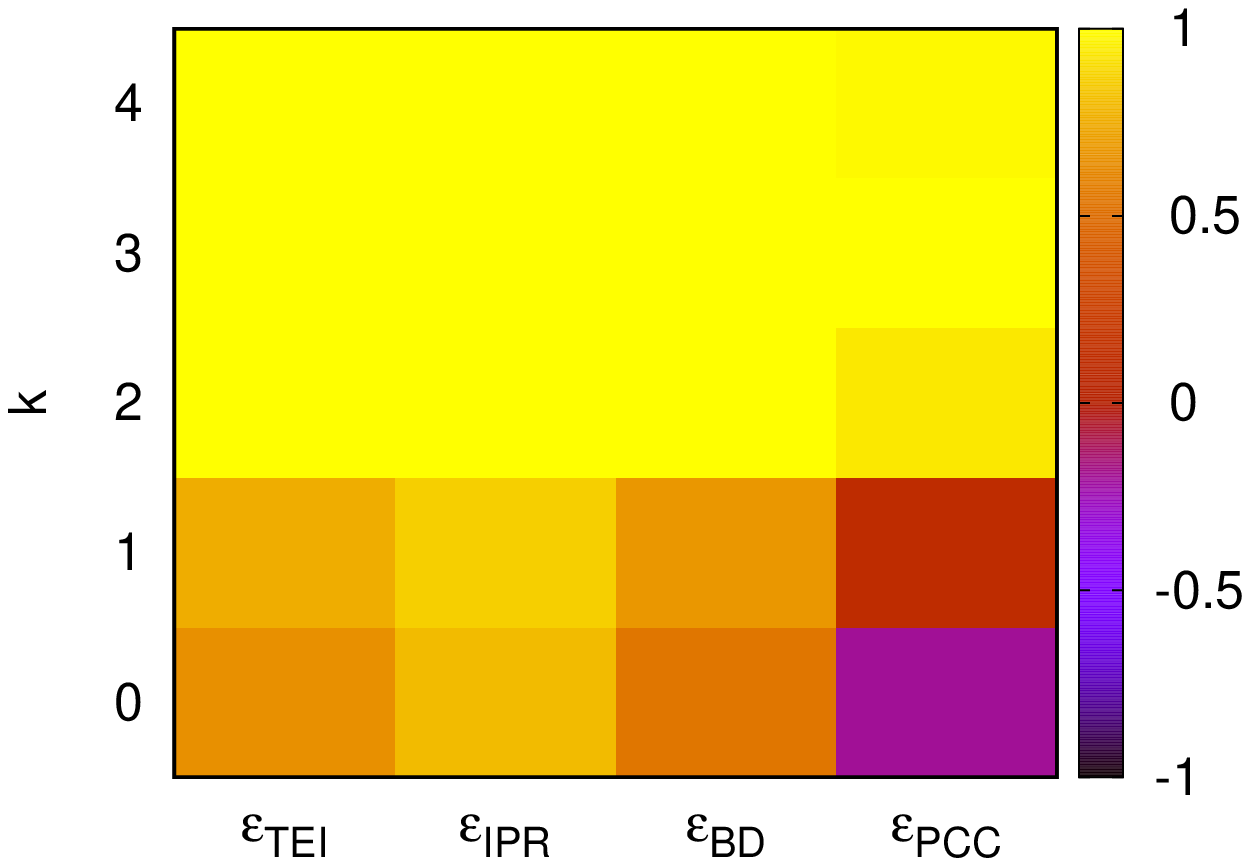}
\caption{Correlation of $\xi_{\textsc{svne}}$ 
with 
 $\xi$-indicators (left),   
 with $\neweps$-indicators for the slice 
 $\thetaa=0,\thetab=\pi/2$ (centre), 
 and with 
 $\neweps$-indicators  for the slice 
 $\thetaa=0,\thetab=0$ (right), for the eigenstates $\ket{\phi_{4,k}}, 
 \,0 \leqslant k \leqslant 4$  in the atom-field interaction model.
 $\omega_{\textsc{f}}=\omega_{\textsc{a}}=
 \gamma=1$.}
\label{fig:CorrPlotN4agarwalVaryLambda}
\end{figure}

In Fig. \ref{fig:CorrPlotN4agarwalVaryLambda}, we plot 
the correlation between various indicators and 
$\xi_{\textsc{svne}}$. 
For  this purpose,  $80$ values of each of the $\xi$-indicators were calculated  
with  $g$ varied in the range $[-1,1.4]$ in  steps of 
$0.03$. Treating these as sets of random numbers, we obtain the PCC between the various indicators and 
$\xi_{\textsc{svne}}$,  as described in the foregoing. 
The performance of the entanglement indicators in this case is similar to that found  in the BEC system. With increase in $\gamma$, the efficacy of all the indicators is marginally decreased.

\section{\label{sec:Ch3work2}Avoided crossings in multipartite HQ systems}
\subsection{The Tavis-Cummings model}

As our third and final example,  we consider hybrid quantum systems comprising several qubits interacting with an external field.  
These systems are described by the class of  Tavis-Cummings models~\cite{tavis} in  a variety of  diverse physical situations  which include inherent field nonlinearities and inter-qubit interactions. The model we consider below is generic, applicable to a system of  several two-level atoms with nearest-neighbour couplings interacting with an external radiation field in the presence of  a   Kerr-like nonlinearity, or to a  chain of $M$ superconducting qubits interacting with a 
microwave field of frequency $\Omega_{\textsc{f}}$. 
In the latter case, the model Hamiltonian (setting $\hbar = 1$) is given  by~\cite{QuantMeta2,QuantMeta} 
\begin{align}
\nonumber H_{\textsc{tc}}=\Omega_{\textsc{f}} 
a^{\dagger} a + \chi a^{\dagger \,2} a^{2}  + \sum_{p=1}^{M} &\Omega_{p} \sigma_{p z} + \Lambda ( a^{\dagger} \sigma_{p}^{-} + a \sigma_{p}^{+})\\
&+ \sum_{p=1}^{M-1} \Lambda_{s} ( \sigma_{p}^{-} \sigma_{(p+1)}^{+} + \sigma_{(p+1)}^{-} \sigma_{p}^{+}).
\label{eqn:HTC}
\end{align}
Here, $\chi$ is the strength of the field nonlinearity, 
$\Lambda$ is the coupling strength between the 
field and each of the $M$ qubits, 
$\sigma_{p}^{\pm}$ are the ladder operators of the 
$p^{\rm th}$ qubit, and $\Lambda_{s}$  is the strength of the interaction between nearest-neighbour qubits. 
$\Omega_{p} = (\Delta_{p}^{2} + \epsilon^{2})^{1/2}$ 
 is the energy difference   between the two levels of the $p^{\rm th}$ qubit, where   $\Delta_{p}$ is the inherent excitation gap and $\epsilon$  is the detuning of the external magnetic flux from the flux quantum 
 $h/(2 e)$. In our numerical computations we have 
 used experimentally  relevant 
 parameter values~\cite{QuantMeta}, namely, $\Omega_{\textsc{f}}/(2 \pi)
 = 7.78 \,{\rm   GHz}$ and  $\epsilon/(2 \pi) =  4.62\,
 {\rm  GHz}$. 
 The  level separations  $\Delta_{p}$  of the individual qubits have been  
 drawn from a  Gaussian  distribution with a mean given by 
  $\aver{\Delta}/(2 \pi) =  5.6\,{\rm  GHz}$ and a standard 
  deviation  $0.2 \,\aver{\Delta}$.

\begin{figure}
\includegraphics[width=0.4\textwidth]{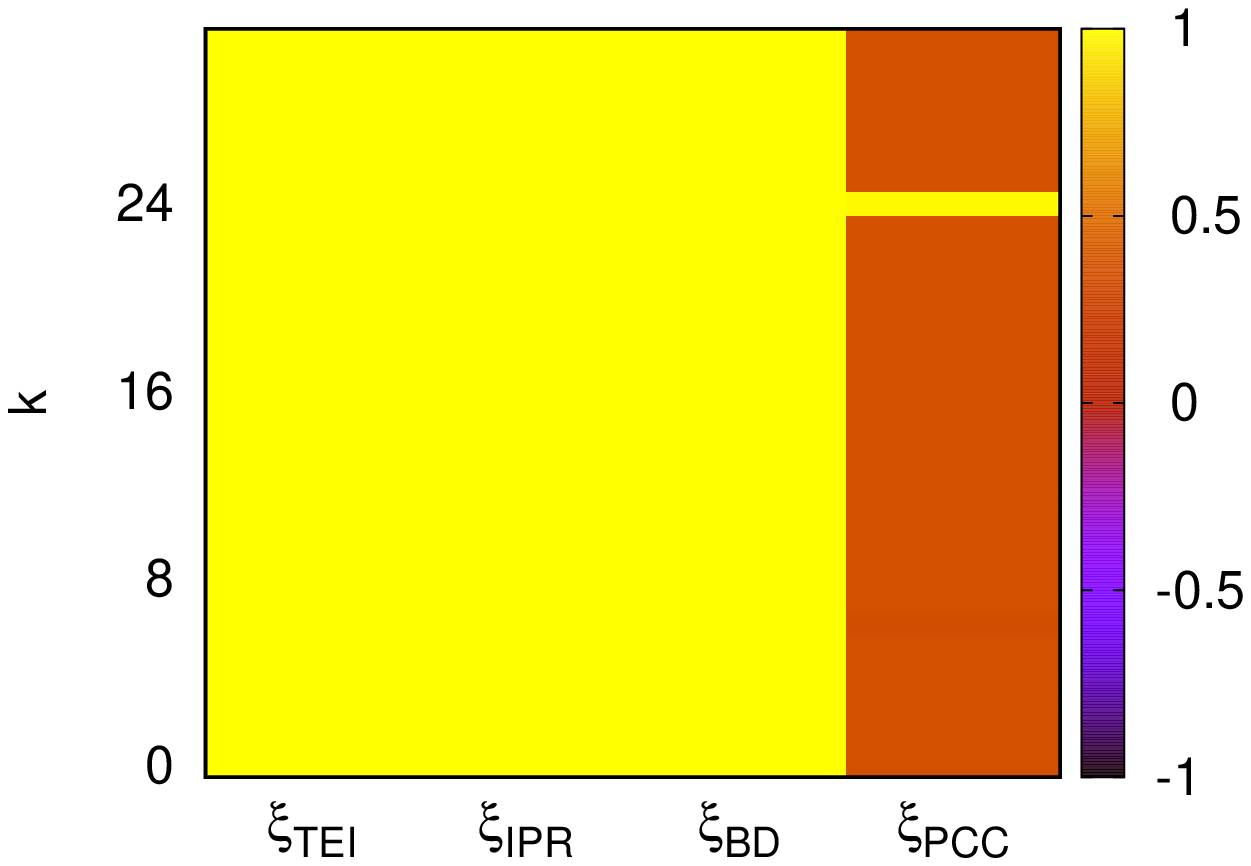}
\includegraphics[width=0.4\textwidth]{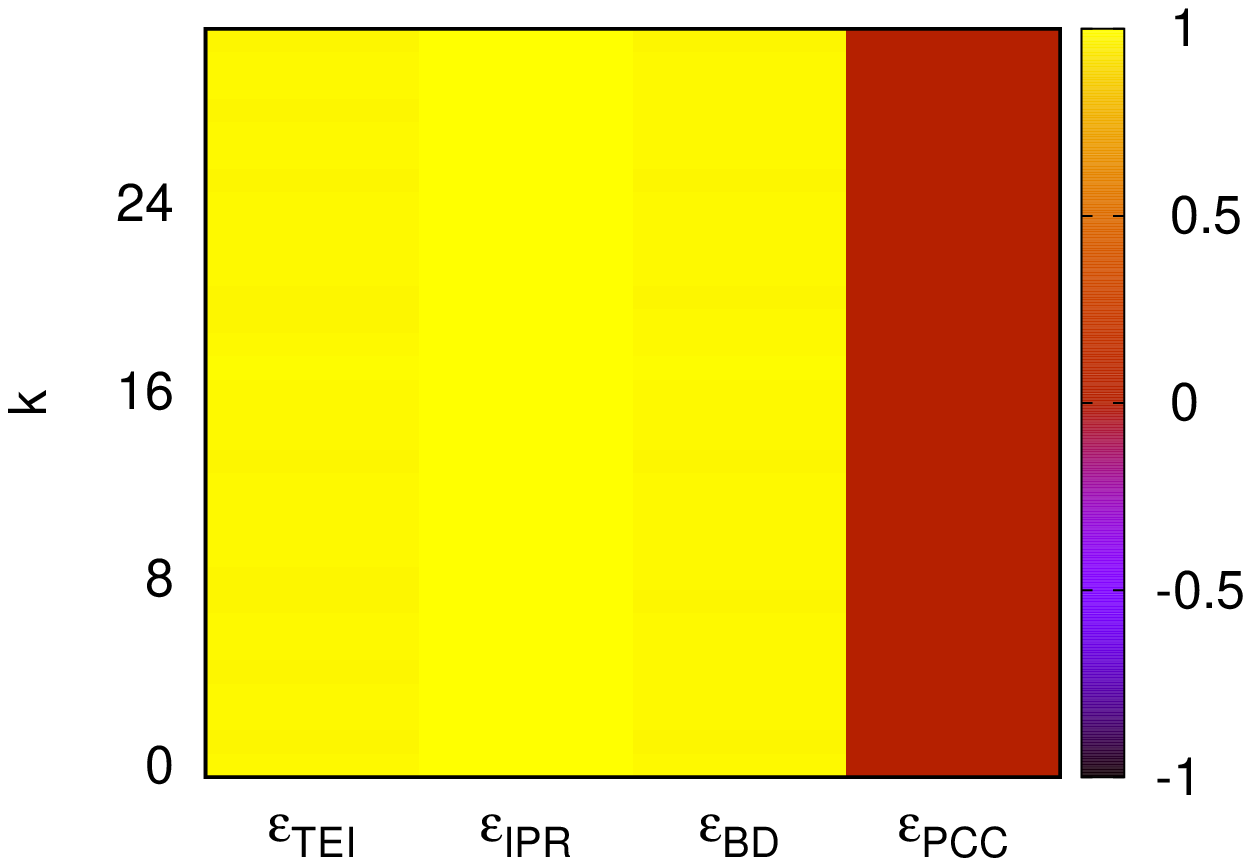}
\caption{Correlation of $\xi_{\textsc{svne}}$ 
with 
 $\xi$-indicators (left) and   
 with $\neweps$-indicators for the slice 
 corresponding to $\theta=\pi/2$ for the field, and 
 the $\sigma_{x}$ basis for each qubit (right) for different values of $\Lambda$.
 The figures are 
 for the eigenstates $\ket{\psi_{5,6,k}}, 
 \,0\leqslant k \leqslant 2^{5}-1$ in Case (i)  in the Tavis-Cummings model.
}
\label{fig:CorrPlotNat5Ntc6TCMVaryg}
\end{figure}

 We  have considered three cases, namely,
(i)\, $\Lambda_{s} = \chi = 0$, \,(ii)\,   
$\Lambda_{s}/(2 \pi) = 1\,{\rm MHz}, \, \chi=0$ 
 and \, (iii)\, $\Lambda_{s}/(2\pi) = 
\chi/(2\pi)  = 1\, {\rm  MHz}$.
In each case, $\Lambda/(2\pi) $ is varied from 
$- 1.2\,{\rm  MHz}$  to $ 1.3\,{\rm  MHz}$  
in steps of $0.025\,{\rm  MHz}$. It is easily 
shown that the total number operator 
\begin{equation*}
\mathcal{N}_{\mathrm{tot}}
=a^{\dagger}a + \sum_{p=1}^{M} \sigma_{p}^{+} \sigma_{p}^{-},
\end{equation*} 
commutes with  $H_{\textsc{tc}}$.  For each value of $\Lambda$ we have numerically solved for the  complete set 
$\lbrace \ket{\psi_{M,N,k}} \rbrace$ 
of common eigenstates of 
 $\mathcal{N}_{\mathrm{tot}}$ and 
 $H_{\textsc{tc}}$, 
where $N = 0,1,\ldots$ is the eigenvalue of 
 $\mathcal{N}_{\mathrm{tot}}$ and 
 $k = 0, 1, \ldots, 2^{M}-1$. 
   Considering the  total system as a bipartite composition of the field subsystem and a subsystem comprising all 
 the qubits, we have computed the entanglement 
 indicators. Figure \ref{fig:CorrPlotNat5Ntc6TCMVaryg} 
 shows   the correlation between the indicators  and 
 $\xi_{\textsc{svne}}$ in Case (i). The associated  
 Pearson correlation coefficients are $0.97$ for 
$\epsarg{tei}$,  $0.99$ for $\epsarg{ipr}$,
$0.97$ for $\epsarg{bd}$, correct to 
two decimal places. 
(The accuracy  of the  $\neweps$-indicators depends, of course,  on the basis chosen.) 
On averaging, we obtain 
the corresponding $\xi$-indicators  
with a PCC equal to $0.99$, showing that these indicators   track  $\xi_{\textsc{svne}}$ very closely. 
We have carried out a similar exercise in Cases 
(ii) and (iii). The results and the inferences drawn from them 
are broadly similar to those 
found in Case (i).  

\begin{figure}
\includegraphics[width=0.32\textwidth]{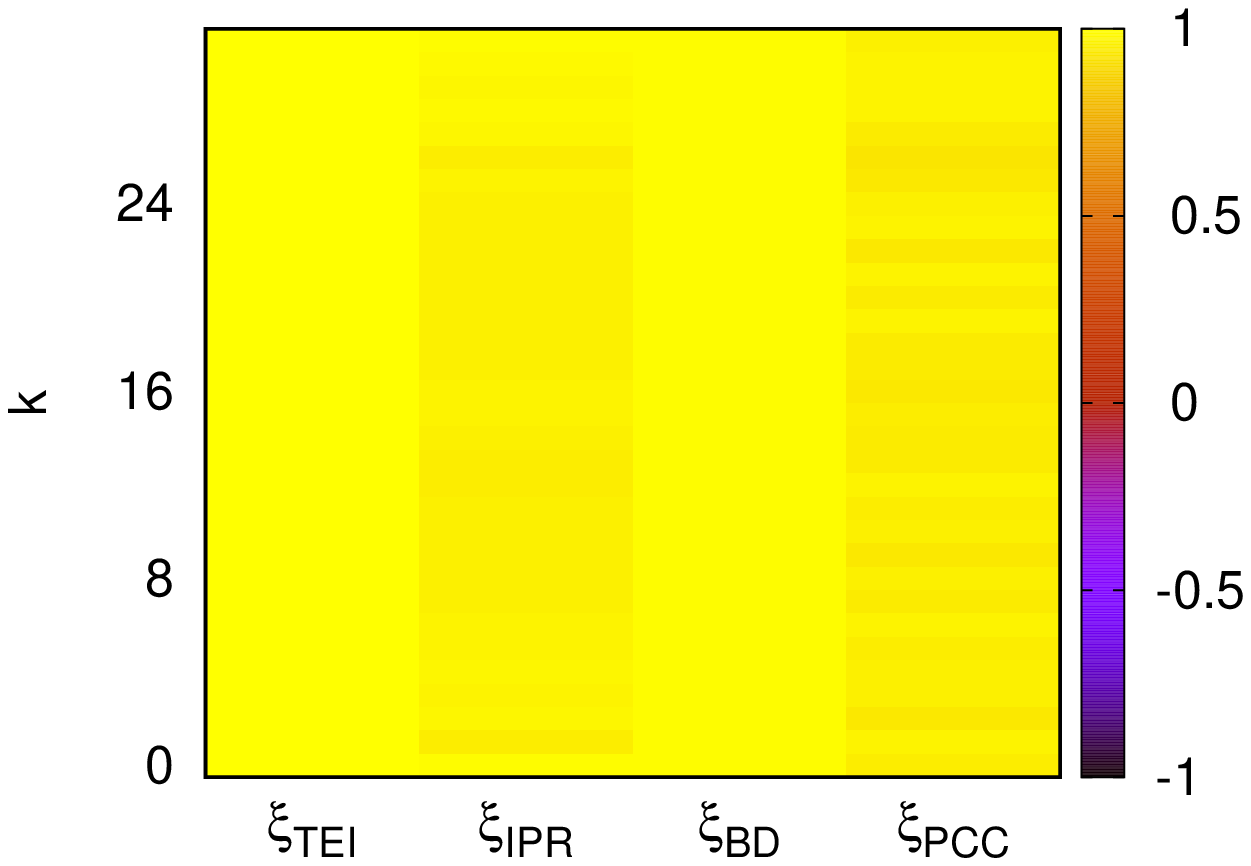}
\includegraphics[width=0.32\textwidth]{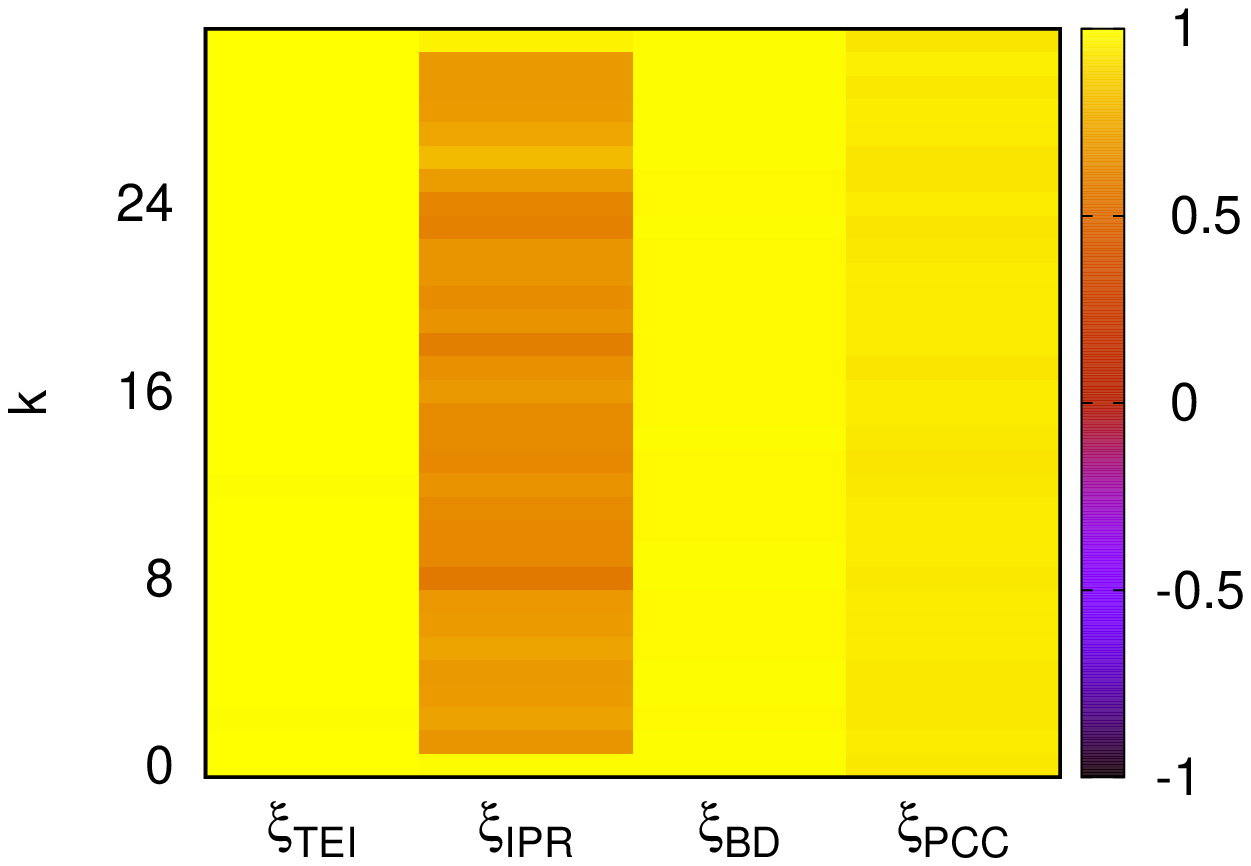}
\includegraphics[width=0.32\textwidth]{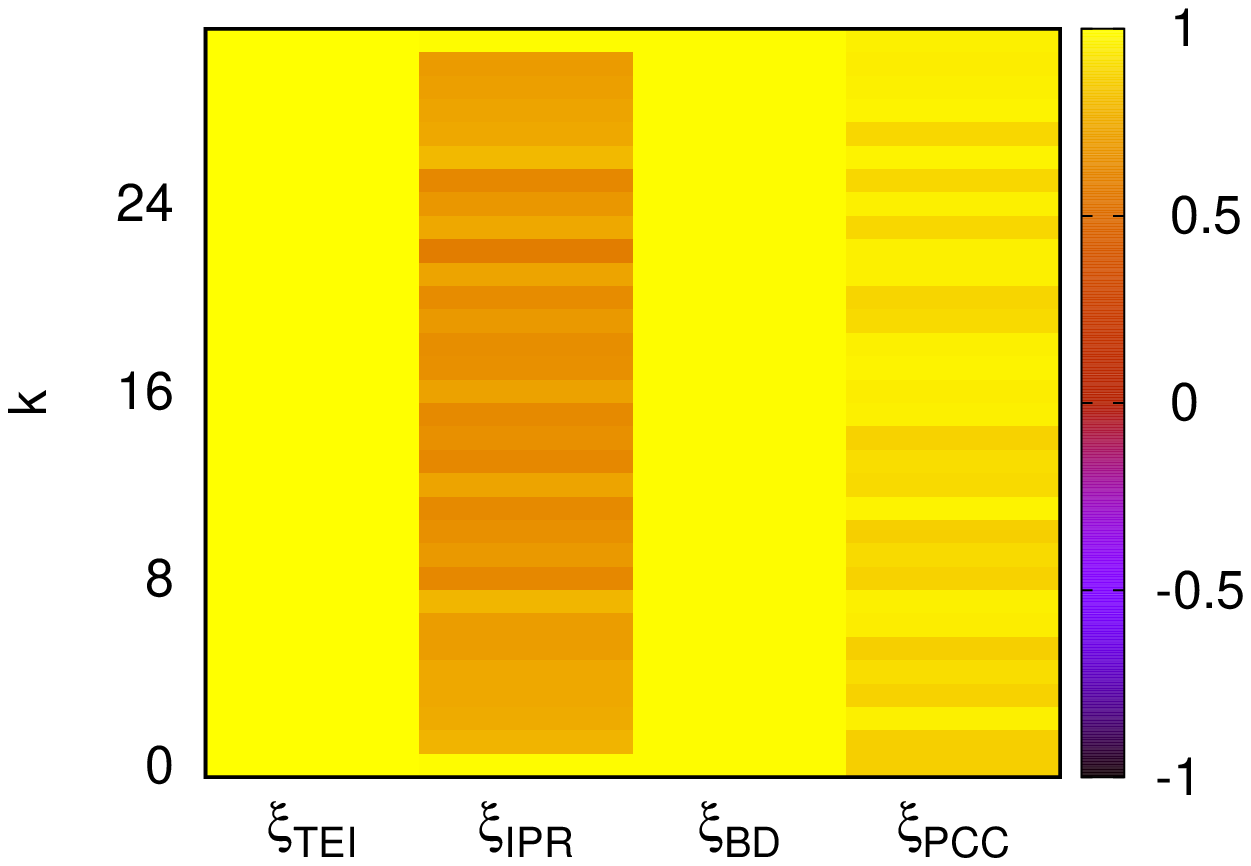}
\caption{Correlation between $\xi_{\textsc{svne}}$ 
and 
 $\xi$-indicators for different values of disorder strength and eigenstates $\ket{\zeta_{5,6,k}}\,0\leqslant k \leqslant 2^{5}-1$. Left to right, Cases (i), (ii) and (iii) respectively in the Tavis-Cummings model.}
\label{fig:CorrPlotNat5Ntc6VaryDis}
\end{figure}

Finally, with  $\Lambda/(2 \pi)$ set equal to  
 $1.2 \,{\rm  MHz}$, we have examined the effect of changing 
 the  strength of the disorder in $\Omega_{p}$ 
  by varying the standard deviation 
  of $\Delta_{p}$ 
   from $0$ to $0.2 \,\aver{\Delta}$ in steps of 
   $2\times 10^{-4}\, \aver{\Delta}$. Calculating the entanglement indicators for each disorder strength in 
   $\Omega_{p}$, we have found the correlations between the $\xi$-indicators and $\xi_{\textsc{svne}}$ in 
   Cases (i), (ii), and (iii). 
     $\xi_{\textsc{tei}}$ and $\xi_{\textsc{bd}}$ turn out to be significantly closer to 
  $\xi_{\textsc{svne}}$, and hence more accurate 
  indicators of entanglement, 
    than $\xi_{\textsc{ipr}}$ and $\xi_{\textsc{pcc}}$
(see Fig. \ref{fig:CorrPlotNat5Ntc6VaryDis}).

\section{\label{sec:Ch3conc}Concluding remarks}
In this chapter, we have considered  generic bipartite continuous-variable systems and hybrid quantum systems in the presence of nonlinearities, and  
tested quantitatively the efficacy of various indicators in estimating entanglement directly from quantum state tomograms close to avoided 
energy-level crossings. We find that the nonlinear correlation between the respective quadratures 
of the  two subsystems reflects very reliably  the extent of entanglement in bipartite CV systems governed by number-conserving Hamiltonians.  We have shown that if the eigenstates of the Hamiltonian are superpositions of Hamming-uncorrelated states, the inverse-participation-ratio-based 
quantifier  $\xi_{\textsc{ipr}}$ is an  excellent indicator of entanglement near avoided  crossings. In fact, even $\epsarg{ipr}$ (the corresponding indicator for a single section of the tomogram) suffices to estimate entanglement reliably. 
The tomographic entanglement indicator 
$\xi_{\textsc{tei}}$ and the Bhattacharyya-distance-based indicator $\xi_{\textsc{bd}}$ are  also good indicators  at avoided crossings, in contrast to the linear correlator $\xi_{\textsc{pcc}}$ which is based on the Pearson correlation coefficient.  
 Entanglement indicators seem to perform better with increasing   
$\aver{N_{\mathrm{tot}}}$.  
The conclusions drawn are both significant and readily applicable in identifying optimal entanglement indicators that are easily obtained from tomograms, without employing state reconstruction procedures.

In the next chapter, we investigate the performance of some of these entanglement indicators during temporal evolution of bipartite CV systems.

%% file: chapter4.tex

\chapter{Assessment and comparison of entanglement indicators in continuous-variable systems}
\label{ch:TEItimeseries}

\section{Introduction}
\label{sec:Ch4intro}

In this chapter, we first examine the entanglement dynamics of bipartite CV systems. We have seen that $\xi_{\textsc{ipr}}$ and $\xi_{\textsc{tei}}$ are good entanglement indicators close to avoided crossings. The entanglement indicators we consider here are $\xi_{\textsc{ipr}}$ and $\xi^{\prime}_{\textsc{tei}}$. The latter is a modification of $\xi_{\textsc{tei}}$. We will show in subsequent sections that it performs as well as $\xi_{\textsc{tei}}$ in the context of entanglement dynamics in CV systems.
The performance of these two indicators are compared at specific instants with both $\xi_{\textsc{svne}}$ and $\xi_{\textsc{sle}}$. We recall that $\xi_{\textsc{svne}}$ is given by $-\mathrm{Tr}\,(\rho_{i} \,\log_{2} \,\rho_{i})$, and the subsystem linear entropy $\xi_{\textsc{sle}}$ is  $1-\mathrm{Tr}\,(\rho_{i}^{2})$ where $\rho_{i}$ is the subsystem density matrix ($i$ denotes subsystems A, B). These two indicators involve both off-diagonal and diagonal elements of the density matrix in any given basis. In contrast, the tomogram only provides information about the diagonal elements, although in several complete bases. It is therefore necessary to carry out a detailed comparative study between the tomographic indicators, on the one hand, and $\xi_{\textsc{svne}}$ and $\xi_{\textsc{sle}}$, on the other, in order to assess their efficacy and limitations.

The systems we consider for our purpose are a multi-level atom interacting with a radiation field~\cite{agarwalpuri}, and the double-well BEC system~\cite{sanz}, examined in Chapter \ref{ch:EIatAC}, in a different context.

We emphasize that $\xi^{\prime}_{\textsc{tei}}$ is not an entanglement \textit{measure} in contrast to $\xi_{\textsc{svne}}$ and $\xi_{\textsc{sle}}$. Keeping this in mind, we have also obtained a long data set of the difference $d_{1}(t)$ between 
$\xi_{\textsc{svne}}$ and $\xi^{\prime}_{\textsc{tei}}$ at various instants of time, and carried out a detailed time-series analysis for several initial states and for 
different strengths of nonlinearity. The ergodicity properties of $d_{1}(t)$ carry information on the performance of $\xi^{\prime}_{\textsc{tei}}$. 

In a recent experiment reported in the literature~\cite{perola}, photon coincidence counts were used to distinguish between two $2$-photon states in a CV bipartite system. Here we have demonstrated that by computing $\epsarg{tei}$ (defined in Eq. \eref{eqn:epsTEI}) from the tomograms corresponding to the two states, we can easily distinguish between them. Through this calculation we have extended our study to chronocyclic tomograms, which are explained in detail in Section~\ref{sec:Ch4chrono}.

The plan of this chapter is as follows. In Section \ref{sec:Ch4_2}, we outline the procedure for obtaining $\xi^{\prime}_{\textsc{tei}}$ from $\epsarg{tei}$. In Section \ref{sec:Ch4models}, we compare $\xi_{\textsc{ipr}}$ and $\xi^{\prime}_{\textsc{tei}}$ with $\xi_{\textsc{svne}}$ and $\xi_{\textsc{sle}}$ during dynamical evolution of both the double-well BEC sysem and the multi-level atom interacting with the radiation field. Section \ref{sec:Ch4NTSA} is devoted to the time-series analysis. In Section \ref{sec:Ch4chrono}, we briefly review the salient features of chronocyclic tomograms, and demonstrate the usefulness of the tomographic indicator in distinguishing between two entangled states. We conclude with brief remarks.

\section{$\xi$-indicators and averaging procedures}
\label{sec:Ch4_2}

\begin{figure}
\centering
\includegraphics[width=0.49\textwidth]{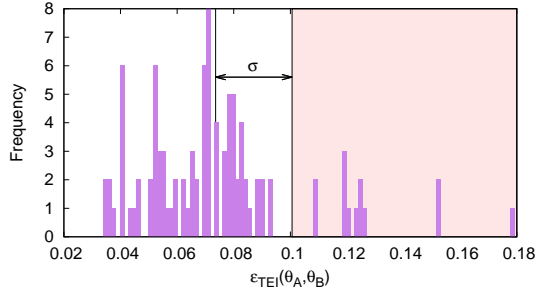}
\caption{Histogram of $\epsarg{tei}(\thetaa,\thetab)$ from $100$ combinations of $\thetaa$ and $\thetab$. The shaded area marks the values of $\epsarg{tei}(\thetaa,\thetab)$ that exceed the mean value $\sim 0.07$ by one standard deviation.}
\label{fig:DTEIexpln}
\end{figure}

The generalised eigenstates of conjugate pairs of quadrature operators constitute a pair of mutually unbiased bases~\cite{xpMUB}, as
\begin{equation}
\left\vert\aver{X_{\theta},\theta|X^{\prime}_{\theta+\pi/2},\theta+\pi/2}
\right\vert = 1/\sqrt{2 \pi \hbar} \,> 0.
\label{eqn:xpMUB}
\end{equation}
We recall that the specific averaging procedure used to obtain any $\xi$-indicator mentioned in Section \ref{sec:Ch3review} involves calculating the corresponding $\neweps$-indicator in several sets of mutually unbiased bases. $\xi^{\prime}_{\textsc{tei}}$ is obtained by averaging \textit{only} over the dominant values of $\epsarg{tei}(\thetaa,\thetab)$. Here we have averaged over $100$ values of $\epsarg{tei}(\thetaa,\thetab)$ to obtain $\xi_{\textsc{tei}}$. $\xi^{\prime}_{\textsc{tei}}$ is calculated only using those values that exceed the mean by one standard deviation. For this purpose we have numerically generated several histograms of $\epsarg{tei}(\thetaa,\thetab)$ for both the atom-field interaction and the double-well BEC models. For a given initial state, the system is unitarily evolved in time, and each histogram corresponds to a specific instant of time.  A typical histogram is presented in  Fig. \ref{fig:DTEIexpln} . The average of $\epsarg{tei}(\thetaa,\thetab)$ over {\em all} the values in the histogram in Fig. \ref{fig:DTEIexpln} has been compared with that obtained by averaging only over the contribution from the shaded portion. The qualitative features were  found to be essentially the same in both cases in all the histograms considered. The results in the following sections are therefore based on the latter averaging procedure, since it is clearly computationally less intensive. We denote the entanglement indicator thus obtained by  $\xi^{\prime}_{\textsc{tei}}$, and in subsequent sections compare $\xi^{\prime}_{\textsc{tei}}$ with $\xi_{\textsc{sle}}$ and $\xi_{\textsc{ipr}}$. 

\section{Entanglement indicators in generic bipartite models}
\label{sec:Ch4models} 

In this section, we examine in detail the performance of $\xi^{\prime}_{\textsc{tei}}$  in the atom-field interaction model \cite{agarwalpuri} and the double-well BEC model \cite{sanz}. 
We consider initial states of the 
total bipartite system that are pure states governed by unitary evolution. 

\subsection{Atom-field interaction model}

The effective Hamiltonian (Eq. \eref{eqn:HAF}), setting $\hbar=1$, is \cite{agarwalpuri}  
\begin{equation}
\nonumber H_{\textsc{af}}=\omega_{\textsc{f}} a^{\dagger} a + \omega_{\textsc{a}} b^{\dagger} b + \gamma b^{\dagger \,2} b^{2} + g ( a^{\dagger} b + a b^{\dagger}).
\end{equation}
We recall that ($a, a^{\dagger}$) are photon  annihilation and creation operators. 
The multi-level atom is modelled as an 
oscillator with harmonic frequency 
$\omega_{\textsc{a}}$ and  ladder operators ($b, b^{\dagger}$).
The anharmonicity of the oscillator is  effectively described by the Kerr-like term in $H_{\textsc{af}}$ with strength $\gamma$.  A  variety of initial states of the total system has been judiciously selected in order to explore the  range of possible nonclassical effects during time evolution.  
The {\em unentangled}  initial states considered correspond to the atom in its ground state $\ket{0}$ and the field in either a CS (e.g., $\ket{\alpha}_{\textsc{f}}$), 
 or an $m$-PACS (e.g., $\ket{\alpha,m}_{\textsc{f}}$). 
We also consider 
two {\em entangled} initial states, namely, the binomial state 
$\ket{\psi_{\rm{bin}}}$ (Eq. \eref{eqn:BS_defn}) and the two-mode squeezed state $\ket{\zeta}$ (Eq. \eref{eqn:sq_state}). 
We recall that the binomial state $\ket{\psi_{\rm{bin}}}$, for a non-negative integer $N$, is given by 
\begin{equation}
\nonumber \ket{\psi_{\rm{bin}}} = 2^{-N/2} \sum_{n=0}^{N} 
{\tbinom{N}{n}}^{1/2}
 \ket{N-n \delim n},
\end{equation}
where $\ket{N-n \delim n} \equiv  \ket{N-n}_{\textsc{f}}\otimes\ket{n}$, 
the product state corresponding to 
the field and the atom  
 in the respective 
 number states $\ket{N-n}$ and $\ket{n}$. 
 Also, $\ket{\zeta} = e^{\zeta^{*} a b - \zeta a^{\dagger} b^{\dagger}} \ket{0 \delim 0},$
($\zeta \in \mathbb{C}$) and  $\ket{0 \delim 0}$ is the 
 product state corresponding to 
 $N = 0, n = 0$. 
 
\begin{figure}[h]
\centering
\includegraphics[width=0.4\textwidth]{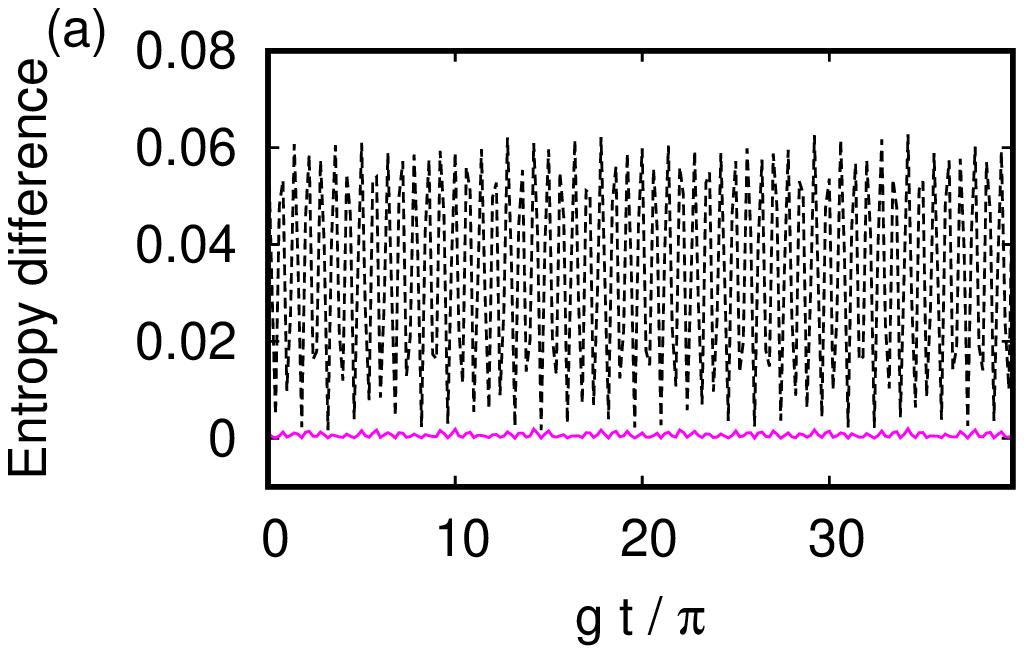}
\includegraphics[width=0.4\textwidth]{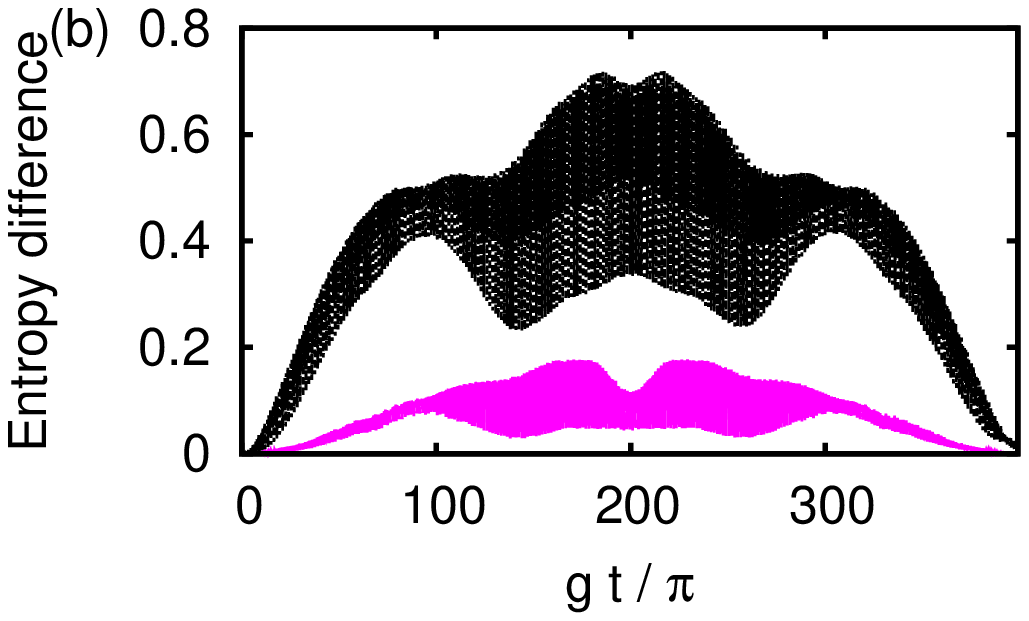}
\caption{$d_{1}(t)$ (black) and $d_{2}(t)$ (pink) vs. scaled time $g t/ \pi$, for $\omega_{\textsc{f}}=\omega_{\textsc{a}}=\gamma=1$ in the atom-field interaction model. (a) $g=0.2$, initial two-mode squeezed state $\ket{\zeta}, \,\zeta = 0.1$ (b) $g=100$, 
initial state $\ket{\alpha}_{\textsc{f}} \otimes \ket{0},  |\alpha|^{2}=1$. }
\label{fig:SLE_vs_SVNE}
\end{figure}

\begin{figure}[h]
\centering
\includegraphics[width=0.4\textwidth]{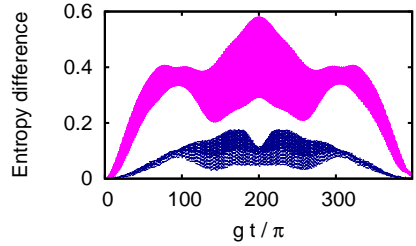}
\caption{$d_{2}(t)$ (blue) and $\Delta(t)$ (pink)
vs. scaled time $g t/\pi$,  with $\omega_{\textsc{f}}=\omega_{\textsc{a}}=\gamma=1$, $g=100$  in the atom-field interaction model. Initial state $\ket{\alpha}_{\textsc{f}} \otimes \ket{0}$, $|\alpha|^{2}=1$.}
\label{fig:comp_diff_SVNE}
\end{figure}

\begin{figure}[h]
\centering
\includegraphics[width=0.4\textwidth]{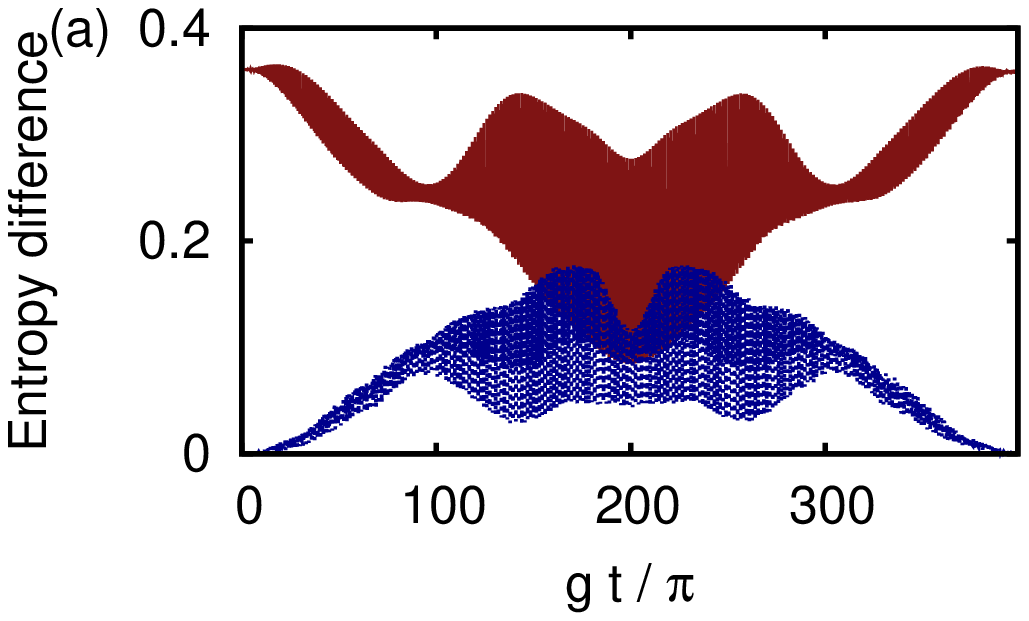}
\includegraphics[width=0.4\textwidth]{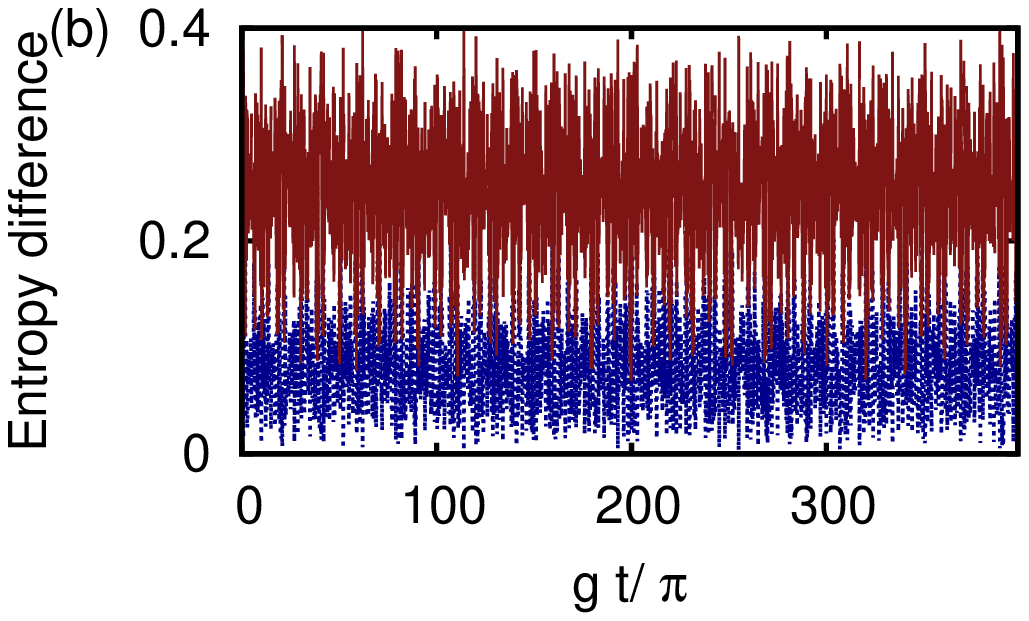}
\includegraphics[width=0.4\textwidth]{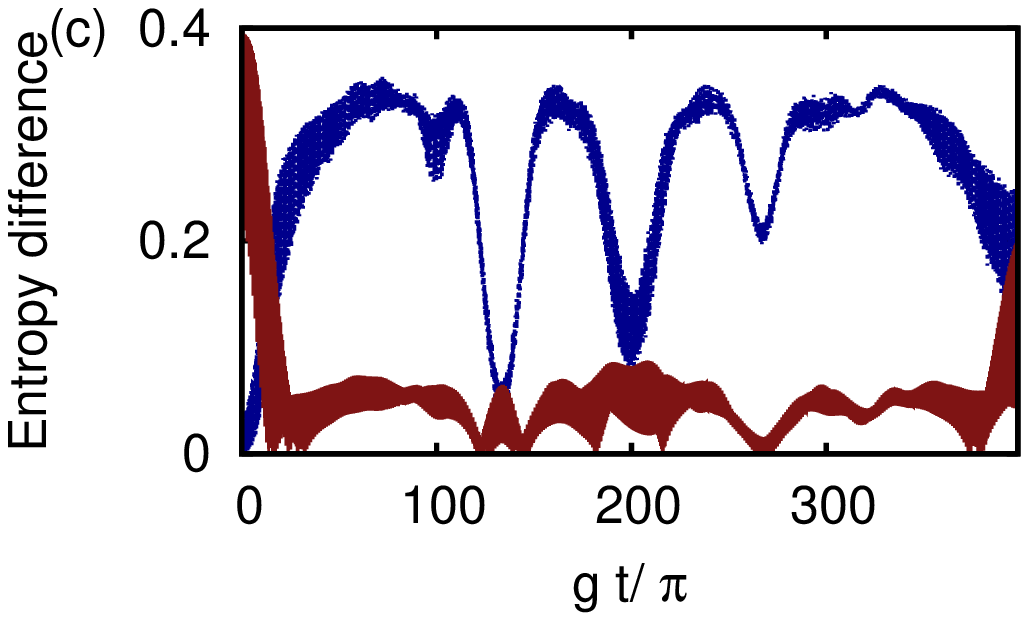}
\includegraphics[width=0.4\textwidth]{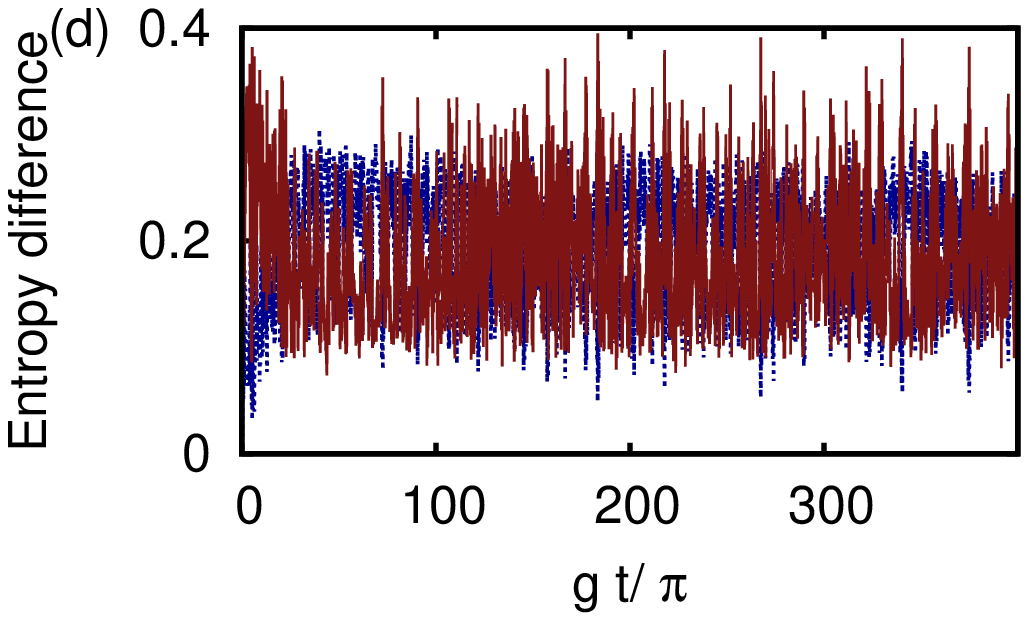}
\caption{$d_{2}(t)$ (blue) and $d_{3}(t)$ (brown) 
vs. scaled time $g t /\pi$, for $\omega_{\textsc{f}}=\omega_{\textsc{a}}=\gamma=|\alpha|^{2}=1$ in the atom-field interaction model. 
(a) and (b): Initial state $\ket{\alpha}_{\textsc{f}} \otimes \ket{0}$, $g=100$ and $0.2$ respectively. (c) and (d): Initial state $\ket{\alpha,5}_{\textsc{f}} \otimes \ket{0}$, $g=100$ and $0.2$ respectively. }
\label{fig:comp_diff_1}
\end{figure}

Corresponding to these initial states we have numerically generated tomograms at approximately 2000 instants, separated by a time step 0.2 $\pi / g$ as the system evolves. From these,  we have obtained $\xi^{\prime}_{\textsc{tei}}$ and  the differences
\begin{equation}
d_{1}(t) = \vert {\xi_{\textsc{svne}}} - \xi^{\prime}_{\textsc{tei}} \vert, \;\;
d_{2}(t) = \vert {\xi_{\textsc{sle}}} - \xi^{\prime}_{\textsc{tei}} \vert.
\label{eqn:d1d2defns}
\end{equation}
These differences are plotted against 
the scaled time $g t / \pi$ 
for an initial two-mode squeezed state (Fig. \ref{fig:SLE_vs_SVNE}(a)), and 
for a factored product of a  CS and atomic ground state $\ket{0}$ (Fig. \ref{fig:SLE_vs_SVNE}(b)). From these plots it is evident that $\xi^{\prime}_{\textsc{tei}}$ is in much better agreement with $\xi_{\textsc{sle}}$ than with $\xi_{\textsc{svne}}$ over the time interval considered, independent of the parameter values and the nature of the initial state.  We therefore choose $\xi_{\textsc{sle}}$ as the reference entanglement indicator.
Next, we compare $d_{2}(t)$ with the difference 
\begin{equation}
\Delta(t) = \vert {\xi_{\textsc{svne}}} - {\xi_{\textsc{sle}}} \vert.
\label{Deltadefn}
\end{equation} 
We have verified that in all the cases considered, 
$\Delta(t) >  d_{2}(t)$ (see, for instance, Fig. \ref{fig:comp_diff_SVNE}). 
In what follows,  we therefore focus only on   $d_{2}(t)$ and the difference 
\begin{equation}
d_{3}(t) = \vert {\xi_{\textsc{sle}}} - \xi_{\textsc{ipr}}\vert.
\label{d3defn}
\end{equation}
This comparison brings out interesting features of both the indicators. When the strength of the nonlinearity is low relative to that of the 
coupling (e.g., $\gamma/g=0.01$), it is known \cite{sudh_rev_moments} 
that  full and fractional revivals occur, and  
entanglement measures may be  expected to display signatures of 
these revival phenomena. From
Fig. \ref{fig:comp_diff_1} (a) we see that at the 
revival time  $g T_{\rm rev}/\pi=400$, $\xi^{\prime}_{\textsc{tei}}$ agrees with $\xi_{\textsc{sle}}$ much more closely  than $\xi_{\textsc{ipr}}$ does. Further, over the entire time interval $(0, T_{\rm rev})$, $d_{2}(t)$ is significantly smaller than $d_{3}(t)$. This feature holds even for larger values of the ratio $\gamma/g$, as can be seen from 
Fig. \ref{fig:comp_diff_1} (b).   
$\xi^{\prime}_{\textsc{tei}}$ is therefore favoured over $\xi_{\textsc{ipr}}$ as an entanglement indicator. The time evolution 
 of the difference $d_{2}(t)$ is drastically different from 
 that of  $d_{3}(t)$ for initial field states that depart from ideal coherence. In this case, over the entire time considered, $\xi_{\textsc{ipr}}$ performs significantly better than $\xi^{\prime}_{\textsc{tei}}$ for small values of $\gamma/g$ (Fig. \ref{fig:comp_diff_1} (c)). As  the value of $\gamma/g$ 
 is increased the two indicators have essentially the same  behaviour (Fig. \ref{fig:comp_diff_1} (d)).

\begin{figure}[h]
\centering
\includegraphics[width=0.4\textwidth]{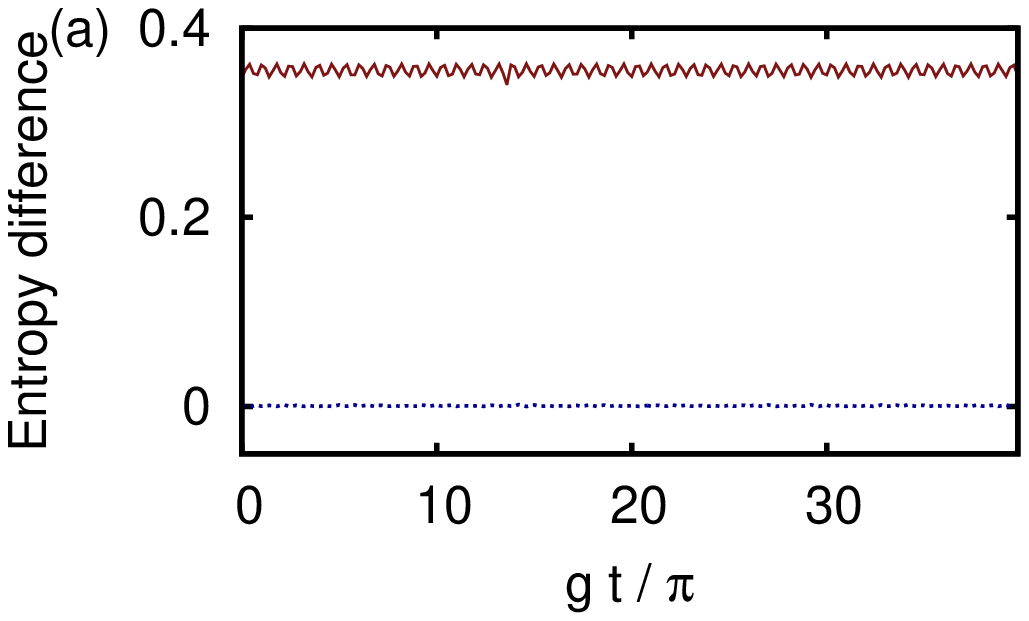}
\includegraphics[width=0.4\textwidth]{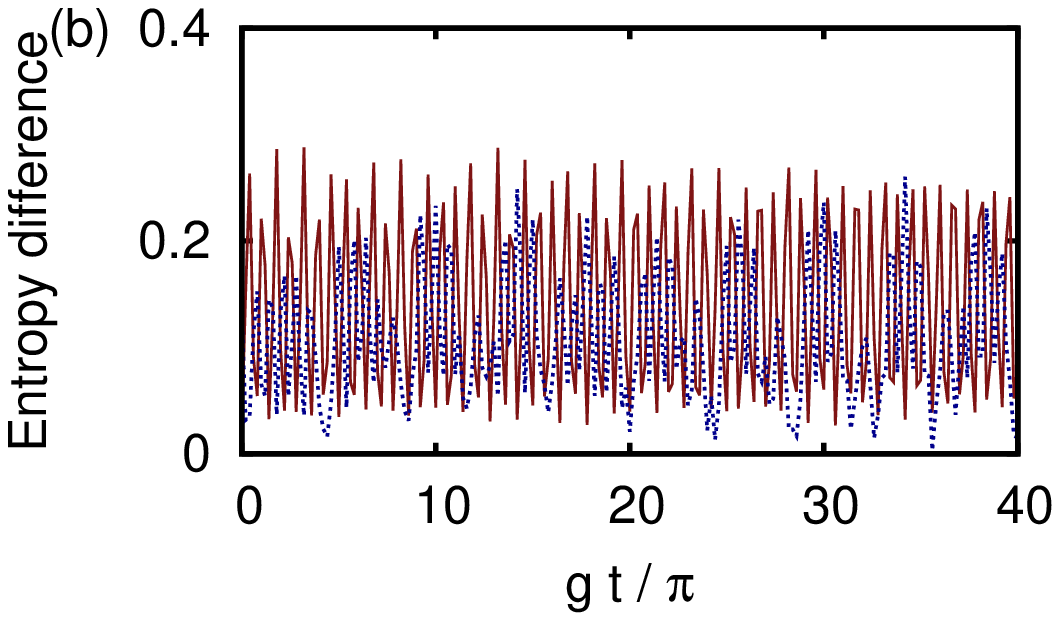}
\caption{$d_{2}(t)$ (blue) and $d_{3}(t)$ (brown) 
vs. scaled time $g t / \pi$, for $\omega_{\textsc{f}}=\omega_{\textsc{a}}=\gamma=1, g=0.2$ in the atom-field interaction model. Initial two-mode squeezed state 
$\ket{\zeta}$, (a) $\zeta=0.1$ and (b) $\zeta=0.7$. }
\label{fig:comp_diff_3}
\end{figure}

We turn now to  entangled initial states. In the case of the two-mode squeezed state $\ket{\zeta}$, we see from Figs. \ref{fig:comp_diff_3} (a) and (b) that  $\xi^{\prime}_{\textsc{tei}}$ fares much better than 
$\xi_{\textsc{ipr}}$ over the entire time interval  considered,  for small values of  $\zeta$.   With an increase in the value of $\zeta$,  both the indicators show comparable departures from $\xi_{\textsc{sle}}$.

\begin{figure}[h]
\centering
\includegraphics[width=0.4\textwidth]{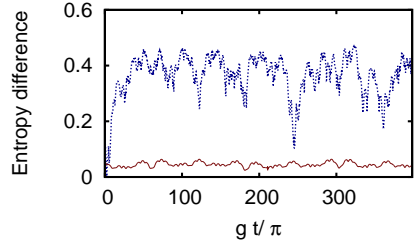}
\caption{$d_{2}(t)$ (blue) and $d_{3}(t)$ (brown) 
vs. scaled time $g t / \pi$,  for $\omega_{\textsc{f}}=\omega_{\textsc{a}}=\gamma=1, 
g=0.2$ in the atom-field interaction model. Initial state $\ket{\psi_{\rm{bin}}}$ with $N=10$.}
\label{fig:comp_diff_4}
\end{figure}

In the case of   an initial binomial state, on the 
other hand,   $\xi_{\textsc{ipr}}$ fares significantly better 
than  $\xi^{\prime}_{\textsc{tei}}$. This can be understood by examining the Hamming distance between  the 
basis states constituting  the binomial state. We define this distance in the context of  continuous variables by extrapolating the idea of Hamming distance for qudits as explained in Section \ref{sec:Ch3work}. As pointed out before, the efficacy of 
$\xi_{\textsc{ipr}}$ 
as an entanglement indicator increases with an 
increase in the Hamming distance.  This indicator is 
therefore especially useful  for superpositions of states that are Hamming-uncorrelated (i.e., separated by a Hamming distance equal to 2 in bipartite states) both in the context of spin systems \cite{ViolaBrown,HamCorBSSV} and CV systems~\cite{sharmila2,arxiv4}.

In the atom-field interaction model considered here, both the subsystems are 
infinite-dimensional. We now examine 
whether the efficacy of  $\xi_{\textsc{ipr}}$ is correlated with 
the Hamming distance in this case as well.  
We note that 
 $\ket{\psi_{\rm{bin}}}$
 can be expanded as a superposition of states which are Hamming-uncorrelated. 
From Fig. \ref{fig:comp_diff_4}, we see that in this case also
 $\xi_{\textsc{ipr}}$ is a significantly better entanglement indicator 
than $\xi^{\prime}_{\textsc{tei}}$.

\subsection{The double-well BEC model}

The effective Hamiltonian (Eq. \eref{eqn:HBEC}) for the system, setting $\hbar = 1$, is given by \cite{sanz}
\begin{equation}
\nonumber H_{\textsc{bec}}
=\omega_{0} N_{\rm{tot}} + \omega_{1} (a^{\dagger} a - b^{\dagger} b) + U  N_{\rm{tot}}^{2} - \lambda (a^{\dagger} b + a b^{\dagger}).
\end{equation}
We recall that $N_{\rm{tot}} = a^{\dagger} a + b^{\dagger} b$. 
Here $(a,a^{\dagger})$ and $(b,b^{\dagger})$ are
 the boson annihilation and creation operators of the atoms in wells A and B respectively. $U$ is the strength of the nonlinearity 
 (both in the individual modes as well as in their interaction),  
 $\lambda$ is the linear interaction strength, and $\omega_{0}, \omega_{1}$ are constants. As in the previous instance,  we select a representative variety of initial states: 
 (i) the unentangled direct product $ \ket{\alpha_{a}, m_{1}} 
 \otimes \ket{\alpha_{b}, m_{2}}$ of boson-added 
 coherent states of atoms in the wells A and B respectively, 
 where $\alpha_{a},\alpha_{b} \in \mathbb{C}$; (ii)  
 the binomial state  $\ket{\psi_{\rm{bin}}}$ (Eq. \eref{eqn:BS_defn}), and (iii) the two-mode squeezed vacuum state $\ket{\zeta}$ (Eq. \eref{eqn:sq_state}),  with the understanding that the basis states are now product states of the species in the two wells. 
 
In each of these cases, we must first obtain  
the state of the system at 
any time $t \geq 0$ as it evolves under the Hamiltonian 
$H_{\textsc{bec}}$. It turns out that, in the case of an initial 
state of type (i) above, the state of the system can be 
calculated  explicitly  as a function of $t$, as outlined in Appendix~\ref{appen:DensMat}. In Cases (ii) and (iii), the state vector at 
time $t$ is computed numerically. 
 Using these, we have generated tomograms at approximately $1000$ instants, separated by a time step 0.001 $\pi/U$.  We have verified that, in this model also $\xi^{\prime}_{\textsc{tei}}$ agrees better with 
$\xi_{\textsc{sle}}$ than with $\xi_{\textsc{svne}}$, and that the difference between $\xi^{\prime}_{\textsc{tei}}$ and $\xi_{\textsc{sle}}$ is smaller than that between 
 $\xi_{\textsc{svne}}$ and $\xi_{\textsc{sle}}$. In what follows, we have therefore
chosen $\xi_{\textsc{sle}}$ as the reference entanglement measure and compared  $d_{2}(t)$ with $d_{3}(t)$.

\begin{figure}[h]
\includegraphics[width=0.32\textwidth]{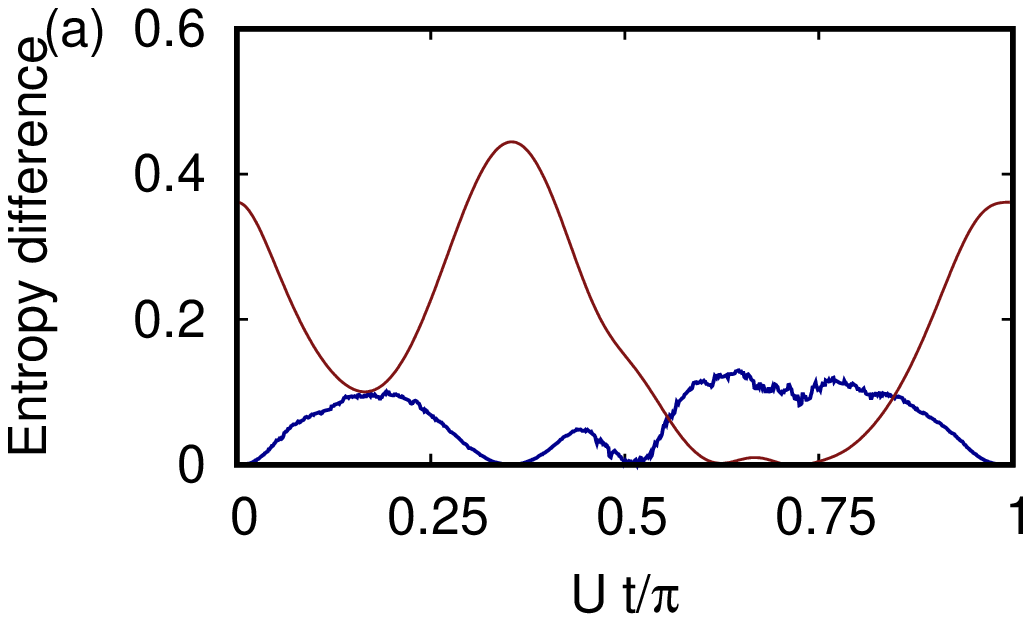}
\includegraphics[width=0.32\textwidth]{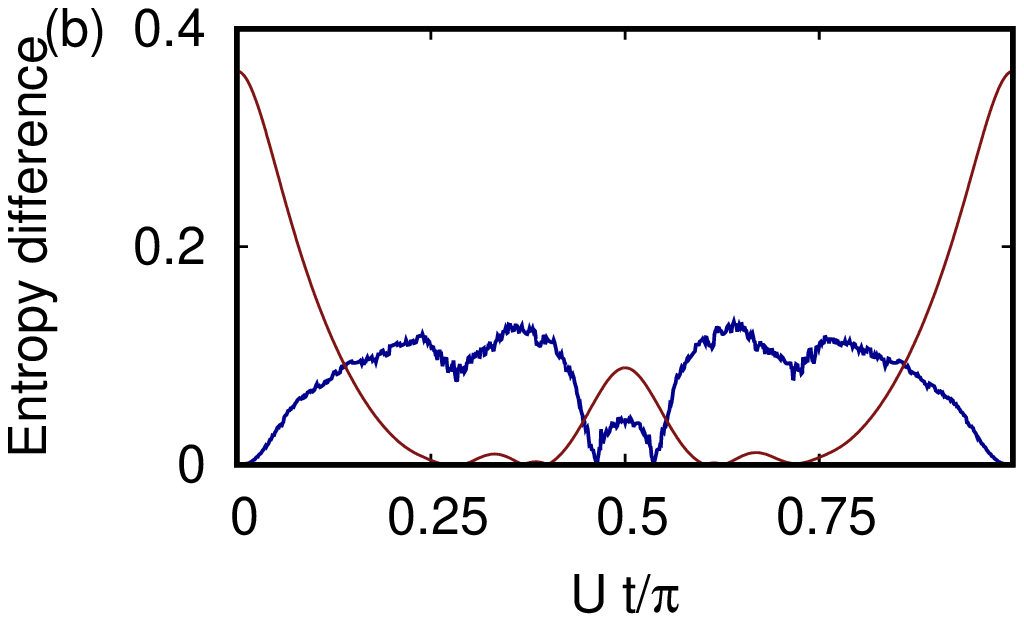}
\includegraphics[width=0.32\textwidth]{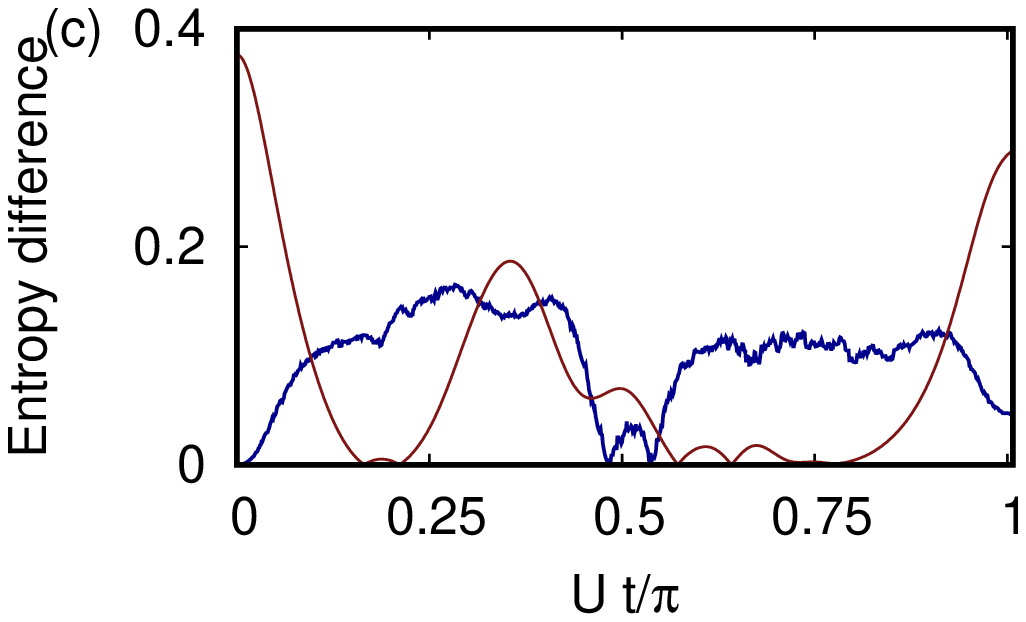}
\caption{$d_{2}(t)$ (blue) and $d_{3}(t)$ (brown) vs. scaled time $U t / \pi$, for $\omega_{0}=U=\vert\alpha\vert^{2} = 1$ in the BEC model. (a) $\omega_{1}=\lambda=1$, initial state $\ket{\alpha} \otimes \ket{\alpha}$; (b) $\omega_{1}=\lambda=0.1$,  initial state $\ket{\alpha} \otimes \ket{\alpha}$; (c) $\omega_{1}=\lambda=1$,  
initial state $\ket{\alpha,1} \otimes \ket{\alpha}$. }
\label{fig:BEC_comp_diff_1}
\end{figure}

\begin{figure}[h]
\centering
\includegraphics[width=0.32\textwidth]{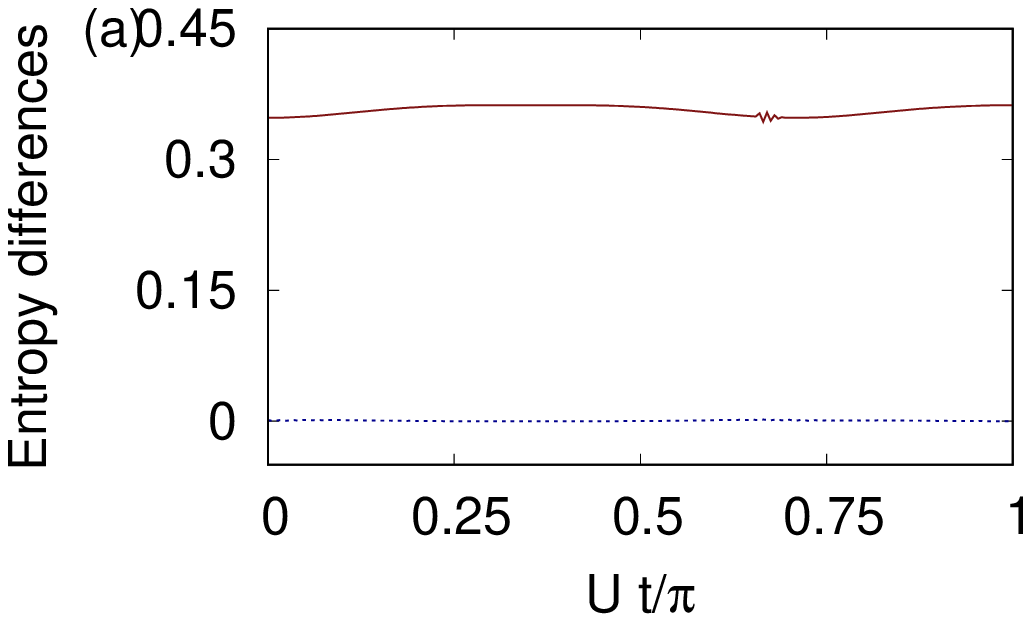}
\includegraphics[width=0.32\textwidth]{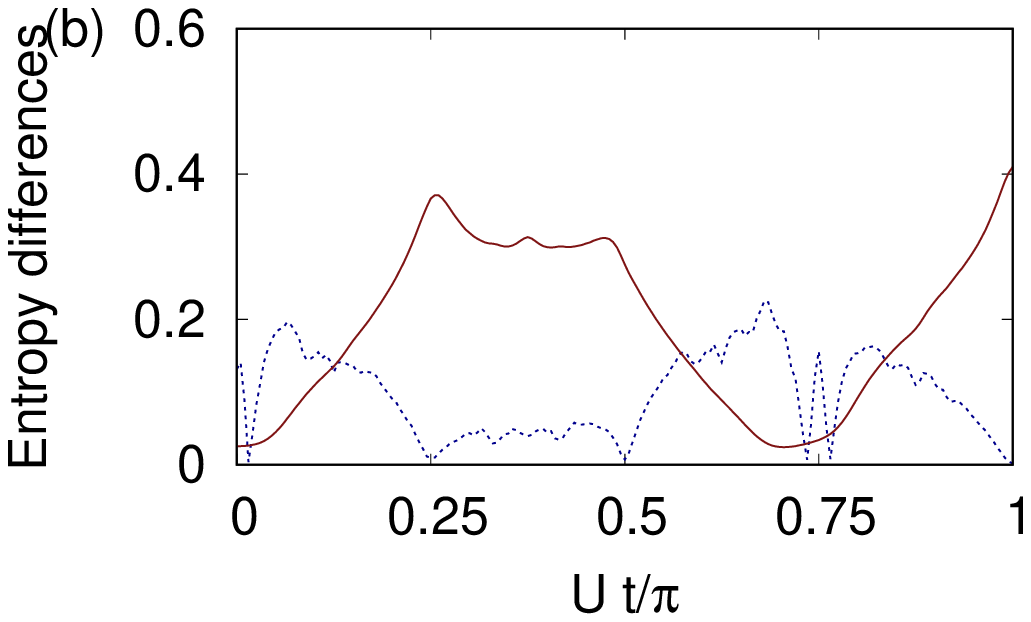}
\includegraphics[width=0.32\textwidth]{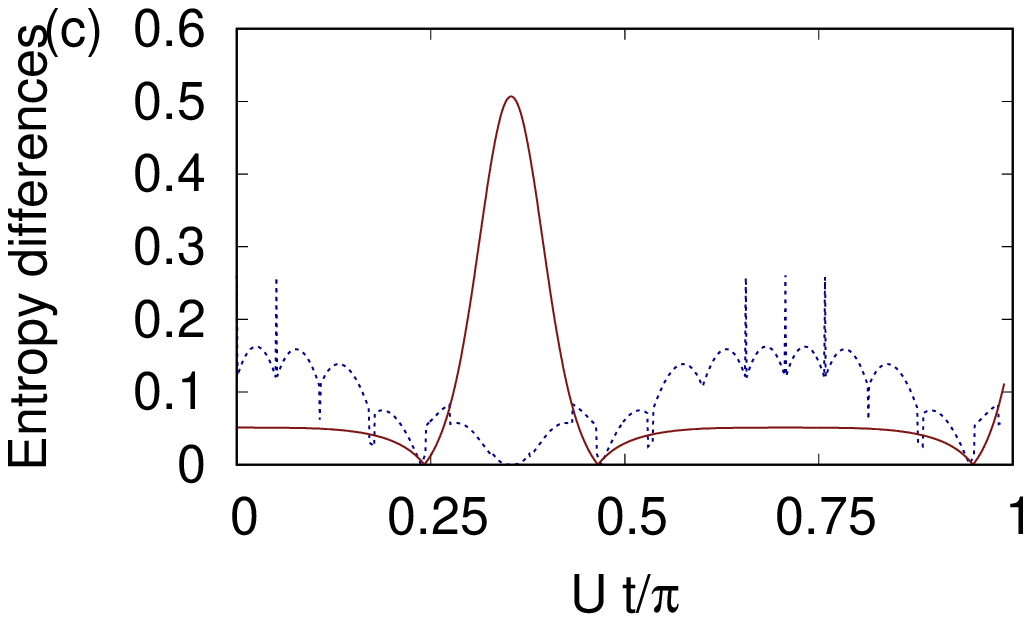}
\caption{$d_{2}(t)$ (blue) and $d_{3}(t)$ (brown) 
vs. scaled time $U t /\pi$, for $\omega_{0}=U=\omega_{1}=\lambda=1$ in the BEC model. (a) and (b): Initial state 
$\ket{\zeta}$, $\zeta=0.1$ and $0.7$ respectively. (c) Initial state $\ket{\psi_{\rm{bin}}}$, $N=10$.}
\label{fig:BEC_comp_diff_2}
\end{figure}

It is evident that the relevant ratio for characterising the dynamics is $U/\lambda_{1}$ where $\lambda_{1} = (\omega_{1}^{2}+ \lambda^{2})^{1/2}$. 
A representative example of the  temporal behaviour of 
 $d_{2}(t)$ and $d_{3}(t)$ is shown in Fig. \ref{fig:BEC_comp_diff_1}.  
 The effect of increasing 
  $U/\lambda_{1}$ can be seen by comparing 
 Figs. \ref{fig:BEC_comp_diff_1} 
 (a) and (b), while that of departure from coherence of the initial state can be seen by comparing  (a) and  (c). 
We have also carried out analogous studies in the case of 
entangled initial states $\ket{\psi_{\rm{bin}}}$ and $\ket{\zeta}$. 
The general trends in the behaviour of  the entanglement 
indicators in these cases are consistent with, and corroborate, those 
found in the atom-field interaction model (see Fig. \ref{fig:BEC_comp_diff_2}). 

\subsubsection{Decoherence effects in the double-well BEC model}
\label{sec:decoherence}

Since decoherence is an important issue in experiments, we now investigate this aspect for an initial state $\ket{\alpha_{a}}\otimes\ket{\alpha_{b}}$, where $\alpha_{a}=\sqrt{0.001}$ and $\alpha_{b}=1$: since $\alpha_{a}$ is sufficiently small in magnitude,  $\ket{\alpha_{a}}$ can be approximated by a superposition of a zero-boson and one-boson states, whereas $\ket{\alpha_{b}}$ is an infinite superposition of the boson number states.  The states get entangled during unitary evolution. We consider the entangled state of the system at  any instant, say $U t = \pi/2$, and apply damping to subsystem A alone for a time interval $\tau$. This toy example suffices to  examine the extent to which an efficient entanglement indicator can be extracted from tomograms, when the system decoheres.
We have set $U$, $\omega_{0}$, $\omega_{1}$ and $\lambda$  equal to $1$, for numerical computation. Decoherence takes place through amplitude damping and phase damping~\cite{agarwalmaster}.  

We recall from Chapter \ref{ch:RevSqueezeOptTomo} that the master equation for amplitude decay (Eq. \eref{eqn:master_eqn}) is given by
\begin{equation}
\nonumber \frac{\rmd \rho}{\rmd \tau} = - \Gamma (a^{\dagger} a \rho - 2 a \rho a^{\dagger} + \rho a^{\dagger} a).
\end{equation}
Here, $\Gamma$ is the rate of loss and $\tau$ the time parameter is reckoned from the instant $U t = \pi/2$. 
The solution to this master equation is ~\cite{agarwalmaster}
\begin{equation}
\rho(\tau) = \sum_{n,m,n',m'=0}^{\infty}\rho_{n,m,n',m'}(\tau) \ket{n \delim m}\bra{n' \delim m'},
\label{eqn:soln_master_bipart}
\end{equation}
where
\begin{align}
\nonumber \rho_{n,m,n',m'}(\tau)=&e^{-\Gamma\tau(n+n')}\sum_{r=0}^{\infty} \sqrt{{n+r \choose r}{n'+r \choose r}} \\
\nonumber &(1-e^{-2\Gamma\tau})^{r} \rho_{n+r,m,n'+r,m'}(\tau=0).
\end{align}
Note that $\rho(\tau=0)$ is chosen to be $\ket{\Psi_{00}(\pi/ 2 U)}\bra{\Psi_{00}(\pi/ 2 U)}$. (The expression for $\ket{\Psi_{00}(t)}$ as a function of time $t$ is given by Eq. \eref{eqn:Psi_t}).

\begin{figure}
\includegraphics[width=0.49\textwidth]{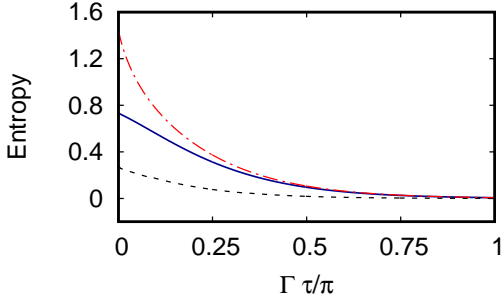}
\caption{$\xi^{\prime}_{\textsc{tei}}$ (black), $\xi_{\textsc{svne}}^{(\textsc{a})}$ (blue) and $\xi_{\textsc{qmi}}$ (red) vs. $\Gamma\tau/\pi$ for amplitude decay with $\Gamma=0.001$ in the BEC model.}
\label{fig:amp_damping}
\end{figure}

As the bipartite state is a mixed state owing to decoherence, $\xi_{\textsc{svne}}$ corresponding to subsystems A and B are not equal to each other and are now denoted by $\xi_{\textsc{svne}}^{(\textsc{a})}$ and $\xi_{\textsc{svne}}^{(\textsc{b})}$, respectively. The  entropy of the full system is denoted by $\xi_{\textsc{svne}}^{(\textsc{ab})}$.  The quantum mutual information 
\begin{equation}
\xi_{\textsc{qmi}}=\xi_{\textsc{svne}}^{(\textsc{a})}+ \xi_{\textsc{svne}}^{(\textsc{b})}- \xi_{\textsc{svne}}^{(\textsc{ab})},
\label{eqn:qmi}
\end{equation}
 is of immediate interest.  At each instant $\tau$, $\xi^{\prime}_{\textsc{tei}}$, $\xi_{\textsc{svne}}^{(\textsc{a})}$ and $\xi_{\textsc{qmi}}$ have been calculated numerically. Since the subsystem entropies are not equal, we expect that (if the tomogram captures decoherence effects well) $\xi^{\prime}_{\textsc{tei}}$ must match $\xi_{\textsc{qmi}}$ during dynamics. 
 This is indeed borne out in Fig. \ref{fig:amp_damping}, where $\xi^{\prime}_{\textsc{tei}}$, $\xi_{\textsc{qmi}}$ and $\xi_{\textsc{svne}}^{(\textsc{a})}$ are compared. We point out that for a bipartite pure state, in the absence of decoherence $\xi_{\textsc{qmi}}=2$ $\xi_{\textsc{svne}}^{(\textsc{a})}=2 \xi_{\textsc{svne}}^{(\textsc{b})}$, and both $\xi_{\textsc{svne}}^{(\textsc{a})}$ and $\xi_{\textsc{svne}}^{(\textsc{a})}$ are denoted by $\xi_{\textsc{svne}}$.

\begin{figure}
\includegraphics[width=0.49\textwidth]{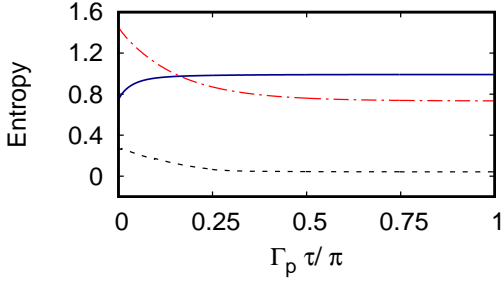}
\caption{$\xi^{\prime}_{\textsc{tei}}$ (black), $\xi_{\textsc{svne}}^{(\textsc{a})}$ (blue) and $\xi_{\textsc{qmi}}$ (red) vs. $\Gamma_{p}\tau/\pi$ for phase damping with $\Gamma_{p}=0.001$ in the BEC model.}
\label{fig:phase_damping}
\end{figure}

Phase damping in A is modelled by (Eq. \eref{eqn:master_eqn_phase})~\cite{agarwalmaster}
\begin{equation}
\nonumber \frac{\rmd \overline{\rho}}{\rmd \tau} = - \Gamma_{p} ((a^{\dagger}a)^{2} \overline{\rho} - 2 a^{\dagger}a \overline{\rho} a^{\dagger}a + \overline{\rho} (a^{\dagger}a)^{2}),
\end{equation}
where $\Gamma_{p}$ is the rate of decoherence. 
The solution  is of the form given in Eq. \eref{eqn:soln_master_bipart}, with
\begin{align}
\nonumber \overline{\rho}_{n,m,n',m'}(\tau)=e^{-\Gamma_{p}\tau(n-n')^{2}} \overline{\rho}_{n,m,n',m'}(\tau=0).
\end{align}
As before, $\xi^{\prime}_{\textsc{tei}}$ mimics $\xi_{\textsc{qmi}}$ reasonably well (Fig. \ref{fig:phase_damping}).

\section{Time-series analysis of $d_{1}(t)$}
\label{sec:Ch4NTSA}

Finally, we turn to an  assessment of 
 the {\em long-time}  behaviour of $\xi^{\prime}_{\textsc{tei}}$ by means of a detailed time-series analysis. 

As we have shown in the foregoing, the deviation of   
$\xi^{\prime}_{\textsc{tei}}$ from $\xi_{\textsc{svne}}$ is much more pronounced than its deviation from $\xi_{\textsc{sle}}$.  It is therefore appropriate to investigate how an initial  difference  between $\xi^{\prime}_{\textsc{tei}}$ and $\xi_{\textsc{svne}}$ changes with time. 
With this in mind, a time series of $d_{1}(t)$  has been obtained 
for both models considered in this chapter, and used to compute local Lyapunov exponents  along the lines customary \cite{AbarbanelBook,NTSA_GP,abarbanel1992} in the study of dynamical systems. This involves  reconstruction of the effective phase space,  estimation of the minimum embedding dimension $d_{\rm{emb}}$ of this space, and  calculation of the exponents themselves. The procedure used is outlined below.

The time series had 20000 data points. The effective phase space was reconstructed using the TISEAN package \cite{tisean}. $100$ different initial values $d_{1}(0)$ were randomly chosen in this phase space. The maximum local Lyapunov exponent corresponding to each $d_{1}(0)$ was computed over the same time interval $L$. (The term `local' refers to the fact that $L$ is much smaller than the time interval over which the maximum Lyapunov exponent $\Lambda_{\infty}$ is obtained in the standard method). The average value $\Lambda_{L}$ of these $100$  maximum local Lyapunov exponents was obtained following the prescription in \cite{abarbanel1992} (see Appendix~\ref{appen:TimeSeries}).   
This procedure was repeated for as many as 14 different values of $L$.  Further, in each case it was verified that,  with an increase in $L$, $\Lambda_{L}$ tends to $\Lambda_{\infty} + (m/L^{q})$, where 
$m$ and $q$ are constants  \cite{abarbanel1992}. We note that two neighbouring initial values of the dynamical variable of interest $d_{1}(t)$ diverges exponentially with $\Lambda_{L}$, in $L$ steps. For completeness, we also present the power spectra corresponding to the various time series. We would expect a broadband spectrum for chaotic data. On the other hand, a `spiky' power spectrum (i.e., a cluster of clearly defined sharp peaks in the power spectrum) points to a possible quasi-periodic behaviour. 
The results are presented below. 
 
\subsection{Atom-field interaction model}

\begin{figure}[h]
\centering
\includegraphics[width=0.4\textwidth]{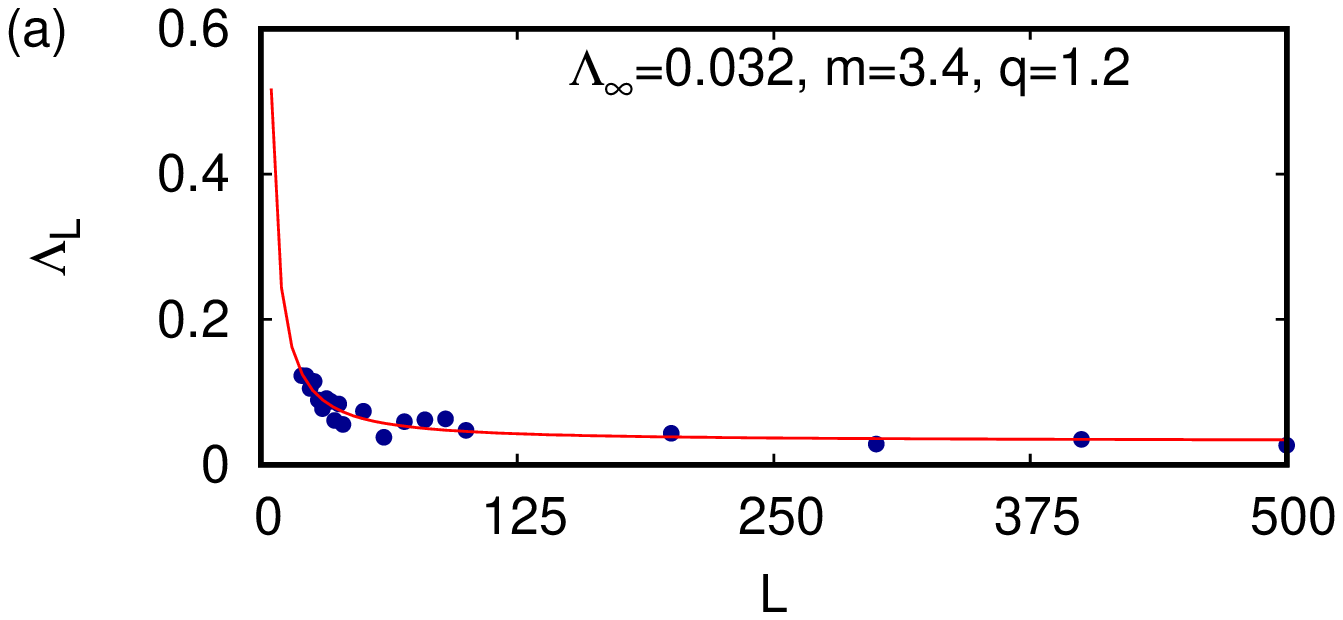}
\includegraphics[width=0.4\textwidth]{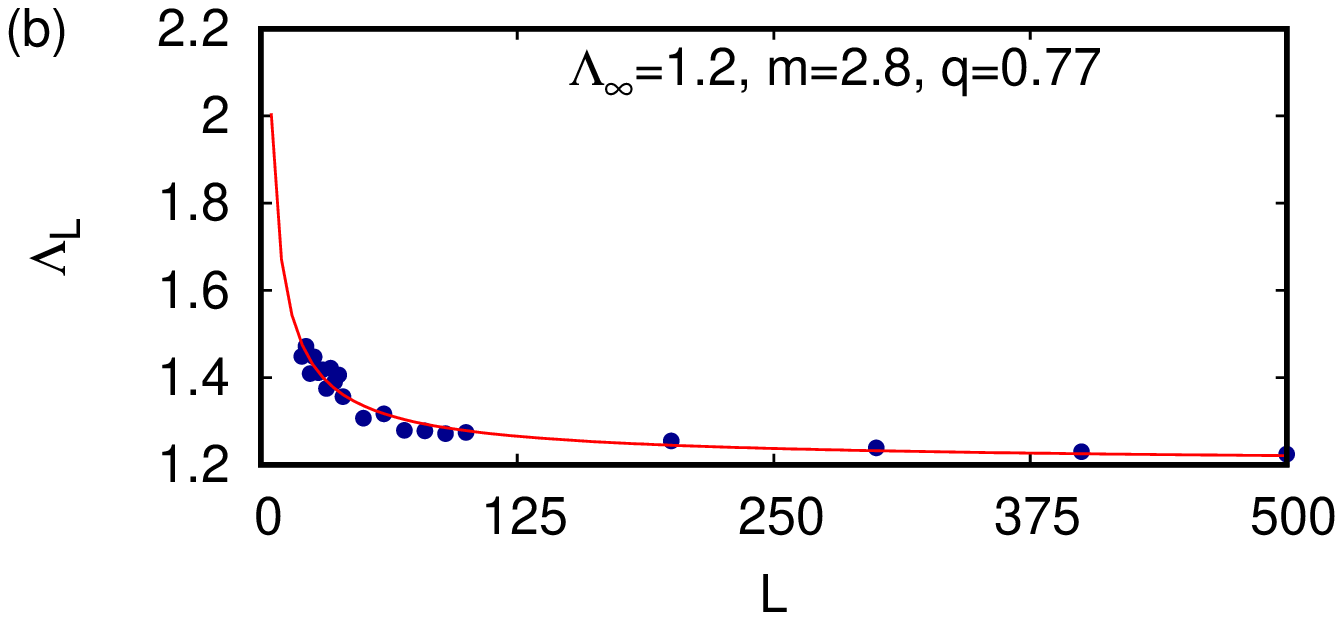}
\includegraphics[width=0.4\textwidth]{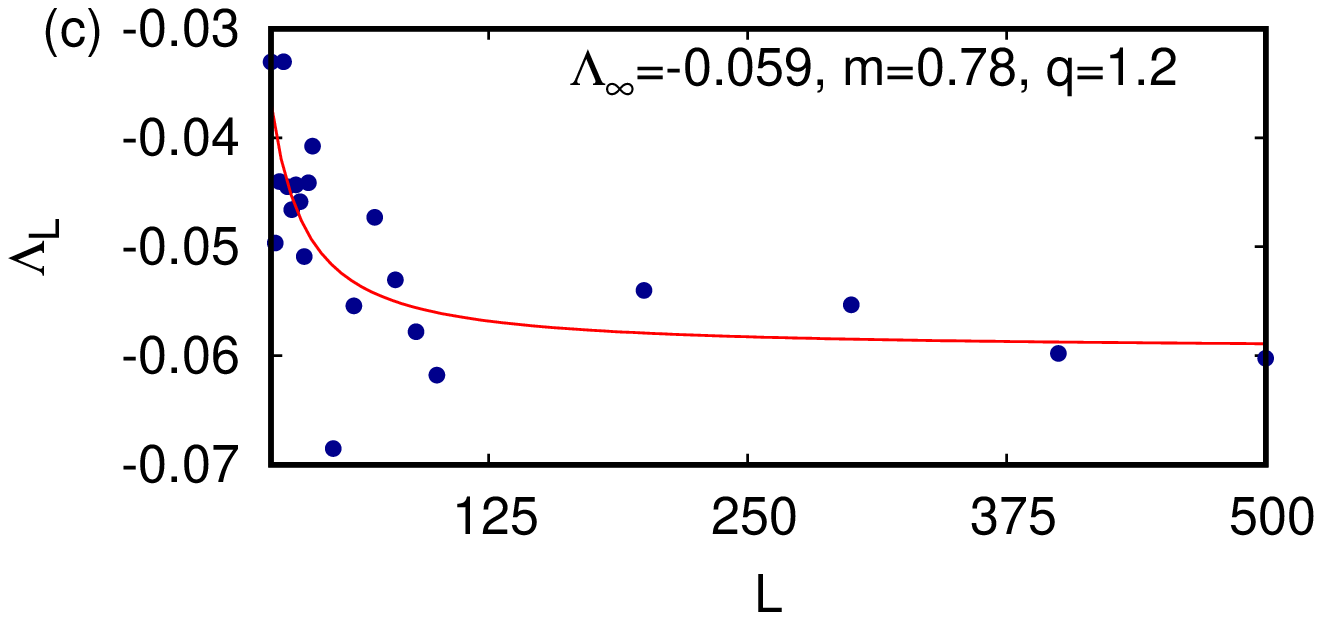}
\includegraphics[width=0.4\textwidth]{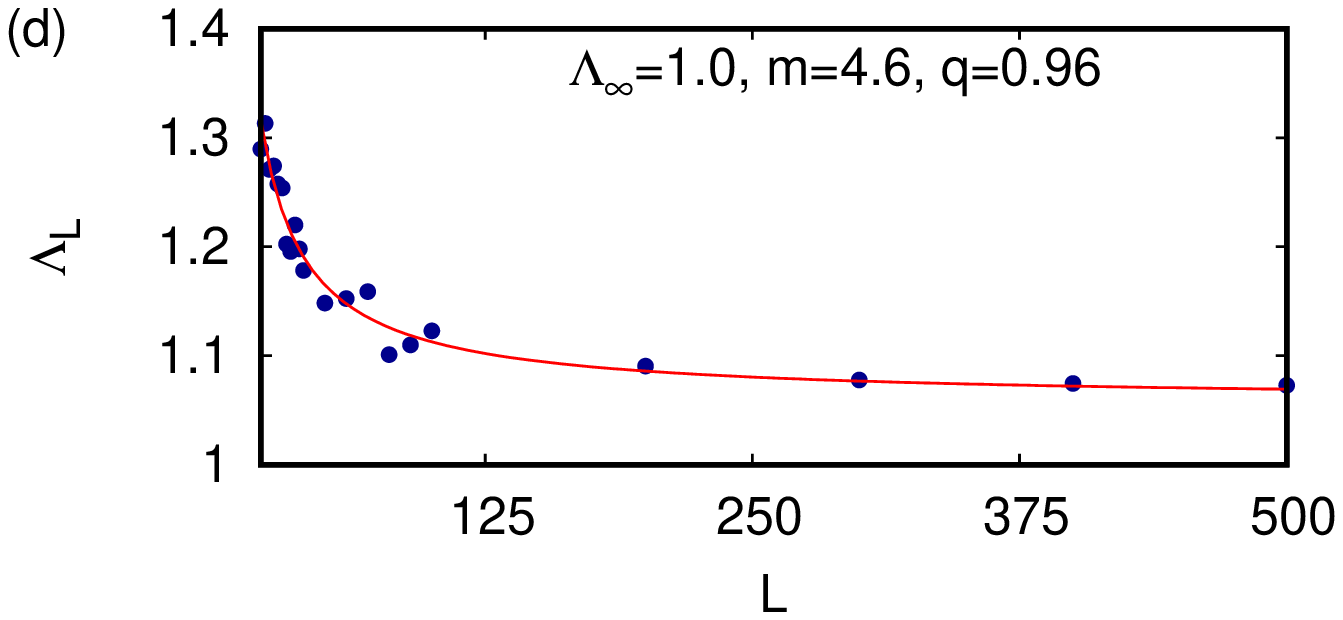}
\caption{$\Lambda_{L}$ obtained from the time series 
of $d_{1}(t)$ (blue) and the fit $\Lambda_{\infty} + (m/L^{q})$ (red)  vs. $L$,  for $\omega_{\textsc{f}}=\omega_{\textsc{a}}=\gamma=1$ in the atom-field interaction model.  Initial state $\ket{\alpha}_{\textsc{f}}\otimes \ket{0}$: (a) $g=100$,  $|\alpha|^{2}=1$ (b) $g=100$,  $|\alpha|^{2}=5$ (c) $g=0.2$,  $|\alpha|^{2}=1$. (d) Initial state 
$\ket{\alpha,5}_{\textsc{f}}\otimes \ket{0}, g=100, |\alpha|^{2}=1$.}
\label{fig:lambda_L_agarwal}
\end{figure}
The difference $d_{1}(t)$ has been obtained at each instant with time step $\delta t = 0.1$ for $20000$ time steps, and the effective phase space has been reconstructed. We see that for both the  initial states $\ket{\alpha}_{\textsc{f}} \otimes \ket{0}$ and  $\ket{\alpha,5}_{\textsc{f}}\otimes \ket{0}$ with $|\alpha^{2}|=1$ and weak nonlinearity ($\gamma/g=0.01$), $\Lambda_{L}$ is positive, and  
both $\Lambda_{L}$ and  $\Lambda_{\infty}$ are 
larger  for the second initial state (compare Figs. \ref{fig:lambda_L_agarwal} (a) and (d)). $\Lambda_{\infty}$ increases with an increase in $|\alpha|^{2}$ for the 
initial state $\ket{\alpha} \otimes \ket{0}$ (compare Figs. \ref{fig:lambda_L_agarwal} (a) and (b)).  In contrast, for strong nonlinearity (e.g., as in Fig. \ref{fig:lambda_L_agarwal} (c),  
$\gamma/g=5$), $\Lambda_{L}$ is negative.

\begin{figure}[h]
\centering
\includegraphics[width=0.4\textwidth]{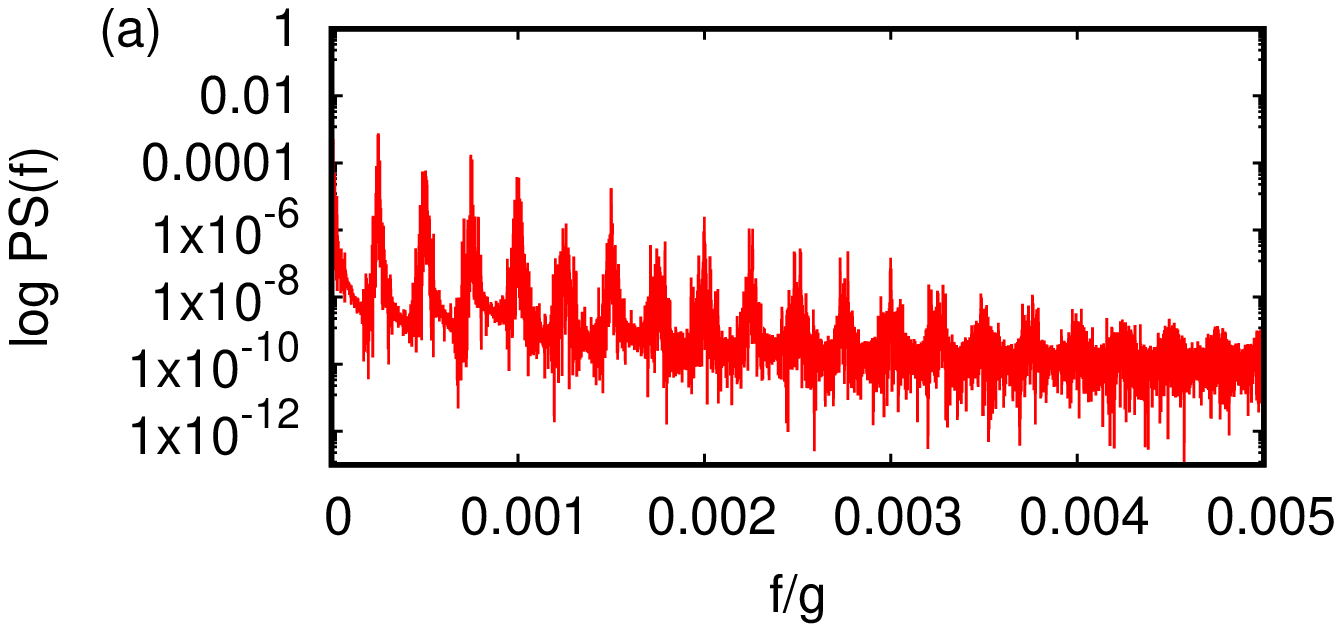}
\includegraphics[width=0.4\textwidth]{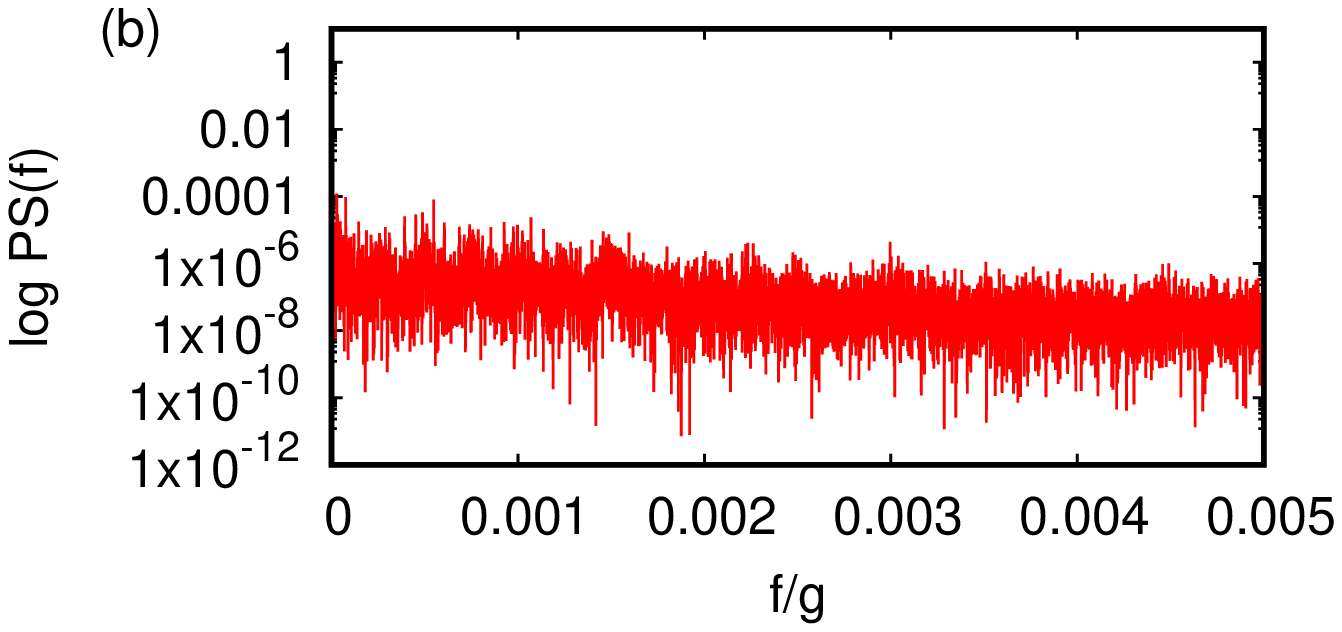}
\includegraphics[width=0.4\textwidth]{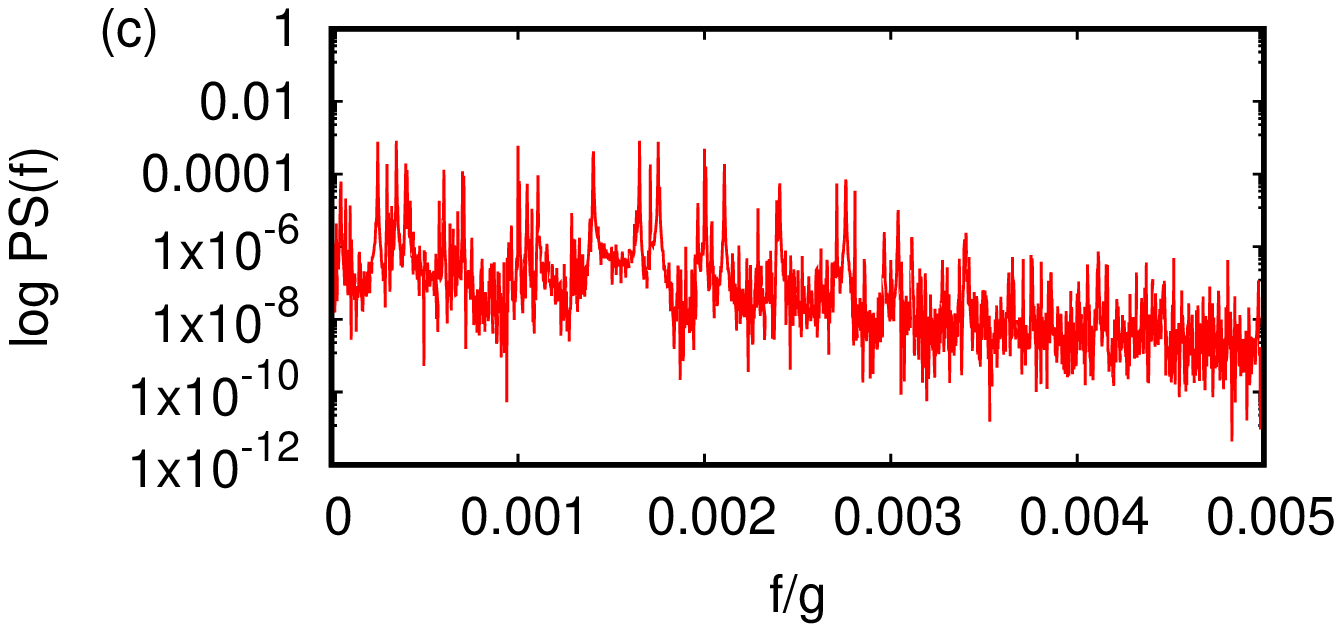}
\includegraphics[width=0.4\textwidth]{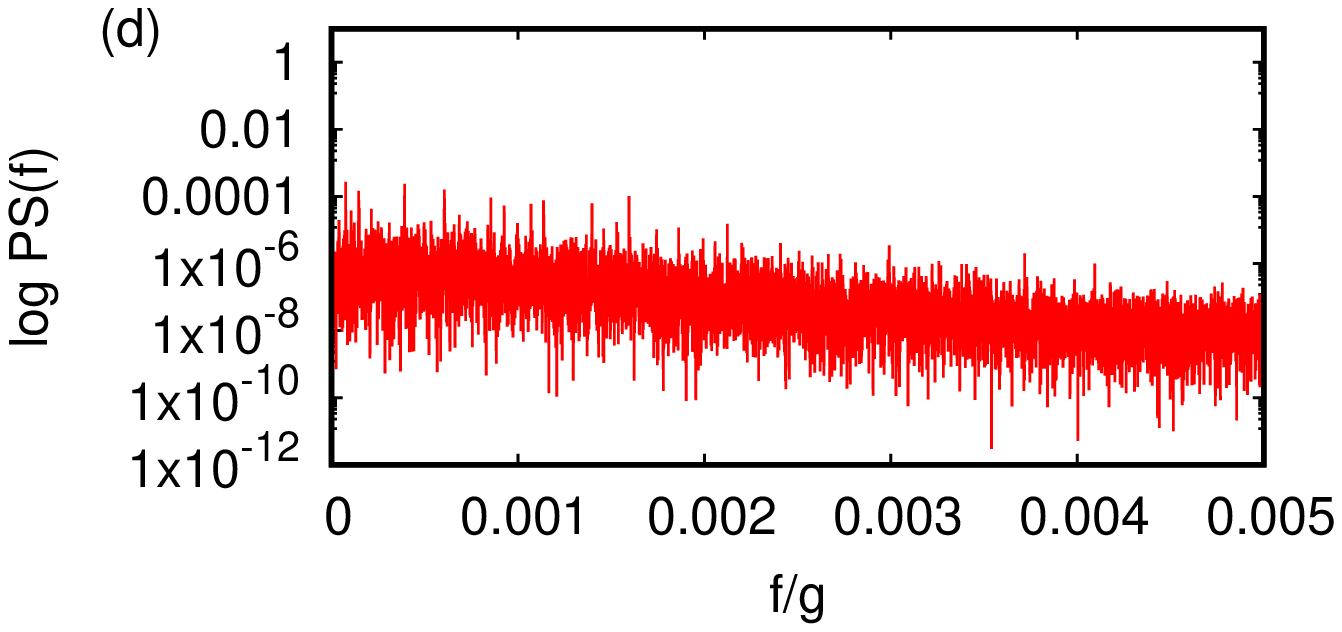}
\caption{Logarithm of $PS(f)$ vs. $f/g$ (red) in the atom-field interaction model, corresponding to the initial states and parameter values in Figs. \ref{fig:lambda_L_agarwal} (a)--(d) respectively.}
\label{fig:powerspec_agarwal}
\end{figure}

For completeness, we present the power spectrum $PS(f)$
of the time series as a function 
of the frequency $f$ in units of $g$, for each of the 
cases corresponding to Figs. 
\ref{fig:powerspec_agarwal} (a) to (d). 
The nearly quasi-harmonic power spectrum for weak nonlinearity 
(Fig. \ref{fig:powerspec_agarwal} (a)) 
changes into a broadband 
spectrum with increasing $|\alpha|^{2}$  
(Fig. \ref{fig:powerspec_agarwal} (b)), while it  loses 
its quasi-harmonicity  without becoming a broadband 
spectrum with increasing nonlinearity
(Fig. \ref{fig:powerspec_agarwal} (c)).  The lack of 
 coherence in the initial state makes the power spectrum 
broadbanded (Fig. \ref{fig:powerspec_agarwal} (d)).  
 When a power spectrum exhibits regular or quasi-regular spikes at different frequencies, it signals quasiperiodicity in the processes generating the time series. A broadband spectrum, on the other hand, is an indicator of non-periodic (including chaotic) time evolution. When the power spectrum of this time series shows quasiperiodic behaviour (Figs. \ref{fig:powerspec_agarwal} (a) and (c)) in the sense above, we see that the local Lyapunov exponent is either negative, or small positive. The two entanglement indicators (namely, $\xi^{\prime}_{\textsc{tei}}$ and $\xi_{\textsc{svne}}$) therefore do not diverge from each other in any significant way. We regard this as a broad corroboration of the validity of $\xi^{\prime}_{\textsc{tei}}$ as an entanglement indicator. In contrast, when the power spectrum is markedly broadband (Figs. \ref{fig:powerspec_agarwal} (b) and (d)), the local Lyapunov exponent is distinctly positive, showing that $\xi^{\prime}_{\textsc{tei}}$ diverges significantly from $\xi_{\textsc{svne}}$. Hence we conclude that $\xi^{\prime}_{\textsc{tei}}$ is not a reliable or satisfactory indicator of entanglement for sufficiently large $|\alpha|^{2}$ or for initial PACS states with significant departure from coherence.

\subsection{The double-well BEC model}

\begin{figure}[h]
\centering
\includegraphics[width=0.4\textwidth]{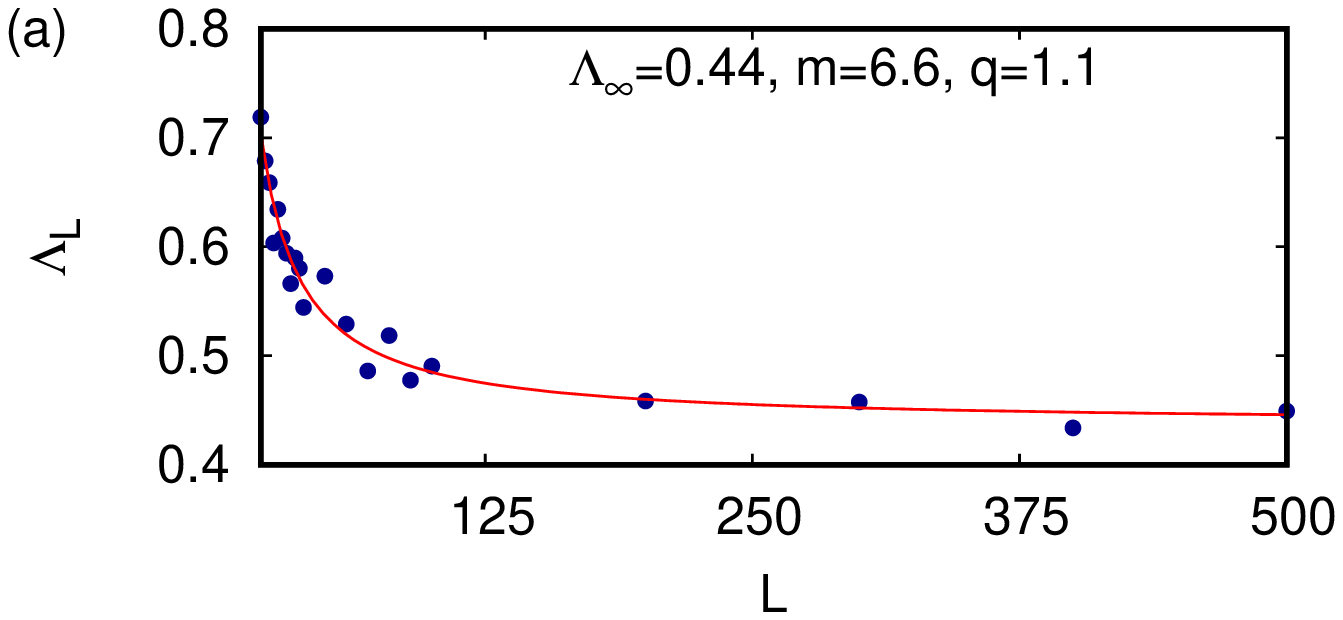}
\includegraphics[width=0.4\textwidth]{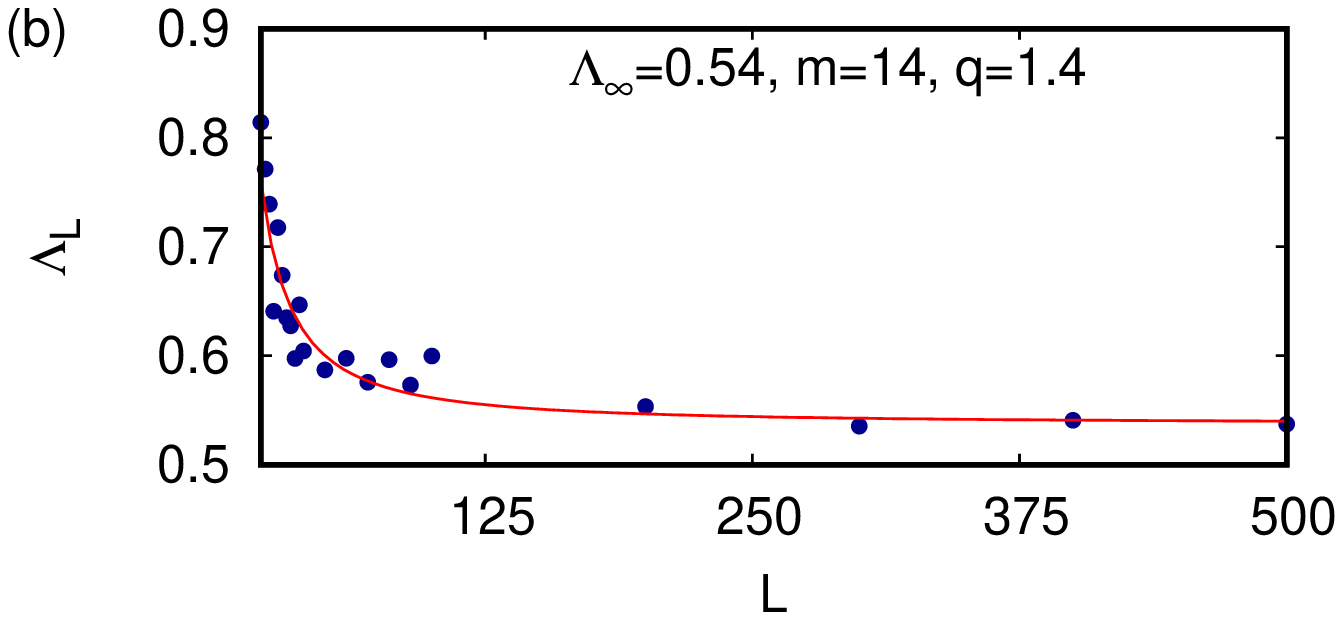}
\includegraphics[width=0.4\textwidth]{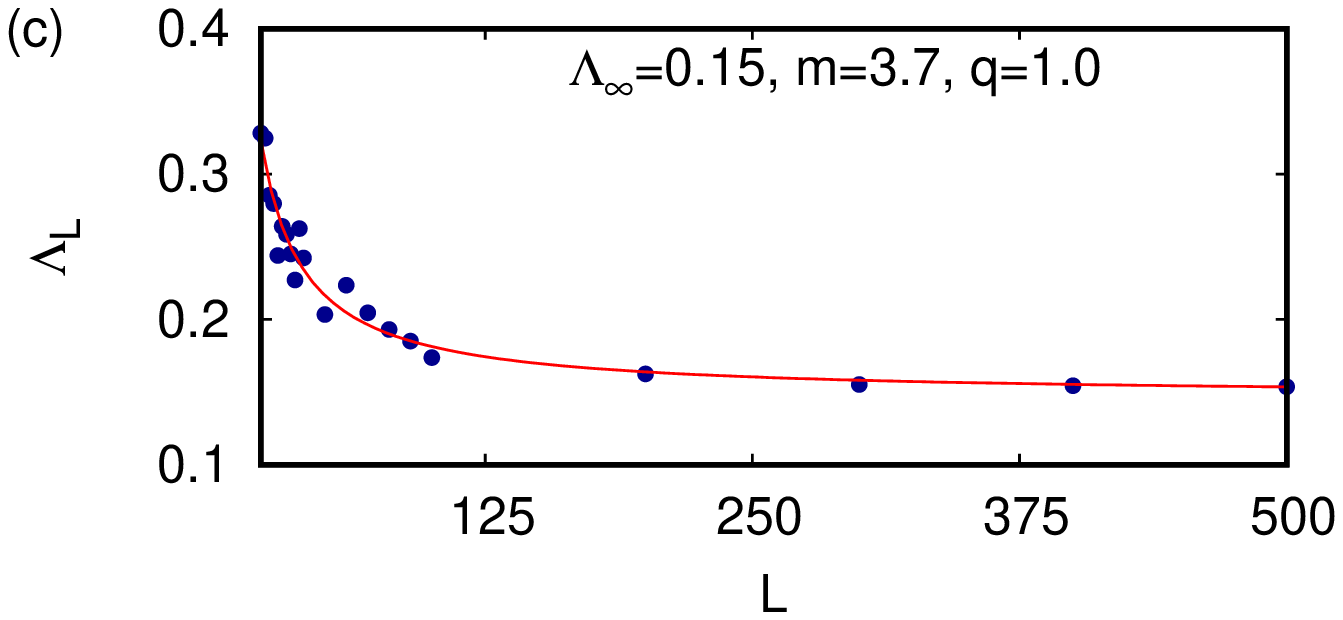}
\caption{$\Lambda_{L}$ obtained from the time series 
of $d_{1}(t)$ (blue) and the fit 
$\Lambda_{\infty} + (m/L^{q})$ (red)  vs. $L$, for 
$\omega_{0}= \omega_{1}=|\alpha|=1$ in the BEC model. 
Initial state $\ket{\alpha,1}\otimes\ket{\alpha}$ 
and (a) hopping frequency 
$\lambda=5$, $U=0.5$ (b) $\lambda=U=1$. 
 (c) Initial state $\ket{\alpha,5}\otimes\ket{\alpha,5}, 
 \lambda=U=1$.}
\label{fig:lambda_L_BEC}
\end{figure}

As in the foregoing,  we generate the 
time series of $d_{1}(t)$ by calculating this  difference 
for $20000$ time steps, in this case with $\delta t = 0.01$. 
As seen in Figs.  \ref{fig:lambda_L_BEC} (a)--(c), 
 in this instance $\Lambda_{L}$ is positive regardless of the degree of coherence of the initial states of the subsystems,  for a wide range of values of the ratio $U/\lambda_{1}$ 
($\lambda_{1}= (\lambda^{2}+\omega_{1}^{2})^{1/2}$). 
 With an increase in $U/\lambda_{1}$,  
  $\Lambda_{\infty}$ increases (Figs. \ref{fig:lambda_L_BEC} (a), (b)). In contrast to the atom-field interaction model, a departure of the initial state from perfect coherence causes  $\Lambda_{\infty}$ 
  to decrease (Figs. \ref{fig:lambda_L_BEC} (a), (c)).

\begin{figure}[h]
\centering
\includegraphics[width=0.4\textwidth]{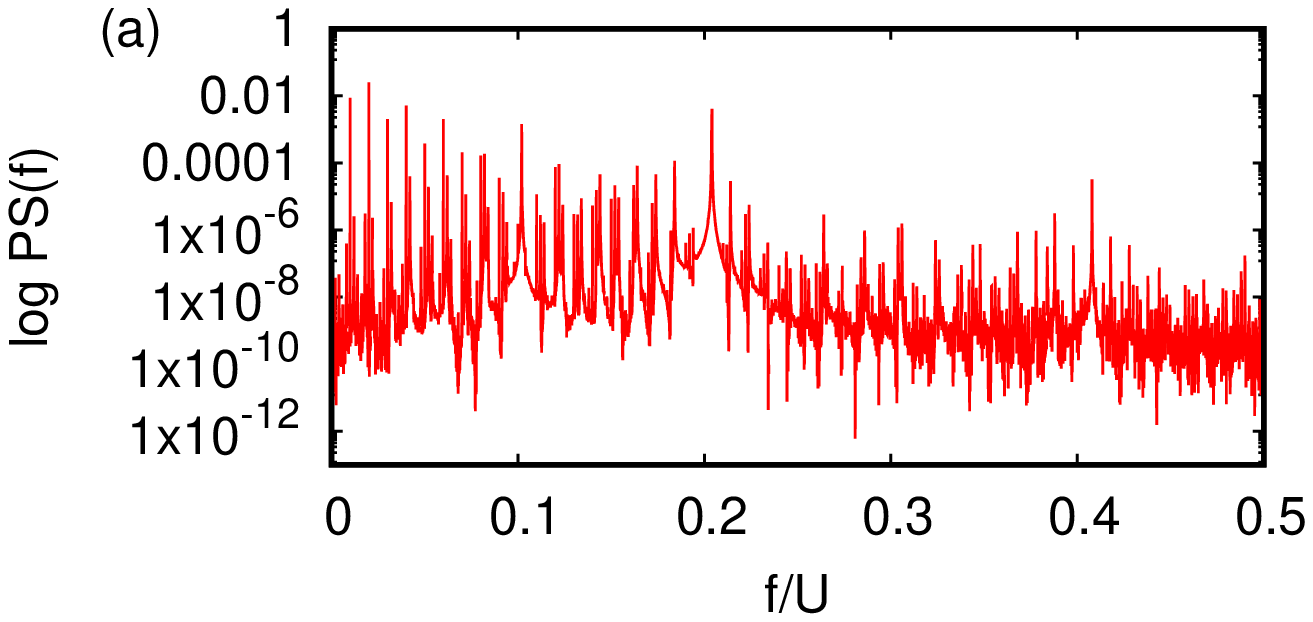}
\includegraphics[width=0.4\textwidth]{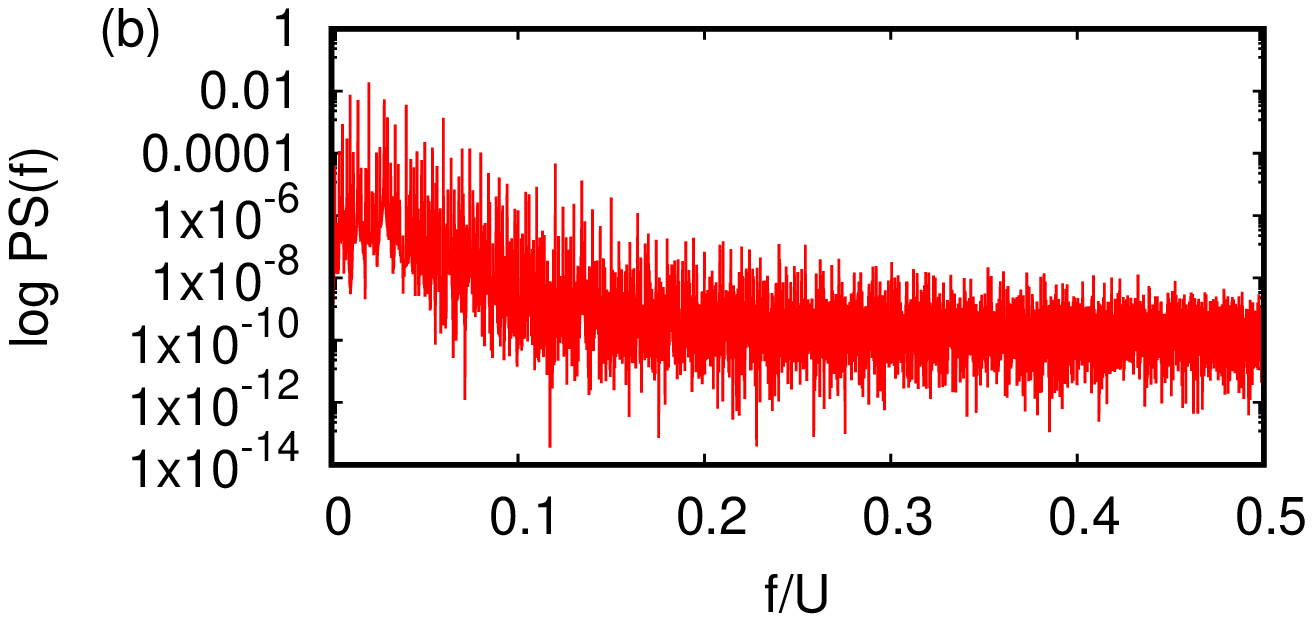}
\includegraphics[width=0.4\textwidth]{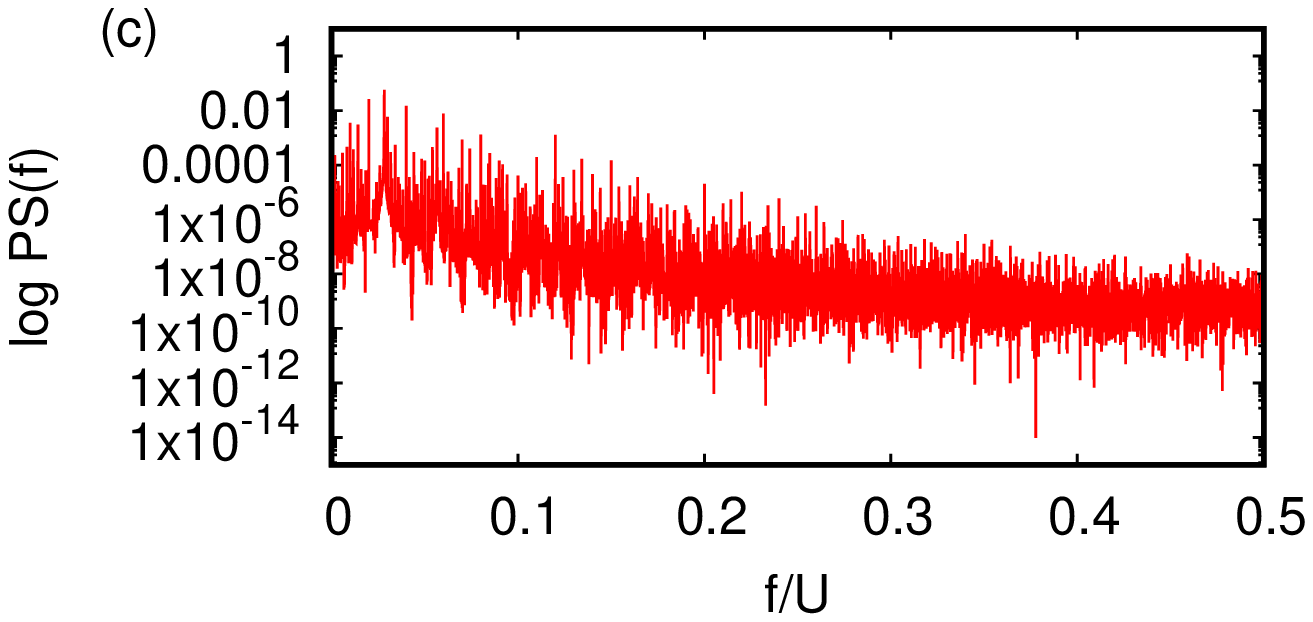}
\caption{Logarithm of $PS(f)$ vs. $f/U$ (red) in the BEC model, corresponding to the initial states and parameter values in 
 Figs. \ref{fig:lambda_L_BEC} (a)--(c) respectively.}
\label{fig:powerspec_BEC}
\end{figure}

The power spectra  corresponding to  the three 
cases in Fig. \ref{fig:lambda_L_BEC} are shown in 
Fig. \ref{fig:powerspec_BEC}.  
When the linear 
part of $H_{\textsc{bec}}$ is dominant  
($\lambda_{1}$ dominates over $U$, 
Fig. \ref{fig:powerspec_BEC} (a)), $PS(f)$ reflects 
a degree of 
quasiperiodicity in the time series. When $U$ becomes 
comparable to $\lambda_{1}$, however, the nonlinearity 
in the Hamiltonian takes over, and $PS(f)$ is a 
broadband spectrum 
(Figs. \ref{fig:powerspec_BEC} (b), (c)).

\section{\label{sec:Ch4chrono}Tomographic signatures in a $2$-photon bipartite system}
As a further illustration of the use of the tomographic 
entanglement indicators, we now assess the performance of $\epsarg{tei}$ which we recall, is an indicator obtained from an appropriately chosen slice of the tomogram. We consider a CV bipartite system that has 
very recently been studied experimentally~\cite{perola}, 
and hence is of particular interest in the present context. 
One of the aims of the experiment was to analyse GKP-like states (Gottesman-Kitaev-Preskill-like states~\cite{gkp} where the `qubit' is encoded in a CV basis, and on which error correction can be implemented with logic gate operations). What is of relevance to us in this context is the distinguishability of the two 
$2$-photon states. We will show that the indicator 
$\epsarg{tei}$ does so unambiguously. 

Some background information is in order. 
The analogy between an ultrashort light pulse and a quantum mechanical wave function 
leads~\cite{paye} to a \textit{chronocyclic} representation for the  study of  ultrashort pulses,  
where the time $t$ and frequency $\omega$ are the conjugate observables.
The state of a single photon of 
frequency $\omega$ is denoted in a spectral representation of infinitely narrow-band pulses
 by $\ket{\omega}$.  The superposed state of a photon that has a frequency $\omega$ with a probability amplitude $\mathcal{S}(\omega)$ is given by $\int \text{d}\omega \, \mathcal{S}(\omega) \ket{\omega}$. In an equivalent temporal representation of infinitely short-duration  pulses $\{\ket{t}\}$, this $1$-photon state is $\int \text{d}t \, \widetilde{\mathcal{S}}(t) \ket{t}$, where $\widetilde{\mathcal{S}}(t)$ is the Fourier transform of $\mathcal{S}(\omega)$.
A family of rotated observables $(\omega \cos \theta + t \sin \theta)$ can then be defined~\cite{beck}, where $\omega$ and $t$ 
have been  scaled by a natural time scale of the 
system to make them dimensionless. Measurements of these rotated observables form the basis of chronocyclic 
tomography, in which  the set of histograms corresponding to these observables gives the tomogram of the state of a $1$-photon system.
In this chronocyclic representation, a $1$-photon state  can  also be described  in the time-frequency `phase space' by a corresponding Wigner function~\cite{paye}. 
Extension to multipartite states corresponding to two or more photons is straightforward. For instance, two photons of frequencies $\omega$ and $\omega^{\prime}$ 
are given by the 
$2$-photon CV  bipartite state $\ket{\omega} \otimes \ket{\omega^{\prime}}$. 

The experiment~\cite{perola} used a 
$2$-photon CV bipartite system to generate 
a pair of distinguishable states that  are frequency 
combs comprising finite-width  peaks.
Despite the finite width, the pair of 2-photon states concerned were shown to be clearly distinguishable 
experimentally. The measured photon coincidence counts were shown to be very different for the two states. Our objective here is to use  chronocyclic tomograms to compute  
the indicator $\epsarg{tei}$ and to show that it 
can also be useful, in principle, to 
distinguish between these $2$-photon states. 

We denote by $\ket{\Psi_{\alpha}}$ 
and $\ket{\Psi_{\beta}}$ these two $2$-photon states.
In the experiment, 
they were  generated  in a setup comprising a cavity with resonant frequency  
$\overline{\omega}$, a photon source using parametric down conversion, and with input photons of frequency $\omega_{p}$.  If  
$\omega_{\textsc{s}}$ and 
$\omega_{\textsc{i}}$  
denote the signal 
and idler frequencies, respectively, 
and $\Omega$ is their difference,  
the first of the two states is
given by 
\begin{align}
\nonumber 
\ket{\Psi_{\alpha}}
= \mathcal{N}_{\alpha}^{-1/2}\int \text{d}\omega_{\textsc{s}} \int \text{d}\omega_{\textsc{i}} &f_{+}(\omega_{\textsc{s}}+\omega_{\textsc{i}}) f_{-}(\Omega)
\times \\
 & f_{\text{cav}}(\omega_{\textsc{s}}) f_{\text{cav}}(\omega_{\textsc{i}}) \ket{\omega_{\textsc{s}}} \otimes \ket{\omega_{\textsc{i}}}.
\label{eqn:Psi_plpl}
\end{align}
Here, 
\begin{equation}
f_{-}(\Omega) = e^{ -(\Omega - \Omega_{0})^{2}/
4 (\vardel \Omega)^{2} },
\label{eqn:fmin}
\end{equation}
where $\Omega_{0}$ 
and $\vardel \Omega$ are the 
mean and standard deviation of $\Omega$. 
$f_{\text{cav}}$ is the Gaussian comb
\begin{equation}
f_{\text{cav}}(\omega)
=\sum_{n} e^{-(\omega -n \overline{\omega})^{2}/
2 (\vardel \omega)^{2}}
\label{eqn:fcav}
\end{equation}
where $\vardel \omega$ is the 
standard deviation of each Gaussian, $\mathcal{N}_{\alpha}$ is the normalisation constant,
and $f_{+}(\omega_{\textsc{s}}+\omega_{\textsc{i}})=\delta(\omega_{p}-\omega_{\textsc{s}}-\omega_{\textsc{i}})$.
 We note that   
 Eq.~\eref{eqn:Psi_plpl} features the product $f_{\text{cav}}(\omega_{\textsc{s}})f_{\text{cav}}(\omega_{\textsc{i}})$, where $f_{\text{cav}}(\omega)$  
 is a superposition of Gaussians corresponding to 
 odd and even values of $n$ such that the two are in phase with each other.

The second $2$-photon state is given by 
\begin{align}
\nonumber \ket{\Psi_{\beta}}=
\mathcal{N}_{\beta}^{-1/2}\int \text{d}\omega_{\textsc{s}} \int \text{d}\omega_{\textsc{i}} &f_{+}(\omega_{\textsc{s}}+\omega_{\textsc{i}}) f_{-}(\Omega)     
\times \\
 & g_{\text{cav}}(\omega_{\textsc{s}}) f_{\text{cav}}(\omega_{\textsc{i}}) \ket{\omega_{\textsc{s}}} \otimes \ket{\omega_{\textsc{i}}}, 
 \label{eqn:Psi_mipl}
\end{align}
where 
\begin{equation}
g_{\text{cav}}(\omega)
=\sum_{n}  (-1)^{n} e^{-(\omega -n \overline{\omega})^{2}/
2 (\vardel \omega)^{2}}.
\label{eqn:gcav}
\end{equation}
Here, $\mathcal{N}_{\beta}$ is the normalisation constant. 
In contrast to $\ket{\Psi_{\alpha}}$, Eq.~\eref{eqn:Psi_mipl} features the product 
$g_{\text{cav}}(\omega_{\textsc{s}}) 
f_{\text{cav}}(\omega_{\textsc{i}})$, 
where $g_{\text{cav}}(\omega)$ 
is a  superposition of Gaussians corresponding to 
odd and even $n$  such that the two are out of phase with each other. In Appendix \ref{appen:2states}, we establish in a straightforward manner that the expressions above in Eqs.~\eref{eqn:Psi_plpl} and \eref{eqn:Psi_mipl} indeed correspond to the two states of concern in the experiment. In what follows, we will use the tomographic approach to distinguish between the two $2$-photon states. Since photon \textit{coincidence} counts were used to experimentally distinguish between the two states, it is reasonable to expect that the time-time slices of the tomograms corresponding to the two $2$-photon states will capture the difference.

The time-time slice of the tomogram corresponding to the state $\ket{\Psi_{\textsc{x}}} (\textsc{x}=\alpha,\beta)$ is given by
\begin{align}
w^{\textsc{x}}(t_{\textsc{s}};t_{\textsc{i}})=\aver{ t_{\textsc{s}};t_{\textsc{i}} \ket{\Psi_{\textsc{x}}}\bra{\Psi_{\textsc{x}}} t_{\textsc{s}};t_{\textsc{i}} },
\label{eqn:tt_slice}
\end{align}
where  $\ket{t_{\textsc{s}};t_{\textsc{i}}}$ 
stands for $\ket{t_{\textsc{s}}}\otimes\ket{t_{\textsc{i}}}$. We point out that this is analogous to the bipartite optical tomogram (Eq.~\eref{eqn:2modeTomoDefn}). 

We work with  the parameter values used 
 in~\cite{perola}, namely,  
$\omega_{p}/(2 \pi)=391.8856$ THz, 
$\overline{\omega}/(2 \pi)=19.2$ GHz, 
$\vardel\omega/(2 \pi)=1.92$ GHz, 
$\Omega_{0}/(2 \pi)=10.9$ THz, and 
$\vardel\Omega/(2 \pi)=6$ THz. 
The time-time slices of the tomograms of 
$\ket{\Psi_{\alpha}}$ and $\ket{\Psi_{\beta}}$ have 
been obtained by 
substituting Eqs.~\eref{eqn:Psi_plpl}  and 
\eref{eqn:Psi_mipl} in turn 
in Eq.~\eref{eqn:tt_slice} and simplifying the 
resulting expressions  
(see Appendix~\ref{appen:ChronoTomo}). 

\begin{figure}
\centering
\includegraphics[width=0.32\textwidth]{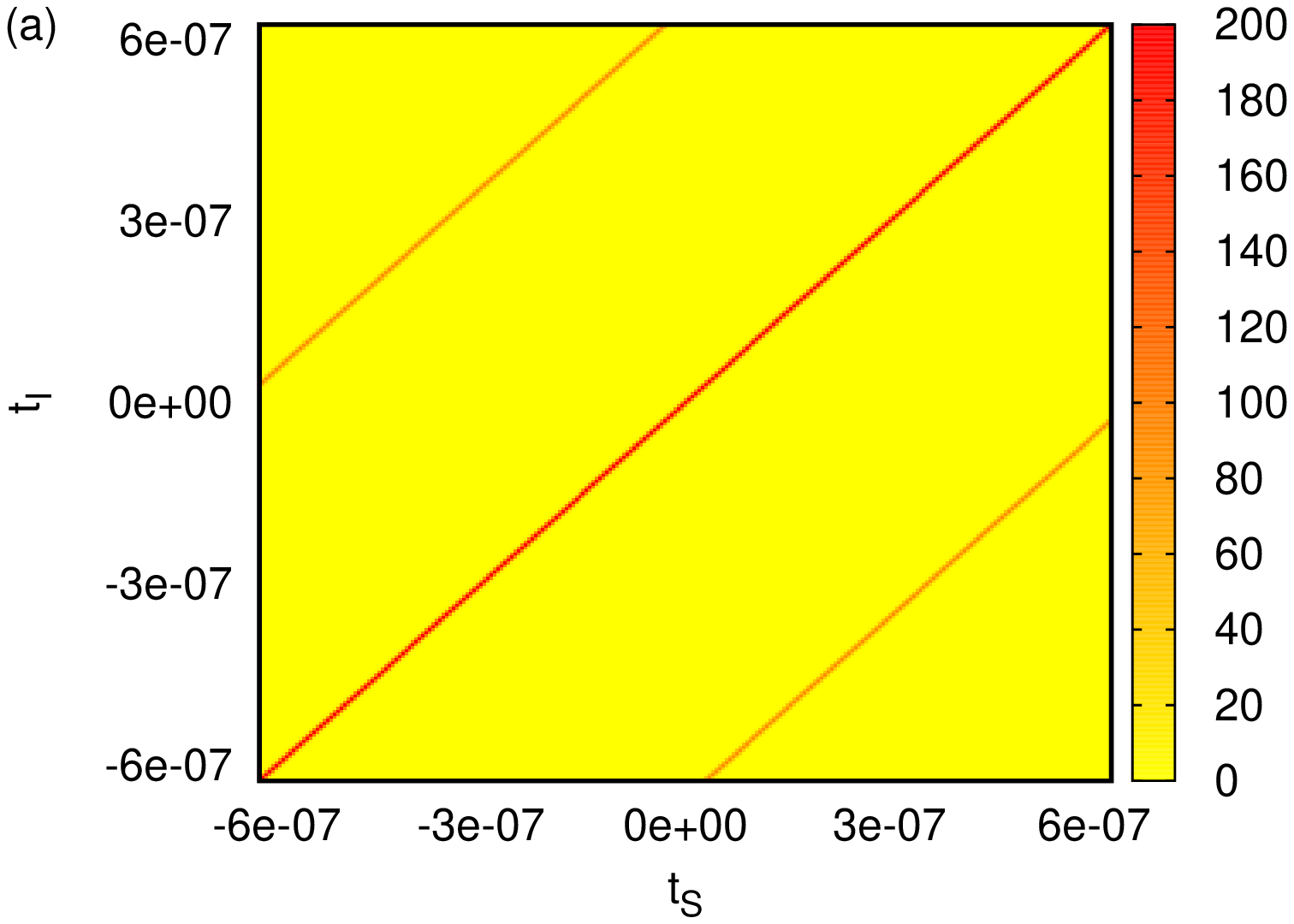}
\includegraphics[width=0.32\textwidth]{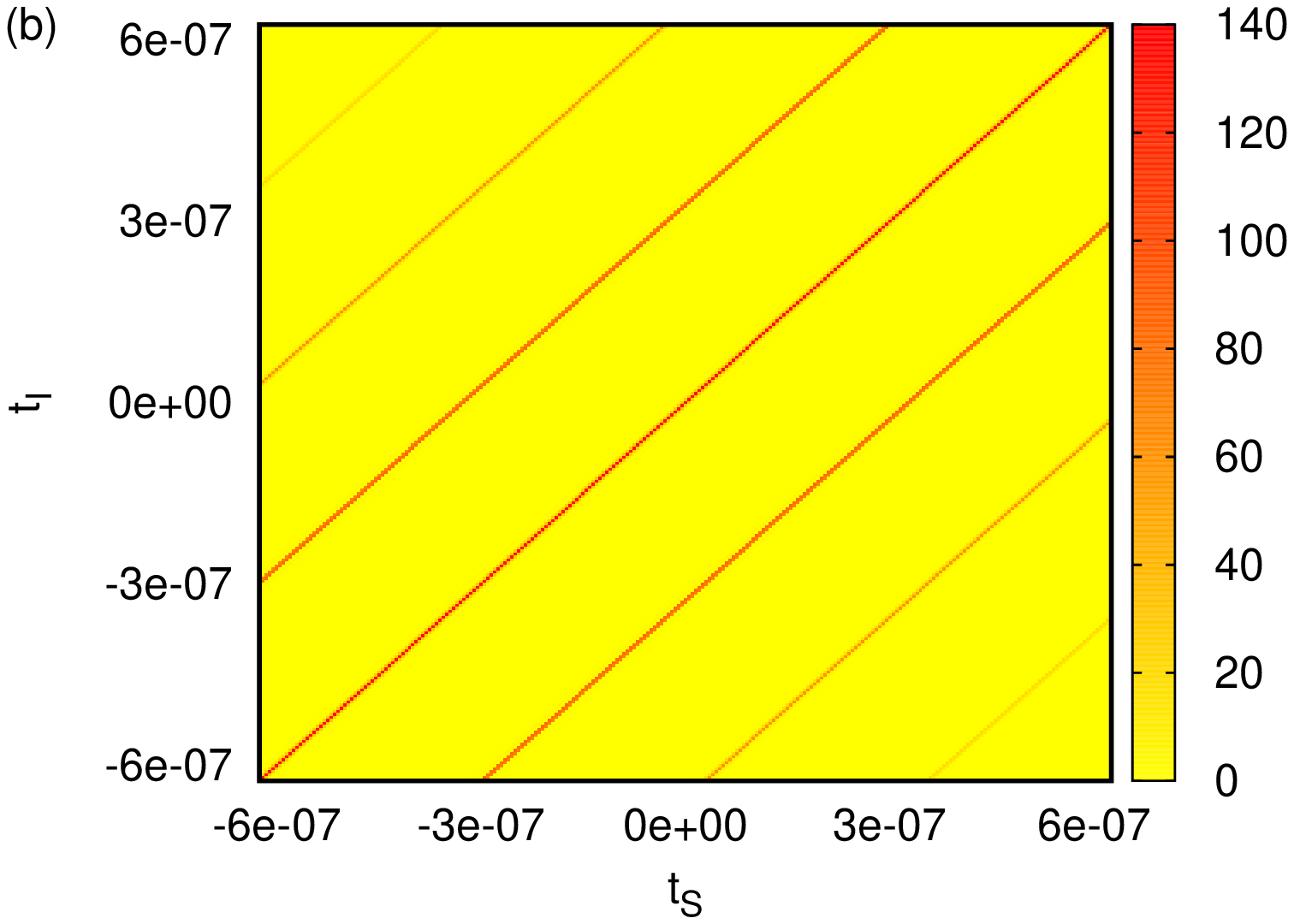}
\includegraphics[width=0.32\textwidth]{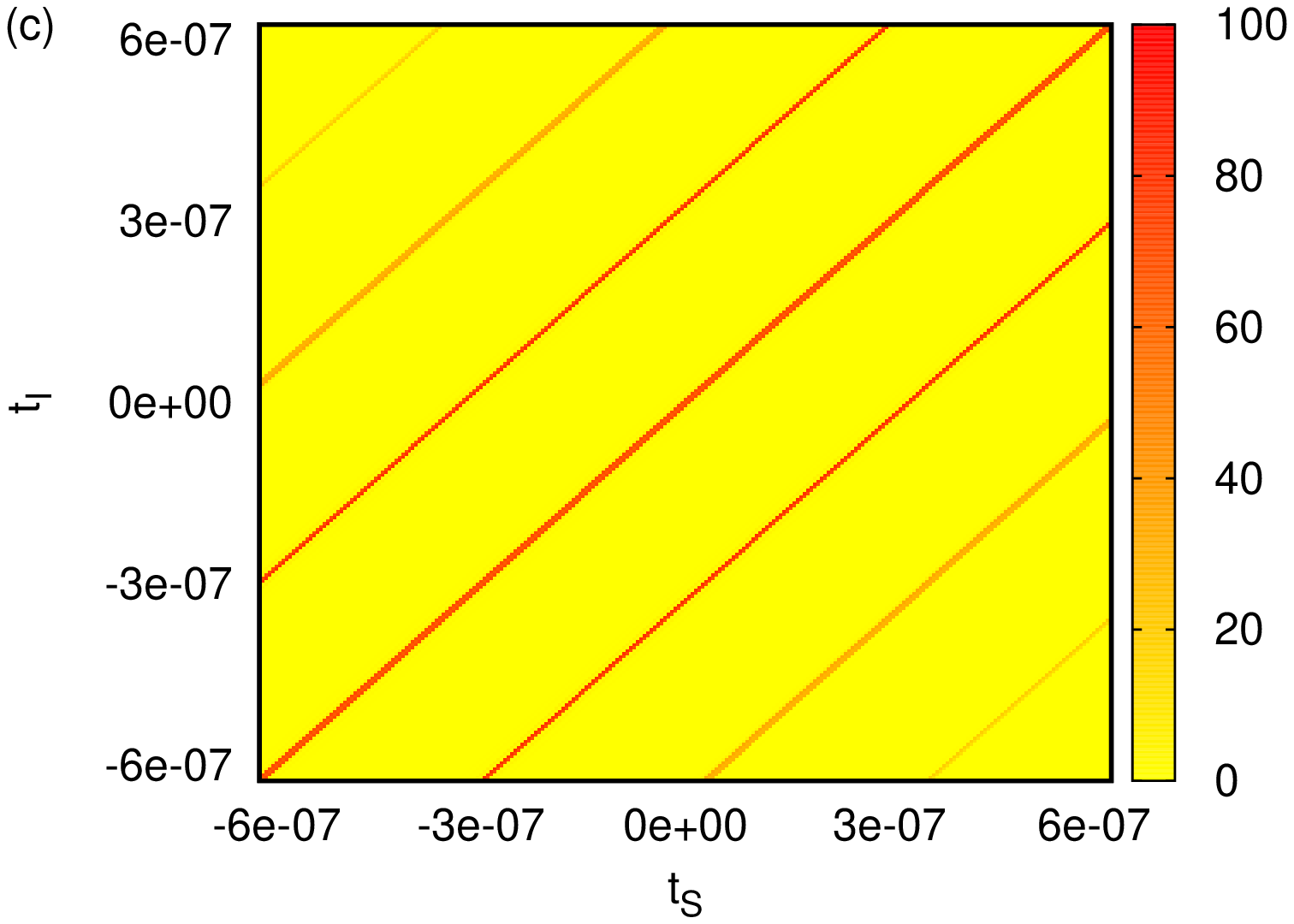}
\caption{Tomographic time-time slice (a) $w^{\alpha}(t_{\textsc{s}};t_{\textsc{i}})$  and (b) $w^{\beta}(t_{\textsc{s}};t_{\textsc{i}})$ vs. $t_{\textsc{s}}$ and $t_{\textsc{i}}$ in seconds. (c) Difference $|w^{\alpha}(t_{\textsc{s}};t_{\textsc{i}})-w^{\beta}(t_{\textsc{s}};t_{\textsc{i}})|$ vs. $t_{\textsc{s}}$ and $t_{\textsc{i}}$ in seconds.}
\label{fig:chrono_tomo}
\end{figure}

As expected, $w^{\alpha}(t_{\textsc{s}};t_{\textsc{i}})$ and $w^{\beta}(t_{\textsc{s}};t_{\textsc{i}})$  are distinctly different 
from each other, as seen in  Figs.~\ref{fig:chrono_tomo} (a)-(c). 
This  difference arises because $\ket{\Psi_{\alpha}}$ and $\ket{\Psi_{\beta}}$ correspond to combs that are clearly displaced with respect to each other, when expressed in the time-time basis.

Next, we calculate  the reduced tomograms 
$w^{\textsc{x}}_{i}(t_{i})$ 
corresponding to subsystem $i$ (where  
$i=\text{S,I}$ and $\textsc{X} =\alpha,\beta$) 
by integrating out the other subsystem,  
as has been done in Eqs.~\eref{eqn:tomo_2_mode_sub_A} and \eref{eqn:tomo_2_mode_sub_B} 
in the case of the optical tomograms. 
(For instance, $w^{\textsc{x}}_{\textsc{s}}(t_{\textsc{s}})=\int \text{d}t_{\textsc{i}} w^{\textsc{x}}(t_{\textsc{s}};t_{\textsc{i}})$.)  
Substituting these full-system and subsystem chronocyclic tomograms in place of the corresponding optical tomograms in Eqs.~\eref{eqn:2modeEntropy}--\eref{eqn:epsTEI}, we obtain the entanglement indicator $\epsarg{tei}$ corresponding to any chosen slice of the chronocyclic tomogram.
(For ease of notation, we have dropped the explicit dependence of 
$\epsarg{tei}$ on the choice of both the tomogram slice and the specific state.) 
In the case of the time-time slice of the tomograms 
we get, finally, the values $\epsarg{tei} = 6.50$ 
for  the state $\ket{\Psi_{\alpha}}$, and 
$\epsarg{tei} = 5.44$ for the state  
$\ket{\Psi_{\beta}}$. Thus,  $\epsarg{tei}$ clearly distinguishes between these two $2$-photon states. 
 We emphasize that the methods used by us could, in principle, provide an alternative approach to the procedure adopted~\cite{perola} in the experiment.


\section{Concluding remarks}
\label{sec:Ch4conclremarks}
Entanglement monotones such as $\xi_{\textsc{svne}}$ and $\xi_{\textsc{sle}}$ can be constructed only if the off-diagonal elements of the density operator are known.
For a single-mode system, the optical tomogram is related to the density matrix $\rho$ by
\begin{equation}
w(X_{\theta},\theta)=\sum_{n,m=0}^{\infty}\bra{X_{\theta},\theta}n\rangle \rho_{nm} \langle m \ket{X_{\theta},\theta},
\label{eqn:tomo_defn_2}
\end{equation}
where $\rho_{nm}=\bra{n}\rho\ket{m}$ and $\lbrace\ket{n}\rbrace$, $\lbrace\ket{m}\rbrace$ constitute the Fock basis. To get $\rho_{nm}$ from the tomogram $w(X_{\theta},\theta)$, we need to invert Eq. \eref{eqn:tomo_defn_2}. But this is essentially the procedure for a full state reconstruction which could be tedious and error-prone. As has been pointed out in \cite{photonnumbertomo2}, even in linear inversion procedures to get $\rho$ from histograms obtained experimentally, small errors in the experimental data get magnified substantially. 
Improved  procedures for minimising errors are of great interest and possibilities for improvement have been explored both experimentally and theoretically (see, for instance,  \cite{photonnumbertomo1,photonnumbertomo3,QubitRecon2016,CVRecon2019}). Experimental challenges are primarily associated with gaining sufficient phase information from more than one  copy of the system. Studies in this regard largely focus on reconstructing single-mode density matrices minimising the errors.  Adapting and extending these ideas to entangled states which is of direct relevance to us, would be a significant step towards extracting entanglement monotones. 

We have demonstrated  that, even without information about the off-diagonal elements of the density matrix, substantial  reproduction of the qualitative aspects of entanglement dynamics can be achieved \textit{using the tomograms alone}. 
The  performance of  the entanglement indicator $\xi^{\prime}_{\textsc{tei}}$ thus obtained, in quantifying the extent of entanglement, has been assessed in this chapter using two model bipartite systems with inherent nonlinearities. We have shown that $\xi^{\prime}_{\textsc{tei}}$ fares significantly better for generic initial states of the system even during temporal evolution, compared to better-known entanglement indicators such as $\xi_{\textsc{ipr}}$. In order to quantify the reliability of the indicator over long intervals of time, the difference between $\xi_{\textsc{svne}}$ and $\xi^{\prime}_{\textsc{tei}}$ has been examined using a time-series analysis. The manner in which this difference $d_{1}(t)$ is sensitive to the nonlinearity of the system, the nature of the interaction, and the precise initial state is revealed by the time-series analysis. The importance and relevance of this investigation lies in the fact that detailed state reconstruction from the tomogram is completely avoided  in identifying an efficient entanglement indicator for generic bipartite systems involving continuous variables.

Further, we have considered a pair of $2$-photon states
 which were experimentally shown to be distinguishable, using the difference in their normalised photon coincidence counts~\cite{perola}. Here we have unambiguously distinguished between these states using the entanglement indicator $\epsarg{tei}$. This demonstrates an alternative procedure using the tomographic approach in an experimentally relevant CV system.

In the next chapter, we extend the investigation to multipartite HQ systems. We shall also demonstrate the utility of the indicator $\xi_{\textsc{tei}}$ with an experimentally-generated tomogram using the IBM quantum computer, and with a spin tomogram computed from experimental data from an NMR experiment~\cite{nmrExpt}.


%% file: chapter5.tex

\chapter{Dynamics of entanglement indicators in hybrid quantum systems and spin systems}
\label{ch:IBM}

\section{\label{sec:Ch5intro} Introduction}
Interesting phenomena such as sudden death and birth of entanglement can arise during temporal evolution in certain HQ systems such as the double Jaynes-Cummings (DJC) model~\cite{eberly}  and the double Tavis-Cummings (DTC) model~\cite{dtcm}. These models of atom-field interactions have been investigated extensively in the literature~\cite{jcm1,jcm2,tcm1}. We examine the performance of $\xi_{\textsc{tei}}$ and $\xi^{\prime}_{\textsc{tei}}$ in these models by comparing them with the quantum mutual information $\xi_{\textsc{qmi}}$ at various instants of time during temporal evolution. $\xi_{\textsc{qmi}}$ is a useful measure of the quantum correlation between any two subsystems of a multipartite system and is defined as (Eq. \eref{eqn:qmi}),
 \begin{equation}
\nonumber \xi_{\textsc{qmi}}=\xi_{\textsc{svne}}^{(\textsc{a})}+\xi_{\textsc{svne}}^{(\textsc{b})}-\xi_{\textsc{svne}}^{(\textsc{ab})}.
 \end{equation}
Here the superscripts (A), (B) and (AB) in the RHS refer to the subsystems A, B and the bipartite subsystem AB respectively. 
The terms, \textit{quantum correlation} and \textit{entanglement} will be used interchangeably in this chapter, as the former sets the upper bound on the latter for any two subsystems.

We have numerically generated tomograms at various instants during temporal evolution in the DJC and the DTC models. Wherever possible, we have compared these tomograms with those obtained from equivalent circuits executed in the IBM quantum computing platform (IBM Q). The tomograms from the IBM quantum computer have been obtained both through experimental runs and through simulations using the IBM open quantum assembly language (QASM) simulator~\cite{ibm_main,qiskit}.

We have also extended our investigations to spin systems. For this purpose, we have used the reconstructed density matrices from the liquid-state NMR experiment, provided to us by the NMR-QIP group in IISER Pune, India. This experiment has been peformed using NMR spectroscopic techniques on $\vphantom{}^{13}\text{C}$, $\vphantom{}^{1}\text{H}$ and $\vphantom{}^{19}\text{F}$ atoms in dibromofluoromethane dissolved in acetone. The extent of entanglement has been quantified using negativity at various instants of time~\cite{nmrExpt}. Since our aim is to assess the usefulness of $\xi_{\textsc{tei}}$, we have numerically calculated the tomograms from these reconstructed density matrices, and compared $\xi_{\textsc{tei}}$ with $\xi_{\textsc{qmi}}$ and negativity.

In Section \ref{sec:Ch5models}, we describe and investigate the DJC and the DTC models. In Section \ref{sec:Ch5NMRexpt}, we analyse tomograms corresponding to the NMR system. We also comment on the spin, higher-order and entropic squeezing properties in this system. We conclude with brief remarks.

\section{\label{sec:Ch5models} Hybrid multipartite models}
 
We now proceed to assess the efficacy of $\xi_{\textsc{tei}}$ and $\xi^{\prime}_{\textsc{tei}}$ in the DJC and DTC models.

\subsection{The double Jaynes-Cummings model}

The model comprises two $2$-level atoms C and D which are initially in an entangled state, with each atom interacting  with  strength 
$g_{0}$ with radiation fields A and B respectively. The effective Hamiltonian (setting $\hbar=1$) is \cite{eberly}  
\begin{align}
\nonumber H_{\textsc{djc}}= &\sum_{j=\textsc{a,b}} 
\chi_{\textsc{f}} a_{j}^{\dagger} a_{j} + \sum_{k=\textsc{c,d}} \chi_{0} \sigma_{k z} + g_{0} \,(a_{\textsc{a}}^{\dagger} \sigma_{\textsc{c} -} + a_{\textsc{a}} \sigma_{\textsc{c} +})\\
&+ g_{0} \,(a_{\textsc{b}}^{\dagger} \sigma_{\textsc{d} -} + a_{\textsc{b}} \sigma_{\textsc{d} +}).
\label{eqn:HDJC}
\end{align}
$a_{j}, a_{j}^{\dagger}$ ($j=\text{A,B}$) are photon  annihilation and creation operators, $\chi_{\textsc{f}}$ is the frequency of the fields, 
and $\chi_{0}$ is the  energy  
difference between the two atomic levels.
In terms of the matrices $\sigma_{x}$, $\sigma_{y}$ and $\sigma_{z}$ (Eq. \eref{eqn:atomops}), the atomic ladder operators for subsystem $k$  
are given by $\sigma_{k \pm}=(\sigma_{k x} \pm i \sigma_{k y})$.
The initial atomic states considered both in the DJC model and the DTC model are of the form
\begin{equation}
 \ket{\psi_{+}}=\big(\ket{g}_{1}\otimes\ket{g}_{2} + \ket{e}_{1}\otimes\ket{e}_{2}\big)/\sqrt{2}
 \label{eqn:psi0_defn}
 \end{equation} 
and 
\begin{equation}
\ket{\phi_{+}}=
\big(\ket{g}_{1}\otimes\ket{e}_{2} + \ket{e}_{1}\otimes\ket{g}_{2}\big)/\sqrt{2}.
 \label{eqn:phi0_defn}
\end{equation}
Here $\ket{g}_{p}$ and  $\ket{e}_{p}$ ($p=1,2$) 
denote the respective ground and excited states 
of atom $p$.  In  the DJC model,  $1$ and $2$ are to be replaced by C and D respectively.  A and B  are initially in the zero-photon states $\ket{0}_{\textsc{a}}$ and $\ket{0}_{\textsc{b}}$. The two initial states of the full system that we consider are $\ket{0}_{\textsc{a}} \otimes \ket{0}_{\textsc{b}} \otimes \ket{\psi_{+}}_{\textsc{cd}} \equiv \ket{0;0;\psi_{+}}$ and 
$\ket{0}_{\textsc{a}} \otimes \ket{0}_{\textsc{b}} \otimes \ket{\phi_{+}}_{\textsc{cd}} \equiv \ket{0;0;\phi_{+}}$.

We have numerically generated tomograms 
at approximately 300 instants of time, 
separated by a time step equal to $0.02$ (in units of 
$\pi/g_{0}$).  From these, we have obtained $\xi_{\textsc{tei}}$ at different instants as the system evolves. It was shown in Chapter \ref{ch:TEItimeseries} that for radiation fields, both $\xi_{\textsc{tei}}$ and $\xi^{\prime}_{\textsc{tei}}$ were in fairly good agreement~\cite{sharmila2}. (We recall that the latter was computed by averaging over only those values of $\epsarg{tei}(\thetaa,\thetab)$ that exceed the mean by one standard deviation).
We now proceed to investigate if this holds even in the case of HQ systems. 

\begin{figure}[h]
\centering
\includegraphics[width=0.32\textwidth]{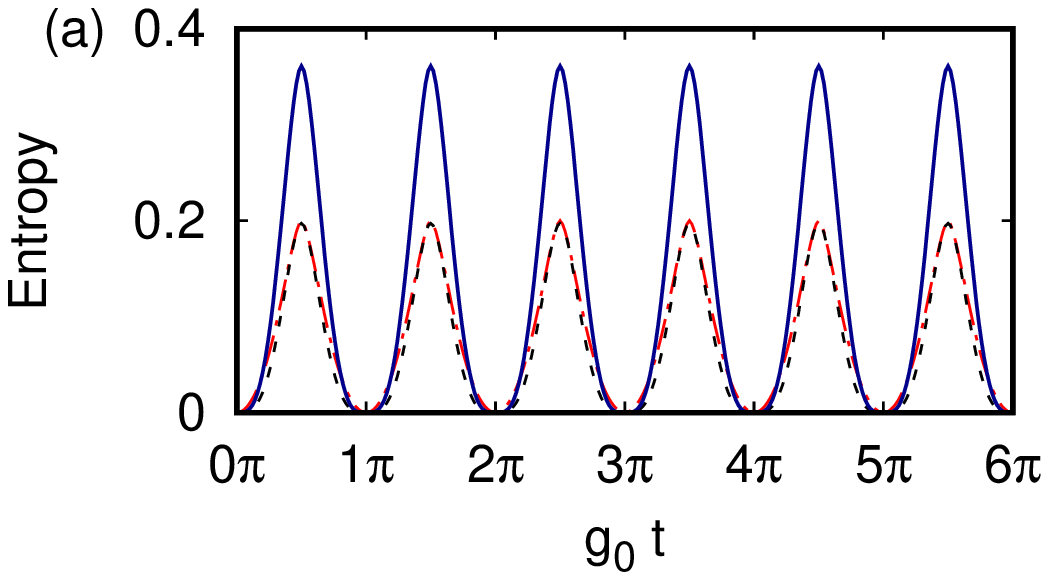}
\includegraphics[width=0.32\textwidth]{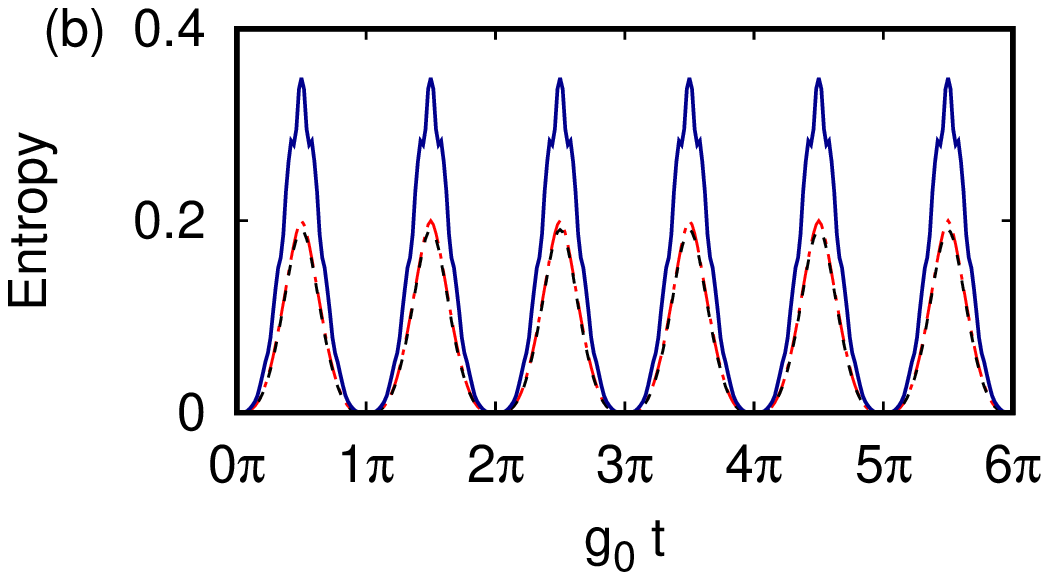}
\includegraphics[width=0.32\textwidth]{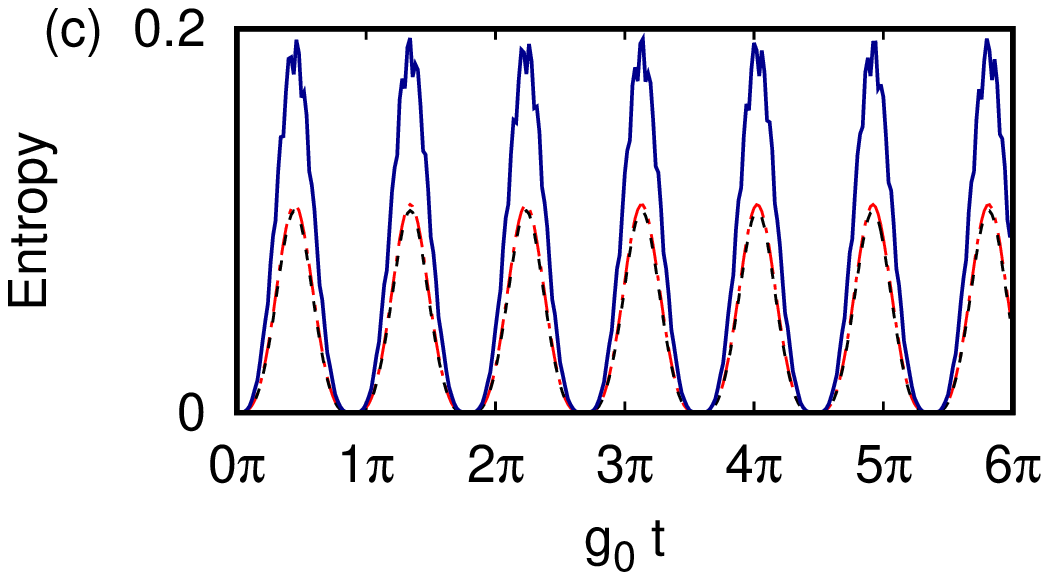}
\caption{$\xi_{\textsc{tei}}$ (black), $\xi^{\prime}_{\textsc{tei}}$ (blue) and $0.1\,\xi_{\textsc{qmi}}$ (red) vs. scaled time $g_{0} t$  for the field subsystem in the DJC model. Initial state (a) $\ket{0;0;\phi_{+}}$ ; (b), (c) $\ket{0;0;\psi_{+}}$. Detuning parameter (a),(b) 0 and (c) 1.}
\label{fig:comp_Neq1_field}
\end{figure}

The entanglement indicators $\xi_{\textsc{tei}}$, $\xi^{\prime}_{\textsc{tei}}$ and $\xi_{\textsc{qmi}}$  are 
plotted against  the scaled time $g_{0} t$ in Figs. 
\ref{fig:comp_Neq1_field} (a)-(c) 
in the case of the field subsystems. The detuning parameter, given by $(\chi_{\textsc{f}} - \chi_{0})$, is $0$ in Figs. \ref{fig:comp_Neq1_field} (a), (b),  
and $1$ in Fig. \ref{fig:comp_Neq1_field} (c). 
The initial states considered are  $\ket{0;0; \phi_{+}}$ 
in Fig. \ref{fig:comp_Neq1_field} (a)
and $\ket{0;0;\psi_{+}}$ in Figs. 
\ref{fig:comp_Neq1_field} (b),(c). 
For ease of comparison, $\xi_{\textsc{qmi}}$ has been  
scaled down by a factor of $10$.  
It is evident that  
$\xi^{\prime}_{\textsc{tei}}$ is a good approximation 
to $\xi_{\textsc{tei}}$ and that both mimic 
$\xi_{\textsc{qmi}}$ closely in all the three cases 
considered. Sensitivity to the precise initial atomic state 
and to the extent of detuning is revealed 
by examining the qualitative features of the indicators in the neighbourhood of their maximum values.

\begin{figure}[h]
\centering
\includegraphics[width=0.32\textwidth]{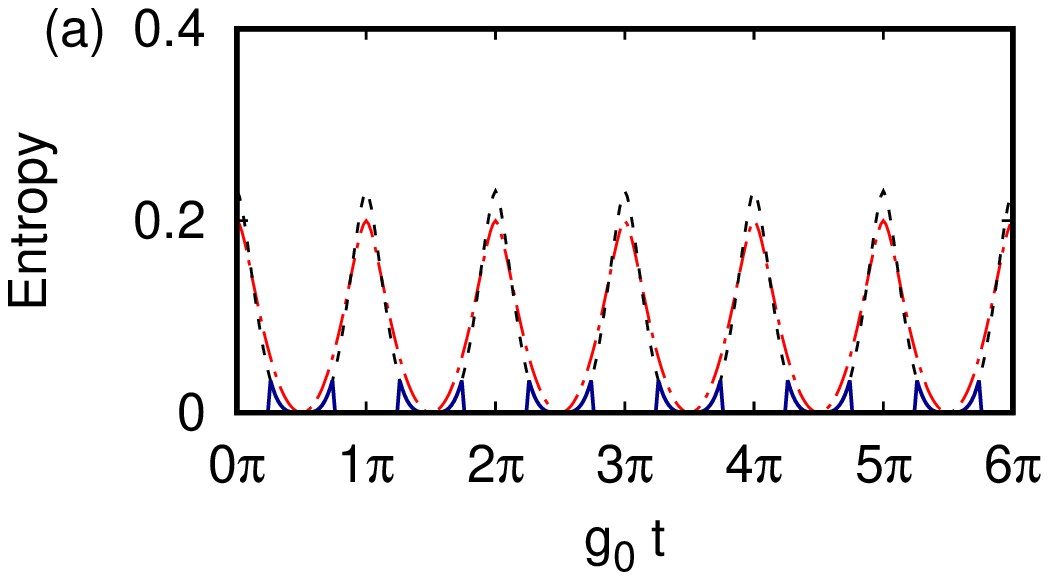}
\includegraphics[width=0.32\textwidth]{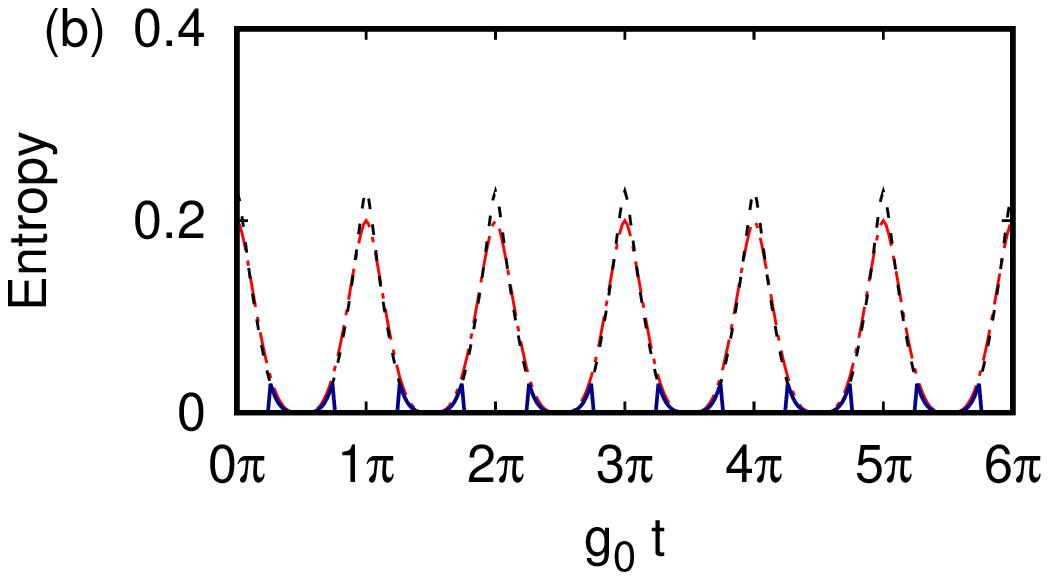}
\includegraphics[width=0.32\textwidth]{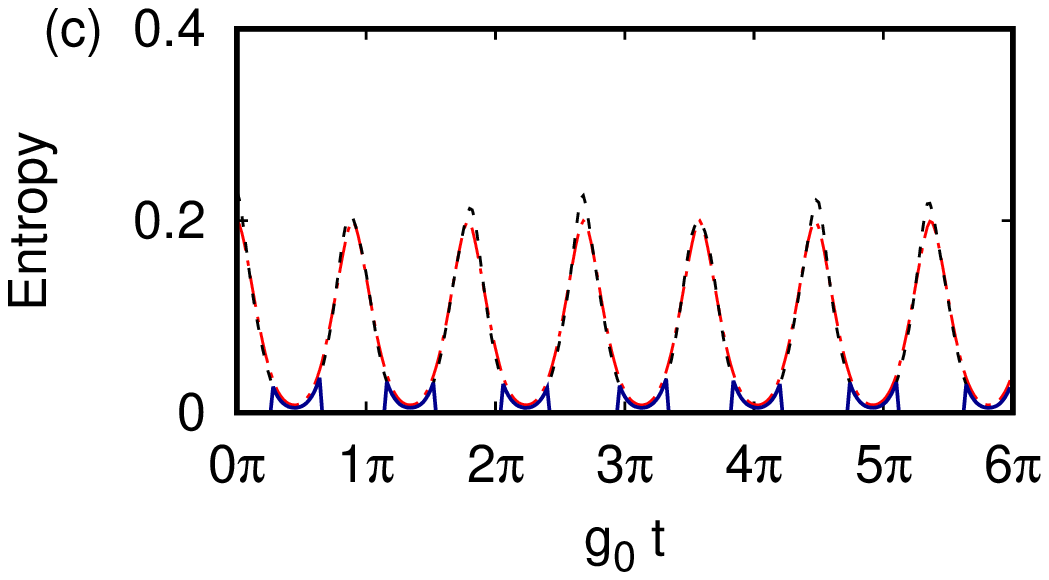}
\caption{$\xi_{\textsc{tei}}$ (black), $\xi^{\prime}_{\textsc{tei}}$ (blue) and $0.1\,\xi_{\textsc{qmi}}$ (red) vs. scaled time $g_{0} t$ for the atomic subsystem in the DJC model.  Initial state (a) $\ket{0;0;\phi_{+}}$ ; (b), (c) $\ket{0;0;\psi_{+}}$. Detuning parameter (a),(b) 0 and (c) 1.}
\label{fig:comp_Neq1_atom}
\end{figure}

Figs. \ref{fig:comp_Neq1_atom} (a)-(c) are plots of 
$\xi_{\textsc{tei}},  \xi^{\prime}_{\textsc{tei}}$ and $\xi_{\textsc{qmi}}$ corresponding to the atomic subsystem for the same set of parameters and initial states as in Figs. \ref{fig:comp_Neq1_field} (a)-(c).  
In this case, although $\xi_{\textsc{tei}}$ is in good 
agreement with  $\xi_{\textsc{qmi}}$  over the time interval considered, $\xi^{\prime}_{\textsc{tei}}$ is not, 
in sharp  contrast to the situation for the field subsystems. 
We note that when the detuning parameter is zero,  
$\xi_{\textsc{qmi}}$ returns to  its initial value of $2$ 
at the instant $g_{0} t = \pi$. 
 We will use this feature in the sequel, when we construct an equivalent circuit for the DJC model. 

\begin{figure}[h]
\centering
\includegraphics[width=\textwidth]{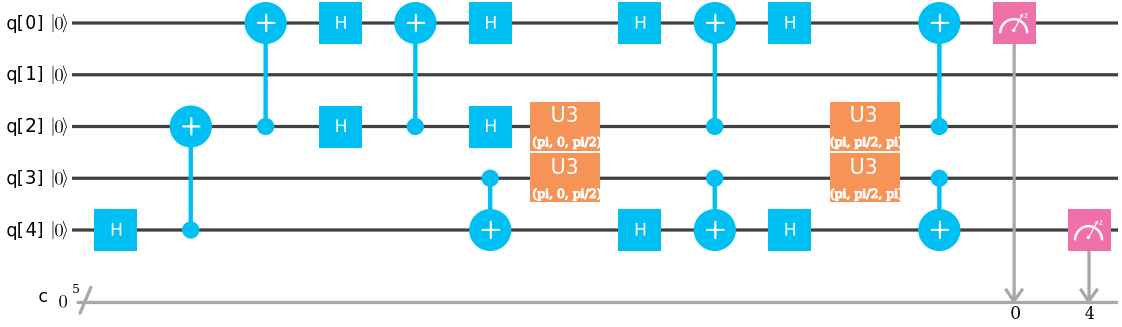}
\vspace*{1 ex}
\caption{Equivalent circuit for the DJC model (created using IBM Q).}
\label{fig:circuit_dia}
\end{figure}

\begin{figure}
\includegraphics[width=0.3\textwidth]{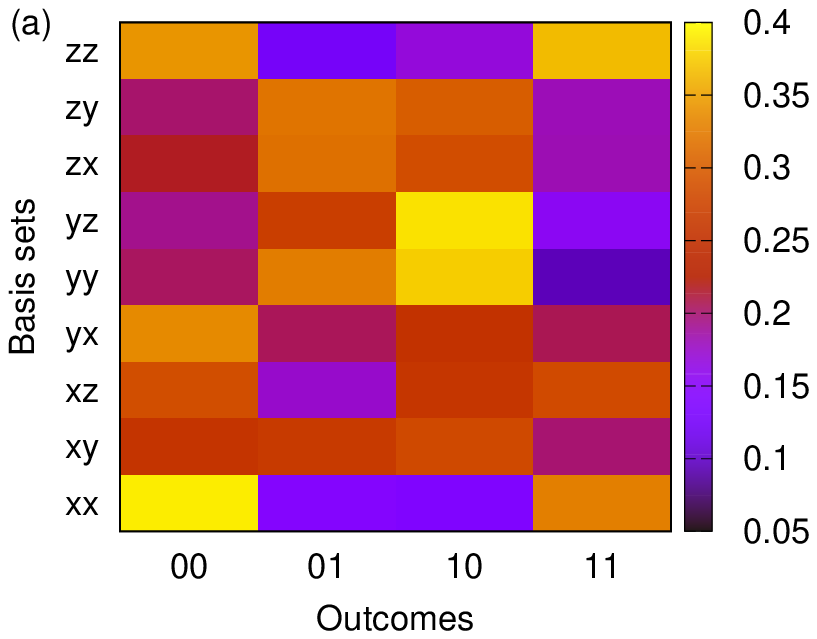}
\includegraphics[width=0.3\textwidth]{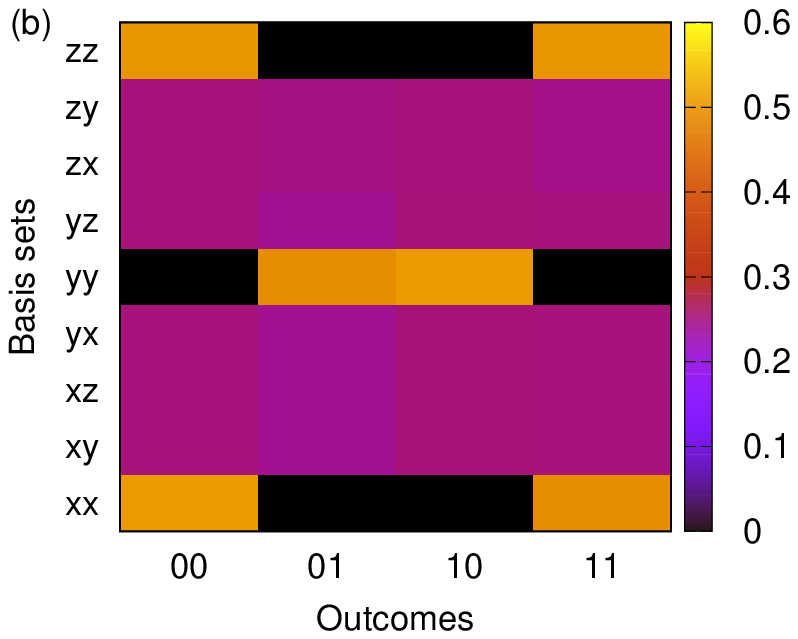}
\includegraphics[width=0.3\textwidth]{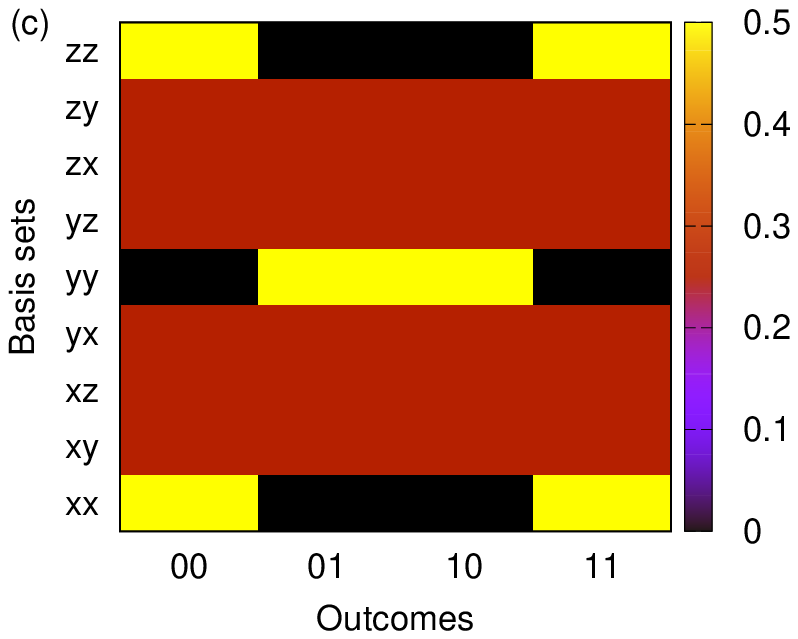}
\caption{Tomograms from (a) IBM Q experiment (b) QASM simulation (c) numerical computations of the DJC model.}
\label{fig:tomograms}
\end{figure}

The equivalent circuit that was provided by us to the IBM Q from which bipartite qubit tomograms are obtained (analogous to the tomograms corresponding to the atomic subsystem of the DJC model)  is shown  in Fig. \ref{fig:circuit_dia}. We use the standard notation  
of the IBM platform \cite{ibm_main}. In the circuit,
$q[0]$ and $q[4]$ are the qubits that follow the dynamics of the atomic subsystem while $q[2]$ and $q[3]$ act as auxiliary qubits to aid the dynamics. Since  transitions between the two energy levels of either atom in the DJC model involve absorption or emission of a single photon,  each auxiliary qubit in the equivalent circuit toggles  between the qubit states $\ket{0}$ and $\ket{1}$ respectively. The operator $U_{3}(\theta^{\prime},\varphi^{\prime},\upsilon)$ in the circuit is given by
\begin{equation}
U_{3}(\theta^{\prime},\varphi^{\prime},\upsilon)=\begin{bmatrix}
\cos\,(\theta^{\prime}/2) \;&\; \;-e^{i \upsilon}\,\sin\,(\theta^{\prime}/2) \\[4pt]
e^{i \varphi^{\prime}}\,\sin\,(\theta^{\prime}/2) \;
& \;\;e^{i (\upsilon+\varphi^{\prime})}\,\cos\,(\theta^{\prime}/2)
\end{bmatrix},
\label{eqn:U3}
\end{equation} 
where  $0\leqslant\theta^{\prime}<\pi, \, 
0\leqslant \varphi^{\prime}<2 \pi$ and $0\leqslant\upsilon<2 \pi$. Each of the four qubits is initially in the qubit state $\ket{0}$. The initial entangled state between $q[0]$ and $q[4]$ (analogous to the initial state $\ket{\psi_{+}}$ of the atomic subsystem) is prepared in the circuit using an Hadamard and a controlled-NOT gate  between $q[4]$ and $q[2]$ and a SWAP gate between $q[2]$ and $q[0]$. Here, $\theta^{\prime}$ is analogous to $g_{0} t$ in the DJC model. We 
choose $\theta^{\prime}=\pi$ so that the extent of entanglement is equal to its initial value ($= 2$), $\varphi^{\prime}=0$ and $\upsilon=\pi/2$. The matrix  
$U_{3}(\pi,\pi/2,\pi)$ which appears in the equivalent circuit is equal to $U_{3}^{\dagger}(\pi,0,\pi/2)$. 
Measurements are carried out in the $x$, $y$ and $z$ bases corresponding to the matrices defined in Eq. \eref{eqn:atomops}. A measurement in the $z$-basis is automatically provided by the IBM platform. A measurement in the $x$-basis is achieved by applying an Hadamard gate followed by a $z$-basis measurement.  Defining the operator
\begin{equation}
S^{\dagger} = \begin{bmatrix}
1 & 0 \\
0 & -i
\end{bmatrix},
\label{eqn:sdg}
\end{equation}
measurement in the $y$-basis is achieved by applying $S^{\dagger}$, then  an Hadamard gate, and  finally a measurement in 
the $z$-basis.
Measurements in the  $x$, $y$ and $z$ bases  are needed for obtaining the spin tomogram, Fig. \ref{fig:tomograms} (a). (This is equivalent to the bipartite atomic tomogram in the DJC model, in the basis sets of $\sigma_{x}$, $\sigma_{y}$ and $\sigma_{z}$). 

These spin tomograms have also been obtained  experimentally 
using the IBM superconducting circuit with appropriate Josephson junctions  (Fig. \ref{fig:tomograms} (a)), and the QASM simulator  provided by IBM. The latter does not take into account losses at various stages of the circuit (Fig. \ref{fig:tomograms} (b)). These tomograms are compared with the atomic tomograms (Fig. \ref{fig:tomograms} (c)) of the DJC model with decoherence effects neglected.
The qualitative features are very similar in Figs. \ref{fig:tomograms} (b) and (c) as the circuit follows the dynamics of the atomic subsystem of the DJC model. As expected, Fig. \ref{fig:tomograms} (a) is distinctly different due to experimental losses.

From these tomograms, $\xi_{\textsc{tei}}$ has been calculated. The values 
obtained from the experiment, simulation and numerical analysis 
are $0.0410 \pm 0.0016$, $0.2311$ and $0.2310$,  respectively. 
Six tomograms were obtained from 
six executions of the experiment. Each execution comprised $8192$ runs over each of the 9 basis sets. The error bar 
was calculated from the standard deviation of $\xi_{\textsc{tei}}$.

It is instructive to estimate the extent of loss in state preparation 
{\em  alone}. For this purpose, an entangled state of two qubits was prepared using an Hadamard and a controlled-NOT gate, to effectively mimic $\ket{\psi_{+}}$. For the initial state $\ket{\psi_{+}}$, the values for $\xi_{\textsc{tei}}$ obtained from the experiment, simulation and the DJC model are $0.0973 \pm 0.0240$, $0.2310$ and $0.2310$ respectively. This demonstrates that substantial losses arise even in state preparation.  In order to examine the  extent to which  an increase in the number of atoms in the system increases these losses, we 
turn to  the DTC model.

\subsection{The double Tavis-Cummings model}

The model comprises  four two-level atoms, $\text{C}_{1}$, $\text{C}_{2}$, $\text{D}_{1}$ and $\text{D}_{2}$, with $\text{C}_{1}$ and $\text{C}_{2}$ (respectively, $\text{D}_{1}$ and $\text{D}_{2}$)  coupled with strength $g_{0}$ to a radiation field A (resp., B) of frequency $\chi_{\textsc{f}}$. The notation used is similar to that in $H_{\textsc{djc}}$ (Eq. \eref{eqn:HDJC}), since the Hamiltonian $H_{\textsc{dtc}}$ can be obtained from the former by appropriate changes. Setting $\hbar=1$, we have~\cite{dtcm},  
\begin{align}
\nonumber H_{\textsc{dtc}}=\sum_{j=\textsc{a,b}} 
&\chi_{\textsc{f}} \,a_{j}^{\dagger} a_{j}  + \sum_{k=1}^{2}
\big\{\chi_{0} \,\sigma_{\textsc{c}_{k} z} + \chi_{0} \,\sigma_{\textsc{d}_{k} z} \\
&+ g_{0} ( a_{\textsc{a}}^{\dagger} \sigma_{\textsc{c}_{k} -} + a_{\textsc{a}} \sigma_{\textsc{c}_{k} +}) + g_{0} ( a_{\textsc{b}}^{\dagger} \sigma_{\textsc{d}_{k} -} + a_{\textsc{b}} \sigma_{\textsc{d}_{k} +})\big\},
\label{eqn:HDTC}
\end{align}
$\text{C}_{1}$ and $\text{D}_{1}$ (respectively, $\text{C}_{2}$ and $\text{D}_{2}$) are in the initial state $\ket{\psi_{+}}$ (Eq. \eref{eqn:psi0_defn}) or $\ket{\phi_{+}}$ (Eq. \eref{eqn:phi0_defn}). Each field is initially in $\ket{0}$. We therefore consider the initial states $\ket{0;0;\psi_{+};\psi_{+}}$, $\ket{0;0;\phi_{+};\phi_{+}}$ and $\ket{0;0;\psi_{+};\phi_{+}}$. The notation $\ket{0;0;\psi_{+};\phi_{+}}$ indicates, for instance, that A and B are in the state $\ket{0}$, the subsystem $(\text{C}_{1}, \text{D}_{1})$ is in the state $\ket{\psi_{+}}$, and the subsystem $(\text{C}_{2}, \text{D}_{2})$ is in the state $\ket{\phi_{+}}$. (We do not consider the initial state $\ket{0;0;\phi_{+};\psi_{+}}$ separately because the results corresponding to that case can be obtained using symmetry arguments from the results for the initial state $\ket{0;0;\psi_{+};\phi_{+}}$). We shall denote $(\text{C}_{1}, \text{C}_{2})$ by C and $(\text{D}_{1}, \text{D}_{2})$ by D.  

\begin{figure}
\includegraphics[width=0.8\textwidth]{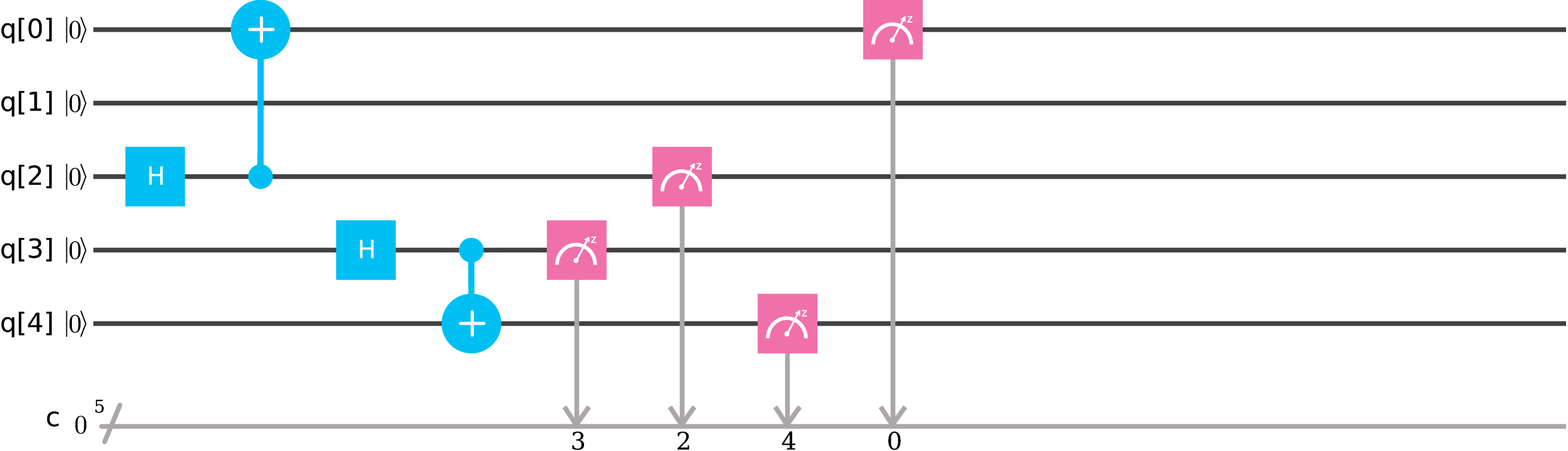}
\caption{Equivalent circuit of the entangled state $\ket{\psi_{+};\psi_{+}}$ in the DTC model (created using IBM Q). }
\label{fig:DTC_circuit}
\end{figure}

An equivalent circuit for the DTC model requires $4$ qubits to mimic the four two-level atoms, and a minimum of $4$ auxiliary qubits to aid the dynamics. In what follows, we assess the extent of losses in  state preparation  {\em alone}. For this purpose, $4$ qubits are prepared in a pairwise entangled state (analogous to the initial state $\ket{\psi_{+};\psi_{+}}$ of the atomic subsystem (C,D)) using $2$ Hadamard and $2$ controlled-NOT gates (Fig. \ref{fig:DTC_circuit}). Here qubits $q[2]$ and $q[3]$ are entangled with qubits $q[0]$ and $q[4]$ respectively. We note that the pair ($q[2]$,$q[3]$) is analogous to subsystem C,  and ($q[0]$,$q[4]$) is analogous to D. The extent of entanglement between C and D is quantified using $\xi_{\textsc{tei}}$. The numerical values obtained from experiment, simulation and the DTC model are $0.2528$, $0.4761$ and $0.4621$, respectively. In this case, the experiment was executed just once. This comprised $8192$ runs~\cite{ibm_main} over each of the $81$ basis sets. Thus the outcome of the experiment, namely $0.2528$, is not prescribed with an error bar. As $4$ qubits are involved in this circuit, the number of possible outcomes is $16$, in contrast to the earlier case which had only $4$ outcomes. Hence, the experimental losses,  as well as the difference between the simulated and the numerically obtained values,  are higher than those obtained for the DJC model. We therefore proceed to investigate the entanglement dynamics in the DTC model numerically in the absence of losses.

\begin{figure}[h]
\centering
\includegraphics[width=0.32\textwidth]{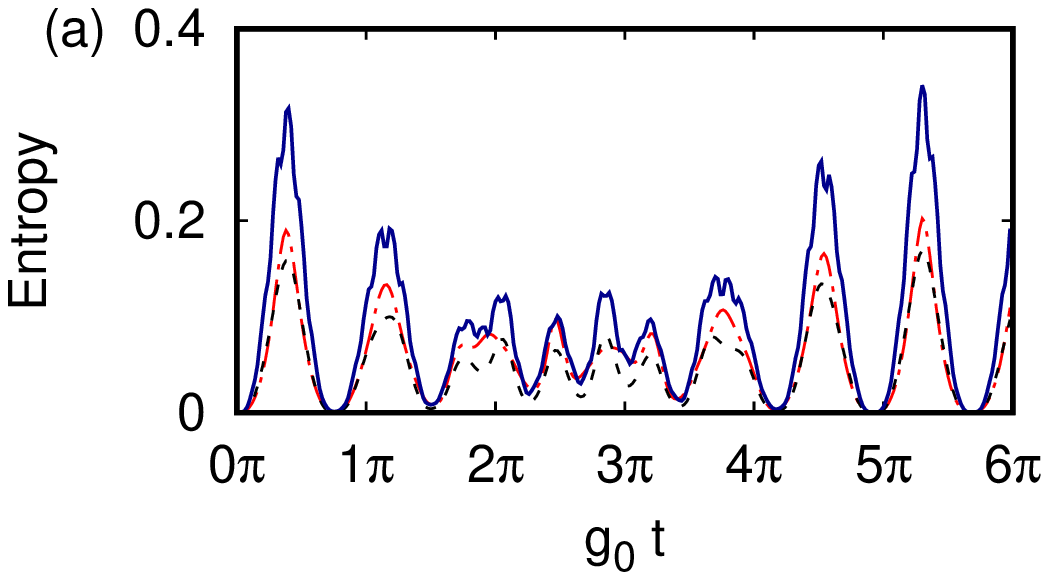}
\includegraphics[width=0.32\textwidth]{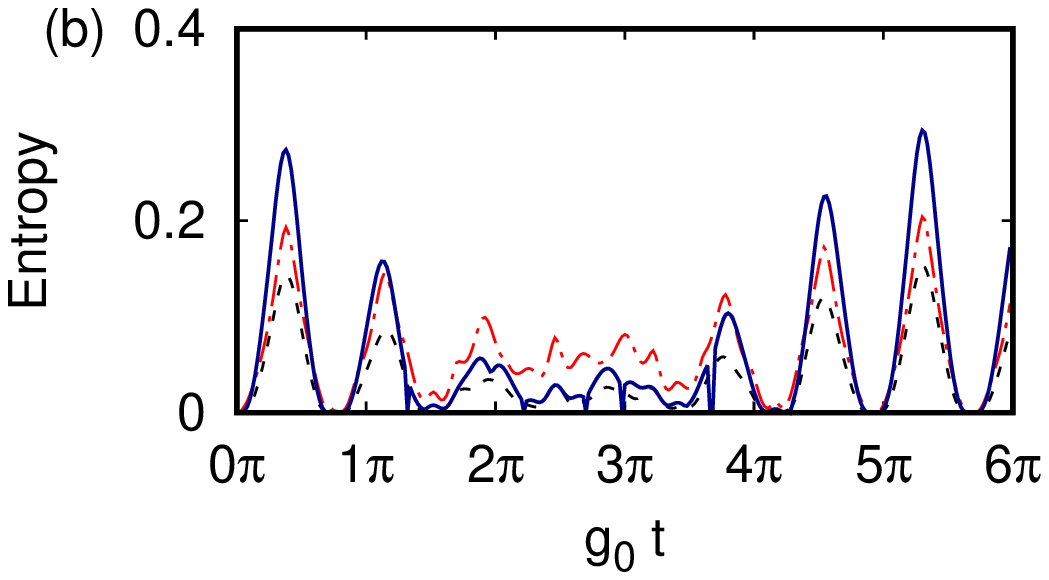}
\includegraphics[width=0.32\textwidth]{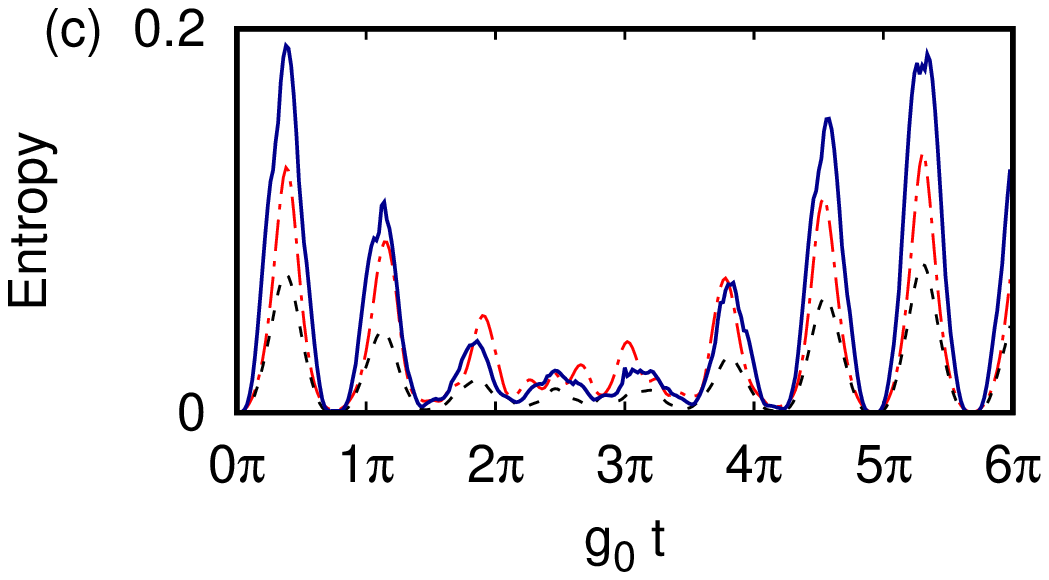}
\caption{$\xi_{\textsc{tei}}$ (black), $\xi^{\prime}_{\textsc{tei}}$ (blue) and $0.1 \,\xi_{\textsc{qmi}}$ (red) vs. scaled time $g_{0} t$ for the field subsystem in the DTC model with zero detuning. Initial field state $\ket{0;0}$, initial atomic state (a) $\ket{\psi_{+}; \psi_{+}}$\, (b) $\ket{\phi_{+};\phi_{+}}$\, (c) $\ket{\psi_{+};\phi_{+}}$.}
\label{fig:comp_Neq2_field}
\end{figure}

\begin{figure}[h]
\centering
\includegraphics[width=0.32\textwidth]{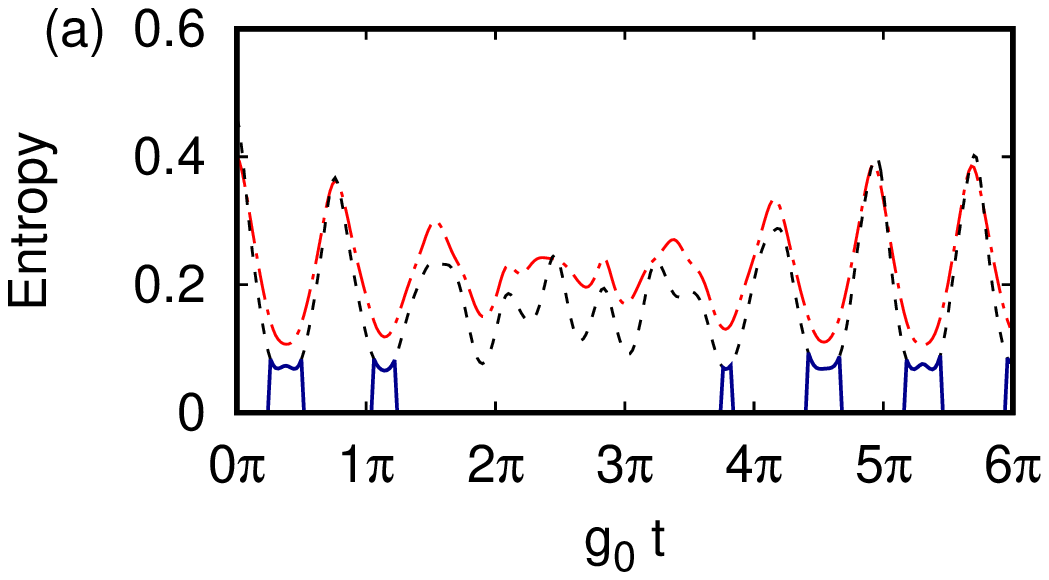}
\includegraphics[width=0.32\textwidth]{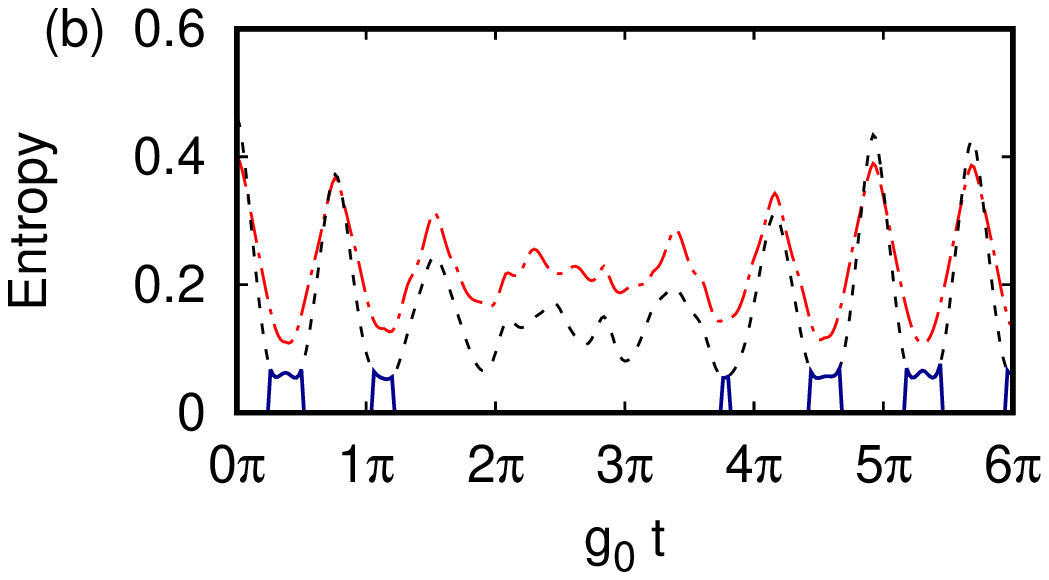}
\includegraphics[width=0.32\textwidth]{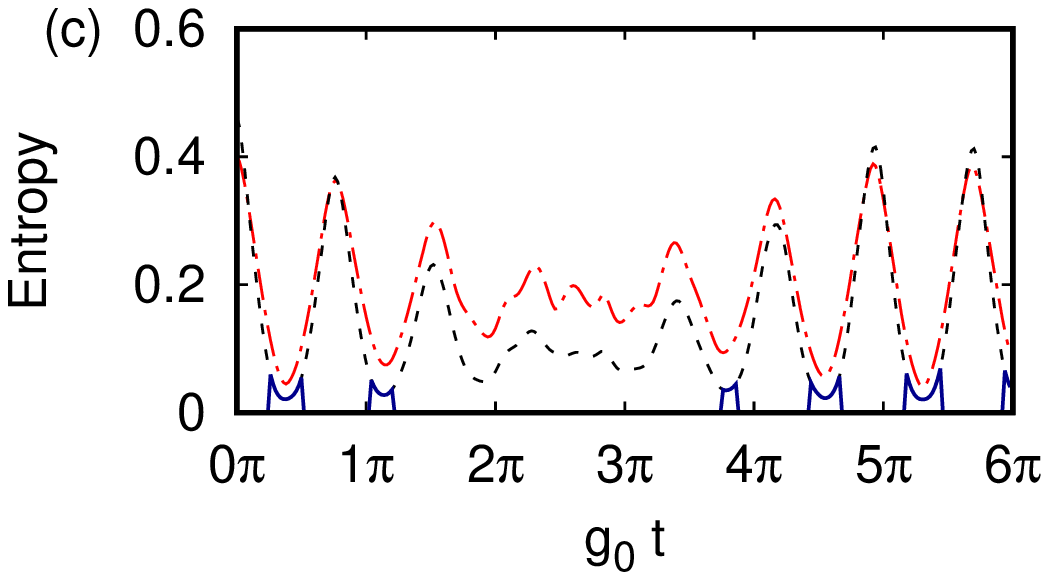}
\caption{$\xi_{\textsc{tei}}$ (black), $\xi^{\prime}_{\textsc{tei}}$ (blue) and $0.1 \,\xi_{\textsc{qmi}}$ (red) vs. scaled time $g_{0} t$ for the bipartite atomic subsystem CD in the DTC model with zero detuning. Initial field state $\ket{0;0}$, initial atomic state (a) $\ket{\psi_{+}; \psi_{+}}$\, (b) $\ket{\phi_{+};\phi_{+}}$\, (c) $\ket{\psi_{+};\phi_{+}}$.}
\label{fig:comp_Neq2_det0_atom}
\end{figure}

We now investigate the extent of entanglement between the field subsystems  A and B, and between  the atomic subsystems C and D, as the system evolves in time.  For this purpose, we have generated tomograms at 300 instants separated by a time step $0.02$ in units of $\pi/g_{0}$ setting the detuning parameter to zero without loss of generality. From these, $\xi_{\textsc{tei}}$ and $\xi^{\prime}_{\textsc{tei}}$ have been obtained. 
Plots of $\xi_{\textsc{tei}}$, 
$\xi^{\prime}_{\textsc{tei}}$ and $\xi_{\textsc{qmi}}$ for the field subsystem is shown in Figs. \ref{fig:comp_Neq2_field} (a)-(c) and for the atomic subsystem in Figs. \ref{fig:comp_Neq2_det0_atom} (a)-(c).
 
As in the DJC model, we see that both $\xi_{\textsc{tei}}$ and $\xi^{\prime}_{\textsc{tei}}$ mimic $\xi_{\textsc{qmi}}$ effectively for the field subsystem, while  $\xi^{\prime}_{\textsc{tei}}$ does not reflect $\xi_{\textsc{qmi}}$ for the atomic subsystem. We therefore do not expect $\xi^{\prime}_{\textsc{tei}}$ to quantify entanglement reasonably in spin systems also. Hence we now proceed to assess the extent of entanglement using $\xi_{\textsc{tei}}$ as the entanglement indicator in a specific spin system. 

\section{\label{sec:Ch5NMRexpt} The spin system}

We now proceed to examine, by means of tomograms,  an entangled system of spins on which an NMR experiment has been performed and reported in the literature~\cite{nmrExpt}. Whereas our investigations on the multipartite HQ systems using the IBM Q platform highlighted the role played by experimental losses which were not taken into account in numerically generated tomograms, our present 
aim  is somewhat different. Here we wish to examine quantitatively the limitations that arise in the tomographic approach  from   neglecting the off-diagonal contributions in the density matrix.

There is also another aspect of interest to us here. In generic multipartite spin systems where the subsystem states are entangled, spin squeezing and entanglement are related to each other~\cite{SpinSqRev}. With this in mind, we will first assess the squeezing properties of the spin system, quantify entanglement with $\xi_{\textsc{tei}}$, and comment on the similarities in the dynamics of spin squeezing and entanglement. As mentioned in Section \ref{sec:Ch5intro}, our starting point is the set of reconstructed density matrices at different instants of time obtained from the experiment. The system of interest comprises $\vphantom{}^{13}\text{C}$ atoms (subsystem M), $\vphantom{}^{1}\text{H}$ atoms (subsystem A) and $\vphantom{}^{19}\text{F}$ 
atoms (subsystem B)  in dibromofluoromethane, evolving in time. Each subsystem is a qubit, and the effective Hamiltonian~\cite{nmrExpt} is
\begin{equation}
H_{\textsc{s}}=4 \chi_{\text{s}} (\sigma_{\textsc{a} x} + \sigma_{\textsc{b} x}) \sigma_{\textsc{m} x},
\label{eqn:Ham_spin}
\end{equation}
where $\chi_{\text{s}}$ is a constant, $\sigma_{x}$ is the usual  spin matrix, and the subscripts A, B and M refer to the corresponding subsystems. The eigenstates of $\sigma_{x}$ are denoted by $\ket{+}$ and $\ket{-}$.

The density matrix at time $t=0$ is 
\begin{equation}
\rho_{\textsc{mab}}(0)=\tfrac{1}{2} \ket{\phi_{+}}_{\textsc{ab \, ab}} \hspace*{-0.25 em}\bra{\phi_{+}} \otimes \rho_{\textsc{m} +} + \tfrac{1}{2} \ket{\psi_{+}}_{\textsc{ab \, ab}} \hspace*{-0.25 em}\bra{\psi_{+}} \otimes \rho_{\textsc{m} -},
\label{eqn:rhozero_defn}
\end{equation}
where $\ket{\psi_{+}}_{\textsc{ab}}$ and $\ket{\phi_{+}}_{\textsc{ab}}$ are defined as in Eqs. \eref{eqn:psi0_defn} and \eref{eqn:phi0_defn} respectively, with the replacement of  
$\ket{e}$ by $\ket{\uparrow}$ and  $\ket{g}$ by $\ket{\downarrow}$. Here, $\ket{\uparrow}$ and $\ket{\downarrow}$ refer to the up and down eigenstates of $\sigma_{z}$ and $\rho_{\textsc{m} +}=\ket{+}_{\textsc{m\, m}} \hspace*{-0.25 em}\bra{+}$, $\rho_{\textsc{m} -}=\ket{-}_{\textsc{m\, m}} \hspace*{-0.25 em}\bra{-}$.
It is straightforward to show that, at any $t> 0$,
\begin{align}
\nonumber\rho_{\textsc{mab}} (t) = &
\tfrac{1}{2}\cos^{2} (2 \chi_{\text{s}} t) 
\big\{\ket{\phi_{+}}_{\textsc{ab \, ab}} \hspace*{-0.25 em}\bra{\phi_{+}} \otimes \rho_{\textsc{m} +} +  \ket{\psi_{+}}_{\textsc{ab \, ab}} \hspace*{-0.25 em}\bra{\psi_{+}} \otimes \rho_{\textsc{m} -} \big\}\\
\nonumber +&\tfrac{1}{2}\sin^{2} (2 \chi_{\text{s}} t)
\big\{ \ket{\psi_{+}}_{\textsc{ab \, ab}} \hspace*{-0.25 em}\bra{\psi_{+}} \otimes \rho_{\textsc{m} +} +  \ket{\phi_{+}}_{\textsc{ab \, ab}} \hspace*{-0.25 em}\bra{\phi_{+}} \otimes \rho_{\textsc{m} -}\big\}\\
+& \tfrac{1}{4} i \sin (4 \chi_{\text{s}} t) 
\big\{
\ket{\phi_{+}}_{\textsc{ab \, ab}} \hspace*{-0.25 em}\bra{\psi_{+}} \otimes \mathbb{I}_{\textsc{m}}  - \ket{\psi_{+}}_{\textsc{ab \, ab}} \hspace*{-0.25 em}\bra{\phi_{+}} \otimes \mathbb{I}_{\textsc{m}}\big\},
\label{eqn:rhot_defn}
\end{align}
where $\mathbb{I}_{\textsc{m}}$ denotes the identity matrix corresponding to subsystem M. The experimentally reconstructed density matrices at different instants have been provided to us by the NMR-QIP group in IISER Pune, India. 
We are concerned with  the reduced  density matrix 
\begin{equation}
\rho_{\textsc{ab}}(t)=\text{Tr}_{\textsc{m}}(\rho_{\textsc{mab}}(t)),
\label{eqn:RhoABnmr}
\end{equation} 
corresponding to the subsystem AB.

\begin{figure}
\centering
\includegraphics[width=0.4\textwidth]{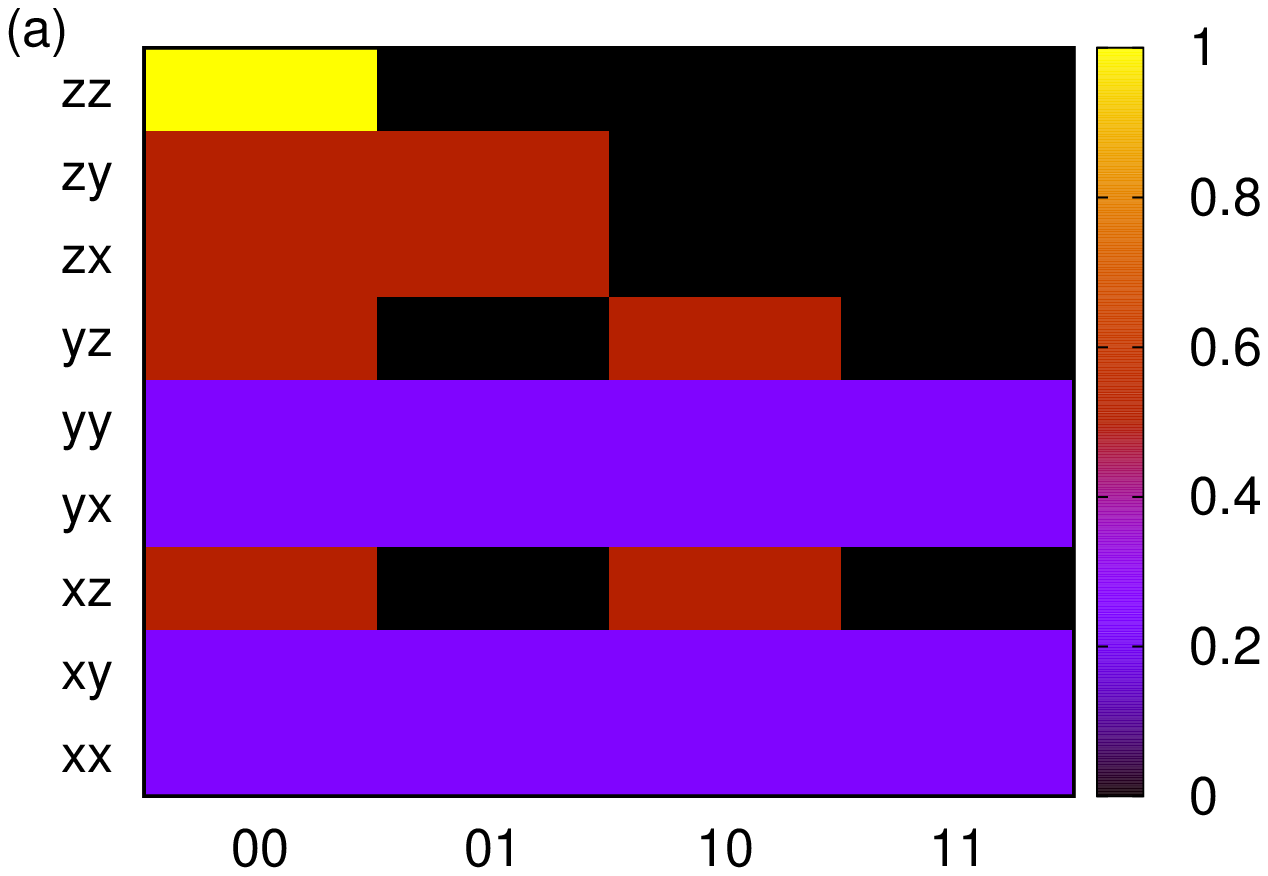}
\includegraphics[width=0.4\textwidth]{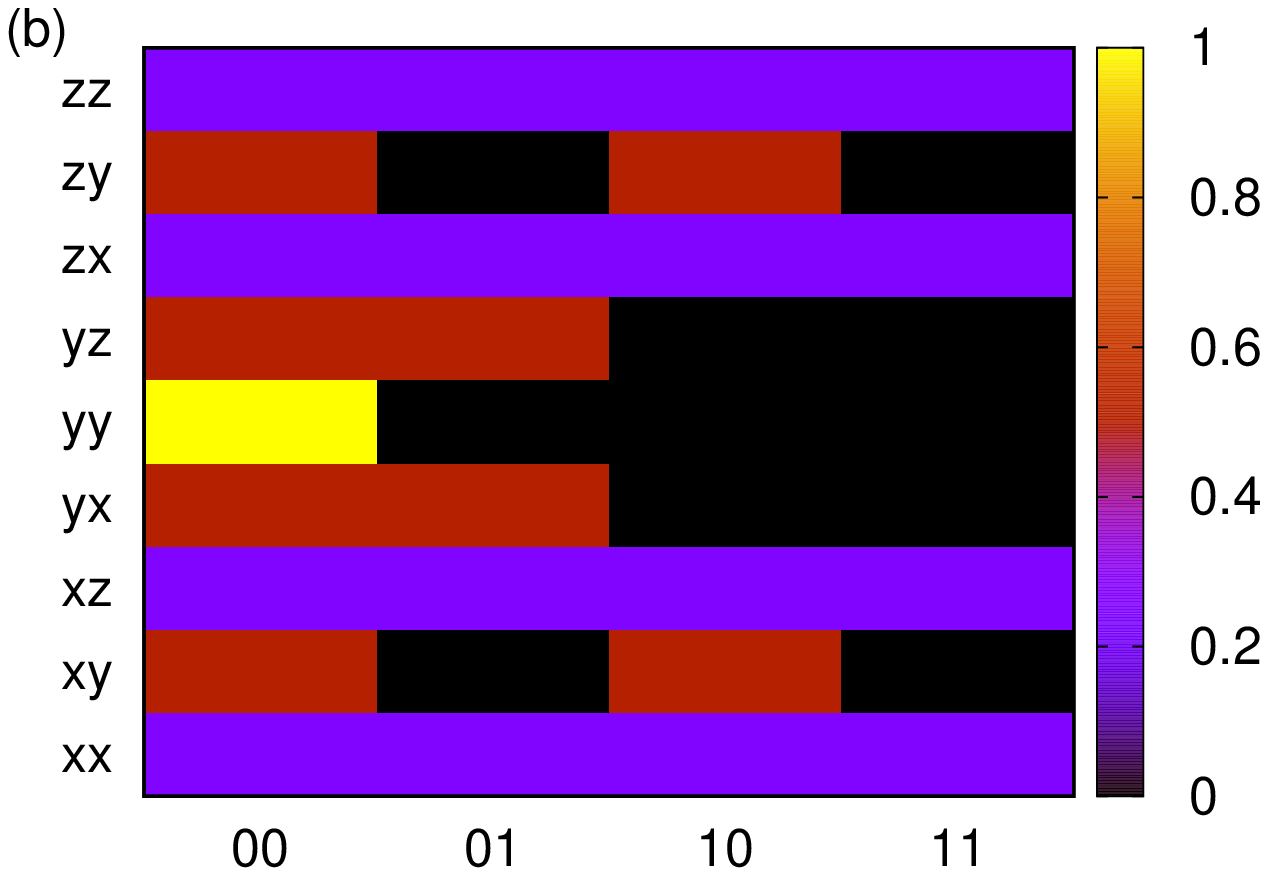}
\includegraphics[width=0.4\textwidth]{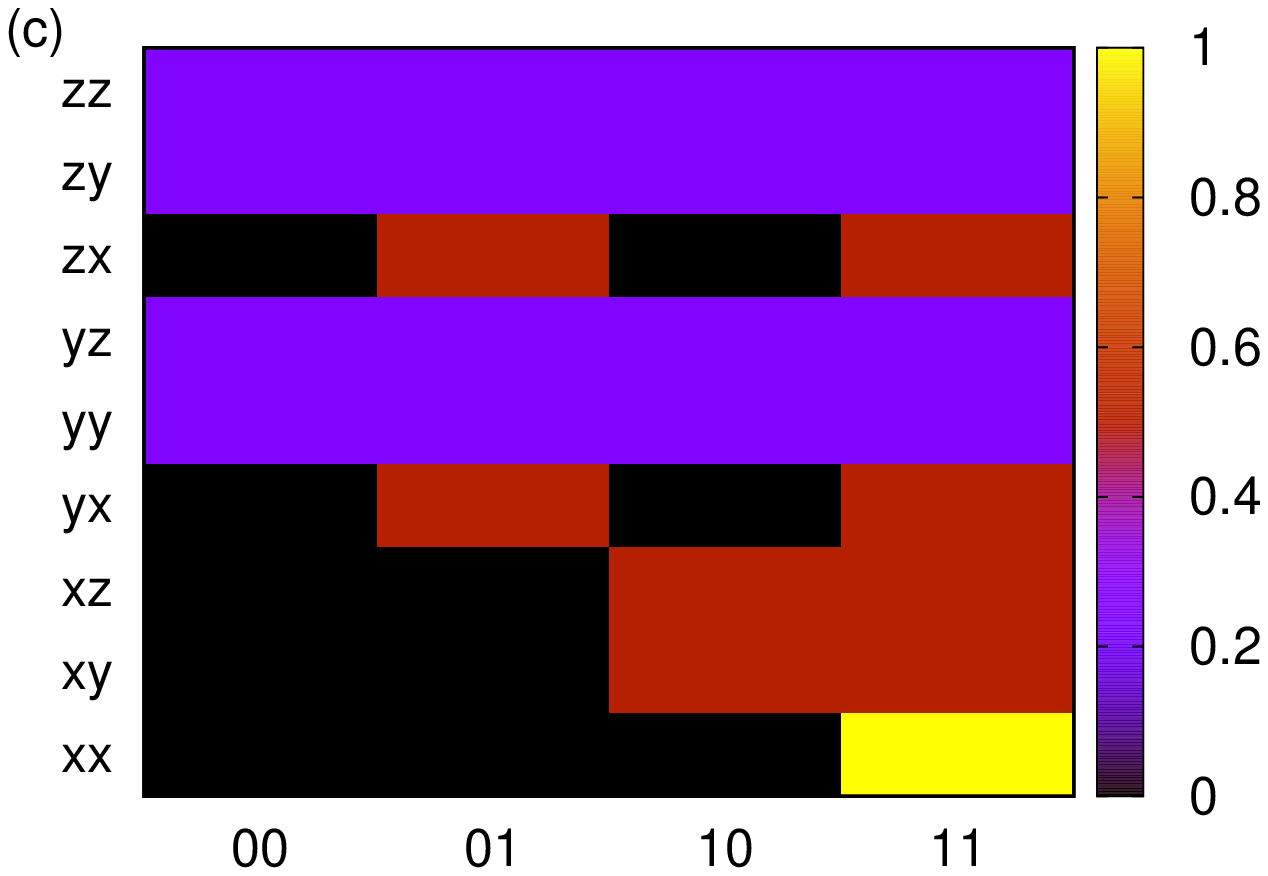}
\includegraphics[width=0.4\textwidth]{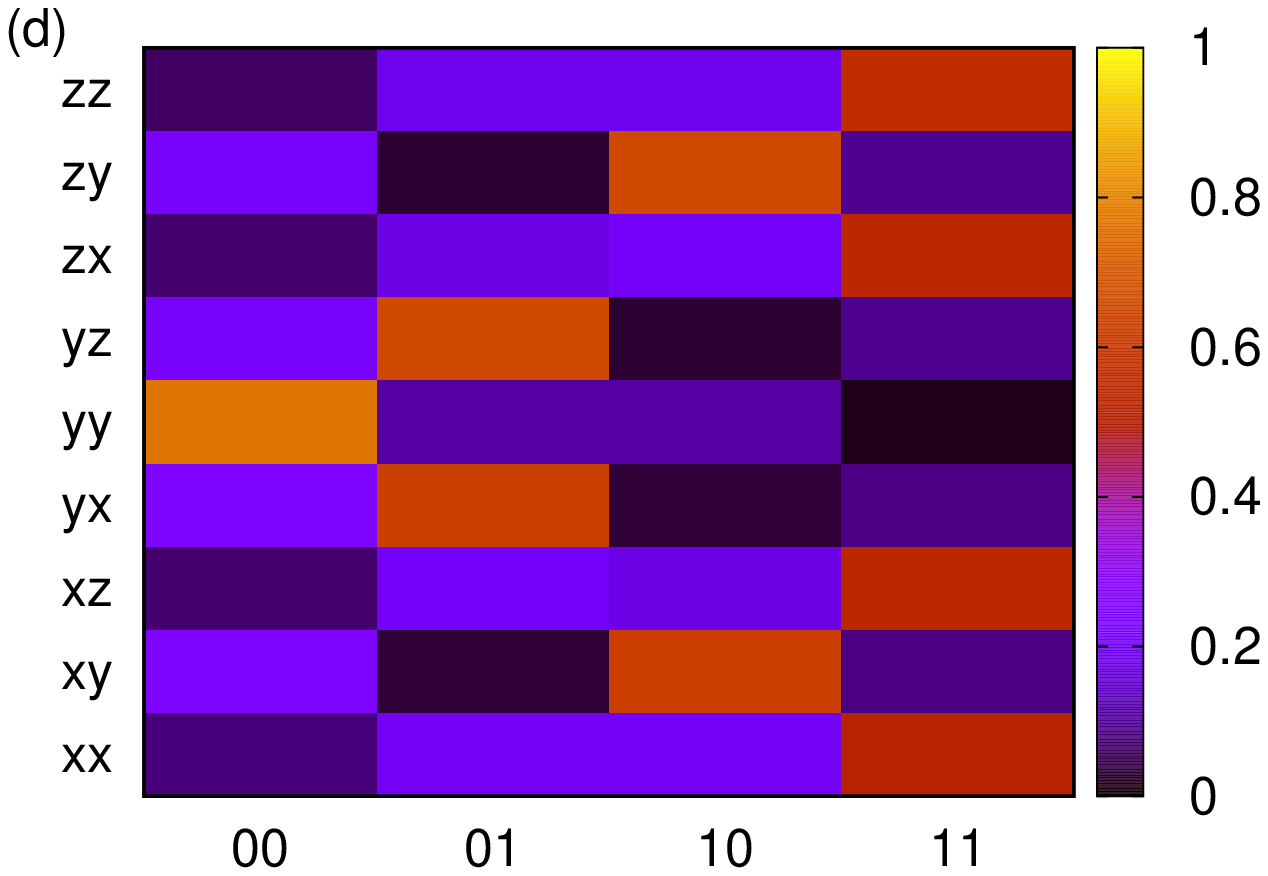}
\caption{Tomograms of spin coherent states with $(\vartheta, \varphi)$ equal to (a) $(0,0)$, (b) $(\pi/4, \pi/2)$, (c) $(\pi/4, 0)$, and (d) $(2 \pi/3, 4 \pi/3)$. The bases are denoted by $x$, $y$ and $z$ and the outcomes by $0$ and $1$.}
\label{fig:spin_CS_tomo}
\end{figure}

In order to  examine its squeezing properties, we use the spin squeezing condition proposed by Kitagawa and 
Ueda~\cite{kitagawa}.
A bipartite system consisting of  two spin-$\tfrac{1}{2}$ subsystems is squeezed, if one of the components normal to the mean spin vector of the bipartite system has a variance less than $0.5$ (the latter being the variance of the 
corresponding spin coherent state). Here, the spin coherent state for the bipartite system AB is defined in terms of the polar and azimuthal angles $(\vartheta,\varphi)$ as 
$\left(\cos\vartheta \ket{\downarrow}_{\textsc{a}} + e^{i \varphi}\,\sin \vartheta  \ket{\uparrow}_{\textsc{a}}\right) \otimes \left(\cos\vartheta \ket{\downarrow}_{\textsc{b}} +e^{i \varphi}\, \sin\vartheta  \ket{\uparrow}_{\textsc{b}}\right)$. For completeness, the tomograms corresponding to spin coherent states for four  different $(\vartheta,\varphi)$ values are shown in Figs. \ref{fig:spin_CS_tomo} (a)-(d). Each of the figures (a), (b), and (c) corresponds to one of the eigenstates of $\sigma_{\textsc{a}z}\, \sigma_{\textsc{b}z}$, $\sigma_{\textsc{a}y}\,\sigma_{\textsc{b}y}$ and $\sigma_{\textsc{a}x}\,\sigma_{\textsc{b}x}$, respectively. The tomogram corresponding to a spin coherent state for $\vartheta=2 \pi/3, \varphi=4 \pi/3$ is plotted in Fig. \ref{fig:spin_CS_tomo} (d).

\begin{figure}
\centering
\includegraphics[width=0.32\textwidth]{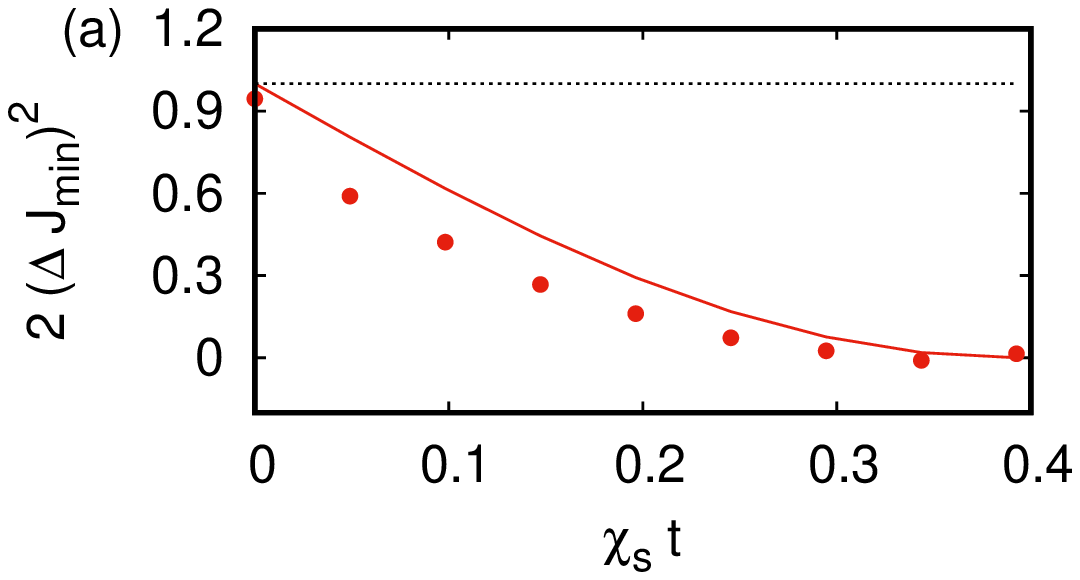}
\includegraphics[width=0.32\textwidth]{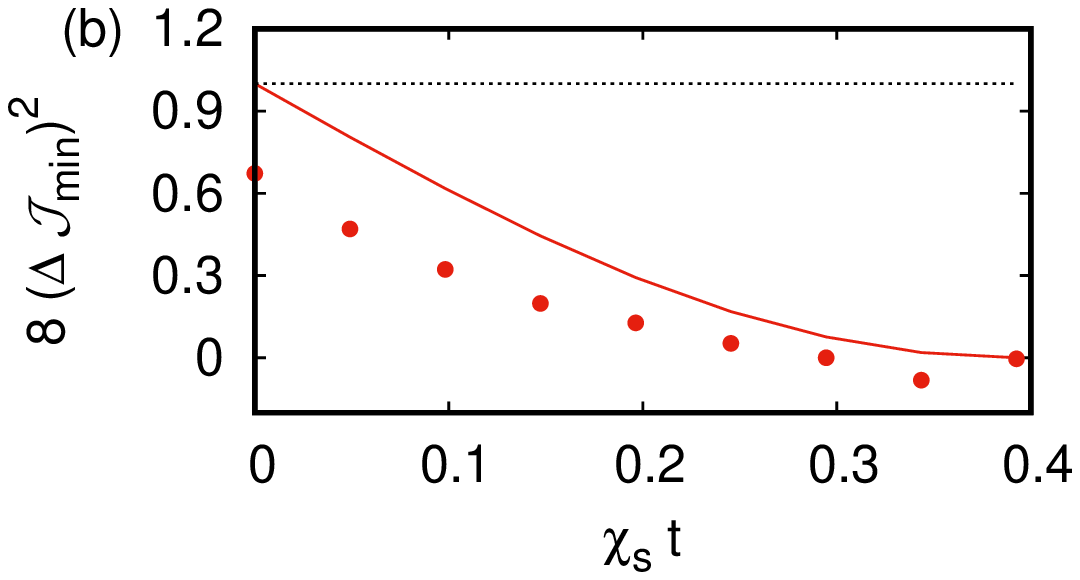}
\includegraphics[width=0.32\textwidth]{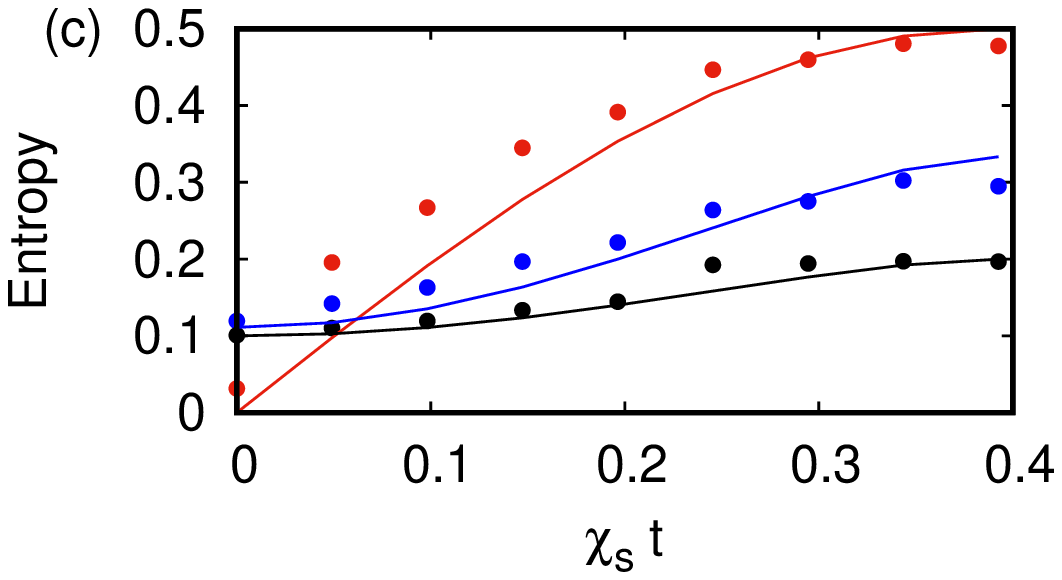}
\caption{(a) $2 (\Delta J_{\text{min}})^{2}$\, (b) $8 (\Delta \mathcal{J}_{\text{min}})^{2}$\, (c) $N(\rho_{\textsc{ab}})$ (red), $\xi_{\textsc{tei}}$ (blue), $0.1 \,\xi_{\textsc{qmi}}$ (black) vs. scaled time $\chi_{\text{s}} t$. The solid curves are computed using Eq. \eref{eqn:RhoABnmr} and the dotted curves from experimental data. The black horizontal line  in (a) 
and  (b) sets the limit below which the state is squeezed.}
\label{fig:squeeze_tei_negativity_compare}
\end{figure}

The mean spin direction is given by 
$\mathbf{v}_{\textsc{s}}(t) = \aver{\mathbf{J}(t)}/|\aver{\mathbf{J}(t)}|$ where 
\begin{equation}
\mathbf{J}= (\sigma_{\textsc{a} x} + \sigma_{\textsc{b} x}) \mathbf{e_{x}} + (\sigma_{\textsc{a} y} + \sigma_{\textsc{b} y}) \mathbf{e_{y}} + (\sigma_{\textsc{a} z} + \sigma_{\textsc{b} z}) \mathbf{e_{z}},
\end{equation}
and $\aver{\mathcal{O}(t)}=\text{Tr}_{\textsc{ab}}(\rho_{\textsc{ab}}(t)\mathcal{O})$ for any operator $\mathcal{O}$. Since $\aver{\sigma_{i x}(t)}$, $\aver{\sigma_{i y}(t)}$ and $\aver{\sigma_{i z}(t)}$ ($i=\text{A,B}$) are 
equal to zero, it follows that $\mathbf{v}_{\textsc{s}}(t)$ is a null vector. Hence, any unit vector $\mathbf{v}_{\perp}$ can be chosen to obtain the required variance. We have calculated 
the variance $(\Delta \; \mathbf{J}\mathbf{\cdot}\mathbf{v}_{\perp})^{2} = \aver{(\mathbf{J}\mathbf{\cdot}\mathbf{v}_{\perp})^{2}}$ as a function of time for $800$ different vectors $\mathbf{v}_{\perp}$ at each instant. From this, we have identified the minimum variance $(\Delta J_{\text{min}})^{2}$ and plotted it as a function of time (Fig.~\ref{fig:squeeze_tei_negativity_compare}(a)). From the figure, it is evident that the variance obtained numerically using  Eq. \eref{eqn:RhoABnmr} and 
that obtained from the experimentally reconstructed density matrices are in good agreement. We also point out that the extent of squeezing,  $[1-2 (\Delta J_{\text{min}})^{2}]$,  increases with time. 

We adapt the Kitagawa-Ueda squeezing condition in the following manner in order to estimate second-order squeezing. By an extension of the preceding argument, we consider the expectation value of the dyad (or 
tensor product) $\mathbf{JJ}$ instead of $\aver{\mathbf{J}}$. In general, $\aver{\mathbf{JJ}}$ is not a null tensor. We may therefore impose  the orthogonality condition  
$\aver{\mathcal{J}}=0$, where 
\begin{equation}
\mathcal{J}=\tfrac{1}{2}\left(\mathbf{v}_{1}\cdot\mathbf{JJ}\cdot \mathbf{v}_{2} + 
\mathbf{v}_{2}\cdot\mathbf{JJ}\cdot \mathbf{v}_{1}\right).
\label{eqn:2orderSqueezeOp}
\end{equation}
$\mathbf{v}_{1}$ and $\mathbf{v}_{2}$ are analogous to the vector $\mathbf{v}_{\perp}$ of  the earlier case.  
The symmetrization with respect to 
$\mathbf{v}_{1}$ and $\mathbf{v}_{2}$ in 
Eq. (\ref{eqn:2orderSqueezeOp}) ensures that  
$\mathcal{J}$ is real. Using Eq. \eref{eqn:RhoABnmr}, we get 
\begin{align}
\nonumber\aver{\mathcal{J} (t)}&=\text{Tr}\,\big(\rho_{\textsc{ab}}(t) \mathcal{J}\big)\\
& = \mathbf{v}_{1 x} \mathbf{v}_{2 x} +
 \tfrac{1}{2} \big\{ \mathbf{v}_{1y} \mathbf{v}_{2 y} + \mathbf{v}_{1z} \mathbf{v}_{2 z} + \sin (4 \chi_{\text{s}} t) \big( \mathbf{v}_{1 y} \mathbf{v}_{2 z} + \mathbf{v}_{2 y} \mathbf{v}_{1 z}\big)\big\}, 
\label{eqn:avercalJ}
\end{align}
where the subscripts $x$, $y$ and $z$ denote the respective components of $\mathbf{v_{1}}$ and $\mathbf{v_{2}}$. (The symmetry between the $y$ and $z$ components in Eq. \eref{eqn:avercalJ} follows from the fact that $[\sigma_{\textsc{a} x}\sigma_{\textsc{b} x},\rho_{\textsc{ab}}(t)]=0$). We have considered a set of $320$ different pairs $(\mathbf{v}_{1},\mathbf{v}_{2})$ for which $\aver{\mathcal{J}(t)}=0$.
For each such pair, the variance $(\Delta\mathcal{J})^{2}$ has been computed, and from this the minimum variance $(\Delta \mathcal{J}_{\text{min}})^{2}$ has been obtained. 
The reference value ($0.125$) below which the state is second-order squeezed is obtained by minimizing the corresponding variance for the spin coherent state with respect to $\vartheta$ and $\varphi$. Plots of $(\Delta \mathcal{J}_{\text{min}})^{2}$ versus time obtained both from the experimentally reconstructed density matrices and from Eq. \eref{eqn:RhoABnmr} are shown in Fig.~\ref{fig:squeeze_tei_negativity_compare}(b). 
The two  curves are in reasonable agreement with each other. As in the earlier case, the measure of second-order squeezing, $[1-8 (\Delta \mathcal{J}_{\text{min}})^{2}]$,  increases with time.
We have verified that neither the state of subsystem A nor that of B displays entropic squeezing~\cite{MassenEUR} at any time.

\begin{figure}
\centering
\includegraphics[width=0.4\textwidth]{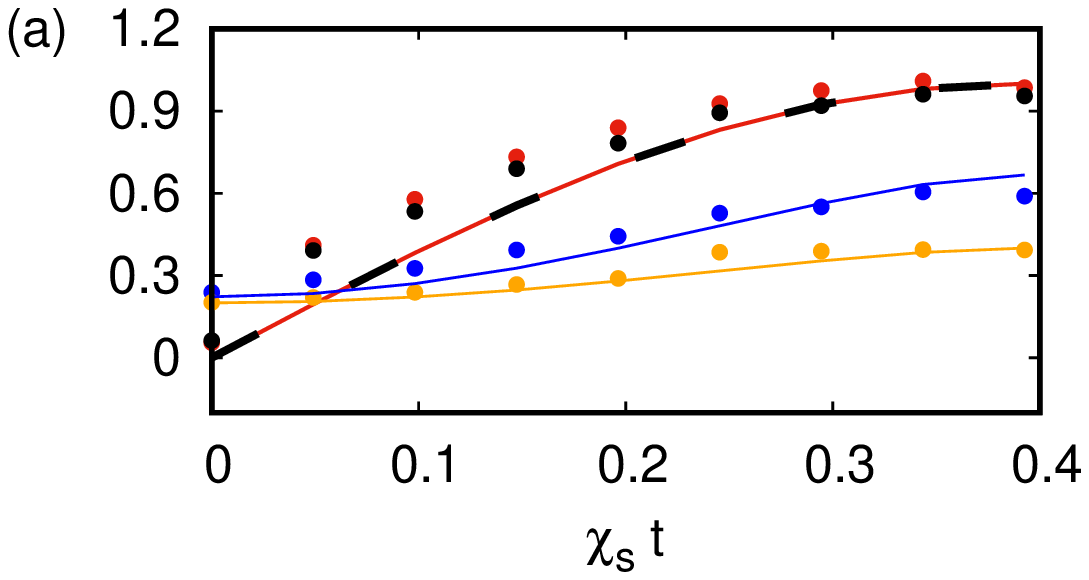}
\includegraphics[width=0.4\textwidth]{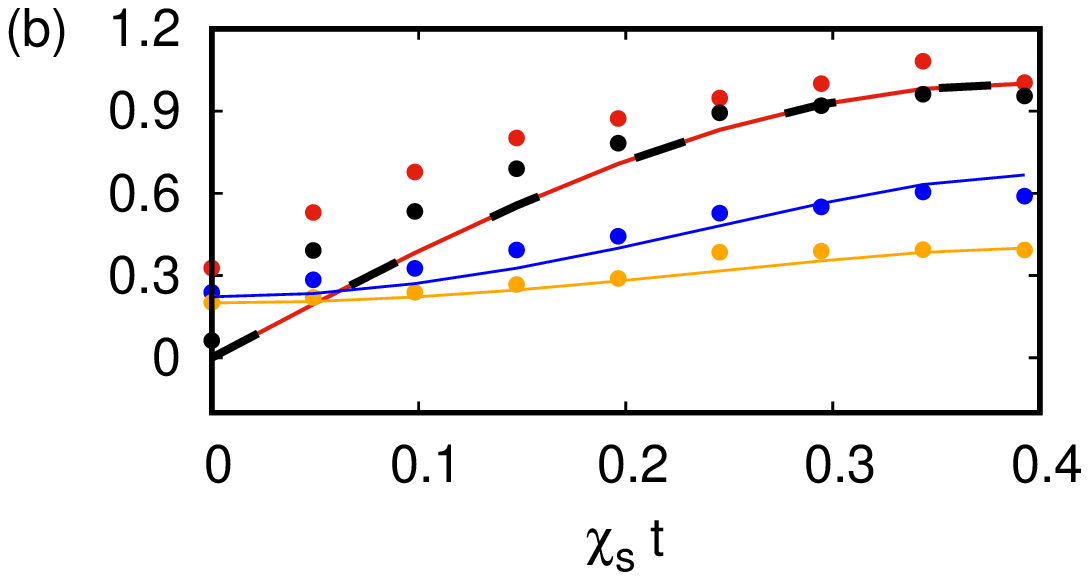}
\caption{$2 \, N(\rho_{\textsc{ab}})$ (black), $2 \, \xi_{\textsc{tei}}$ (blue), $0.2 \,\xi_{\textsc{qmi}}$ (orange), and (a) $[1 - 2 (\Delta J_{\text{min}})^{2}]$ (red) (b) $[1 - 8 (\Delta \mathcal{J}_{\text{min}})^{2}]$ (red) vs. scaled time $\chi_{\text{s}} t$. The solid curves are computed using Eq. \eref{eqn:RhoABnmr} and the dotted curves from experimental data.}
\label{fig:squeeze_extent_entang_indics_compare}
\end{figure}

We now turn to the entanglement dynamics in the NMR experiment~\cite{nmrExpt}. We have computed $\xi_{\textsc{tei}}$ from tomograms obtained both from Eq. \eref{eqn:RhoABnmr} and from the experimentally reconstructed density matrices at different instants of time. This is compared with two indicators, $\xi_{\textsc{qmi}}$ and the negativity $N(\rho_{\textsc{ab}})$. The latter has been reported in the experiment, and is defined as  $N(\rho_{\textsc{ab}})=
\tfrac{1}{2}\sum_{i} \left(|\mathcal{L}_{i}|-\mathcal{L}_{i}\right)$. 
Here $\lbrace\mathcal{L}_{i}\rbrace$ is the set 
of  eigenvalues of $\rho^{T_{\textsc{a}}}
_{\textsc{ab}}$, 
the partial transpose of $\rho_{\textsc{ab}}$ with respect to the subsystem A. (Equivalently, 
the  partial transpose  
 $\rho^{T_{\textsc{b}}}_{\textsc{ab}}$  
 may be used.)  
Although all three entanglement indicators 
are in agreement  in their gross features (Fig.~\ref{fig:squeeze_tei_negativity_compare} (c)), $\xi_{\textsc{tei}}$ is closer to $\xi_{\textsc{qmi}}$ 
owing to the similarity in their definitions. 
The difference between $\xi_{\textsc{tei}}$ and negativity is clearly due to the limitations that arise in the tomographic approach because the off-diagonal contributions of the density matrix have been neglected. We  emphasise, however, 
 that $\xi_{\textsc{tei}}$ is still a good entanglement indicator which compares favourably with $\xi_{\textsc{qmi}}$. Figures \ref{fig:squeeze_extent_entang_indics_compare} (a) and (b) facilitate comparison between $[1-2 (\Delta J_{\text{min}})^{2}]$, $[1-8 (\Delta \mathcal{J}_{\text{min}})^{2}]$, $N(\rho_{\textsc{ab}})$, $\xi_{\textsc{tei}}$, and $\xi_{\textsc{qmi}}$.
It is clear that $N(\rho_{\textsc{ab}})$ characterises the degree  of squeezing and higher-order squeezing extremely well, and that $\xi_{\textsc{tei}}$ and $\xi_{\textsc{qmi}}$ are reasonable quantifiers of squeezing properties.

\section{Concluding remarks}
\label{sec:Ch5conclremarks}

In this chapter, we have compared the performance of $\xi_{\textsc{tei}}$ and $\xi^{\prime}_{\textsc{tei}}$ with standard entanglement indicators in multipartite HQ models (DJC and DTC), and we have calculated $\xi_{\textsc{tei}}$ in the context of an NMR experiment.
In the former, entanglement within the field subsystem and within the atomic subsystem have been examined. While $\xi^{\prime}_{\textsc{tei}}$ satisfactorily mimics $\xi_{\textsc{qmi}}$ for the field subsystem, it does not fare well in the atomic subsystem. 
In contrast, $\xi_{\textsc{tei}}$ performs well for both subsystems.
 An equivalent circuit for the DJC model was both experimentally run and numerically simulated to obtain $\xi_{\textsc{tei}}$. This facilitates estimation of experimental losses. Our results show that the IBM simulation agrees well with numerical simulation of the DJC model. Further, we have established that in multipartite HQ models, the  difference between $\xi_{\textsc{tei}}$ obtained experimentally and numerically increases significantly with an increase in the number of atoms. 
Corresponding to the NMR experiment, we have shown that $\xi_{\textsc{tei}}$ is in fair agreement with both the negativity $N(\rho_{\textsc{ab}})$ and $\xi_{\textsc{qmi}}$. Further, it gives a reasonable estimate of squeezing and second-order squeezing. We therefore conclude that the entanglement indicator $\xi_{\textsc{tei}}$ obtained \textit{directly} from tomograms is a convenient and practical quantifier of entanglement both in multipartite HQ systems and in spin systems.


%% file: chapter6.tex

\chapter{Conclusion}
\label{ch:conc}

In this thesis, we have carried out a detailed investigation of the manner in which nonclassical effects such as  wave packet revival phenomena, squeezing properties of different states, and quantum entanglement in bipartite systems, can be identified and quantitatively estimated directly from tomograms, without detailed state reconstruction. We have examined both single-mode and bipartite CV systems with more than one time scale, and investigated full and fractional revivals of quantum wave packets governed by nonlinear Hamiltonians. A range of squeezing phenomena such as quadrature squeezing, higher-order Hong-Mandel and Hillery-type squeezing, and entropic squeezing, have been examined in various 
quantum systems evolving in time. 
Several entanglement indicators that can be  obtained from appropriate tomograms have been proposed, and their performance assessed through detailed comparison with standard measures of entanglement. We have examined CV, multipartite HQ, and spin systems in this thesis. In each of these cases, we have studied the nonclassical properties captured through numerically generated tomograms of quantum states obtained from known initial states evolving under  nonlinear Hamiltonians. We have  also worked with experimental data generated elsewhere (and in the IBM quantum computing platform). Thus comparison between experimentally obtained tomograms and numerically generated tomograms was feasible in certain cases. We have examined  optical tomograms, qubit (spin) tomograms and chronocyclic tomograms in different contexts. We have used a range of initial states in our  numerical computations.  This includes the standard CS, PACS (equivalently, boson-added CS), direct products of these states, binomial states, and two-mode squeezed states. 

Several bipartite entanglement indicators have been analysed 
both at avoided energy-level crossings of specific systems, and during temporal evolution of certain Hamiltonian systems. We have commented on decoherence effects wherever possible. Some of these indicators are inherently quantum mechanical in nature 
(such as 
$\xi_{\textsc{tei}}$, $\xi^{\prime}_{\textsc{tei}}$ and 
$\xi_{\textsc{ipr}}$) while others ($\xi_{\textsc{bd}}$ and  
$\xi_{\textsc{pcc}}$) are inspired by classical tomography. The extent to which nonlinear correlations are captured by these indicators has been assessed in detail. Our main results are the following.

\noindent(1) We have established that, as an entanglement indicator and quantifier, $\xi_{\textsc{bd}}$ performs as well as $\xi_{\textsc{tei}}$ close to avoided energy-level crossings in both CV and HQ systems.

\noindent(2) If a bipartite state is a superposition of states that are Hamming-uncorrelated, $\xi_{\textsc{ipr}}$ is a better entanglement indicator than $\xi_{\textsc{tei}}$. In fact, we find that even $\epsarg{ipr}$ (the corresponding indicator from a single section of the tomogram) suffices to estimate entanglement reliably near avoided energy-level crossings for superpositions of Hamming-uncorrelated states such as the eigenstates of certain number-conserving Hamiltonians.

\noindent(3) In the context of bipartite CV systems $\xi^{\prime}_{\textsc{tei}}$ (obtained by averaging only over the dominant $\epsarg{tei}$) is found to be a reasonably good entanglement indicator. However, the extent to which it agrees with $\xi_{\textsc{svne}}$ is very sensitive to the precise initial state, the nature of the interaction, and the inherent nonlinearities in the system. 

\noindent(4) In contrast to the foregoing inference, $\xi^{\prime}_{\textsc{tei}}$ is a poor entanglement indicator for qubit systems, and $\xi_{\textsc{tei}}$ has to be used for such systems.

The tomographic approach to deriving  entanglement indicators and distinguishing between states acquires  further  
validation by its application  to experiments on both CV and spin systems reported in the literature. 
In the former case, two $2$-photon states generated 
experimentally were 
distinguished from each other 
using interferometric methods~\cite{perola}. We have shown, by examining the entanglement indicator $\epsarg{tei}$ corresponding to a judiciously selected slice of the bipartite chronocyclic tomograms concerned, that these states can be distinguished clearly from each other. In the latter case, by analysing the experimental data~\cite{nmrExpt} pertaining to the spin system of interest  and  constructing the corresponding qubit tomograms, we have established that $\xi_{\textsc{tei}}$ is a reasonably good bipartite entanglement indicator in this context. 
The limitations that arise from  neglecting the contributions of off-diagonal elements of the density matrix (in computing the tomographic entanglement indicators) are clearly seen in this context. 
Further, using the IBM quantum computing platform we have shown that, even in multipartite HQ systems comprising radiation fields interacting with atomic subsystems, 
$\xi_{\textsc{tei}}$ extracted from appropriate qubit tomograms is a reasonably good bipartite entanglement indicator. 
Here, we have compared indicators computed from experimentally obtained tomograms, on the one hand,  and those computed from numerically generated tomograms which do not account for such losses, on the other. This facilitates understanding the role of experimental losses. 
The thesis also highlights how entanglement indicators which are not monotones (and hence not {\it bona fide}   entanglement 
{\em measures}) 
can still mirror the qualitative features of bipartite  entanglement quite  well, with the added advantage  that they can be computed from tomograms in a straightforward manner. 

Finally, we mention some avenues 
for future research. The extension of our studies on nonclassical effects to multipartite systems, including the computation of multipartite entanglement indicators from tomograms, is an open question. In this context, a detailed assessment of the advantages, if any, of the tomographic approach over state reconstruction programmes needs to be carried out. Further, comparison with other quantifiers of entanglement in multipartite systems would facilitate a deeper understanding of the tomographic approach. The manner in which the nonclassicality of a state is mirrored in the tomogram (similar to negative values of the Wigner function) also merits  examination  in greater detail.


%% file: appendix_TomoSinModExpr.tex
\chapter{Eigenvectors of rotated quadrature operators}
\label{appen:TomoSinModExpr}

We first obtain a useful expression~\cite{MankoFockStates} for the single-mode tomogram in terms of the Hermite polynomials. We also indicate the procedure to obtain eigenvectors of rotated quadrature operators~\cite{barnett}.

The overlap between a photon number state $\ket{n}$ and the field quadrature state $\ket{X}$ can be expressed in terms of the Hermite polynomials $H_{n}(X)$ as, 
\begin{align*}
\aver{ X \vert n } = (e^{-X^{2}/2}H_{n}(X))/(\pi^{1/4} \sqrt{n! 2^{n}}).
\end{align*}
Using this, and Eq. \eref{eqn:phaseshift}, we see that
\begin{equation}
\aver{ X_{\theta} ,\theta \vert n }= e^{-X_{\theta}^{2}/2} \;\frac{e^{-i n \theta} H_{n}(X_{\theta})}{\pi^{1/4}\, \sqrt{n!}\; 2^{\frac{n}{2}}}.
\label{eqn:XthetaNoverlap}
\end{equation}
For a pure state $\ket{\psi}$, $w(X_{\theta},\theta)=|\aver{X_{\theta},\theta|\psi}|^{2}$. Expanding $\ket{\psi}$ in the photon number basis $\lbrace \ket{p} \rbrace$ as $\sum\limits_{p=0}^{\infty} c_{p} \ket{p}$, the corresponding tomogram is given by~\cite{MankoFockStates} 
\begin{align*}
w(X_\theta, \theta) = \frac{e^{-X_{\theta}^{2}}}{\sqrt{\pi}} \left| \sum_{n=0}^{\infty} \frac{c_{n} e^{-i n \theta} }{\sqrt{n!} 2^{\frac{n}{2}} } H_{n}(X_{\theta})  \right|^{2}.
\end{align*}
The above expression can be extended in a straightforward manner to mixed states.

Using Eq. \eref{eqn:XthetaNoverlap} and $\ket{X_{\theta},\theta}=\sum\limits_{p=0}^{\infty} \ket{p} \aver{p|X_{\theta},\theta}$, we get
\begin{align}
\nonumber \ket{X_{\theta},\theta}&=\sum\limits_{n=0}^{\infty} \frac{e^{-X_{\theta}^{2}/2}}{\pi^{1/4}} \; H_{n}(X_{\theta}) \frac{e^{i n \theta} }{\sqrt{n!}\; 2^{\frac{n}{2}}}  \ket{n} \\
&= \frac{e^{-X_{\theta}^{2}/2}}{\pi^{1/4}} \sum\limits_{n=0}^{\infty} \; \frac{H_{n}(X_{\theta})}{n!} \left(\frac{e^{i \theta} a^{\dagger} }{\sqrt{2}}\right)^{n} \ket{0}, 
\label{eqn:XthetaIntermedStep}
\end{align}
where $\ket{0}$ is the zero-photon state.
The generating function for the Hermite polynomials is 
\begin{equation}
\sum\limits_{p=0}^{\infty} H_{p}(x) \frac{t^{p}}{p!}=e^{2 x t - t^{2}}.
\label{eqn:HermiteGenFn}
\end{equation}
Substituting Eq. \eref{eqn:HermiteGenFn} in Eq. \eref{eqn:XthetaIntermedStep}, we get~\cite{barnett}
\begin{equation}
\ket{X_\theta, \theta} =\frac{1}{\pi^{1/4}} \exp\left(-\:\frac{X_{\theta}^{2}}{2} +\: 2 \, X_{\theta} \frac{a^{\dagger} e^{i \theta}}{\sqrt{2}} \, -\: \left(\frac{a^{\dagger} e^{i \theta} }{\sqrt{2}}\right)^{2} \,   \right)\ket{0}.
\label{eqn:xthetaDefn}
\end{equation}
 The procedure outlined above can be extended to obtain the corresponding expressions for multimode tomograms.


%% file: appendix_NormalOrderedMom.tex
\chapter{Normal-ordered moments from optical tomograms}
\label{appen:NormOrdMom}

We outline the procedure   \cite{wunsche} for obtaining normal-ordered moments for infinite-dimensional single-mode systems from optical tomograms.  We expand any normal-ordered operator $F$ in the form
\begin{equation}
F=\sum_{k,l=0}^{\infty} F_{k, l} a^{\dagger k} a^{l}.
\label{eqn:expn_normal_order}
\end{equation}
Here the coefficients
\begin{align}
F_{k, l}=\sum_{s=0}^{\lbrace k,l \rbrace} \frac{(-1)^{s}}{s! \sqrt{(k-s)! (l-s)!}} \bra{k-s}F \ket{l-s},
\label{eqn:F_klmn}
\end{align}
and $\lbrace k,l \rbrace$ is used to denote min($k$,$l$). 
(Equation \eref{eqn:F_klmn} can be verified as follows. We first obtain $\bra{p}F\ket{q}$ using Eq. \eref{eqn:expn_normal_order}. Substituting this in Eq. \eref{eqn:F_klmn}, we get
\begin{align}
F_{k, l} = \sum_{s=0}^{\lbrace k,l \rbrace} \sum_{u=0}^{\lbrace k-s,l-s \rbrace} \frac{(-1)^{s}}{s! u!} F_{k-s-u, l-s-u}.
\label{eqn:intermed_chk}
\end{align}
Now replacing $u$ with $u'=s+u$, changing the order of the sums, and using
\begin{equation}
\sum_{s=0}^{k}\frac{(-1)^{s}}{s! (k-s)!} = \delta_{k,0},
\label{eqn:binomial}
\end{equation}
it follows that  the expansion of $F$ above is valid.)

We write the projection operator 
\begin{align}
\ket{k}\bra{l}= (k! l!)^{-1/2} \sum_{u=0}^{\infty} \frac{(-1)^{u}}{u!} a^{\dagger k+u} a^{l+u} \label{eqn:projection}.
\end{align}
Using $F=\sum_{k,l=0}^{\infty} \ket{l}\bra{k} \mathrm{Tr}(\ket{k}\bra{l}F)$ and Eq. \eref{eqn:projection}, we see that
\begin{align}
F=\sum_{k,l=0}^{\infty} A_{k, l} \mathrm{Tr}(a^{\dagger k} a^{l} F).
\label{eqn:op_normal_mom}
\end{align}
Here 
\begin{align}
A_{k, l}=\sum_{s=0}^{\lbrace k,l \rbrace} &\frac{(-1)^{s}}{s! \sqrt{(k-s)! (l-s)!}} \ket{l-s}\bra{k-s}.
\label{eqn:A_klmn}
\end{align}
We now consider the  special case when $F$ is the density operator $\rho$.
Using Eqs. \eref{eqn:op_normal_mom}, \eref{eqn:A_klmn}, 
\begin{align}
\langle X_{\theta},\theta \vert m \rangle \langle n \vert X_{\theta},\theta \rangle = \frac{e^{-X_{\theta}^{2}}}{\sqrt{\pi}} \frac{e^{- i (m-n)\theta}}{\sqrt{m! n!}\sqrt{ 2^{m+n}}} H_{m} (X_{\theta}) H_{n} (X_{\theta}),
\label{eqn:tomo_gen}
\end{align}
and the following property of the Hermite polynomials
\begin{align}
H_{k+l}(X_{\theta})=\sum_{s=0}^{\lbrace k,l \rbrace} \frac{(-2)^{s} k! l! H_{k-s}(X_{\theta}) H_{l-s}(X_{\theta})}{s! (k-s)! (l-s)!},
\label{eqn:Hermite_prop}
\end{align}
we can show that 
\begin{align}
\nonumber &w(X_{\theta},\theta) = \bra{X_{\theta},\theta}\rho\ket{X_{\theta},\theta}\\
&= \frac{e^{-X_{\theta}^{2}}}{\sqrt{\pi}} \sum_{k,l=0}^{\infty} \frac{e^{i (k-l)\theta}}{\sqrt{2^{k+l}} k! l!} H_{k+l}(X_{\theta}) \mathrm{Tr}(a^{\dagger k} a^{l} \rho).
\label{eqn:tomo_normal_mom}
\end{align}
Using the orthonormality  property of the Hermite polynomials together with the expression 
\begin{equation}
\sum_{u=0}^{n} \exp(2 \pi i u s / (n+1))=(n+1)\delta_{s,0},
\label{eqn:unity_root_sum}
\end{equation}
in Eq. \eref{eqn:tomo_normal_mom} gives
\begin{align}
\label{eqn:appen_wunsche}
\nonumber \aver{a^{\dagger k} a^{l}} &= \mathrm{Tr}(a^{\dagger k} a^{l} \rho)\\
\nonumber &= C_{k l} \sum_{m=0}^{k+l} \exp \left(-i(k-l)\left(\frac{m\pi}{k+l+1}\right)\right) \\
& \int_{-\infty}^{\infty} \rmd X_\theta  \ w\left(X_\theta, \ \frac{m\pi}{k+l+1}\right) H_{k+l}\left( X_{\theta}  \right) ,
\end{align}
where
\begin{equation}
\nonumber C_{k l}=\frac{k! l!}{(k+l+1)! \sqrt{2^{k+l}}}.
\end{equation}
This procedure can be extended in a straightforward manner to multimode systems.


%% file: appendix_DensMat.tex
\chapter{Numerical computation of the time-evolved density matrix of the bipartite BEC system}
\label{appen:DensMat}

We outline the essential steps in computing the time-evolved density matrix of the double-well BEC system, for initial states  $\ket{\Psi_{m_{1}m_{2}}} = \ket{\alpha_{a},m_{1}} \otimes \ket{\alpha_{b},m_{2}}$ with Hamiltonian $H_{\textsc{bec}}$ (Eq. \eref{eqn:HBEC}). We denote the time-evolved density matrix by $\rho_{m_{1},m_{2}}(t)$.

The procedure for obtaining $\rho_{0,0}(t)$ corresponding to the initial state $\ket{\alpha_{a}} \otimes \ket{\alpha_{b}}$ is outlined in \cite{sanz}.  We obtain~\cite{sharmila} $\rho_{m_{1},m_{2}}(t)=\ket{\Psi_{m_{1}m_{2}}(t)}\bra{\Psi_{m_{1}m_{2}}(t)}$ from $\rho_{0,0}(t)$ through appropriate transformations.  
We first write
\begin{equation}
\ket{\Psi_{m_{1}m_{2}}(t)}=M_{m_{1},m_{2}}(t) \ket{\Psi_{00}(t)},
\end{equation} 
where
\begin{equation}
M_{m_{1},m_{2}}(t)=\frac{1}{\mu}\exp(-i H_{\textsc{bec}} t) a^{\dagger m_{1}} b^{\dagger m_{2}} \exp(i H_{\textsc{bec}} t), 
\end{equation}
and $\mu$ is given in terms of the Laguerre polynomials as $\sqrt{m_{1}! L_{m_{1}}(-{|\alpha_{a}|}^{2}) m_{2}! L_{m_{2}}(-{|\alpha_{b}|}^{2})}$.  In order to recast $M_{m_{1},m_{2}}(t)$ in a simpler form we introduce the operator $V=\exp(\kappa (a^{\dagger}b-b^{\dagger}a)/2)$, where $\kappa=\tan^{-1}(\lambda/\omega_{1})$~\cite{sanz}. Consequently, $H_{\textsc{bec}}$ can be written as  $V \widetilde{H}_{\textsc{bec}} V^{\dagger}$, where 
\begin{equation}
\widetilde{H}_{\textsc{bec}}=\omega_{0} N_{\text{tot}} + \lambda_{1} (a^{\dagger}a - b^{\dagger}b) + U N_{\text{tot}}^{2},
\label{eqn:VHVdag}
\end{equation}
and $\lambda_{1}=(\lambda^{2}+\omega_{1}^{2})^{1/2}$.  Hence,
\begin{align}
\nonumber M_{m_{1},m_{2}}(t)=\frac{1}{\mu} V \exp(-i \widetilde{H}_{\textsc{bec}} t) V^{\dagger}& a^{\dagger m_{1}} b^{\dagger m_{2}}\\
& V \exp(i \widetilde{H}_{\textsc{bec}} t) V^{\dagger}.
\end{align}
This expression can now be simplified using the following identities which can be obtained in a straightforward manner by using the Baker-Hausdorff lemma.
\begin{align}
V^{\dagger} a^{\dagger} V = a^{\dagger} \cos(\kappa / 2) + b^{\dagger} \sin(\kappa / 2),\\
V^{\dagger} b^{\dagger} V = b^{\dagger} \cos(\kappa / 2) - a^{\dagger} \sin(\kappa / 2),\\
V a^{\dagger} V^{\dagger} = a^{\dagger} \cos(\kappa / 2) - b^{\dagger} \sin(\kappa / 2),\\
V b^{\dagger} V^{\dagger} = b^{\dagger} \cos(\kappa / 2) + a^{\dagger} \sin(\kappa / 2),\\
\nonumber \exp(-i \lambda_{1} (a^{\dagger}a -b^{\dagger}b) t) a^{\dagger p} b^{\dagger q} \exp(i \lambda_{1} (a^{\dagger}a -b^{\dagger}b) t)\\
= a^{\dagger p} b^{\dagger q} \exp(-i (p-q) \lambda_{1} t), \\
\nonumber \exp(-i \omega_{0} N_{\text{tot}} t) a^{\dagger p} b^{\dagger q} \exp(i \omega_{0} N_{\text{tot}} t)\\
= a^{\dagger p} b^{\dagger q} \exp(-i (p+q) \omega_{0} t),\\
\nonumber \exp(-i U N_{\text{tot}}^{2} t) a^{\dagger p} b^{\dagger q} \exp(i U N_{\text{tot}}^{2} t) \\
=a^{\dagger p} b^{\dagger q} \exp(-i U t (p+q) (2 N_{\text{tot}} + p + q)).
\end{align}
Further, using binomial expansions for the two commuting operators $a^{\dagger}$ and $b^{\dagger}$ and defining $p_{max}=(k+m_{2}-l)$ and $q_{max}=(l+m_{1}-k)$, we arrive at the following simplified expression for $M_{m_{1},m_{2}}(t)$.
\begin{align}
\nonumber &M_{m_{1},m_{2}}(t)= \frac{1}{\mu} \biggl[\sum_{k=0}^{m_{1}}\sum_{l=0}^{m_{2}}\sum_{p=0}^{p_{max}}\sum_{q=0}^{q_{max}}(-1)^{k-p} {m_{1} \choose k} {m_{2} \choose l} \\
\nonumber &\hspace{2 em} {p_{max} \choose p} {q_{max} \choose q} \exp(-i\lambda_{1} t (2(k-l)+m_{2}-m_{1}))\\ 
\nonumber &\hspace{2 em} (\cos(\kappa/2))^{(k+l+p+q)} (\sin(\kappa/2))^{(2(m_{1}+m_{2})-(k+l+p+q))}\\
\nonumber &\hspace{2 em} a^{\dagger (p+q_{max}-q)} b^{\dagger (q+p_{max}-p)} \biggr] \exp(-i\omega_{0} t(m_{1}+m_{2}))\\
& \times \exp(-i U t (m_{1}+m_{2}) (2 N_{\text{tot}} + m_{1}+m_{2})).
\label{eqn:appen_intermed_rho_numerics}
\end{align}
The density matrix can now be expressed in terms of $M_{m_{1},m_{2}}(t)$ and $ \rho_{0,0} (t)$ as
\begin{equation}
\rho_{m_{1},m_{2}} (t) = M_{m_{1},m_{2}}(t) \rho_{0,0}(t) M^{\dagger}_{m_{1},m_{2}}(t).
\label{eqn:appen_rho_numerics}
\end{equation}


%% file: appendix_TimeSeries.tex
\chapter{Estimation of the maximal local Lyapunov exponent $\Lambda_{L}$}
\label{appen:TimeSeries}

We outline the procedure to obtain the maximal local Lyapunov exponent $\Lambda_{L}$ from a time series $\lbrace y(t_{n}) \rbrace$ ($n=1,2,\dots,N_{0}$). The underlying dynamical process is inherently nonlinear. The basic premise is that the ergodicity properties pertaining to the dynamics can be captured through a detailed time-series analysis. The first step is to reconstruct an effective phase space of dimension $d_{\rm{emb}}$ (the embedding dimension), in which the dynamics will be examined. 
For this purpose, a time delay $\tau_{\text{d}}$ is obtained such that data points separated by a time interval $\geq \tau_{\text{d}}$ can be treated as independent dynamical variables. 
This is done as follows. We note that for two data points $y(t_{n})$ and $y(t_{n}+T)$ ($T$ is any time duration such that $t_{n}+T \leq t_{N_{0}}$), the information about $y(t_{n})$ that can be deduced from the measured value  of $y(t_{n} + T)$ should tend to zero for sufficiently large~$T$.
Hence, if $p(y(t_{n}))$ and $p(y(t_{n}+T))$ are the individual probability densities for obtaining the values $y(t_{n})$ and $y(t_{n}+T)$ respectively, and $p(y(t_{n}), y(t_{n}+T))$ is the corresponding joint probability density, the mutual information
\begin{equation}
I(T) = \sum_{y(t_{n}),\, y(t_{n}+T)} p(y(t_{n}),\, y(t_{n}+T))\, \log_{2} \left\{ \frac{p(y(t_{n}),\,y(t_{n}+T))}{p(y(t_{n}))\, p(y(t_{n}+T))} \right\}
\label{eqn:mutual_info_timeseries}
\end{equation}
should tend to zero,  for sufficiently large $T$. A frequently used prescription is to choose $\tau_{\text{d}}$ to be the  first minimum of $I(T)$~\cite{TimeDelay}.

Using $\tau_{\text{d}}$, $d_{\rm{emb}}$ is determined as follows. The next step is to examine the correlation integral
\begin{equation}
C(r)=\lim_{M_{0}\rightarrow\infty} \frac{1}{M_{0}^{2}} \sum\limits_{\substack{i,j=0\\i\neq j}}^{M_{0}} \Theta\left(r- \mu_{ij} \right)=\int_{0}^{r} \rmd^{d}r' c(\mathbf{r'}),
\label{eqn:corr_integ}
\end{equation} 
where,
\begin{equation}
\nonumber \mu_{ij}=\left(\sum\limits_{p=0}^{d-1} \left(y(t_{i} + p \tau_{\text{d}}) - y(t_{j} + p \tau_{\text{d}})\right)^{2} \right)^{1/2},
\end{equation}
and $\Theta(r-\mu_{ij})$ is the Heaviside function. $C(r)$ gives an estimate of the average correlation between $M_{0}$ points in a $d$-dimensional phase space and the integrand $c(\mathbf{r'})$ is the standard correlation function. Here $M_{0}$ is fixed such that $t_{M_{0}}+(d-1)\tau_{\text{d}}\leq t_{N_{0}}$. If $r$ is sufficiently small, $C(r) \sim r^{\ell}$~\cite{grassberger}. $C(r)$ is calculated for various values of $d$. The embedding dimension $d_{emb}$ is determined by the requirement that for any $d \geq d_{\rm emb}$,  $\ell$ remains constant. 

Vectors in the reconstructed phase space are denoted by 
\begin{equation}
\mathbf{y}(n)=[ y(t_{n}), y(t_{n}+\tau_{\text{d}}),\dots,y(t_{n}+(d_{\rm{emb}}-1)\tau_{\text{d}})].
\end{equation}
The basic assumption is that the dynamics is guided by a map $\mathbf{F}$ such that 
\begin{equation}
\mathbf{y}(n+1)=\mathbf{F(y}(n)\mathbf{)}.
\label{eqn:FmapDefn}
\end{equation}
Of immediate relevance to us is the Jacobian 
\begin{equation}
\mathbf{DF}_{ij}\mathbf{(y}(n)\mathbf{)}=\frac{\partial \mathbf{F}_{i}\mathbf{(y}(n)\mathbf{)}}{\partial \mathbf{y}_{j}(n)},
\end{equation}
where $i,j=0,1,\dots,d_{\rm{emb}}-1$. Since the expression for $\mathbf{F}$ is not a priori known explicitly, the Jacobian is obtained from the time series as follows~\cite{abarbanel1992}.
For a given $\mathbf{y}(n)$, we find $k$ nearest neighbours in the reconstructed phase space. They are denoted by $\lbrace \mathbf{y}^{(1)}(n),\mathbf{y}^{(2)}(n),\dots, \mathbf{y}^{(k)}(n)\rbrace$. Defining $\mathbf{Dy}^{(p)}(n)=\mathbf{y}^{(p)}(n)- \mathbf{y}(n)$ ($p=1,2,\dots,k$), it can be shown that,
\begin{align}
\nonumber \mathbf{F}_{i}\mathbf{(y}^{(p)}(n)\mathbf{)}- \mathbf{F}_{i}\mathbf{(y}(n)\mathbf{)}= &\mathbf{DF}_{ij}\mathbf{(y}(n)\mathbf{)} \mathbf{Dy}_{j}^{(p)}(n) \\
& + \mathbf{Q}_{ilm}\mathbf{(y}(n)\mathbf{)} \mathbf{Dy}_{l}^{(p)}(n) \mathbf{Dy}_{m}^{(p)}(n) + \cdots
\label{eqn:JacobSLE}
\end{align}
where to second-order, $\mathbf{DF}_{ij}\mathbf{(y}(n)\mathbf{)}$ and $\mathbf{Q}_{ilm}\mathbf{(y}(n)\mathbf{)}$ need to be determined. We can numerically find them by solving the set of simultaneous linear equations \eref{eqn:JacobSLE}. We note that the LHS of Eq. \eref{eqn:JacobSLE} can be determined using Eq. \eref{eqn:FmapDefn}. We point out that $k$ is fixed such that this set of equations is uniquely solved.

We now define,
\begin{equation}
\mathbf{DF}^{L}(n)=\mathbf{DF}\mathbf{(y}(n+L-1)\mathbf{)}\cdot \mathbf{DF}\mathbf{(y}(n+L-2)\mathbf{)}\cdots \mathbf{DF}\mathbf{(y}(n)\mathbf{)}.
\end{equation}
The Oseledec matrix $\mathbf{M_{\text{os}}}$ can be expressed in terms of $\mathbf{DF}^{L}(n)$ as
\begin{equation}
\mathbf{M_{\text{os}}}(n,L)=\left( (\mathbf{DF}^{L}(n))^{\text{T}} \cdot \mathbf{DF}^{L}(n) \right)^{1/(2L)}.
\label{eqn:OSL}
\end{equation}
The spectrum of local Lyapunov exponents corresponding to $L$ steps of time are the eigenvalues of $\mathbf{M_{\text{os}}}$. Of direct relevance to us are the ergodicity properties of $d_{1}(t)$ (Eq. \ref{eqn:d1d2defns}) that can be obtained from the corresponding time series. Hence, replacing $y(t_{n})$ by $d_{1}(t_{n})$ and setting $N_{0}=20000$, we have obtained $\tau_{\text{d}}$ and $d_{\rm{emb}}$ from the time series using the TISEAN package~\cite{tisean}. Using the procedure outlined above, the local Lyapunov exponents have been calculated. This procedure has been repeated for $100$ randomly chosen values of $n$ in the time series and the average $\Lambda_{L}$ of the corresponding maximal local Lyapunov exponents has been calculated.

The global Lyapunov exponents are the eigenvalues of $\lim\limits_{L\to\infty}\mathbf{M_{\text{os}}}(n,L)$. We denote the maximal global Lyapunov exponent by $\Lambda_{\infty}$. Then, for generic dynamical systems, it has been shown~\cite{abarbanel1991, abarbanel1992} that $\Lambda_{L}$ tends to $\Lambda_{\infty} + (m/L^{q})$ ($m,\,q$: constants), for sufficiently large $L$.


%% file: appendix_2states.tex
\chapter{The 2-photon frequency combs}
\label{appen:2states}

In Section \ref{sec:Ch4chrono}, the expressions for the two $2$-photon states $\ket{\Psi_{\alpha}}$ and $\ket{\Psi_{\beta}}$ which are distinguished from each other using tomograms are given in Eqs. \eref{eqn:Psi_plpl} and \eref{eqn:Psi_mipl} respectively. For convenience, we give the expressions below.
\begin{align*}
\ket{\Psi_{\alpha}}
= \mathcal{N}_{\alpha}^{-1/2}\int \text{d}\omega_{\textsc{s}} \int \text{d}\omega_{\textsc{i}} f_{+}(\omega_{\textsc{s}}+\omega_{\textsc{i}}) f_{-}(\Omega)
  f_{\text{cav}}(\omega_{\textsc{s}}) f_{\text{cav}}(\omega_{\textsc{i}}) \ket{\omega_{\textsc{s}}} \otimes \ket{\omega_{\textsc{i}}},
\end{align*}
where $\mathcal{N}_{\alpha}$ is the normalisation constant. Here, $f_{-}(\Omega)$ and $f_{\text{cav}}(\omega)$
are defined in Eqs. \eref{eqn:fmin} and \eref{eqn:fcav} respectively. 
The other $2$-photon state
\begin{align*}
\ket{\Psi_{\beta}}=
\mathcal{N}_{\beta}^{-1/2}\int \text{d}\omega_{\textsc{s}} \int \text{d}\omega_{\textsc{i}} f_{+}(\omega_{\textsc{s}}+\omega_{\textsc{i}}) f_{-}(\Omega)     
 g_{\text{cav}}(\omega_{\textsc{s}}) f_{\text{cav}}(\omega_{\textsc{i}}) \ket{\omega_{\textsc{s}}} \otimes \ket{\omega_{\textsc{i}}}, 
\end{align*}
where $g_{\text{cav}}(\omega)$ is defined in Eq. \eref{eqn:gcav}, 
and $\mathcal{N}_{\beta}$ is the normalisation constant.

In what follows, we outline the procedure to show that these two states are indeed the two states which were shown to be distinguishable using photon coincidence counts, in the experiment reported in \cite{perola}. 

In the experimental setup, the sum of the frequencies of the signal and the idler photons matches the pump frequency, i.e., $(\omega_{\textsc{s}}+\omega_{\textsc{i}}=\omega_{\textsc{p}})$. Hence, as stated in the supplementary material of \cite{perola}, $f_{+}(\omega_{\textsc{s}}+\omega_{\textsc{i}})$ can be replaced by $\delta(\omega_{\textsc{s}}+\omega_{\textsc{i}}-\omega_{p})$. Integrating over the variable $\Omega_{+}\:(=\omega_{\textsc{s}}+\omega_{\textsc{i}})$, appropriately changing the integration variables, noting that $\Omega=\omega_{\textsc{s}}-\omega_{\textsc{i}}$, and dropping the normalisation constant, we get
\begin{equation}
\ket{\Psi_{\alpha}}
= \int \text{d}\Omega f_{-}(\Omega) f_{\text{cav}}(\omega_{\textsc{s}}) f_{\text{cav}}(\omega_{\textsc{i}}) \ket{\omega_{\textsc{s}}} \otimes \ket{\omega_{\textsc{i}}}.
\label{eqn:B19perola}
\end{equation}
This can be identified as one of the states considered in the experiment, namely, Eq.~(B19) in \cite{perola}, on changing the notation from $\Omega$, $\omega_{\textsc{s}}$, $\omega_{\textsc{i}}$ in Eq. \eref{eqn:B19perola} to $\omega_{-}$, $\omega_{s}$, $\omega_{i}$ respectively.

We now proceed to establish that the other $2$-photon state $\ket{\Psi_{\beta}}$ considered by us, is the same as the state $\ket{\psi^{\prime}}\:(=C^{\prime} Z_{t_{\textsc{s}}} \ket{\widetilde{+}}_{\omega_{\textsc{s}}} \otimes \ket{\widetilde{+}}_{\omega_{\textsc{i}}})$ defined in \cite{perola}. Here, $C^{\prime}\ket{t_{\textsc{s}};t_{\textsc{i}}}=\ket{t_{\textsc{s}}+t_{\textsc{i}};t_{\textsc{s}}-t_{\textsc{i}}}$ where, for instance, $\ket{t_{\textsc{s}}}\otimes\ket{t_{\textsc{i}}}$ is denoted by $\ket{t_{\textsc{s}};t_{\textsc{i}}}$ with $t_{\textsc{s}}$ and $t_{\textsc{i}}$ being the time variables associated with the signal and the idler photons respectively, and $Z_{t_{\textsc{s}}}\ket{\widetilde{+}}_{\omega_{\textsc{s}}}=\ket{\widetilde{-}}_{\omega_{\textsc{s}}}$. 
It is convenient to express $\ket{\widetilde{+}}_{\omega_{\textsc{x}}}$ and $\ket{\widetilde{-}}_{\omega_{\textsc{x}}}$ $(\textsc{x}=\textsc{s,i})$ as
\begin{equation}
\ket{\widetilde{+}}_{\omega_{\textsc{x}}}=\int \text{d}\omega_{\textsc{x}}\, \int \text{d}t_{\textsc{x}}\, \,\exp\left(-\frac{t_{\textsc{x}}^{2}}{2 \kappa_{\textsc{x}}^{2}} - \frac{\omega_{\textsc{x}}^{2}}{2 (\Delta \omega)^{2}}\right)\, \sum_{n}\, e^{i (\omega_{\textsc{x}} + n \overline{\omega}) t_{\textsc{x}}} \,  \ket{\omega_{\textsc{x}} + n \overline{\omega}}, 
\label{eqn:tildePl}
\end{equation}
and,
\begin{equation}
\ket{\widetilde{-}}_{\omega_{\textsc{x}}}=\int \text{d}\omega_{\textsc{x}}\, \int \text{d}t_{\textsc{x}}\, \,\exp\left(-\frac{t_{\textsc{x}}^{2}}{2 \kappa_{\textsc{x}}^{2}} - \frac{\omega_{\textsc{x}}^{2}}{2 (\Delta \omega)^{2}}\right)\, \sum_{n} (-1)^{n}\, e^{i (\omega_{\textsc{x}} + n \overline{\omega}) t_{\textsc{x}}} \,  \ket{\omega_{\textsc{x}} + n \overline{\omega}}. 
\label{eqn:tildeMin}
\end{equation}
These expressions follow from the properties of the displacement operator, and Eqs.~(B1), (B2), (B7) given in \cite{perola}. Here, $\kappa_{\textsc{x}}$ $(\textsc{x}=\textsc{s,i})$ is the standard deviation in $t_{\textsc{x}}$.
It follows from Eqs. \eref{eqn:tildePl} and \eref{eqn:tildeMin} that
\begin{align}
\nonumber\ket{\psi^{\prime}}=\int \text{d}t \, \int \text{d}t^{\prime} \, \int \text{d}\omega \, \int \text{d}\omega^{\prime} \, \exp \left(- \frac{ t^{2} (\Delta \omega_{p})^{2} + t^{\prime 2} (\Delta \Omega)^{2}}{2}-\frac{\omega^{2} + \omega^{\prime 2}}{2 (\Delta \omega)^{2}}\right) \\
\sum_{n,m} (-1)^{n} e^{i (n \overline{\omega} + \omega) (t+t^{\prime})} e^{i (m \overline{\omega} + \omega^{\prime}) (t-t^{\prime})} \ket{n \overline{\omega} + \omega}\otimes\ket{m \overline{\omega} + \omega^{\prime}},
\label{eqn:psiPrime}
\end{align}
where $\Delta\omega_{p}$ is the standard deviation in $\omega_{p}$. Integrating over the time variables $t$ and $t^{\prime}$, writing $(\omega_{\textsc{s}}=n \overline{\omega} + \omega)$, $(\omega_{\textsc{i}}=m \overline{\omega} + \omega^{\prime})$ where $n$, $m$ are integers, and using the fact that $f_{+}$ is a Gaussian function with a standard deviation $\Delta\omega_{p}$ ($\Delta \omega_{p} \ll \Delta \Omega$), it is straightforward to see that Eq. \eref{eqn:psiPrime} can be expressed as $\ket{\Psi_{\beta}}$ (Eq. \eref{eqn:Psi_mipl}) unnormalised. 


%% file: appendix_ChronoTomo.tex
\chapter{Expressions for the chronocyclic tomograms}
\label{appen:ChronoTomo}

We are interested in the time-time slice of the tomograms corresponding to $\ket{\Psi_{\alpha}}$ (Eq. \eref{eqn:Psi_plpl}) and $\ket{\Psi_{\beta}}$ (Eq. \eref{eqn:Psi_mipl}). As a first step, we calculate the explicit expressions for the states $\ket{\Psi_{\alpha}}$ and $\ket{\Psi_{\beta}}$ in the Fourier transform basis (i.e., time-time basis) using $f_{+}(\omega_{\textsc{s}}+\omega_{\textsc{i}})= \delta(\omega_{\textsc{s}}+\omega_{\textsc{i}}-\omega_{p})$ in Eqs.~\eref{eqn:Psi_plpl} and \eref{eqn:Psi_mipl}. The $2$-photon state $\ket{\Psi_{\alpha}}$ in the time-time basis is given by
\begin{align}
\nonumber \ket{\Psi_{\alpha}}= \frac{1}{\sqrt{\mathcal{M}_{\alpha}\tau_{\textsc{p}}}} \,\int \text{d}t_{\textsc{s}} \,\int \text{d}t_{\textsc{i}} \, \exp \left( -\frac{(t_{\textsc{i}}-t_{\textsc{s}})^{2} \, (\vardel\omega)^{2} \, (\vardel\Omega)^{2}}{4\: ((\vardel\omega)^{2} + (\vardel\Omega)^{2})} \right) \\
\times \left[\mathcal{F}(t_{\textsc{i}} - t_{\textsc{s}})\right]^{2} \ket{t_{\textsc{s}};t_{\textsc{i}}},
\label{eqn:Psi_plpl_tt}
\end{align}
and the time-time slice $w^{\alpha}(t_{\textsc{s}};t_{\textsc{i}})$ corresponding to $\ket{\Psi_{\alpha}}$ is
\begin{align*}
w^{\alpha}(t_{\textsc{s}};t_{\textsc{i}}) = \frac{1}{\mathcal{M}_{\alpha}\tau_{\textsc{p}}} \, \exp \left( -\frac{(t_{\textsc{i}}-t_{\textsc{s}})^{2} \, (\vardel\omega)^{2} \, (\vardel\Omega)^{2}}{2\: ((\vardel\omega)^{2} + (\vardel\Omega)^{2})} \right) 
 \Big\vert\mathcal{F}(t_{\textsc{i}} - t_{\textsc{s}}) \Big\vert^{4},
\end{align*}
where $\tau_{\textsc{p}}=1$s (introduced for dimensional purposes), 
\begin{align*}
\mathcal{F}(t_{\textsc{i}} - t_{\textsc{s}}) = \sum\limits_{n} \exp \left( \frac{i (t_{\textsc{i}} - t_{\textsc{s}})\, n \, \overline{\omega} (\vardel\Omega)^{2}}{2 ((\vardel\omega)^{2} + (\vardel\Omega)^{2})} \right),
\end{align*}
and the normalisation constant is
\begin{align}
\mathcal{M}_{\alpha}= \frac{\pi}{\mu_{0}} \sum_{m,n,m',n'} \exp \left( -\frac{(n-n'+m'-m)^{2}\, \overline{\omega}^{2}\, (\vardel\Omega)^{2}}{2\, (\vardel\omega)^{2}\: ((\vardel\omega)^{2} + (\vardel\Omega)^{2})} \right),
\label{eqn:NormConst2Alph}
\end{align}
where $\mu_{0}=\left(\frac{\pi (\vardel \Omega)^{2} (\vardel \omega)^{2}}{2 ((\vardel \Omega)^{2} + (\vardel \omega)^{2})}\right)^{1/2}$.

Similarly, the other $2$-photon state $\ket{\Psi_{\beta}}$ in the time-time basis is given by
\begin{align}
\nonumber \ket{\Psi_{\beta}}= \frac{1}{\sqrt{\mathcal{M}_{\beta}\tau_{\textsc{p}}}} \,\int \text{d}t_{\textsc{s}} \,\int \text{d}t_{\textsc{i}} \, \exp \left( -\frac{(t_{\textsc{i}}-t_{\textsc{s}})^{2} \, (\vardel\omega)^{2} \, (\vardel\Omega)^{2}}{4\: ((\vardel\omega)^{2} + (\vardel\Omega)^{2})} \right)\\
\times \mathcal{G}(t_{\textsc{i}} - t_{\textsc{s}}) \,
\mathcal{F}(t_{\textsc{i}} - t_{\textsc{s}}) \ket{t_{\textsc{s}};t_{\textsc{i}}},
\label{eqn:Psi_mipl_tt}
\end{align}
and the time-time slice corresponding to $\ket{\Psi_{\beta}}$ is
\begin{align*}
w^{\beta}(t_{\textsc{s}};t_{\textsc{i}}) = \frac{1}{\mathcal{M}_{\beta}\tau_{\textsc{p}}} \, \exp \left( -\frac{(t_{\textsc{i}}-t_{\textsc{s}})^{2} \, (\vardel\omega)^{2} \, (\vardel\Omega)^{2}}{2\: ((\vardel\omega)^{2} + (\vardel\Omega)^{2})} \right)  \Big\vert \mathcal{G}(t_{\textsc{i}} - t_{\textsc{s}}) \, \mathcal{F}(t_{\textsc{i}} - t_{\textsc{s}}) \Big\vert^{2},
\end{align*}
where,
\begin{align*}
\mathcal{G}(t_{\textsc{i}} - t_{\textsc{s}}) = \sum\limits_{n} (-1)^{n} \exp \left( \frac{i (t_{\textsc{i}} - t_{\textsc{s}})\, n \, \overline{\omega} (\vardel\Omega)^{2}}{2 ((\vardel\omega)^{2} + (\vardel\Omega)^{2})} \right),
\end{align*} 
and the normalisation constant $\mathcal{M}_{\beta}$ is essentially the same as $\mathcal{M}_{\alpha}$ with an extra factor of $(-1)^{n+n'}$ within the summation in Eq.~\eref{eqn:NormConst2Alph}.
